\documentclass[letterpaper, 11pt]{article}

\usepackage{mathtools}
\usepackage{array}
\usepackage{multirow}
\usepackage[T1]{fontenc}
\usepackage[utf8]{inputenc}
\usepackage{helvet}
\usepackage{enumitem}
\usepackage{amsmath,amsfonts,amssymb,bm}
\usepackage{titlesec}
\usepackage{pgfgantt,rotating}
\usepackage{float}
\usepackage{graphicx}
\usepackage{sidecap}
\usepackage{wrapfig}
\usepackage{bold-extra}
\usepackage{authblk}
\usepackage{appendix}
\usepackage{amsfonts}
\usepackage{amsmath}
\usepackage{upgreek}
\usepackage{newtxmath}
\usepackage[utf8]{inputenc}
\usepackage{color}
\usepackage{rotating}
\usepackage{cite}

\usepackage[margin=1in]{geometry}

\DeclareMathOperator{\tr}{tr}
\DeclareMathOperator{\KL}{KL}

\newcommand{\beq}{\begin{equation}}
\newcommand{\eeq}{\end{equation}}
\newcommand{\bgqar}{\begin{eqnarray}}
\newcommand{\enqar}{\end{eqnarray}}
\newcommand{\bgqarn}{\begin{eqnarray*}}
\newcommand{\enqarn}{\end{eqnarray*}}
\newcommand{\bgary}{\begin{array}}
\newcommand{\enary}{\end{array}}

\newcommand{\etal}{{\it et al. }}

\graphicspath{{figures/}}

\title{Gaussian Process Regression for Active Sensing Probabilistic Structural Health Monitoring: Experimental Assessment Across Multiple Damage and Loading Scenarios}

\author{Ahmad Amer}
\author{Fotis Kopsaftopoulos\footnote{Corresponding author.}}

\affil{\small Intelligent Structural Systems Laboratory (ISSL) \\ Department of Mechanical, Aerospace and Nuclear Engineering \\ Rensselaer Polytechnic Institute, Troy, NY, USA \\ Email: \{amera2,kopsaf\}@rpi.edu}

\date{\today}

\begin{document}

\maketitle


\begin{abstract}

In the near future, Structural Health Monitoring (SHM) technologies for aircraft will be capable of overcoming the drawbacks in the current maintenance and life-cycle management paradigms, namely: cost, increased downtime, less-than-optimal safety management paradigm and the limited applicability of fully-autonomous operations. In the context of SHM for aircraft structures, one of the most challenging tasks is structural damage quantification. Currently-utilized quantification techniques face accuracy and/or robustness issues when it comes to the varying operating and environmental conditions involved in day-to-day operations. In addition, the damage/no-damage paradigm of current industrial frameworks does not offer much information to maintainers on the ground for proper decision-making. In this study, a novel structural damage quantification framework is proposed based on the widely-used Damage Indices (DIs) and Gaussian Process Regression Models (GPRMs) in order to overcome the aforementioned shortcomings. The proposed framework takes a simple approach to the damage quantification problem by using DI values for training, and provides confidence bounds on the quantified states using a novel state prediction technique. The novelty in this approach to state quantification lies in calculating the probability of an incoming test DI point originating from a specific state, which allows for probability-educated decision-making. In addition, the proposed methods are shown to quantify multiple structural states simultaneously from incoming DI test points. This framework is applied to three test cases: a Carbon Fiber-Reinforced Plastic (CFRP) coupon with attached weights as simulated damage, an aluminum coupon with a notch, and an aluminum coupon with attached weights as simulated damage under varying loading states. The novel state prediction method presented herein is applied to single-state quantification in the first two test cases, as well as the third one assuming the loading state is known. Finally, the proposed method is applied to the third test case assuming neither the damage size nor the load is known in order to predict both simultaneously from incoming DI test points. In applying this framework, two forms of GPRMs (standard and variational heteroscedastic) are used in order to critically assess their performances with respect to the three test cases.

\end{abstract}

\newpage\pagebreak 

\tableofcontents 

\section{Introduction} \label{sec:intro}

In the context of engineering structures, structural safety, maintenance and life-cycle management processes are a major factor in sustainability \cite{Biondini-Frangopol16,Frangopol-Soliman16}.  In particular, the aerospace industry is one that depends heavily on schedule-based procedures in order to sustain proper life cycle management, ensure safety, and improve performance \cite{Frangopol-Maute03,Gomes-etal18,Jordan-etal18}. Most of such procedures include some form of Non-destructive Evaluation (NDE) techniques, in which aircraft need to be inspected on a regular basis on the ground before operations can be resumed regardless of structural state \cite{Jordan-etal18,Dong-Kim18,Frangopol-Maute03}. This framework, although very effective in the sustainability efforts of the aerospace industry, suffers from a number of drawbacks; the most economically-prominent of which are cost, increased downtime, less-than-optimal safety management paradigm (damage can occur and grow between scheduled procedures) and the limited applicability of fully-autonomous operations \cite{Frangopol-Maute03,Frangopol-Soliman16,Dong-Kim18,Davis-etal15}. As such, research endeavors in the past 40 years have been directed towards developing sustainability efforts that can be applied online (limiting downtime and increasing safety) and in an automated fashion (limiting the need for costly man hours, and also allowing for autonomous operation) within the frameworks of Structural Health Monitoring (SHM) \cite{Cawley18}. Because of the complexity of aircraft operations, manifested in multiple operational cycles, and, within each cycle, the varying operational and environmental conditions, the aerospace industry poses as a very rich arena for development of SHM techniques \cite{Dong-Kim18}.

By convention, a complete SHM system is one that can implement all 4 levels of SHM with high accuracy and robustness, namely: damage detection, localization, quantification and remaining-useful-life estimation \cite{Qiu-etal13,Romano-etal19,Janapati-etal16,Das-Saha18}. When it comes to SHM in aerospace structures today, the most common metric used for interrogating structural health is composed of one or more Damage Indices (DIs). These DIs compare specific features of the sensor response signals for the unknown structural state to those features coming from the healthy structure \cite{Ihn-Chang08,Giurgiutiu11}. The features can be the amplitude, energy or phase of the signals, and they have been extensively used in the literature for damage detection \cite{Jin-etal18,Xu-etal13,Ihn-Chang08,Janapati-etal16,Giurgiutiu11,Nasrollahi-etal18} and damage quantification \cite{Lim-etal11,Soman-etal18,Yan-etal19,Dragan-etal13}. From an industrial point of view, one of the major advantages of DIs is the simplicity of application and interpretation of results, which facilitates decision-making by maintainers on the ground \cite{Jin-etal18}. When it comes to active-sensing SHM in particular, however, the currently-utilized DIs face a number of challenges, the most prominent of which is their deterministic nature, i.e. they do not account for operational and environmental uncertainties \cite{Farrar-Worden07,Amer-Kopsaftopoulos19a,Ahmed-Kopsaftopoulos19a,Kopsaftopoulos-etal-MSSP18}. This is particularly important with the growing need for online damage detection and quantification owing to the urban air mobility ``revolution'' \cite{Dutta20}, where the SHM metric being implemented must be robust enough to the different sources of uncertainty pertaining to flight operations, whilst sensitive enough to detect and quantify damage accurately \cite{Amer-etal20}. In addition, the damage/healthy paradigm of DI frameworks does not provide much useful information to maintainers, since it depends on user expertise \cite{Qiu-etal16,Wang-etal18}, and is thus subject to error, especially under the varying environmental and operational conditions natural to aircraft operations.



Accordingly, many researchers sought to developing active-sensing, guided-wave SHM metrics that are more robust and accurate. Within the field of damage quantification in particular, which is more involved than damage detection \cite{Farrar-Worden07}, researchers used conventional/non-conventional DIs (for instance, see \cite{Lim-etal11,Soman-etal18,Yan-etal19,Dragan-etal13}), advanced signal processing techniques (see \cite{Patra-Banerjee17a,Amjad-etal15,Das-etal09}), advanced statistical inference/modelling techniques (see \cite{Zhao-etal07,2013-23,2016-9}), and analytical guided-wave models (see for instance \cite{Banerjee-Ahmed13,Borkowski-Chattopadhyay14,Srivastava-Scalea10}) in order to quantify damage within an active-sensing, guided-wave framework. Other techniques involved the use of image processing of guided-wave maps (such as in \cite{Muller-etal19}). Similar to the case of damage detection \cite{Amer-Kopsaftopoulos21a}, all of these techniques either suffer from an incomplete treatment of the challenges mentioned above, which face damage detection and quantification alike, or increased model complexity. The following paragraphs briefly outline some of the more important studies in the literature dealing with damage quantification.

Being a relatively-simple metric to calculate, many researchers sought for non-conventional DIs in order to obtain a robust and accurate damage quantification metric. Wang and coworkers \cite{Wang-etal15} compared 4 parametric models which were linear combinations of different signal features (such as time of flight and amplitude) in quantifying the size of a crack on an aircraft riveted lap joint. Reynolds and Chattopadhyay \cite{Reynolds-Chattopadhyay08} modelled the system under interrogation using a number of transfer functions in order to decouple the effect of structural anomalies and that of the sensor interaction with the structure on the received signals. Then, they proposed a DI based on the attenuation of signal intensity with damage, relating that DI to the transfer functions of the system in order to corner down the effect of damage on the received signals and thus quantify damage. Observing that DIs based on guided waves alone might not meet the required level of accuracy in aerospace applications, Vanniamparambil \etal \cite{Vanniamparambil-etal12} proposed a data fusion technique, in which information from acoustic emission, guided waves, and digital image correlation is used to indicate crack growth in an Al 2024 Alloy. 

Because of the drawbacks of DIs, many researchers turned their attention to advanced signal processing techniques in order to quantify damage in an active-sensing framework. A common approach in this family of techniques involved the use of wavelet transforms, measuring the change in wavelet energies as indication of damage size \cite{Rizzo-Scalea05,Rizzo-etal09,Wang-etal19}. Other techniques involved the use of Hilbert Huang transforms \cite{Amjad-etal15}. While all of these techniques indeed show superiority to approaches utilizing DIs, they still lack the proper identification of quantification confidence bounds, and are thus not suitable for stochastic systems without further development. 

In order to tackle this problem, some researchers utilized advanced modelling techniques where signal features are fitted using different probability distributions in order to properly extract confidence intervals on the quantification decision \cite{Zhao-etal07,2017-5}. For instance, Yang \etal \cite{2016-9,2017-5} used Bayesian updating for crack size quantification in 6 Al plates having notches with increasing lengths within an active-sensing approach. Normalized amplitude and phase changes in the first-arrival wave mode (S0) in the signals were used as the parameters in a linear model that predicted damage size. Having initially calculated the parameter values from the experiments, Monte Carlo simulations were then used to fit the parameters to a probability distribution to be able to proceed with Bayesian updating of the notch size model. Similar strategies have been demonstrated for estimating delamination propagation in composites \cite{2013-23}. Other endeavors included the use of Gaussian mixture models, \cite{Qiu-etal16}, neural networks \cite{Su-Ye04}, or Principal Component Analysis \cite{Tibaduiza-etal16}. In general, these techniques lead to an accurate and robust damage quantification process (same can be said for detection,) and many of them fall under the umbrella of statistical/probabilistic SHM. However, these approaches either involve complicated steps (such as the case of Gaussian mixture models \cite{Qiu-etal16}), or require many data sets for model training and building processes (such as the case of the matching pursuit decomposition \cite{Das-etal09} or Bayesian updating \cite{2016-9}). Thus, there still lies the need for developing damage quantification techniques that can overcome such shortcomings of currently-used approaches. These challenges can be summarized as follows:

\begin{itemize}
\item The lack of robustness of DI-based quantification techniques under varying operational and/or environmental conditions.
\item The need for user expertise for interpreting DI results.
\item The inability of most of the other approaches to provide quantification confidence intervals for proper decision-making without sacrificing simplicity in defining/calculating the damage quantification metric.
\end{itemize}

One promising approach, which also falls under probabilistic SHM techniques, is Gaussian Process Models (GPMs). GPMs, being probabilistic machine learning models, have recently seen interest within the vibration-based SHM community \cite{Abdessalem-etal17,Bull-etal19,Gonzaga-etal19,Avendano-etal17-GPRM,Valencia-etal18}. This interest originates from the fact that GPMs model uncertainty in the available data regardless of whether the uncertainty sources are known or not. Furthermore, GPMs can extract confidence bounds on state predictions \cite{Lazaro-Gredilla-etal13}. Most importantly, with the availability of data under different conditions, the effect of any condition can be conveniently modelled using GPMs. Consequently, the authors have recently utilized GP Regression Models (GPRMs) \cite{Amer-Kopsaftopoulos19a,Amer-Kopsaftopoulos19b} and GP Classification Models (GPCMs) \cite{Amer-Kopsaftopoulos20} for the task of probabilistic damage quantification in active-sensing, guided-wave SHM. In addition, GPRMs were also used by the authors to predict damage size and load simultaneously either by themselves \cite{Amer-etal20,Amer-etal21a}, or with the assistance of physics-based models \cite{Amer-etal21a}. 

The \textbf{aim} of the present study is to expand on previous work using data-based GPRMs for damage quantification within active-sensing, guided-wave SHM in order to critically assess the usefulness of such models for different types of coupons/damages/states. DIs are utilized in order to train GPRMs for predicting damage sizes for a Carbon fiber-reinforced composite (CFRP) with simulated damage and an Al coupon with a notch, as well as damage size and/or load state for another Al coupon with simulated damage. In addition, this study assesses two types of GPRMs: standard homoscedastic GPRMs (SGPRMs), and the Variational Heteroscedastic GPRMs (VHGPRMs) presented by Lazaro-Gredilla and Titsias \cite{Lazaro-Gredilla-Titsias11} in order to compare the performances of both approaches with respect to accuracy and robustness. The novelty in this work can be summarized through the following points:

\begin{itemize}
\item The application of SGPRMs and VHGPRMs for damage quantification in active-sensing, guided-wave SHM.
\item The use of conventional DIs in training accurate and robust GPRMs, which allows for proper modelling of uncertainties using a well-known, widely-used SHM metric.
\item The accurate simultaneous prediction of damage size and load states within an active-sensing framework.
\item The extraction of confidence bounds over damage size and/or load prediction for probability-educated decision-making, which contrasts with the damage/healthy paradigm of conventional DIs.
\end{itemize}

The paper is organized as follows: Section \ref{Sec:background} presents the overall background of the methods used herein, including the utilized state-of-the-art DI formulae (Section \ref{Sec:ref_DI}), the SGPRM (Section \ref{Sec:SGPRM}) and VHGPRM (Section \ref{Sec:VHGPRM}) formulations, and the state prediction framework used in this study (Section \ref{sec:prediction}). Then, the following sections present the application of the proposed framework to the three coupons: the Al coupon with a notch (Section \ref{Sec:notched_Al}), the CFRP coupon with simulated damage (Section \ref{Sec:cfrp}), and the Al coupon with simulated damage under different loading conditions (Section \ref{Sec:instron_Al}). Finally, Section \ref{Sec:conc} draws the most important conclusions and presents the areas of development within the proposed framework.

\section{Background} \label{Sec:background}

\subsection{Reference Damage Indices} \label{Sec:ref_DI}

In this study, the proposed methods were compared with two formulations for damage indices. The first one was the basic Root-Mean-Square Deviation (RMSD), which reads:

\beq
DI = \sqrt{\frac{\sum_{t=1}^{N}{(y_0[t]-y_u[t])^2}}{N}}
\eeq

In the above equation, $y_0[t]$ and $y_u[t]$ are the signals coming from the healthy and unknown states of the system, indexed with normalized discrete time $t$ ($t=1,2,3,\ldots, N$ where $N$ here is 2500, which includes roughly the first $100 \mu s$ of the signals). The second DI used herein was formulated by Janapati \etal \cite{Janapati-etal16}. This DI was selected in particular due to its high sensitivity to damage size, and low sensitivity to changes in the material properties of the different components of the sensor network (coupon, sensors, and adhesive layers). The formulation is as follows:
\begin{equation}
Y_{u}^n[t] =\frac{y_u[t]}{\sqrt{\sum_{t=1}^{N}{y^2_u[t]}}},
\quad Y_{0}^n[t]=\frac{\sum_{t=1}^{N}{(y_0[t]\cdot Y_{u}^n[t])}}{y_0[t]\cdot \sum_{t=1}^{N}{y_0^2[t]}},
\quad DI=\sum_{t=1}^{N}{(Y^n_{u}[t] - Y^n_{0}[t])}
\end{equation}
such that $Y^n_0[t]$ and $Y^n_u[t]$ are normalized baseline (healthy) and unknown (test/inspection) signals, respectively.

\subsection{Standard Gaussian Process Regression Models (SGPRMs)} \label{Sec:SGPRM}

\subsubsection{Formulation}

In this section, a brief introduction of standard GPRMs will be provided. For a full treatment, the reader is directed to reference \cite{Rasmussen-Williams06}, given a training data set $\mathcal{D}$ containing $n$ inputs-observation pairs \{$\mathbf{x}_i \in \mathbb{R}^D ,y_i \in \mathbb{R},\ i=1,2,3,\ldots,n$\}, a standard (homoscedastic) GPRM can be formulated as follows:
\begin{equation}
    y=f(\mathbf{x})+\epsilon
    \label{eq:SGPRM}
\end{equation}
where, in a Bayesian setting, a GP prior with mean $m(\mathbf{x})$ and covariance  $k(\mathbf{x},\mathbf{x'})$ is placed on the latent function $f(\mathbf{x})$, and an independent, identically-distributed ($\mathit{iid}$), zero-mean Gaussian prior with variance $\sigma_n^2$ is placed on the noise term $\epsilon$, that is:

\begin{equation}
\begin{array}{lr}
f(\mathbf{x}) \sim \mathcal{GP}(m(\mathbf{x}),k(\mathbf{x},\mathbf{x'})), \, & \, \epsilon \sim iid \, \mathcal{N}(0,\sigma_n^2)
\end{array}
\end{equation}

As is common in the GPRM literature, $m(\mathbf{x})$ is set to zero, and the squared exponential covariance function (kernel) is used for the latent function GP, owing to its ability to monotonically decrease as input values go farther from each other, which allows for similar latent function values for close input points, and vice versa:

\begin{equation}
 k(\mathbf{x},\mathbf{x'})=\sigma_0^2\exp(-\frac{1}{2}(\mathbf{x}-\mathbf{x'})^T\Lambdaup^{-1}(\mathbf{x}-\mathbf{x'}))
 \label{eq:SE_kernel}
\end{equation}

In equation \ref{eq:SE_kernel}, $\sigma_0^2$ is the output variance, and $\Lambdaup^{-1}$ is the inverse of a diagonal matrix of the characteristic input length scales corresponding to each dimension ($D$ i.e each covariate) in the input data. For a single-input dimension (i.e. $D=1$), the entries along the diagonal of $\Lambdaup^{-1}$ will be identical; otherwise, there will be a separate input length scale for every covariate in the training input data. 

\subsubsection{Training}

Training of the GPRM involves optimizing the hyperparameters ($\theta \equiv {\sigma_0^2,\Lambdaup, \sigma_n^2}$), which is typically done \textit{via} Type II Maximum Likelihood \cite[Chapter 5, pp. 109]{Rasmussen-Williams06}, whereas the marginal likelihood (evidence) of the training observations is maximized (or its negative log is minimized for reasons related to computational stability). That is, the following expression is minimized with respect to $\theta$:
\begin{subequations}
\begin{eqnarray}
%
-\log p(\mathbf{y}|X,\theta) = -\log \mathcal{N}(\mathbf{y}|\mathbf{0},K_{XX}+\sigma_n^2\mathbb{I}) \\
= -\frac{1}{2}\mathbf{y}^{T}(K_{XX}+\sigma_n^2\mathbb{I})^{-1}\mathbf{y}-\frac{1}{2}\log |K_{XX}+\sigma_n^2\mathbb{I}|-\frac{n}{2}\log2\pi
\end{eqnarray}
\end{subequations}

In the expression above, $K_{AB}$ denotes $K(A,B)$ (covariance matrix), and $\mathbb{I}$ the identity matrix.

\subsubsection{Prediction} \label{sec:pred_SGPRM}

Prediction can be done by assuming joint Gaussian distribution between the training observations ($\mathbf{y}$), and a test observation ($y_\ast$ - to be predicted) at the set of test inputs ($\mathbf{x}_\ast$) as follows:
\begin{equation}
\left[\begin{array}{c} \mathbf{y} \\ y_\ast \end{array} \right] = \mathcal{N}\left[\mathbf{0},\begin{array}{cc} K_{XX}+\sigma_n^2\mathbb{I} & \mathbf{k}_{X\mathbf{x}_\ast} \\ \mathbf{k}_{\mathbf{x}_\ast X}&k_{\mathbf{x}_\ast\mathbf{x}_\ast}+\sigma_n^2\mathbb{I} \end{array} \right]
\end{equation}
where $\mathbf{k}_{X\mathbf{x}_\ast}$ is the vector of covariances between $X$ and $\mathbf{x}_\ast$. By invoking the properties of multivariate Gaussian distributions \cite{Rogers-etal20}, the predictive distribution over $y_\ast$ can be defined as follows:
\begin{subequations}
\begin{eqnarray}
p(y_\ast|\mathbf{x}_\ast,X,\mathbf{y})=\mathcal{N}(\mathbb{E}\{y_\ast\}, \mathbb{V}\{y_\ast\}) \\ 
\mathbb{E}\{y_\ast\} = \mathbf{k}_{\mathbf{x}_\ast X}(K_{XX}+\sigma_n^2\mathbb{I})^{-1}\mathbf{y} \\
\mathbb{V}\{y_\ast\} = k_{\mathbf{x}_\ast\mathbf{x}_\ast}-\mathbf{k}_{\mathbf{x}_\ast X}(K_{XX}+\sigma_n^2\mathbb{I})^{-1}\mathbf{k}_{X\mathbf{x}_\ast}+\sigma_n^2
\end{eqnarray}
\end{subequations}
such that $\mathbb{E}\{y_\ast\}$ and $\mathbb{V}\{y_\ast\}$ are the predictive mean and variance, respectively, at the set of test inputs.


\subsection{Variational Heteroscedastic Gaussian Process Regression Models (VHGPRMs)} \label{Sec:VHGPRM}

\subsubsection{Formulation}

One of the inherent drawbacks of using standard (homoscedastic) GPRMs is the assumption of a fixed noise variance throughout the input space, which, in many real-life applications, is impractical. Thus, a number of modifications have been put forward to allow for the noise variance to vary with the input (that is, an input-dependent noise variance) \cite{Lazaro-Gredilla-etal13,Rogers-etal20}. This is to say that the GPRM formulation in equation \ref{eq:SGPRM} would become:
\begin{equation}
y=f(\mathbf{x})+\epsilon(\mathbf{x})
    \label{eq:HGPRM}
\end{equation}
with the noise prior defined as
\begin{equation}
    \epsilon \sim \mathcal{N}(0,r(\mathbf{x}))
\end{equation}
One of the most common strategies is to treat this input-dependent noise as a GP itself, which was first put forward by Goldberg \textit{et al.} \cite{Goldberg-etal98}, that is
\begin{subequations}
\begin{eqnarray}
    r(\mathbf{x}) = \exp(g(\mathbf{x})) \label{eq:var_prior} \\
    g(\mathbf{x})\sim\mathcal{GP}(\mu_0,k_g(\mathbf{x},\mathbf{x}'))
\end{eqnarray}
\end{subequations}
such that $\mu_0$ and $k_g(\mathbf{x},\mathbf{x}')$ are the mean and covariance for the GP prior on the noise variance function, respectively, where an exponential function is used in order to ensure that the noise variance stays positive \cite{Lazaro-Gredilla-Titsias11}. It is worth noting that the subscript $g$ was introduced to differentiate between the covariance function of the noise GP and that of the noise-free process. At this point, it is useful to introduce shorthand notations for the covariance functions as follows:
\begin{eqnarray}
    K_j(X,X) \equiv{K_j} \nonumber \\
    K_j(\mathbf{x}_\ast X) \equiv{K_{j\ast}}\nonumber \\
    K_j(\mathbf{x}_\ast \mathbf{x}_\ast) \equiv{K_{j\ast \ast}}\nonumber 
\end{eqnarray}
Such that $j$ can be $f$ or $g$. Although the formulation in equation \ref{eq:var_prior} provides a better treatment of data with hetersocedastic noise, the added complexity results in making the marginal likelihood and the predictive distribution over unknown observations not analytically-tractable. One of the proposed approaches to approximate them was put forward by Lazaro-Gredilla and Titsias \cite{Lazaro-Gredilla-Titsias11}, which is based on variational approximations. Briefly, their approach is based on approximating $p(f,g|D)$ by $q(f)q(g)$ \textit{via} the minimization of the Kullback-Leibler divergence between them, where the $q(f)$ and $q(g)$ are the variational probability densities (arbitrary density functions) over sets of $f$ and $g$, respectively. The resulting marginal variational (MV) bound ($M$) becomes:
\begin{equation}
M(\mathbf{\mu},\Sigmaup) = \log \mathcal{N}(\mathbf{y}|\mathbf{0},K_f+R)-\frac{1}{4}\tr(\Sigmaup)-\KL(\mathcal{N}(g|\mathbf{\mu},\Sigmaup)||\mathcal{N}(g|\mu_0\mathbf{1},K_g))
\label{eq:MV}
\end{equation}

In equation \ref{eq:MV}, the mean $\mathbf{\mu}$ and covariance $\Sigmaup$ come from the restricting $q(g)$ to be $\mathcal{N}(g|\mathbf{\mu},\Sigmaup)$, the $\KL(\cdot)$ term is the Kullback Leibler divergence between the GP prior on $g$ and the aforementioned restriction on $q(g)$, $\tr(\cdot)$ denotes the trace of the enclosed matrix, $\mathbf{1}$ indicates a vector of ones, and $R$ is a diagonal matrix with elements:
\begin{equation}
    R_{ii} = \exp(\mathbf{\mu}_i-\frac{\Sigmaup_{ii}}{2})
\end{equation}
where $i$ is as defined before. 

\subsubsection{Training}

In equation \ref{eq:MV}, the number of free parameters to be determined becomes n+n(n+1)/2, which would make the training process much more computationally exhaustive. Thus, Lazaro-Gredilla and Titsias \cite{Lazaro-Gredilla-Titsias11} proposed a reparametrization of $\mathbf{\mu}$ and $\Sigmaup$ at the maxima of the marginal variational bound into the following:

\begin{equation}
\mathbf{\mu}=K_g(\Lambdaup-\frac{1}{2}\mathbb{I})\mathbf{1}+\mu_0\mathbf{1}, \, \, \, \Sigmaup^{-1}=K_g^{-1}+\Lambdaup
\end{equation}
such that $\Lambdaup$ is a positive semi-definite diagonal matrix of the variational parameters (to be determined through optimization).

\subsubsection{Prediction} \label{sec:pred_VHGPRM}

Based on the formulation presented by Lazaro-Gredilla and Titsias \cite{Lazaro-Gredilla-Titsias11}, the predictive distribution for a new test point $y_\ast$ can be simplified to:
\begin{eqnarray}
    q(y_\ast) = \int\int p(y_\ast|f_\ast,g_\ast)q(f_\ast)q(g_\ast)df_\ast dg_\ast \nonumber \\
    = \int \mathcal{N}(y_\ast|a_\ast,c_\ast^2+\exp(g_\ast))\mathcal{N}(g_\ast|\mu_\ast,\sigma_\ast^2)dg_\ast
\end{eqnarray}
with:
\begin{subequations}
\begin{eqnarray}
a_\ast = \mathbf{k}_{f\ast}(K_{f}+R)^{-1}\mathbf{y} \\
c_\ast^2 = k_{f\ast\ast}-\mathbf{k}_{f\ast}^T(K_{f}+R)^{-1}\mathbf{k}_{f\ast} \\
\mu_\ast = \mathbf{k}_{g\ast}^T(\Lambdaup-\frac{1}{2}\mathbb{I})\mathbf{1}+\mu_0 \\
\sigma_\ast^2=k_{g\ast\ast}-\mathbf{k}_{g\ast}^T(K_{g}+\Lambdaup^{-1})^{-1}\mathbf{k}_{g\ast}
\end{eqnarray}
\end{subequations}

Although the integration above does not have an analytical solution, the first two moments of the predictive distribution can be calculated analytically as follows:
\begin{subequations}
\begin{eqnarray}
 \mathbb{E}\{y_\ast|\mathbf{x}_\ast,\mathcal{D}\} = a_\ast  \\
 \mathbb{V}\{y_\ast|\mathbf{x}_\ast,\mathcal{D}\} = c_\ast^2+\exp(\mu_\ast+\frac{\sigma_\ast^2}{2})
\end{eqnarray}
\end{subequations}
Thus, under the assumption of a Gaussian predictive distribution over the unknown test observation $y_\ast$, which is not necessarily true, the variance of the test observation can be predicted along with the test observation itself, which allows for a more flexible model compared to standard GPRMs \cite{Rogers-etal20}. 


\subsection{Damage State Quantification} \label{sec:prediction}

\begin{figure}[t!]
\centering
\includegraphics[scale=0.55]{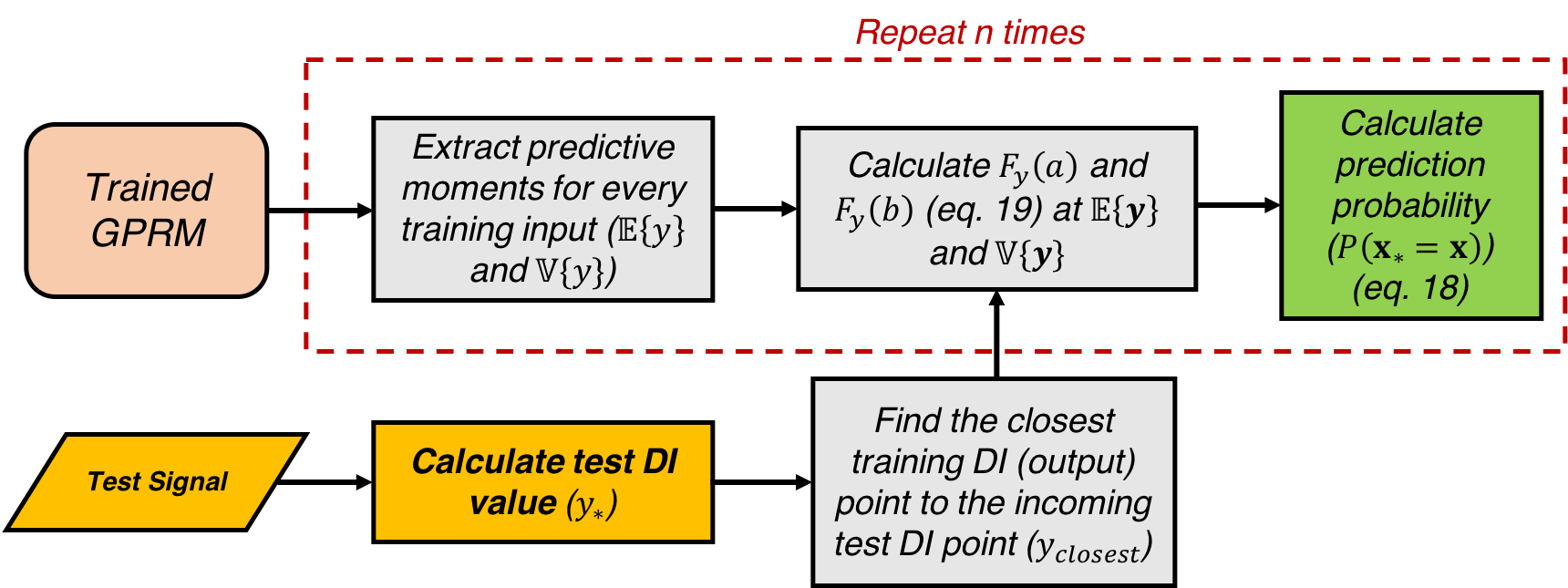}
\caption{A schematic flow diagram showing the steps taken in this study to calculate the state prediction probability.}
\label{fig:prob} \vspace{10pt}
\end{figure}

In many practical cases, such as the case in this study, the test observations at which prediction should take place are actually the GPRM targets, not the inputs i.e. in this study, a test DI ($y_\ast$ - target) would be available, and the damage size and/or load state ($\mathbf{x_\ast}$ - inputs) would be estimated by the GPRM. In this work, the proposed method of estimating the inputs from the test target values is based on the probability that a test target value $y_\ast$ belongs to a specific damage size and/or load state $\mathbf{x}$. Figure \ref{fig:prob} schematically outlines the steps to calculate the state (input) prediction probability in this case. As shown, this probability ($P(\mathbf{x_\ast}=\mathbf{x})$) can be estimated from the Cumulative Distribution Function of the targets as follows:

\beq P(\mathbf{x}_\ast=\mathbf{x}) = F_{y}(b;{E}\{y\},\mathbb{V}\{y\})-F_{X}(a;{E}\{y\},\mathbb{V}\{y\})     \label{eq:CDF1} \eeq
such that,
\begin{subequations}
    \begin{eqnarray}
        F_{y}(s;\mu,\sigma)=\frac{1}{\sigma\sqrt{2\pi}}\sum_{t=-\infty}^{s}{e^{\frac{-(t-\mu)^2}{2\sigma^2}}} \\
        a=y_{\ast}-2 \sqrt{\mathbb{V}\{y_{closest}\}} \\
        b=y_{\ast}+2 \sqrt{\mathbb{V}\{y_{closest}\}}
    \end{eqnarray}
    \label{eq:CDF2}
\end{subequations}
In the above equations, $y_{closest}$ is the closest training DI value to the value of the incoming test DI point ($y_\ast$), and $\mathbb{V}\{y_{closest}\}$ is the corresponding GPRM predictive variance. Also, $\mathbb{E}\{y\}$ and $\mathbb{V}\{y\}$ are the GPRM predictive mean and variance, respectively, at the training input $\mathbf{x}$. Finally, $F_{y}(\cdot)$ is the value of the Cumulative Distribution Function of the targets at the enclosed point and distribution. This probability is calculated for every set of inputs (states) in the training data. As shown in this expression, information from the uncertainty in the GPRM (predictive variance) corresponding to the closest training DI value to the test one is leveraged in order to properly estimate the probability of damage size and/or load state. In addition, this framework can be easily extended to VHGPRMs by replacing the the SGPRM predictive means and variances by VHGPRM ones. Although this framework works for both single-state (damage size only) and two-state (load and damage size) predictions, the readers are directed to Section \ref{sec:2-state} for more information on the latter. Finally, it is worth noting that, although the present treatment of the available DI data dictates the use of the state prediction framework presented in this section, it might also be of interest in using more elaborate multi-output, noisy-input GPRMs in which the DI values are the inputs and the states are the outputs. In such a case, GPRM prediction methodologies, such as those presented in Sections \ref{sec:pred_SGPRM} but suited to multi-output, noisy-input GPRMs, can be used for state quantification directly.

\subsubsection{Notes on two-state prediction using DI-trained GPRMs} \label{sec:2-state}

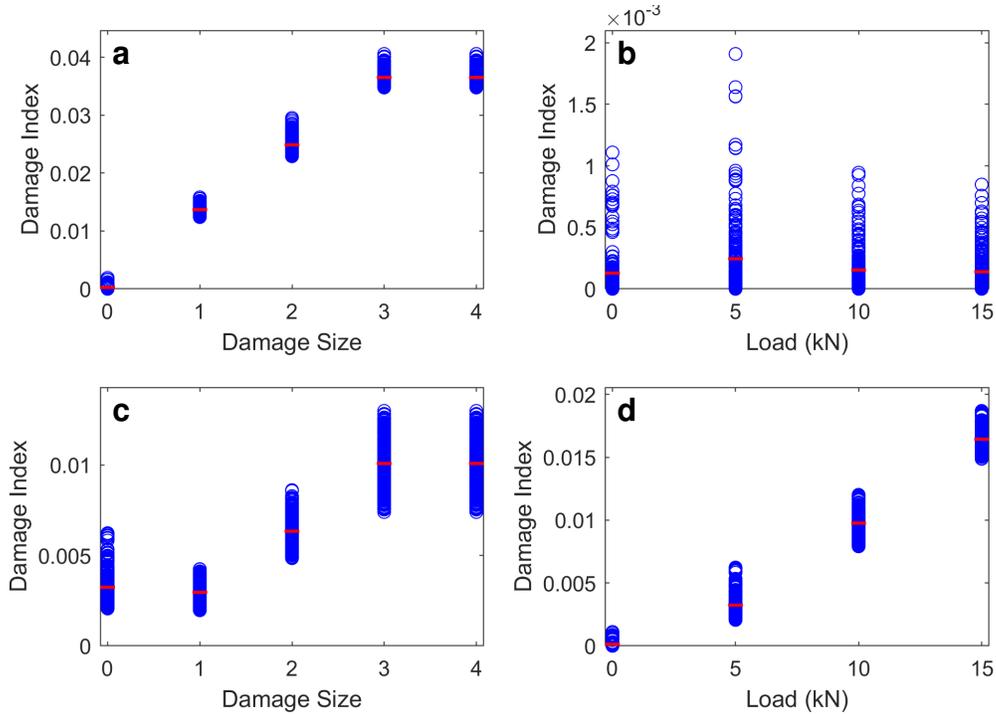
\begin{figure}[t!]
    \centering
    \begin{picture}(400,300)
    \put(0,40){\includegraphics[trim = 20 0 20 15,clip,scale=0.8]{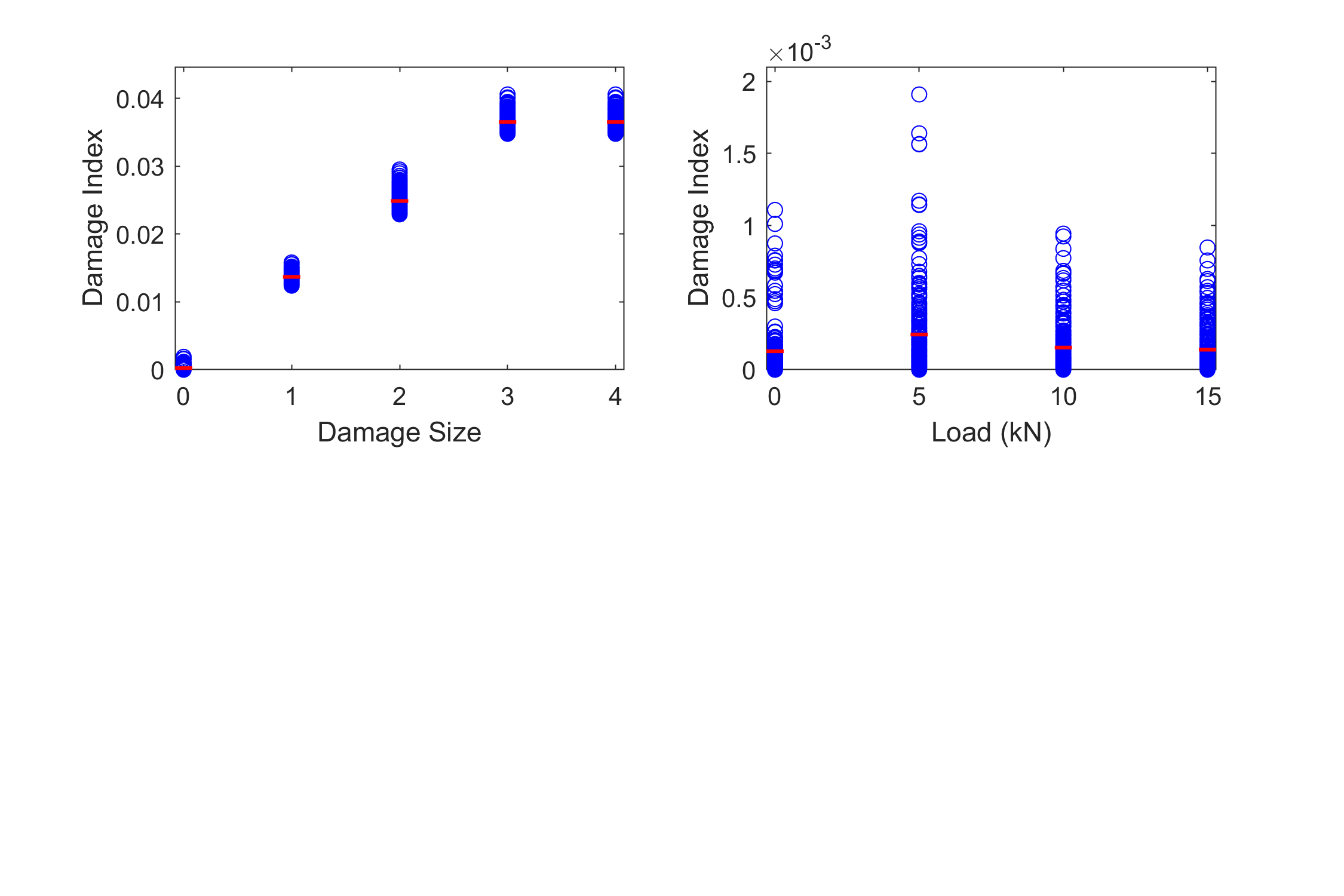}}
    \put(0,-95){\includegraphics[trim = 20 0 20 15,clip,scale=0.8]{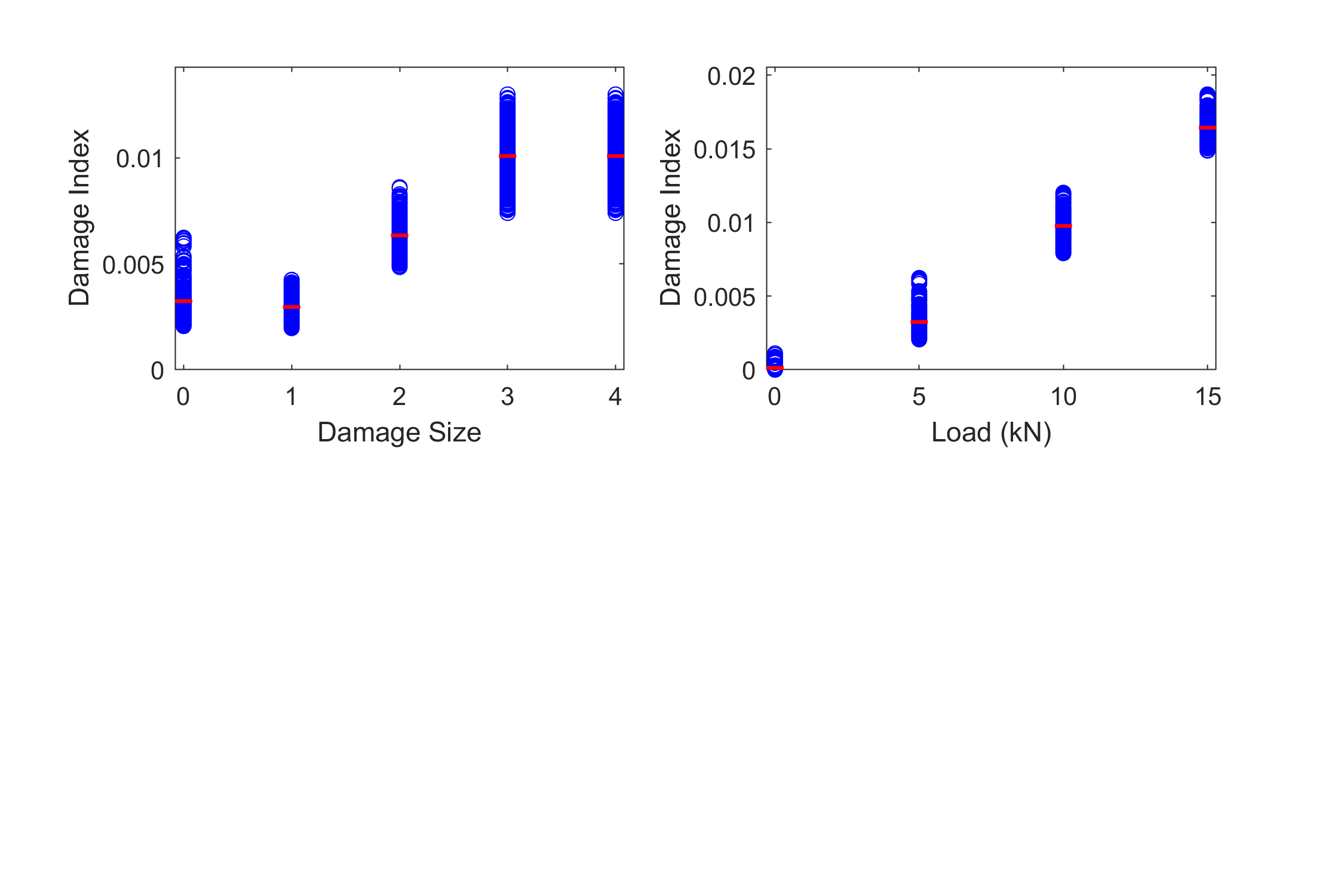}}
   \put(45,295){\color{black} \large {\fontfamily{phv}\selectfont \textbf{a}}}
    \put(236,295){\large {\fontfamily{phv}\selectfont \textbf{b}}}
   \put(45,160){\large {\fontfamily{phv}\selectfont \textbf{c}}} 
   \put(236,160){\large {\fontfamily{phv}\selectfont \textbf{d}}} 
    \end{picture} \vspace{-55pt}
    \caption{Indicative DI plots from path 1-6 in the third test case in this study (Al coupon with simulated damage) showing the evolution of DIs calculated with respect to damage size and load using both classes of reference signals: (a) evolution of DI values calculated using class 1 reference signals with damage size at a load of 5 kN; (b) evolution of DI values calculated using class 1 reference signals with load in the healthy state; (c) evolution of DI values calculated using class 2 reference signals with damage size at a load of 5 kN; (d) evolution of DI values calculated using class 2 reference signals with load in the healthy state. The red dots indicate the means of the DI values at every state.} 
\label{fig:example_DI} \vspace{0pt}
\end{figure}

As aforementioned, the trained models in this study are used for predicting damage size at a known load, or predicting both damage size and load state simultaneously. Since GPRMs are trained herein using DI values, which depend heavily on the reference signal(s) being used for their calculations, in order to train GPRMs to predict both load state and damage size accurately, the models must be trained with two classes of reference signals. The first class comprises the signals coming from the healthy structure at various loading states, which would result in DI values that show a uniform evolution with damage size, but not necessarily with load since the reference for each loading state (and damage size) would be the healthy signal at that specific load, not at the unloaded state. This can be evidently seen in Figure \ref{fig:example_DI} panels a and b, which show indicative DI plots from the third test case in this study (the Al coupon with simulated damage) when the DI's are calculated using reference signals from the aforementioned class of signals (class 1). The second class of reference signals contains the signals coming form the unloaded case at various damage sizes. This latter class would result in DI values that evolve nicely with load states, but fail to properly follow the effects of damage evolution since the DI values at different damage sizes would be referenced to the unloaded case at the corresponding damage size, not the healthy case. Figure \ref{fig:example_DI} panels c and d show the evolution of the DIs calculated using this class of signals (class 2). Thus, in order to obtain accurate predictions of both damage size and loading state, GPRMs need to be trained with DI values referenced to both classes of reference signals. 

\begin{table}[b]
\centering
\caption{The reference signals encompassed by each of the two classes of reference signals used in this study for calculating the DI values for the purposes of training multi-input GPRMs in the third test case presented herein.}\label{tab:switch_covariate}
\renewcommand{\arraystretch}{1.2}
{\footnotesize
\begin{tabular}{|c|c|c|} 
\hline
Class & Reference Signal State & Switch Covariate \\ 
\hline
 & Healthy @0 kN & \\
1 & Healthy @5 kN & 1 \\
 & Healthy @10 kN & \\
 & Healthy @15 kN & \\
\hline
 & Healthy @0 kN & \\
 & 1 weight @0 kN & \\
2 & 2 weights @0 kN & 2\\
 & 3 weights @0 kN & \\
 & 4 weights @0 kN & \\
\hline
\end{tabular}} 
\end{table}
Table \ref{tab:switch_covariate} presents the set of reference signals used in calculating the DI values in each class. Also shown in Table \ref{tab:switch_covariate} is the value of the so-called ``switching covariate'', which is simply a third covariate in the input space of the trained GPRMs ($\mathbf{x}_i \in \mathbb{R}^3$, damage size and load being the first two covariates) that acquires a value of 1 or 2 to identify each of the two different classes of DI reference signals in the training space. That is, the training inputs and outputs can be defined as follows: 
\beq X \in \{\mathbf{x}_{1}^1 \; \mathbf{x}_{2}^1 \; \ldots \; \mathbf{x}_{n}^1 \; \mathbf{x}_{1}^2 \; \mathbf{x}_{2}^2 \; \ldots \; \mathbf{x}_{n}^2\}^T, \mathbf{y} \in \{y_{1}^1 \; y_{2}^1 \; \ldots \; y_{n}^1 \; y_{1}^2 \; y_{2}^2 \; \ldots \; y_{n}^2\}^T  \eeq
where the superscript denotes the class of reference signals to which each target (DI value) is referenced. Note that for the inputs, this superscript only denotes the value of the third covariate (the switch covariate) and it does not affect the values of the first two covariates (damage size and load). It is worth noting here that it is also possible to reformulate this problem using a multi-output GPRM approach with a 2-dimensional input space. However, this 3-covariate framework was chosen herein for convenience. 

\begin{figure}[t!]
\centering
\includegraphics[scale=0.5]{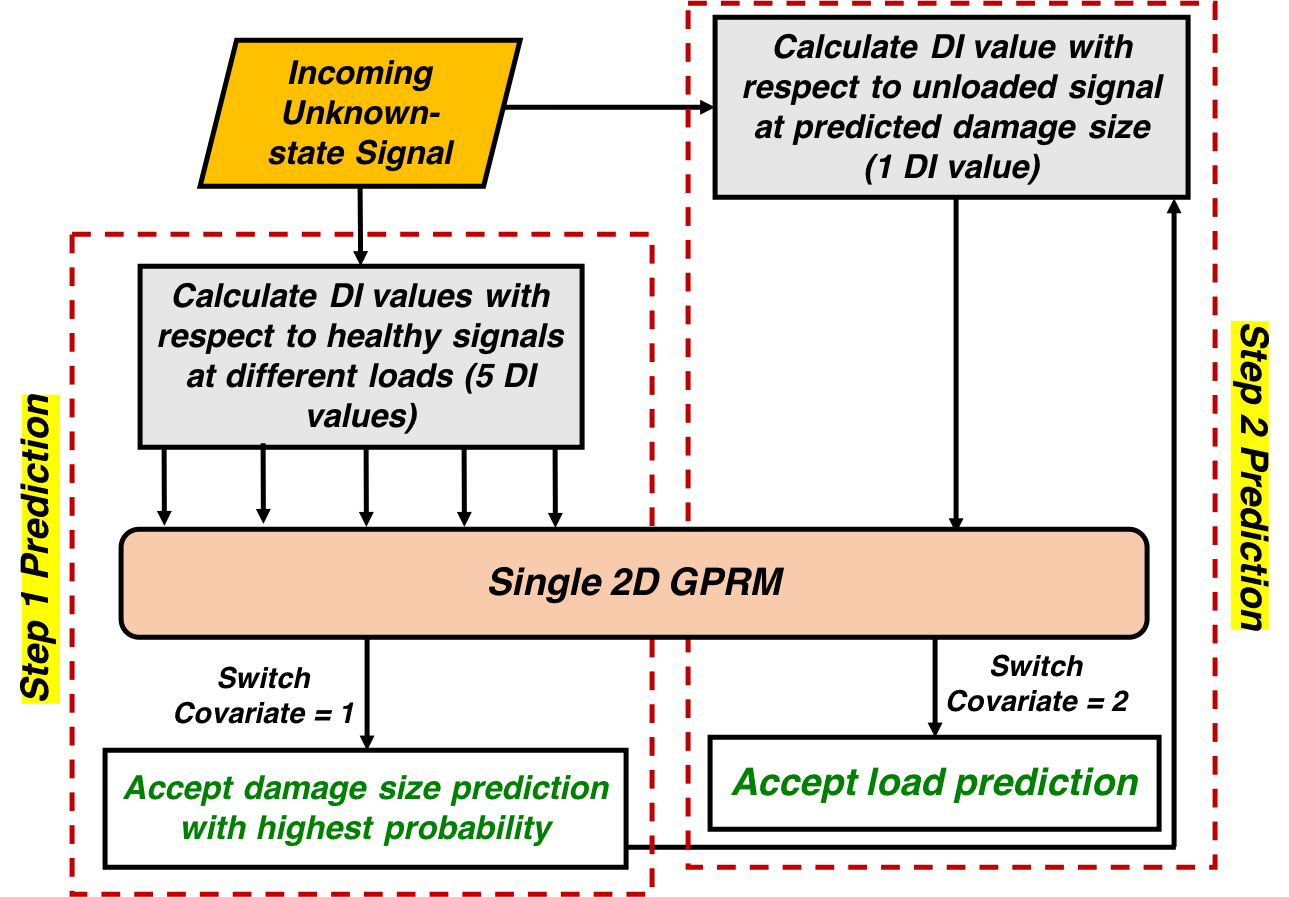}
\caption{A schematic showing the 2-step prediction process followed in simultaneously predicting damage size and load in the Al coupon with simulated damage.}
\label{fig:2-step_process} \vspace{10pt}
\end{figure}

Figure \ref{fig:2-step_process} schematically outlines the two-state prediction process adopted here. As shown, with properly trained GPRMs, two-state predictions are made in a two-step process. In the first step of this prediction process, the DI values of the incoming test signal is first calculated with respect to the reference signals in class 1 (healthy signals at various loading states). Each of these test DI values (5 values in this study) is fed into the prediction process outlined in Figure \ref{fig:prob}, and a probability is calculated for the damage size and load, repeating this process for each of the DI values calculated with respect to class 1 reference signals, and choosing the pair of states with highest probability. It is important to note that, in this step, the switching covariate is set to 1 in order to only use the predictive means and variances related to that part of the trained GPRM. This step generally leads to an accurate damage size prediction, but a poor load prediction, again owing to the poor evolution of class 1 DIs with loads. Thus, only damage size prediction is accepted in this step. In the second step of this prediction process, a new test DI value of the incoming test signal is calculated, this time with respect to the reference signal from the class 2 signals that corresponds to the damage size predicted in the first step. Then, knowing the damage size, this test DI value is fed into the prediction process outlined in Figure \ref{fig:prob} with the switching covariate set to 2 in order to calculate the values of the CDF using the proper predictive moments from class 2 DIs. This process leads to an accurate load size prediction, since it is implemented at the known damage size predicted in the first step.

\section{Test Case I: Al Coupon with Varying Notch Sizes} \label{Sec:notched_Al}

\subsection{Test Setup}

\begin{figure}[t!]
\centering
\includegraphics[scale=0.35]{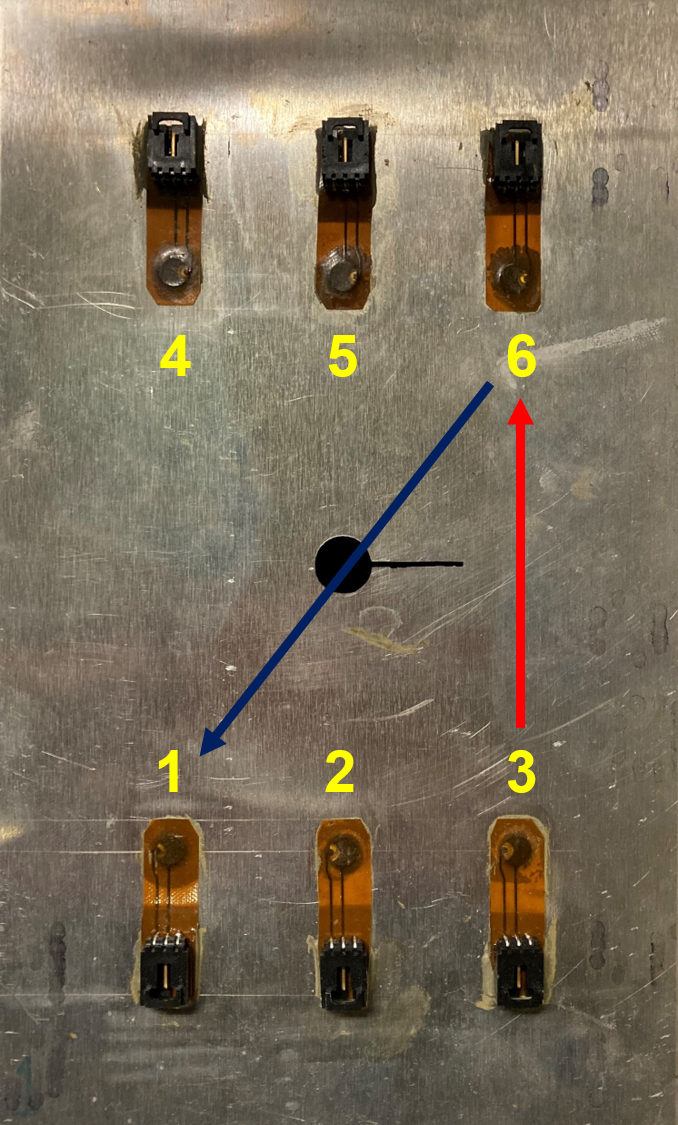}
\caption{The notched Al coupon used in this study shown here with a 20-mm notch (largest damage size). The arrows show the signal paths presented herein.}
\label{fig:Al_coupon} \vspace{10pt}
\end{figure}

The first coupon used in this study was a 6061 Aluminum $152.4 \times  254$ mm ($6 \times 10$ in) coupon ($2.36$ mm/$0.093$ in thick) (McMaster Carr) with a 12-mm (0.5-in) diameter hole in the middle, as shown in Figure \ref{fig:Al_coupon}. Six PZT (Lead Zirconate Titanate) SMART Layers type PZT-5A (with thickness $0.2$ mm/$0.008$ in and diameter $3.175$ mm/$1/8$ in; Acellent Technologies, Inc) were attached to the CFRP coupon using Hysol EA 9394 adhesive. In order to simulate damage, up to six three-gram weights were sequentially attached to the surface of the plate using tacky tape. Damage was simulated by cutting a notch starting from the middle hole using an end-mill and a $0.8128$-mm ($0.032$-in) hand saw, with varying lengths between $2$ and $20$ mm, in $2$-mm increments.

In order to interrogate the coupon, each sensor was actuated, in a consecutive manner, using 5-peak tone bursts (5-cycle Hamming-filtered sine waves) with 90 V peak-to-peak amplitude and various center frequencies. 20 response signals per structural case were collected at each sensor at a sampling rate of 24 MHz using a ScanGenie III data acquisition system (Acellent Technologies, Inc). Preliminary analysis for the best separation between the first two wave packets in various response signal paths wa done, and a center frequency of 250 kHz was chosen for the analysis presented herein. For the GPRMs, the normalized mean squared error (NMSE) presented in ref. \cite{Lazaro-Gredilla-Titsias11} and the residual sum of squares divided by the sample sum of squares (RSS/SSS) were used for assessing the trained models, and all values are reported herein with respect to the validation (test) data that was not used in the training phase. All analysis was done in Matlab.\footnote{Matlab version R2020a, GPRM training and prediction: the different functions within the GPML package available at http://www.gaussianprocess.org/gpml/code/matlab/doc/index.html. VHGPRM training and prediction: http://www.tsc.uc3m.es/~miguel.}

\subsection{Results \& Discussion}

\begin{figure}[t!]
    \centering
    \begin{picture}(400,300)
    \put(0,40){\includegraphics[trim = 20 0 20 12,clip,scale=0.8]{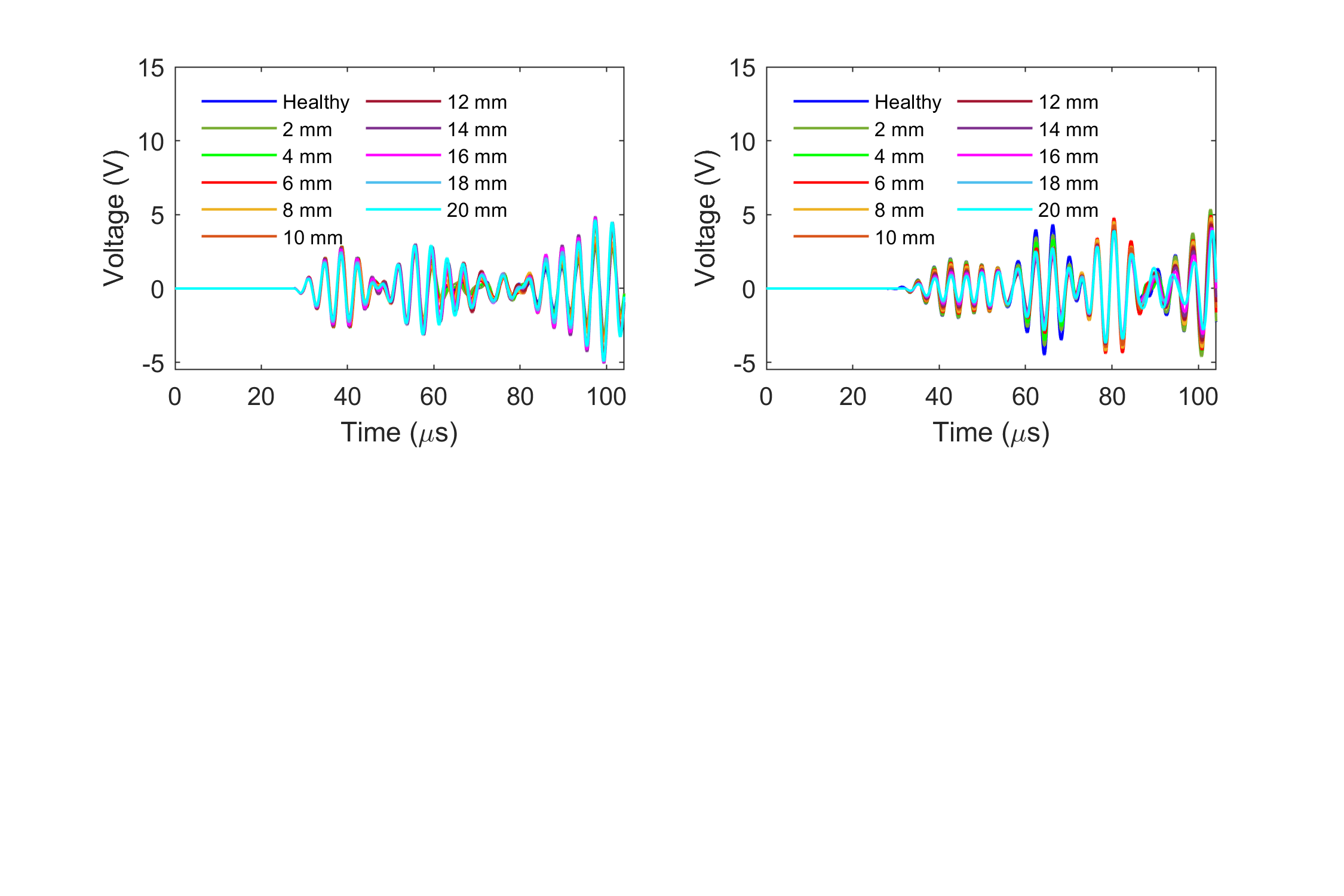}}
    \put(0,-95){\includegraphics[trim = 20 0 20 15,clip,scale=0.8]{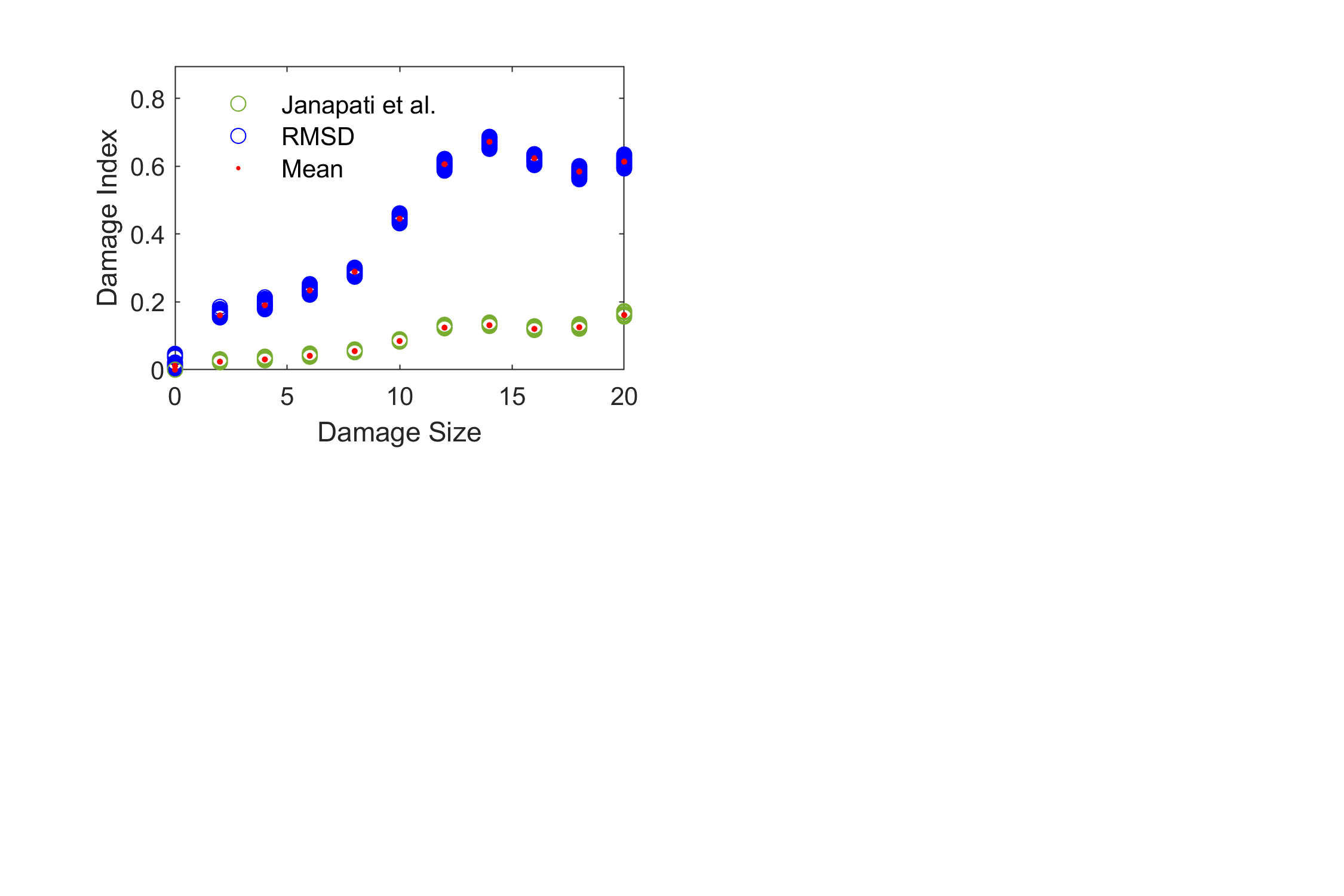}}
    \put(190,-95){\includegraphics[trim = 20 0 20 15,clip,scale=0.8]{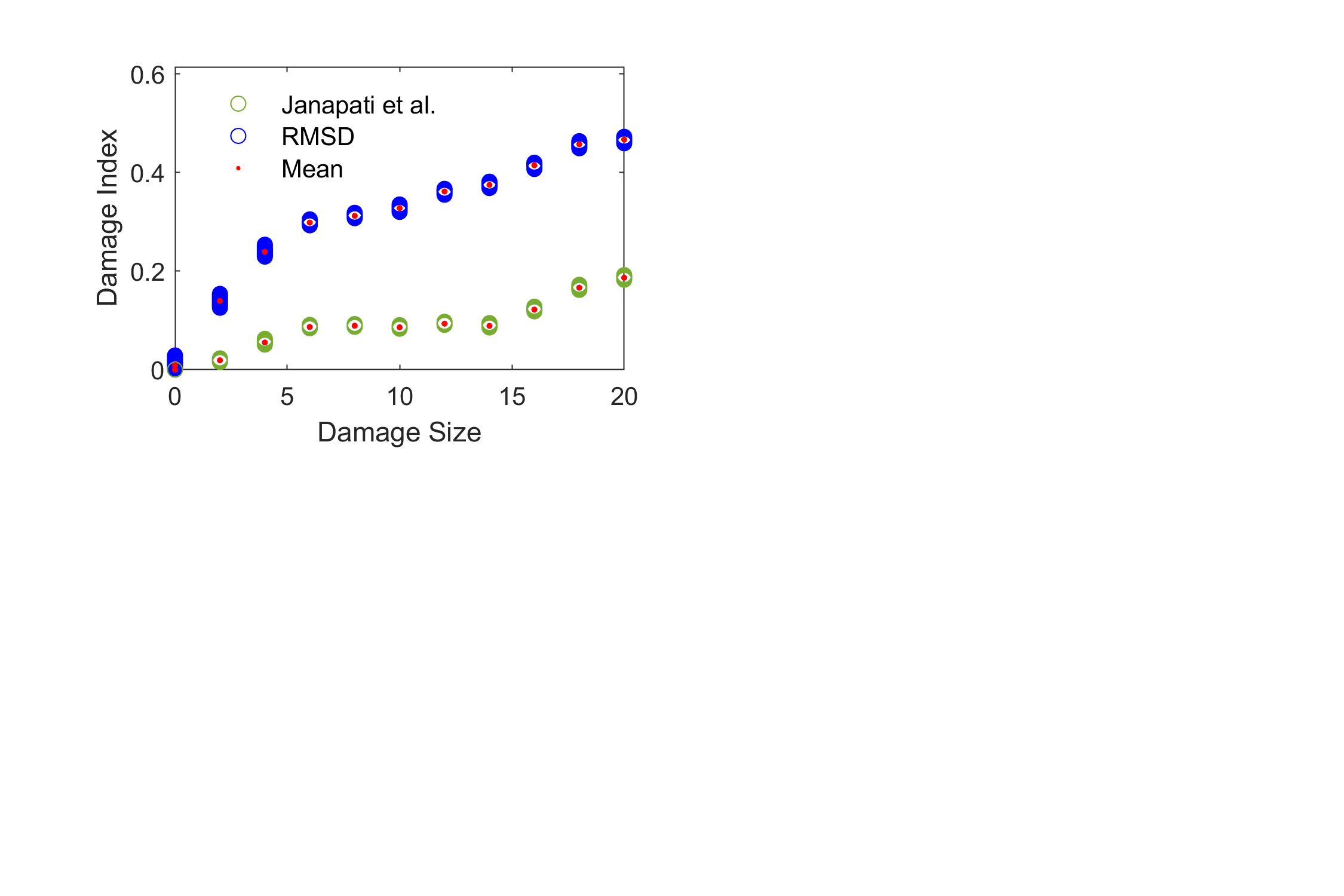}}   \put(45,295){\color{black} \large {\fontfamily{phv}\selectfont \textbf{a}}}
    \put(235,295){\large {\fontfamily{phv}\selectfont \textbf{b}}}
   \put(45,160){\large {\fontfamily{phv}\selectfont \textbf{c}}} 
   \put(235,160){\large {\fontfamily{phv}\selectfont \textbf{d}}} 
   \put(120,130){\color{red}\framebox(70,25){}}
   \put(122,80){\color{red}\framebox(55,20){}}
   \put(265,80){\color{red}\framebox(70,20){}}
    \end{picture} \vspace{-55pt}
    \caption{Notched Al coupon: indicative signals and DI plots: (a) signals from path 3-6; (b) signals from path 6-1; (c) DI values from path 3-6; (d) DI values from path 6-1.} 
\label{fig:notch_signals} \vspace{-12pt}
\end{figure}
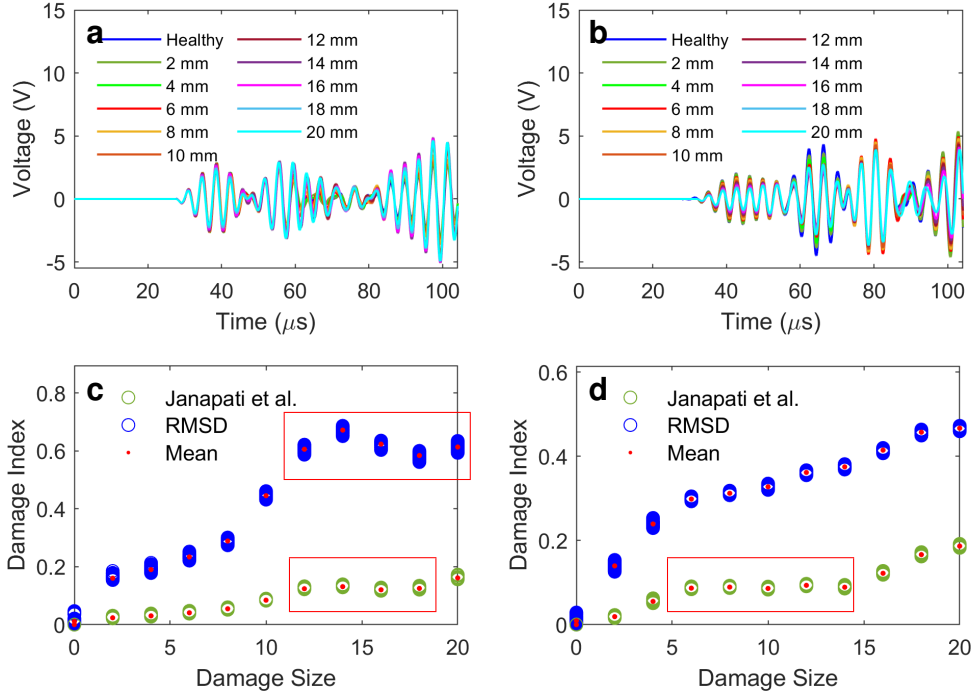

Figure \ref{fig:notch_signals} panels a and b show the signals from the actuator-sensor paths 3-6 and 6-1, respectively, under varying notch sizes (see Figure \ref{fig:Al_coupon} for sensor numbering). Figure \ref{fig:notch_signals} panels c and d show the corresponding DI evolution from both DI formulations used in this study. As shown in the latter two panels, aside from the difference in amplitude, both DI formulations show similar trends with respect to damage size. Also, it is worth noting that the saturation phenomenon (indicated by red frames in the panels) seems to occur under different damage sizes for both paths. As will be shown later, as these DI values are used in training GPRMs, damage size predictions in the areas of saturation might be challenging since the models would not be able to differentiate between the different damage sizes having similar DI values. Finally, because path 6-1 intersects the notch, the DI evolution using both formulations in this path is more uniform with notch size compared to that for path 3-6.

\begin{figure}[t!]
    \centering
    \begin{picture}(400,300)
    \put(0,40){\includegraphics[trim = 20 0 20 15,clip,scale=0.8]{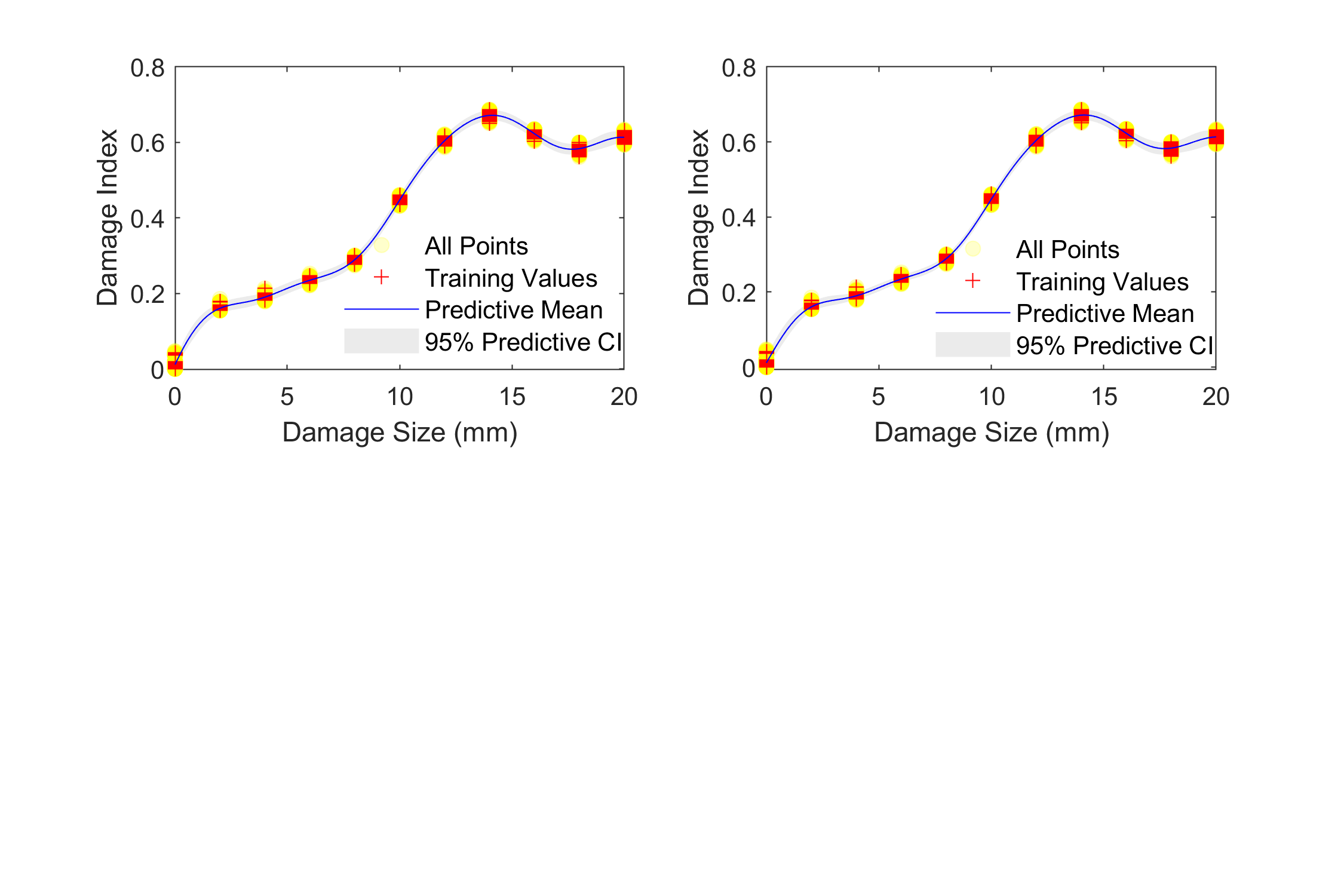}}
    \put(0,-95){\includegraphics[trim = 20 0 20 15,clip,scale=0.8]{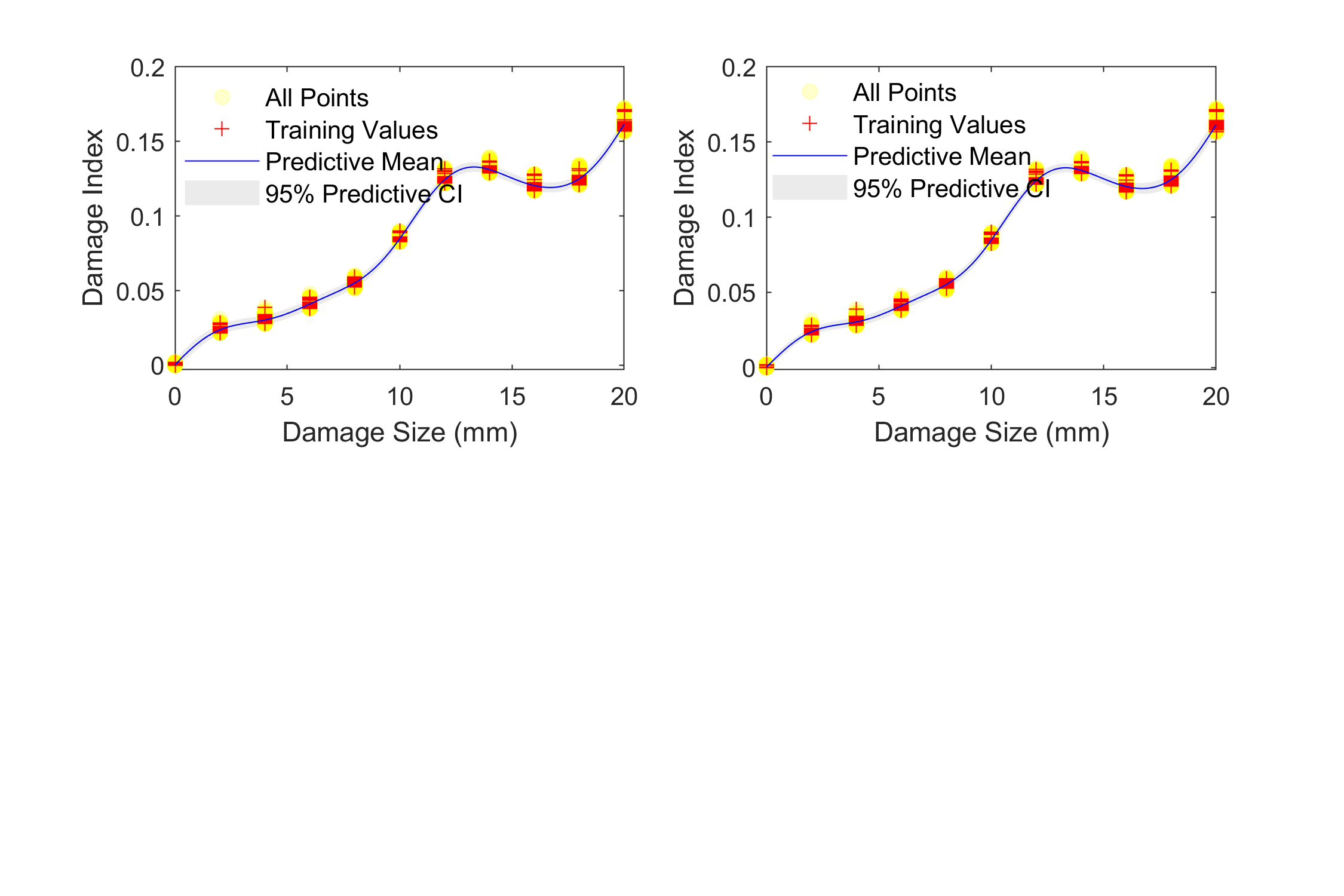}}
    \put(45,295){\color{black} \large {\fontfamily{phv}\selectfont \textbf{a}}}
    \put(235,295){\large {\fontfamily{phv}\selectfont \textbf{b}}}
    \put(45,160){\large {\fontfamily{phv}\selectfont \textbf{c}}} 
    \put(235,160){\large {\fontfamily{phv}\selectfont \textbf{d}}}
    \end{picture} \vspace{-55pt}
    \caption{Notched Al coupon: GPRM predictive mean and variance for path 3-6: (a) SGPRM results based on the RMSD DI; (b) VHGPRM results based on the RMSD DI; (c) SGPRM results based on the DI formulation from \cite{Janapati-etal16}; (d) VHGPRM results based on the DI formulation from \cite{Janapati-etal16}.}
\label{fig:notch_gprm_3-6} \vspace{0pt}
\end{figure}
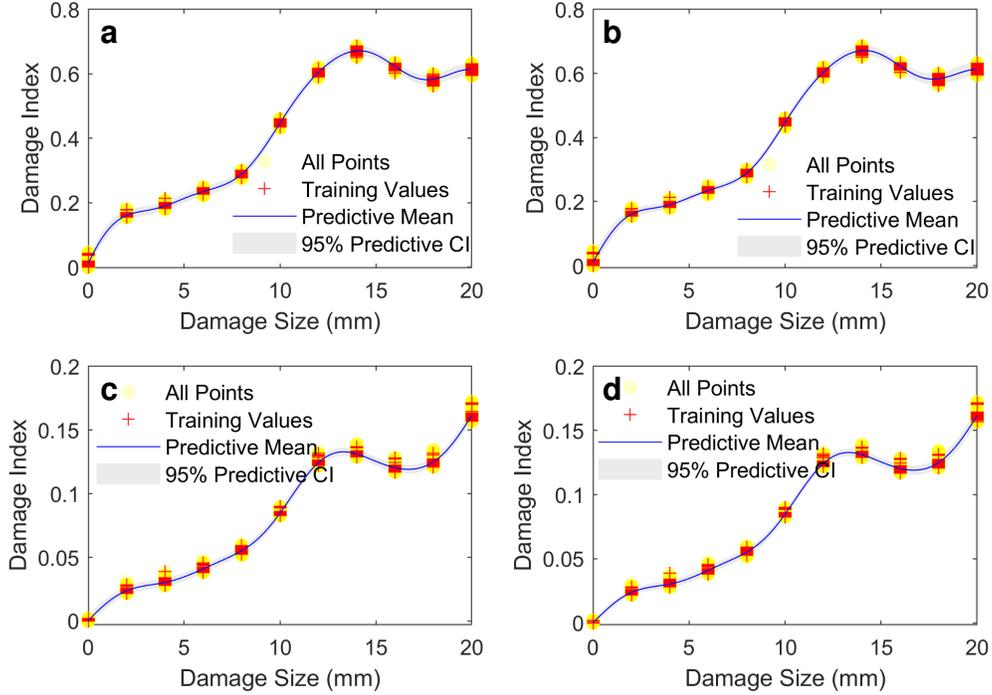

\begin{figure}[t!]
    \centering
    \begin{picture}(400,300)
    \put(0,40){\includegraphics[trim = 20 0 20 15,clip,scale=0.8]{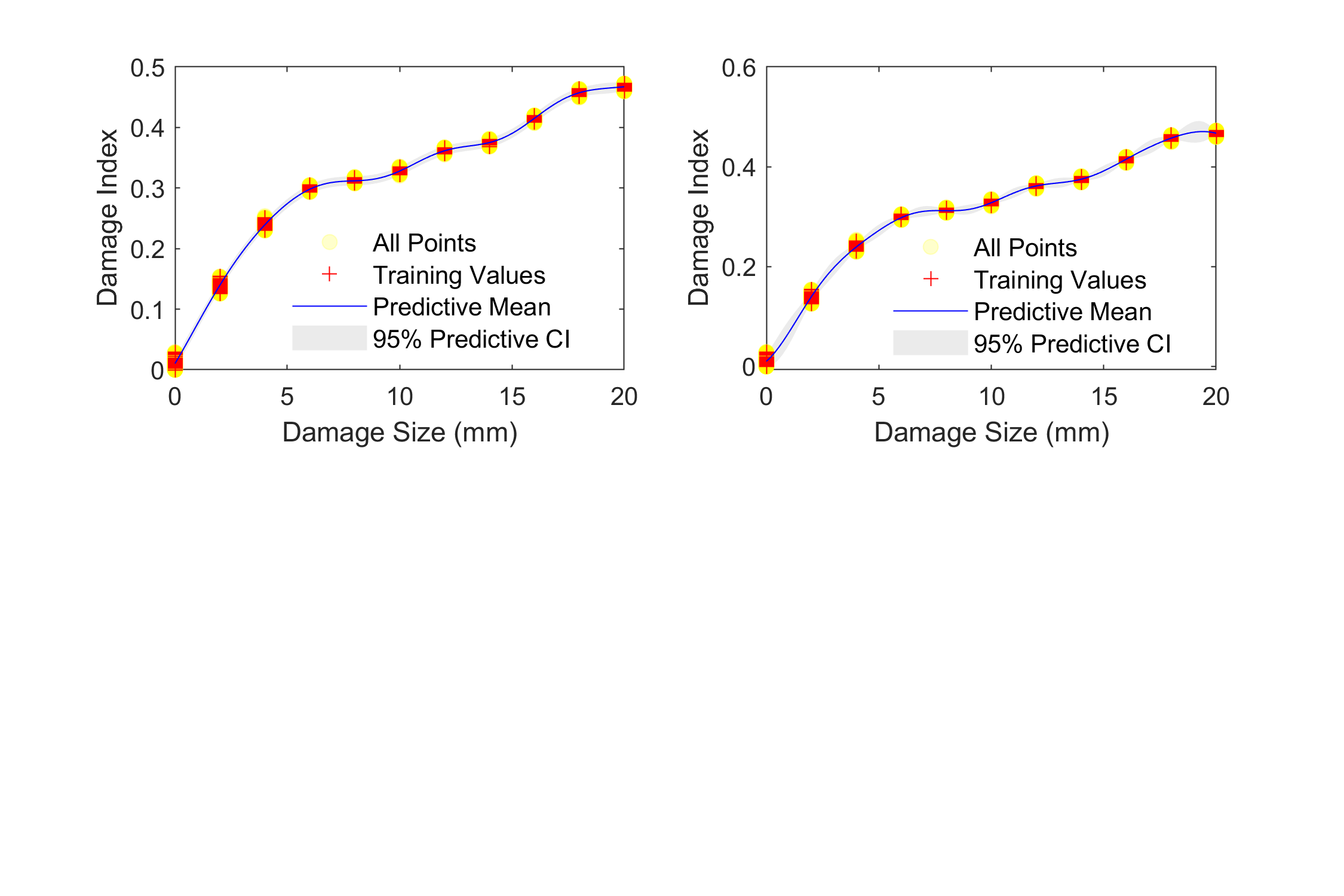}}
    \put(0,-95){\includegraphics[trim = 20 0 20 15,clip,scale=0.8]{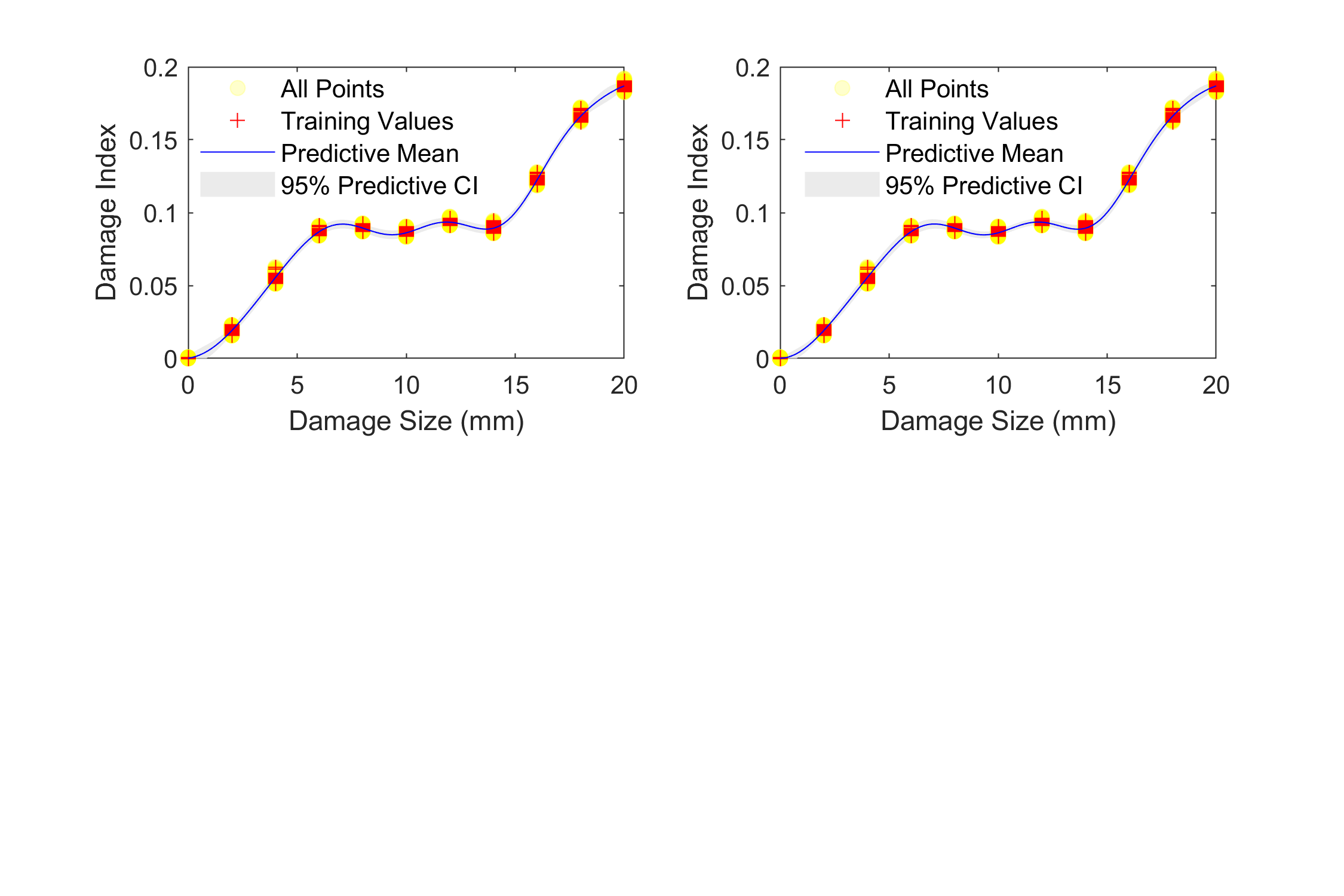}}
    \put(45,295){\color{black} \large {\fontfamily{phv}\selectfont  \textbf{a}}}
    \put(235,295){\large {\fontfamily{phv}\selectfont \textbf{b}}}
    \put(45,160){\large {\fontfamily{phv}\selectfont \textbf{c}}} 
    \put(235,160){\large {\fontfamily{phv}\selectfont \textbf{d}}} 
    
    
    \end{picture} \vspace{-55pt}
    
    \caption{Notched Al coupon: GPRM predictive mean and variance for path 6-1: (a) SGPRM results based on the RMSD DI; (b) VHGPRM results based on the RMSD DI; (c) SGPRM results based on the DI formulation from \cite{Janapati-etal16}; (d) VHGPRM results based on the DI formulation from \cite{Janapati-etal16}.} 
\label{fig:notch_gprm_6-1} \vspace{0pt}
\end{figure}
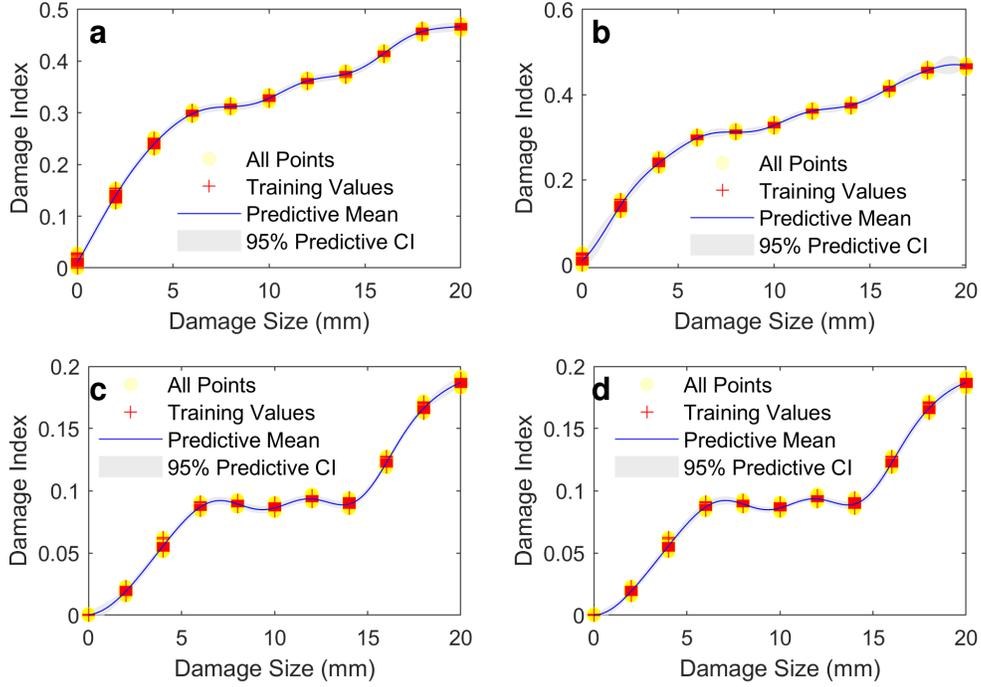

SGPRMs and VHGPRMs were trained for both paths in order to model the evolution of both DI formulations with notch size, whiles modelling the sources of uncertainty in the data. Tables \ref{tab:notch_rmsd} and \ref{tab:notch_janapati} show model training and testing information for both DI formulations. briefly, about $50\%$ of the available DI data points for each formulation were used to train the models, while the remaining $50\%$ was left for testing the models. As shown in the tables, the difference in model performance between SGPRM and VHGPRM is minimal, although the VHGPRM training process takes well over double the time required for training the SGPRM for the reasons previously noted in Section \ref{Sec:VHGPRM}. 
\begin{table}[b]
\centering
\caption{Summary of GPRM$^*$ information$^\dagger$ for the notched Al coupon based on the RMSD DI formulation.}\label{tab:notch_rmsd}
\renewcommand{\arraystretch}{1.2}
{\footnotesize
\begin{tabular}{|c|c|c|c|c|c|c|c|c|} 
\hline
Signal & \multicolumn{2}{c}{NMSE} & \multicolumn{2}{|c|}{RSS/SSS (\%)} & \multicolumn{2}{c}{Training Time (s)} & \multicolumn{2}{|c|}{Prediction Time (s)} \\ 
\cline{2-9}
 Path & SGPRM & VHGPRM & SGPRM & VHGPRM & SGPRM & VHGPRM & SGPRM & VHGPRM \\
\hline
3-6 & 8.87E-4 & 8.88E-4 & 0.026 & 0.026 & 8.8 & 21.5 & 0.1577 & 0.1751 \\  
\hline
6-1 & 7.780E-4 & 7.781E-4 & 0.012 & 0.012 & 8.8 & 23.7 & 0.1630 & 0.1753 \\  
\hline
\multicolumn{7}{l}{$^*$22.5\% (990 points) of the data was used for training each model.} \\
\multicolumn{9}{l}{$^\dagger$Numbers approximated to the last quoted decimal place, and times estimated based on an Intel Core i3 laptop} \\
\multicolumn{8}{l}{  with 4 Gb of RAM.}
\end{tabular}} 
\end{table}
\begin{table}[b]
\centering
\caption{Summary of GPRM$^*$ information$^\dagger$ for the notched Al coupon based on the DI formulation from \cite{Janapati-etal16}}\label{tab:notch_janapati}
\renewcommand{\arraystretch}{1.2}
{\footnotesize
\begin{tabular}{|c|c|c|c|c|c|c|c|c|} 
\hline
Signal & \multicolumn{2}{c}{NMSE} & \multicolumn{2}{|c|}{RSS/SSS (\%)} & \multicolumn{2}{c}{Training Time (s)} & \multicolumn{2}{|c|}{Prediction Time (s)} \\ 
\cline{2-9}
 Path & SGPRM & VHGPRM & SGPRM & VHGPRM & SGPRM & VHGPRM & SGPRM & VHGPRM \\
\hline
3-6 & 0.0011 & 0.0011 & 0.032 & 0.032 & 8.9 & 22.9 & 0.1704 & 0.1734 \\
\hline
6-1 & 8.035E-4 & 8.033E-4 & 0.024 & 0.024 & 8.9 & 23.6 & 0.1670 & 0.1822 \\  
\hline
\multicolumn{7}{l}{$^*$22.5\% (990 points) of the data was used for training each model.} \\
\multicolumn{9}{l}{$^\dagger$Numbers approximated to the last quoted decimal place, and times estimated based on an Intel Core i3 laptop} \\
\multicolumn{8}{l}{  with 4 Gb of RAM.}
\end{tabular}} 
\end{table}
Figure \ref{fig:notch_gprm_3-6} shows the predictive means and confidence intervals for SGPRMs and VHGPRMs trained using both DI data sets for path 3-6. Figure \ref{fig:notch_gprm_6-1} shows the corresponding plots for path 6-1. As shown in both figures, the GPRMs closely follow the evolution of the DI in all cases and conveniently model the uncertainty in the data sets through the underlying assumptions in the models. In addition, both SGPRMs and VHGPRMs perform just as well, with a slight difference in the evolution of the confidence intervals in the VHGPRM plot (for instance, see Figure \ref{fig:notch_gprm_6-1}b towards the lower notch sizes). This similarity between both types of models stems from the lack of changing variance with damage size in the data sets recorded from this Al coupon.

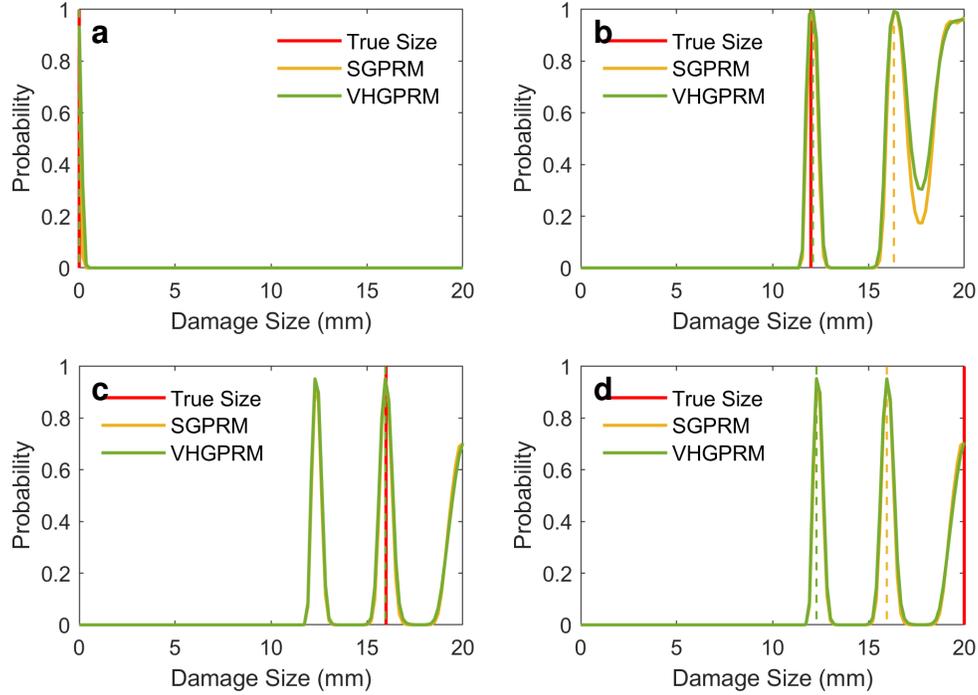
\begin{figure}[t!]
    \centering
    \begin{picture}(400,300)
    \put(0,40){\includegraphics[trim = 20 0 20 15
    ,clip,scale=0.8]{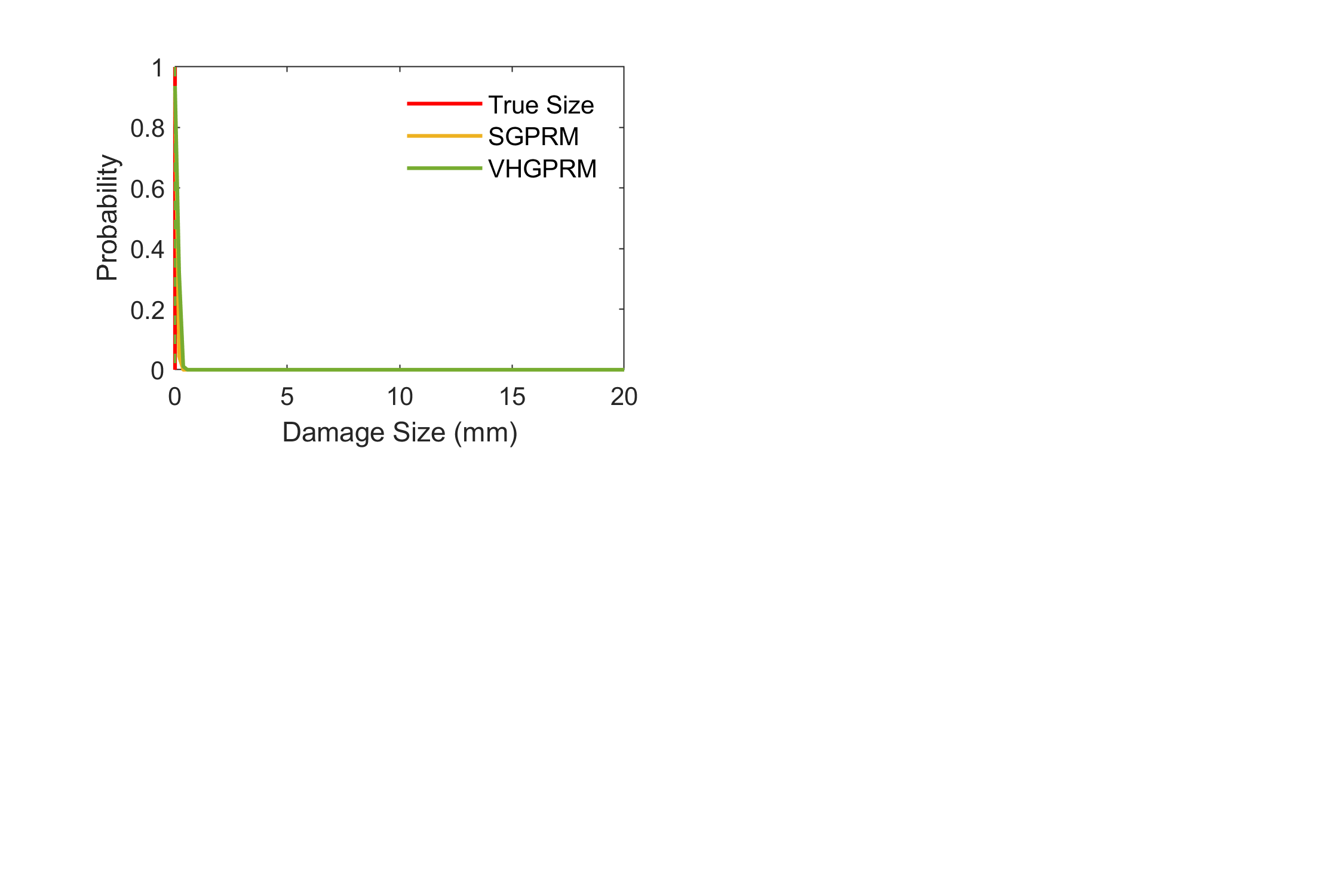}}
    \put(190,40){\includegraphics[trim = 20 0 20 15,clip,scale=0.8]{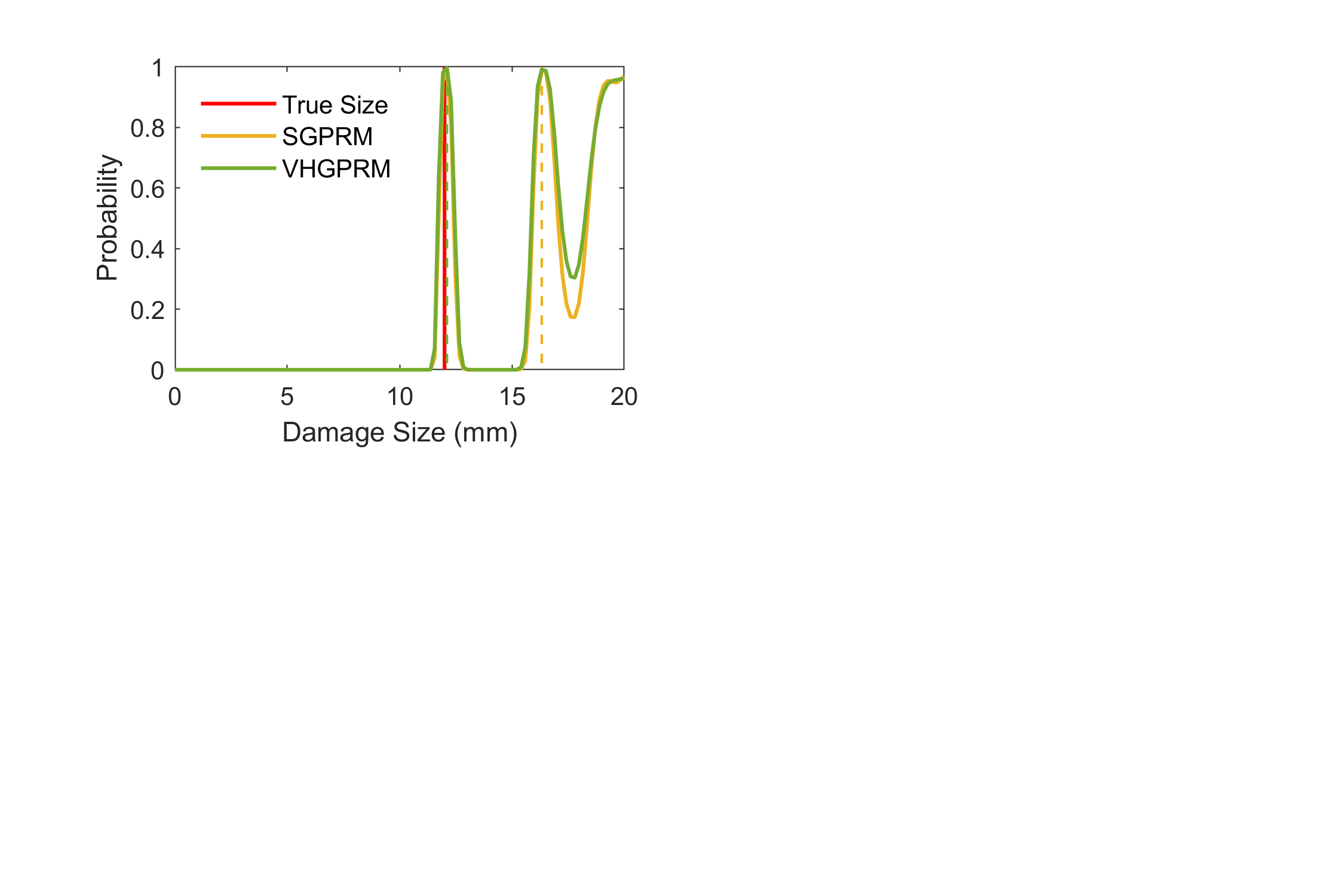}}
    \put(0,-95){\includegraphics[trim = 20 0 20 15,clip,scale=0.8]{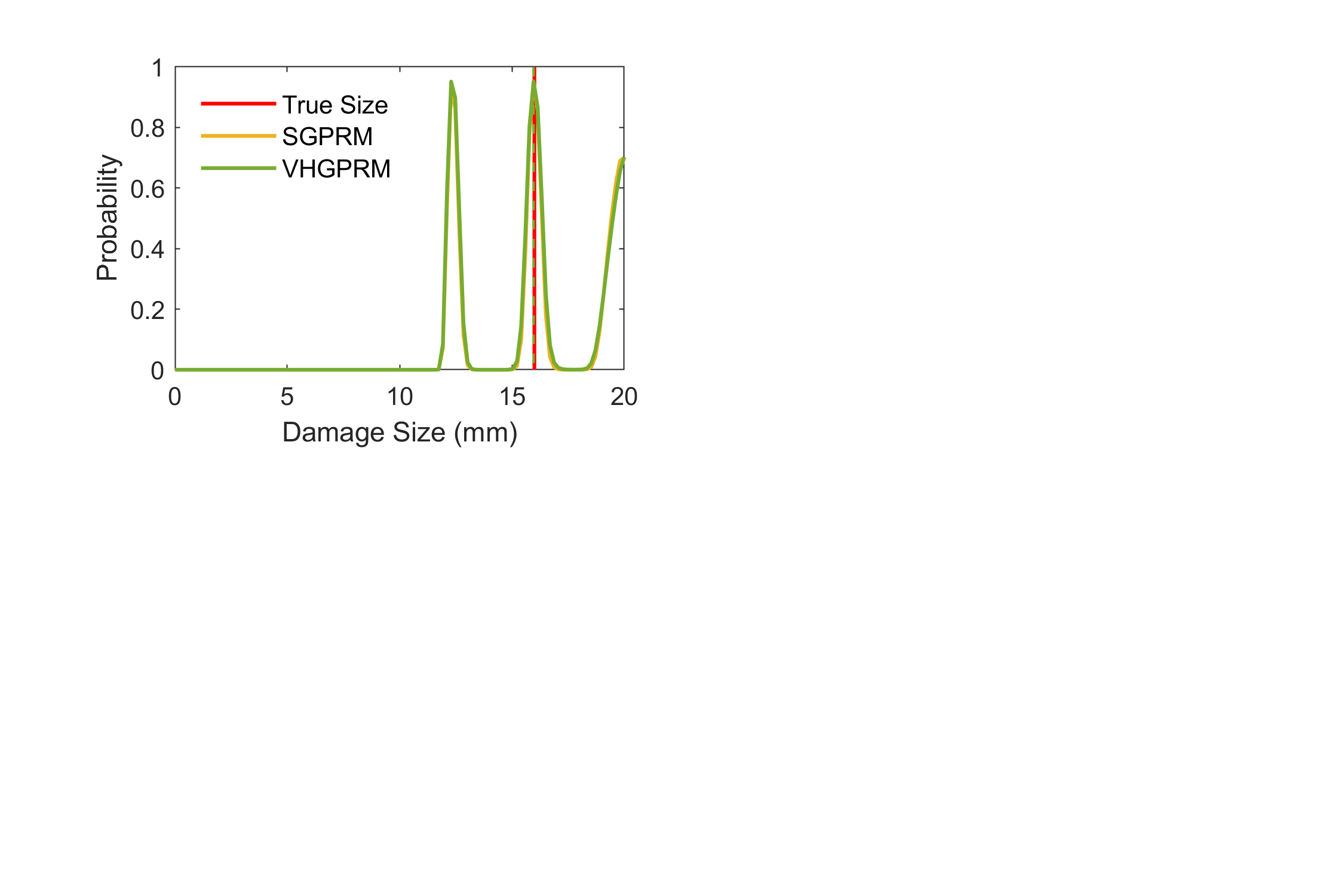}}
   \put(190,-95){\includegraphics[trim = 20 0 20 15,clip,scale=0.8]{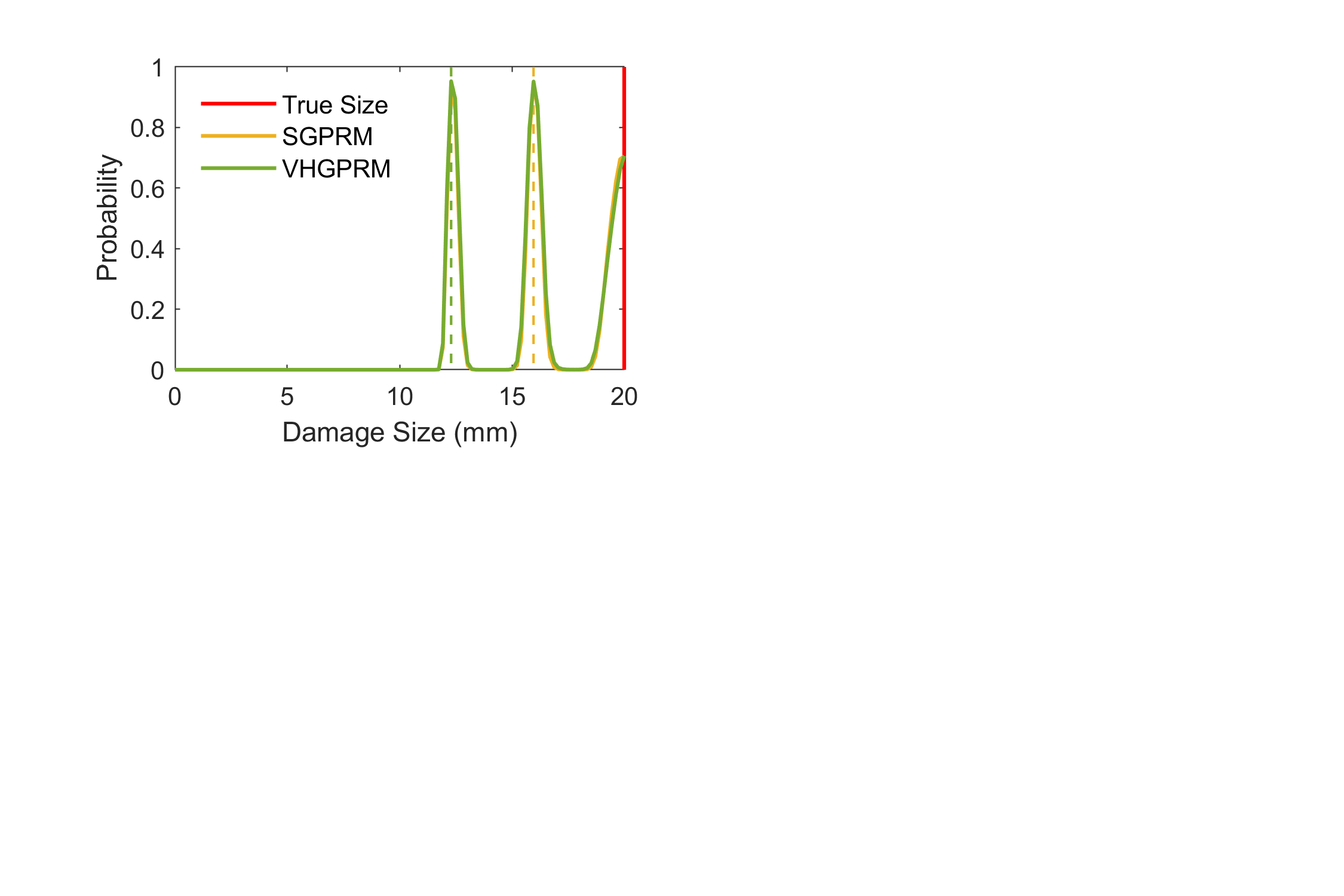}}
   \put(45,295){\color{black} \large {\fontfamily{phv}\selectfont \textbf{a}}}
    \put(235,295){\large {\fontfamily{phv}\selectfont \textbf{b}}}
   \put(45,160){\large {\fontfamily{phv}\selectfont \textbf{c}}} 
   \put(235,160){\large {\fontfamily{phv}\selectfont \textbf{d}}} 
    \end{picture} \vspace{-55pt}
    \caption{Notched Al coupon: damage size prediction results for path 3-6 based on the RMSD DI: (a) prediction probabilities for the healthy case; (b) prediction probabilities for a notch size of 12 mm; (c) prediction probabilities for a notch size of 16 mm; (d) prediction probabilities for a notch size of 20 mm. Dashed vertical lines indicate the maximum probability corresponding to each model.} 
\label{fig:notch_prob_3-6_rmsd} \vspace{0pt}
\end{figure}

In order to use the trained models for damage quantification, the DI data points not used in training were considered test points coming off the coupon with an unknown damage size. Then, following the damage size prediction methodology outlined in Section \ref{sec:prediction}, the probability of each DI test point coming from each damage state in the coupon was calculated. Figure \ref{fig:notch_prob_3-6_rmsd} shows prediction probabilities from both model types using the RMSD DI for path 3-6 for 4 indicative test points. The shown points were selected from the healthy case, as well as three damage cases where the original RMSD DI values (framed area in Figure \ref{fig:notch_signals}c) were very close just to examine whether the models have the capability to differentiate between the different cases based on the training data, and whether there exist any differences in quantification performance between SGPRMs and VHGPRMs for this coupon. As shown, although both models fail to properly predict the damage size of $20$ mm, the VHGPRM outperforms the SGPRM in quantifying the damage size of $12$ mm. In addition, both models can accurately predict the damage size of $16$ mm, as well as the healthy case. Furthermore, both models can also accurately predict almost all other damage sizes (not shown here), as will be indicated later on when analyzing the summary prediction results.

\begin{figure}[t!]
    \centering
    \begin{picture}(400,300)
    \put(0,40){\includegraphics[trim = 20 0 20 15,clip,scale=0.8]{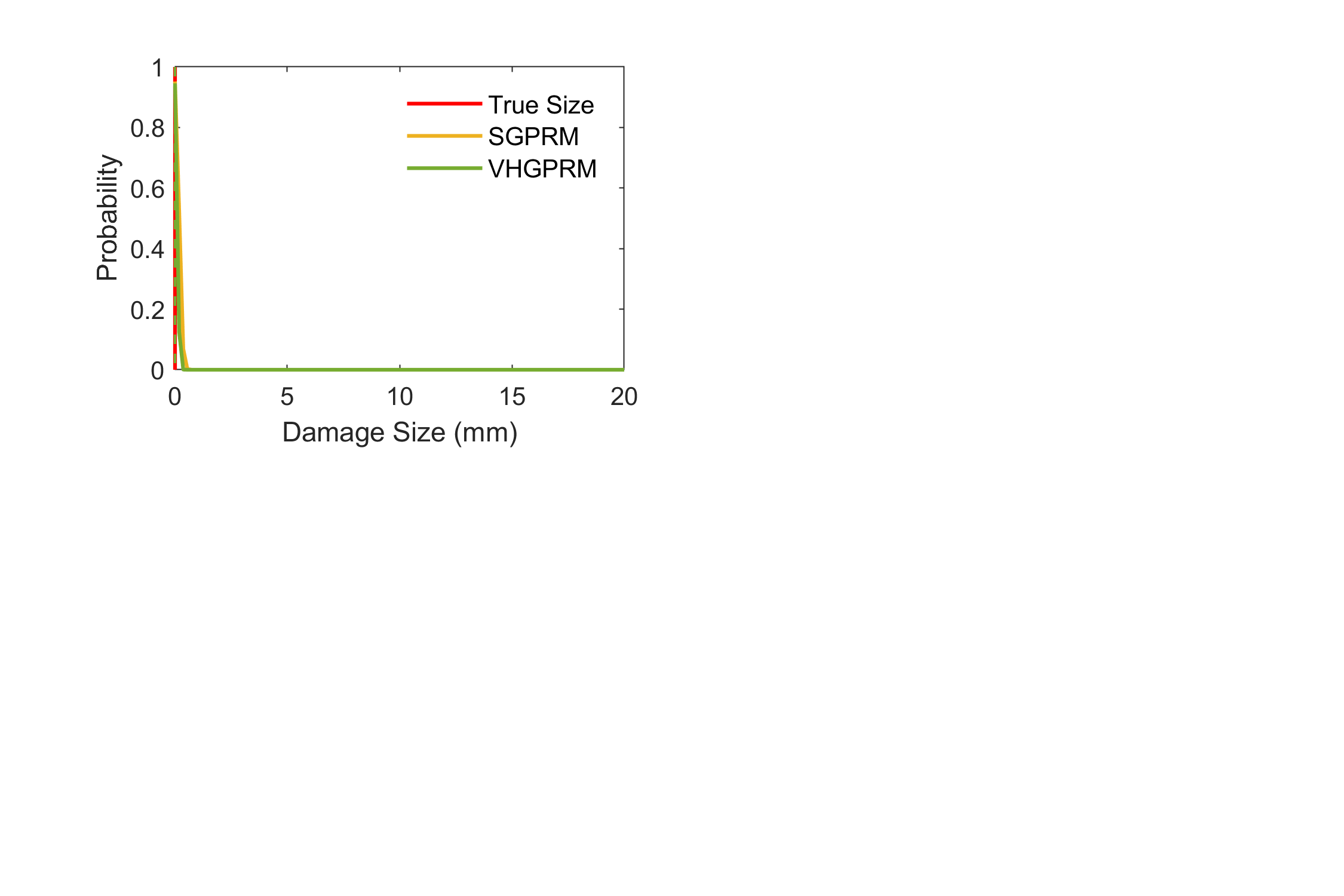}}
    \put(190,40){\includegraphics[trim = 20 0 20 15,clip,scale=0.8]{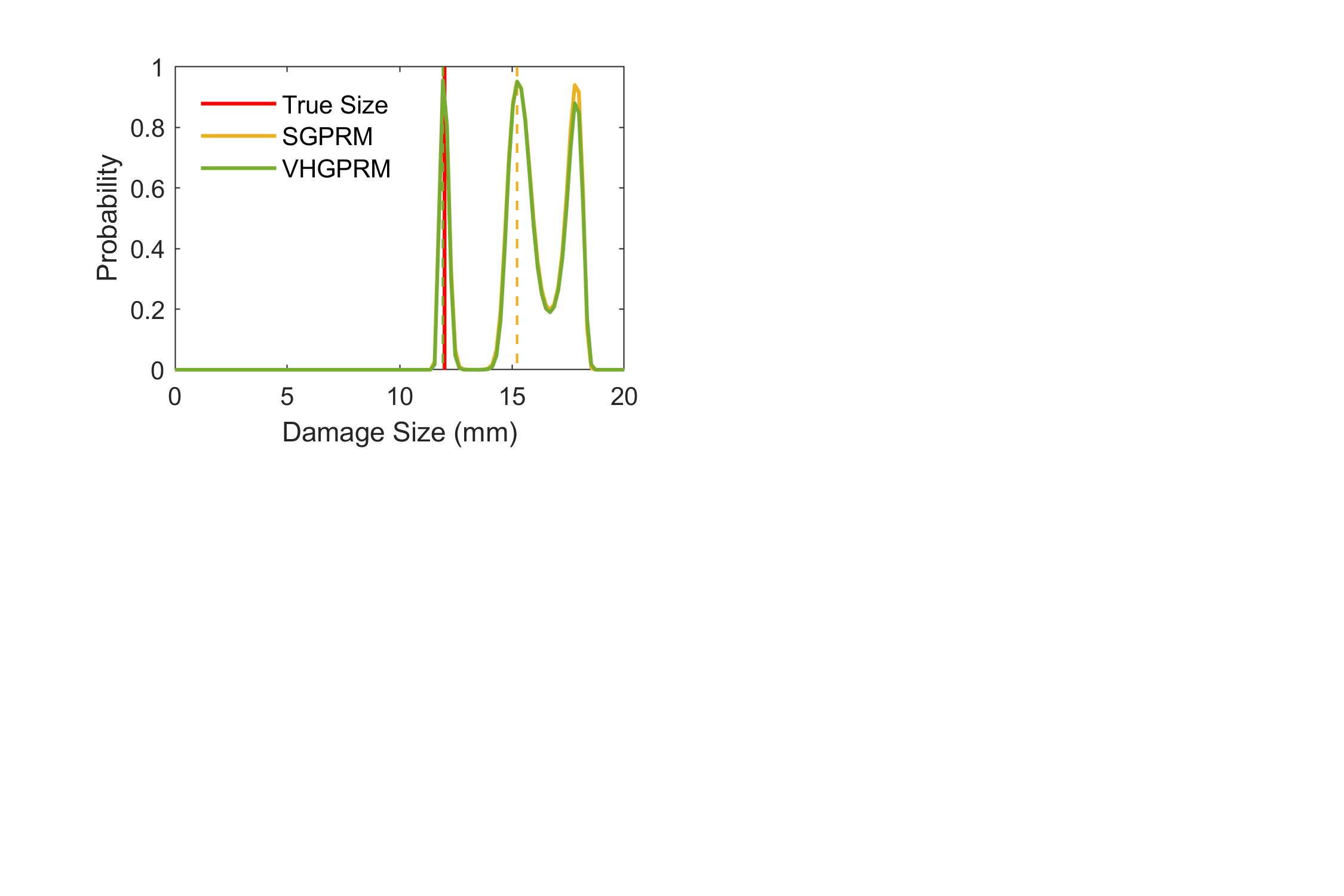}}
    \put(0,-95){\includegraphics[trim = 20 0 20 15,clip,scale=0.8]{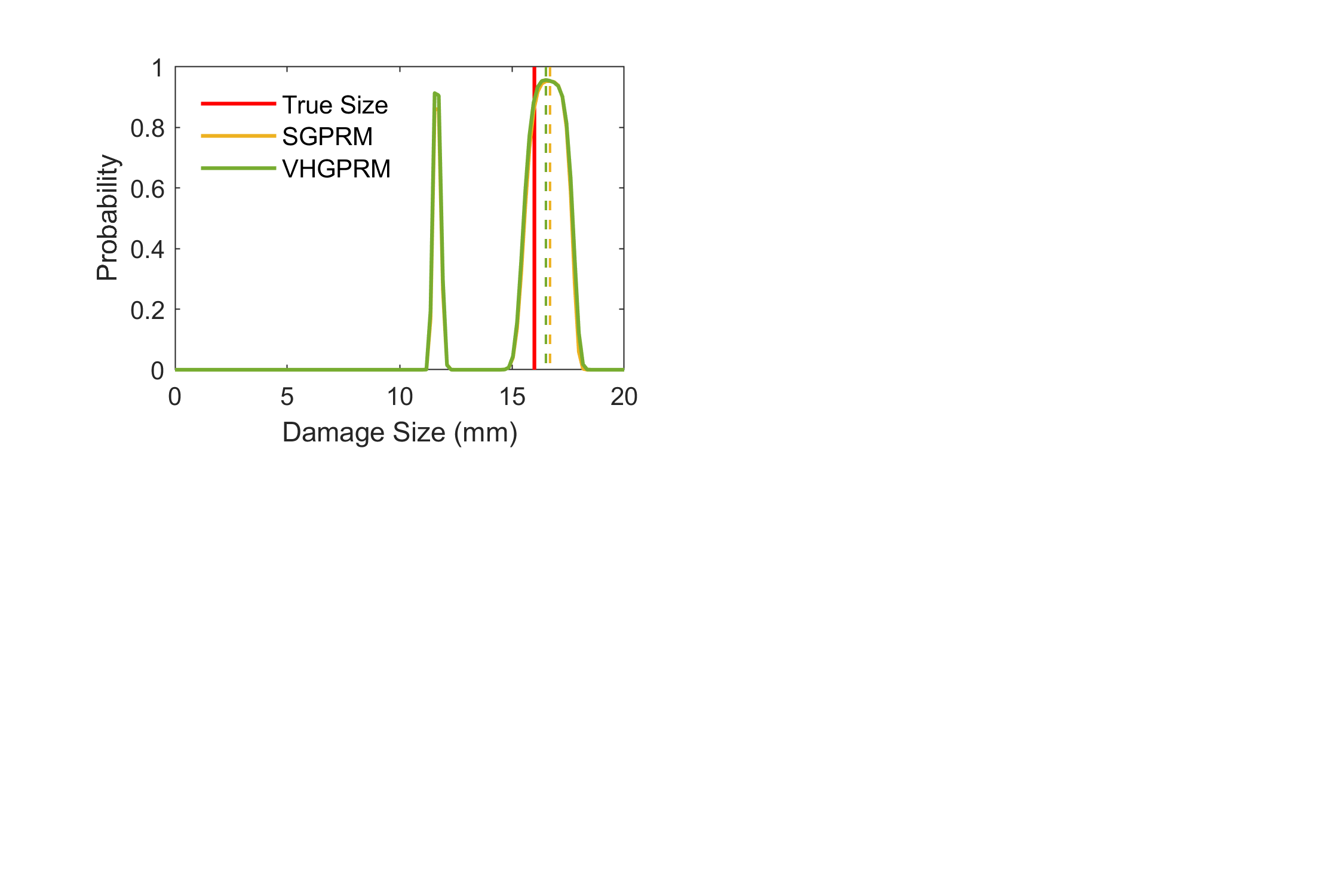}}
   \put(190,-95){\includegraphics[trim = 20 0 20 15,clip,scale=0.8]{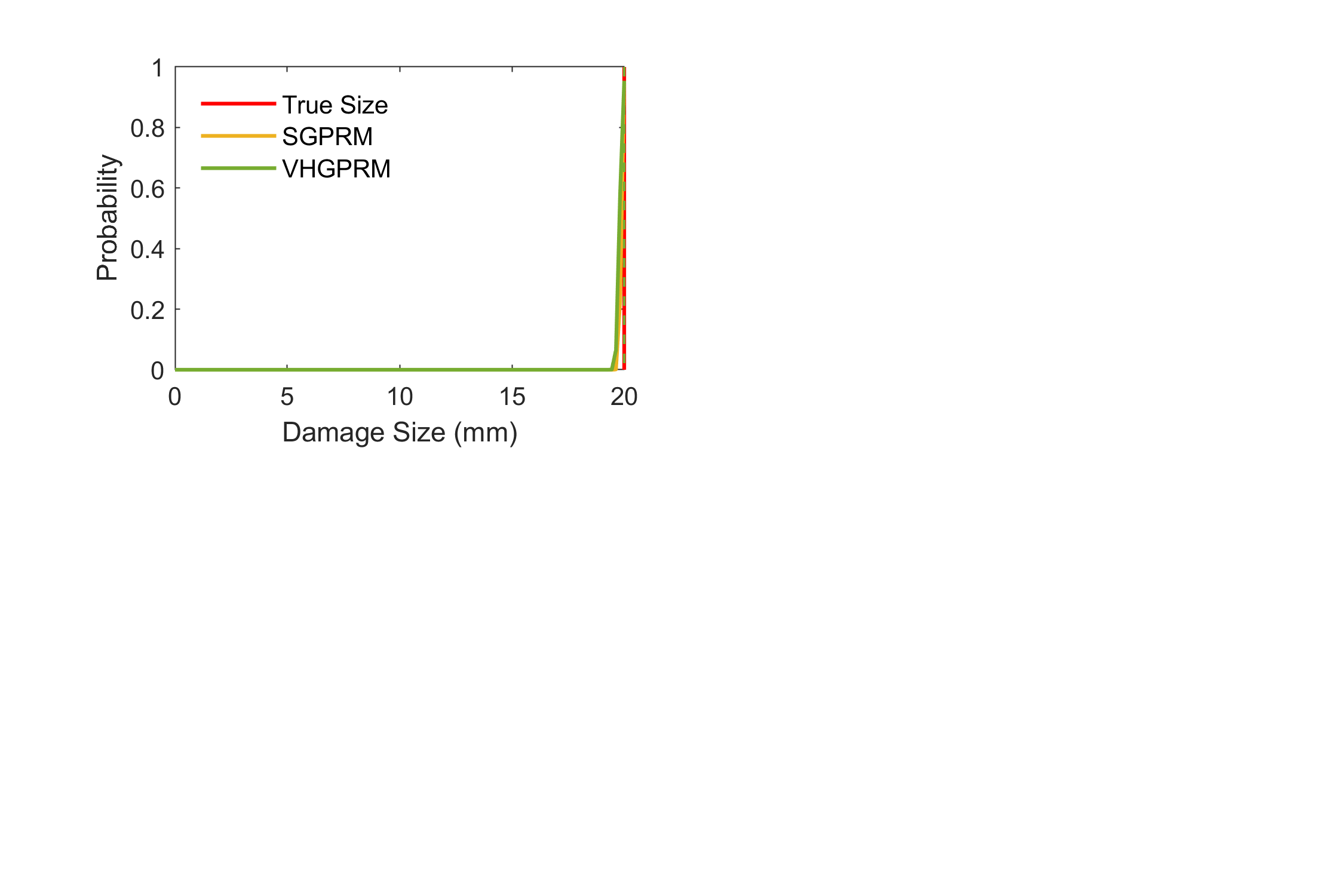}}
   \put(45,295){\color{black} \large {\fontfamily{phv}\selectfont \textbf{a}}}
    \put(235,295){\large {\fontfamily{phv}\selectfont \textbf{b}}}
   \put(45,160){\large {\fontfamily{phv}\selectfont \textbf{c}}} 
   \put(235,160){\large {\fontfamily{phv}\selectfont \textbf{d}}} 
    \end{picture} \vspace{-55pt}
    \caption{Notched Al coupon: damage size prediction results for path 3-6 based on the DI formulation from \cite{Janapati-etal16}: (a) prediction probabilities for the healthy case; (b) prediction probabilities for a notch size of 12 mm; (c) prediction probabilities for a notch size of 16 mm; (d) prediction probabilities for a notch size of 20 mm. Dashed vertical lines indicate the maximum probability corresponding to each model.} 
\label{fig:notch_prob_3-6_janapati} \vspace{10pt}
\end{figure}

Figure \ref{fig:notch_prob_3-6_janapati} shows the corresponding prediction probabilities from the GPRMs trained using the DI formulation presented in \cite{Janapati-etal16}. As shown, both models perform accurately in all selected test points except for the damage size of $12$ mm, where the VHGPRM again outperforms the SGPRM in damage quantification. The reason why the models trained using data sets from the second DI formulation perform better under a damage size of $20$ mm can be attributed to the way this DI formulation responds to that damage size, as can be seen in Figure \ref{fig:notch_signals}c and Figure \ref{fig:notch_gprm_3-6} panels c and d. As shown there, unlike the RMSD DI, the DI values from the second formulation clearly has a value relatively far from DI values at other notch sizes, thus making it easy for the GPRMs to accurately quantify this damage size. It is important to note here that the predictions shown in Figures \ref{fig:notch_prob_3-6_rmsd} and \ref{fig:notch_prob_3-6_janapati} are only predictions at 8 indicative DI points, and are not necessarily representing the prediction capability of each model across all $1000+$ test points in each DI data set. 

\begin{figure}[t!]
    \centering
    \begin{picture}(400,300)
    \put(0,40){\includegraphics[trim = 20 0 20 15,clip,scale=0.8]{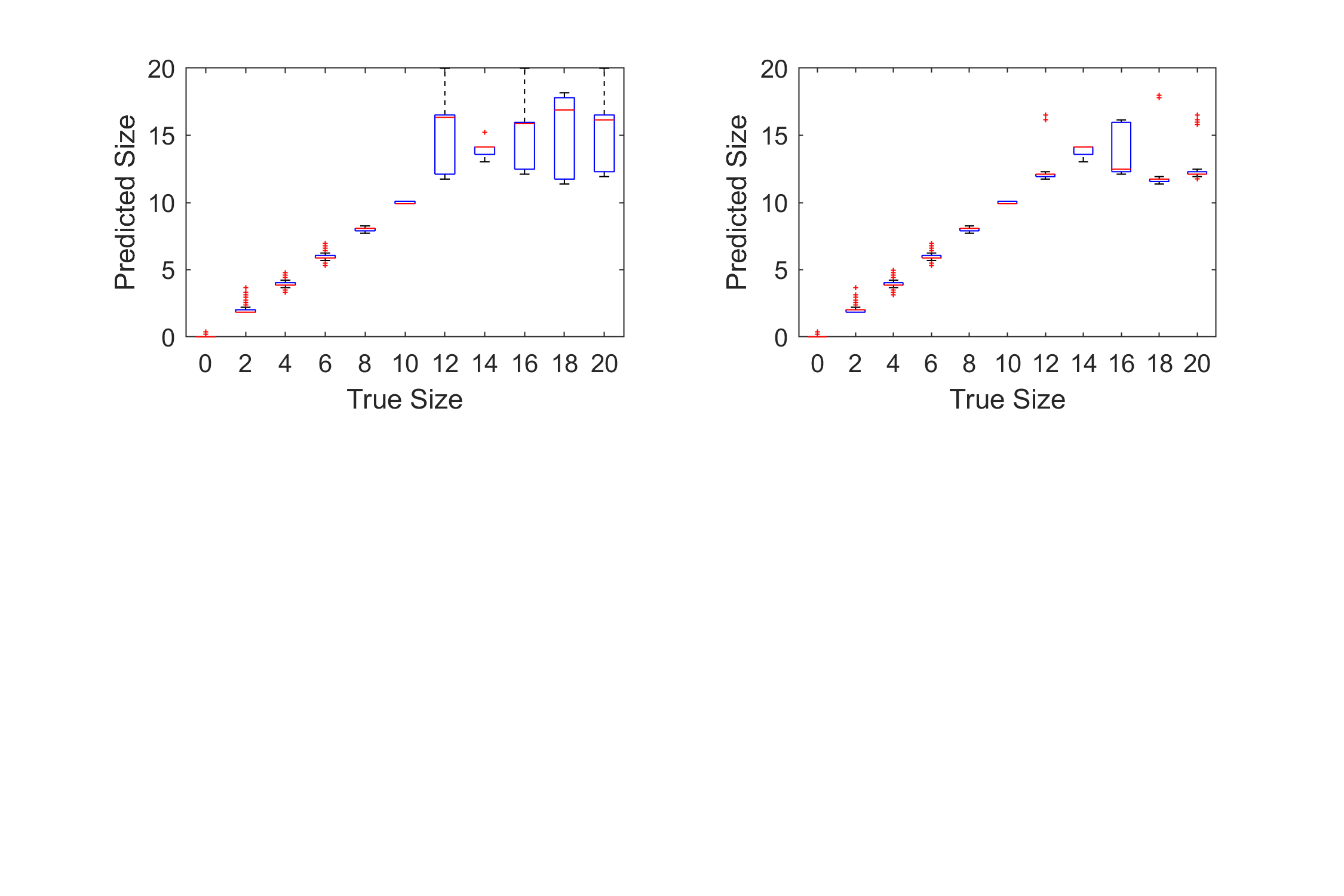}}
    \put(0,-85){\includegraphics[trim = 20 0 20 15,clip,scale=0.8]{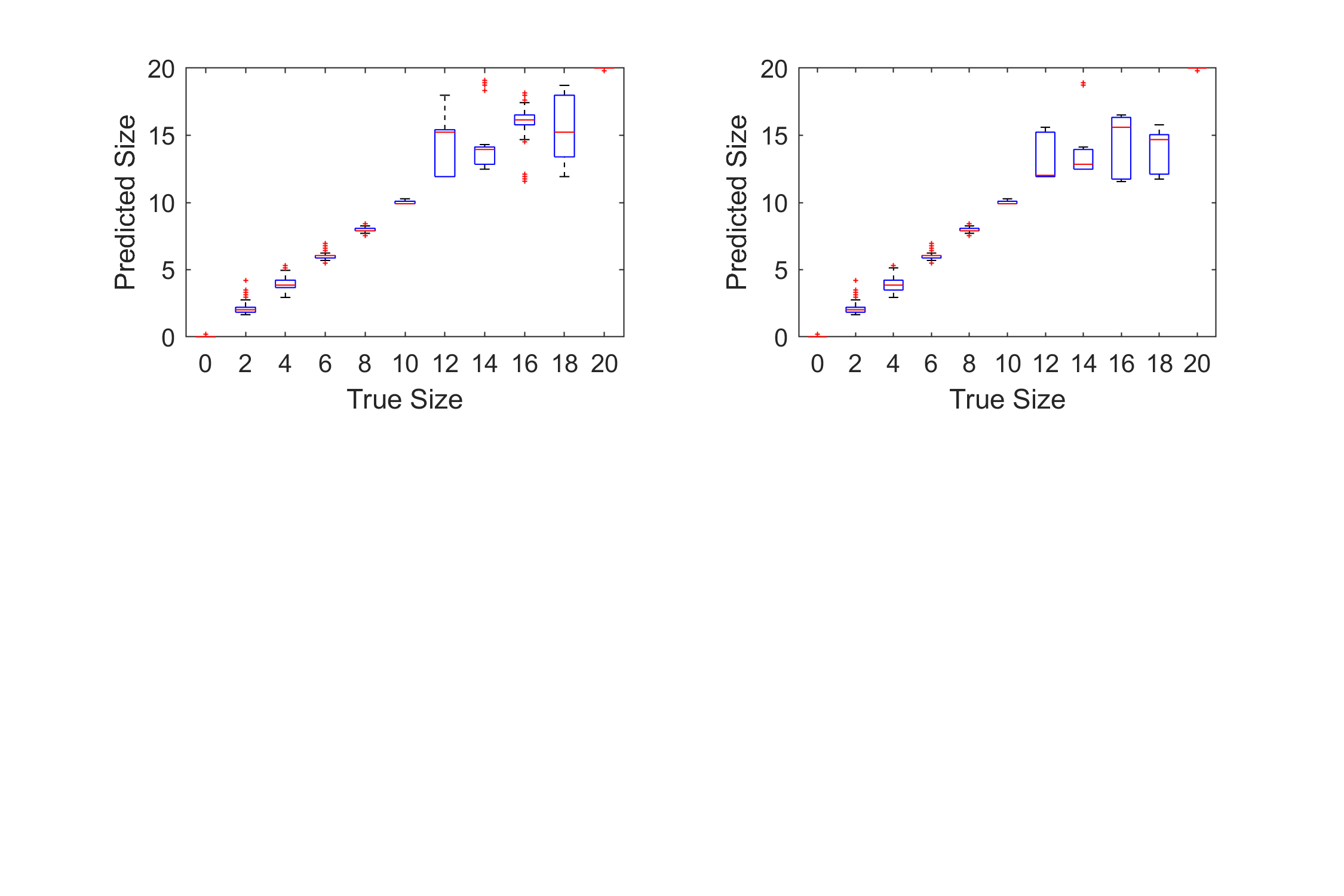}}
   \put(47,295){\color{black} \large {\fontfamily{phv}\selectfont \textbf{a}}}
    \put(245,295){\large {\fontfamily{phv}\selectfont \textbf{b}}}
   \put(47,170){\large {\fontfamily{phv}\selectfont \textbf{c}}} 
   \put(245,170){\large {\fontfamily{phv}\selectfont \textbf{d}}} 
    \end{picture} \vspace{-75pt}
    \caption{Notched Al coupon: true/predicted damage size boxplots for path 3-6: (a) SGPRM state prediction based on the RMSD DI; (b) VHGPRM state prediction based on the RMSD DI; (c) SGPRM state prediction based on the DI formulation from \cite{Janapati-etal16}; (d) VHGPRM state prediction based on the DI formulation from \cite{Janapati-etal16}.} 
\label{fig:notch_boxplot_3-6} \vspace{10pt}
\end{figure}

In order to present prediction results from all DI test points, summary box-plots were created presenting the models' predictions (point of maximum probability under each damage size) against the true damage size corresponding to each test DI point. Figure \ref{fig:notch_boxplot_3-6} panels a and b show the summary results from the RMSD DI-trained models, while panels c and d show the corresponding results for the second DI formulation. In these plots, each box contains a red horizonal line indicating the median, with the top and bottom edges of the box indicating the $75^{th}$ and $25^{th}$ percentiles, respectively. the small red crosses indicate outliers in the data. Closely examining these plots, a few interesting observations come up. The first one is that the VHGPRM based on the RMSD DI values (Figure \ref{fig:notch_boxplot_3-6}b) outperforms the corresponding SGPRM (Figure \ref{fig:notch_boxplot_3-6}a) in the cases of $12$ and $14$ mm, and shows a narrower prediction window in the case of $16$ mm. Conversely, the SGPRM shows a median closer to the true notch size in the case of an $18$-mm notch. Another observation pertaining to the GPRMs based on the DI formulation from \cite{Janapati-etal16} is that the VHGPRM again shows a more accurate prediction at the notch size of $12$ mm, as indicated by the median predictions in Figure \ref{fig:notch_boxplot_3-6} panels c and d. finally, it seems that the SGPRM generally outperforms the VHGPRM in predicting damage sizes above $12$ mm, where either most of the predictions are confined around the true damage size (notch size of $16$ mm), or are spread out with the median prediction very close to the true notch size (notch size of $14$ mm), or the range of predictions it self (between the $25^{th}$ and $75^{th}$ percentiles) actually encompasses the true damage size (notch size of $18$ mm).

\begin{figure}[t!]
    \centering
    \begin{picture}(400,300)
    \put(0,40){\includegraphics[trim = 20 0 20 15
    ,clip,scale=0.8]{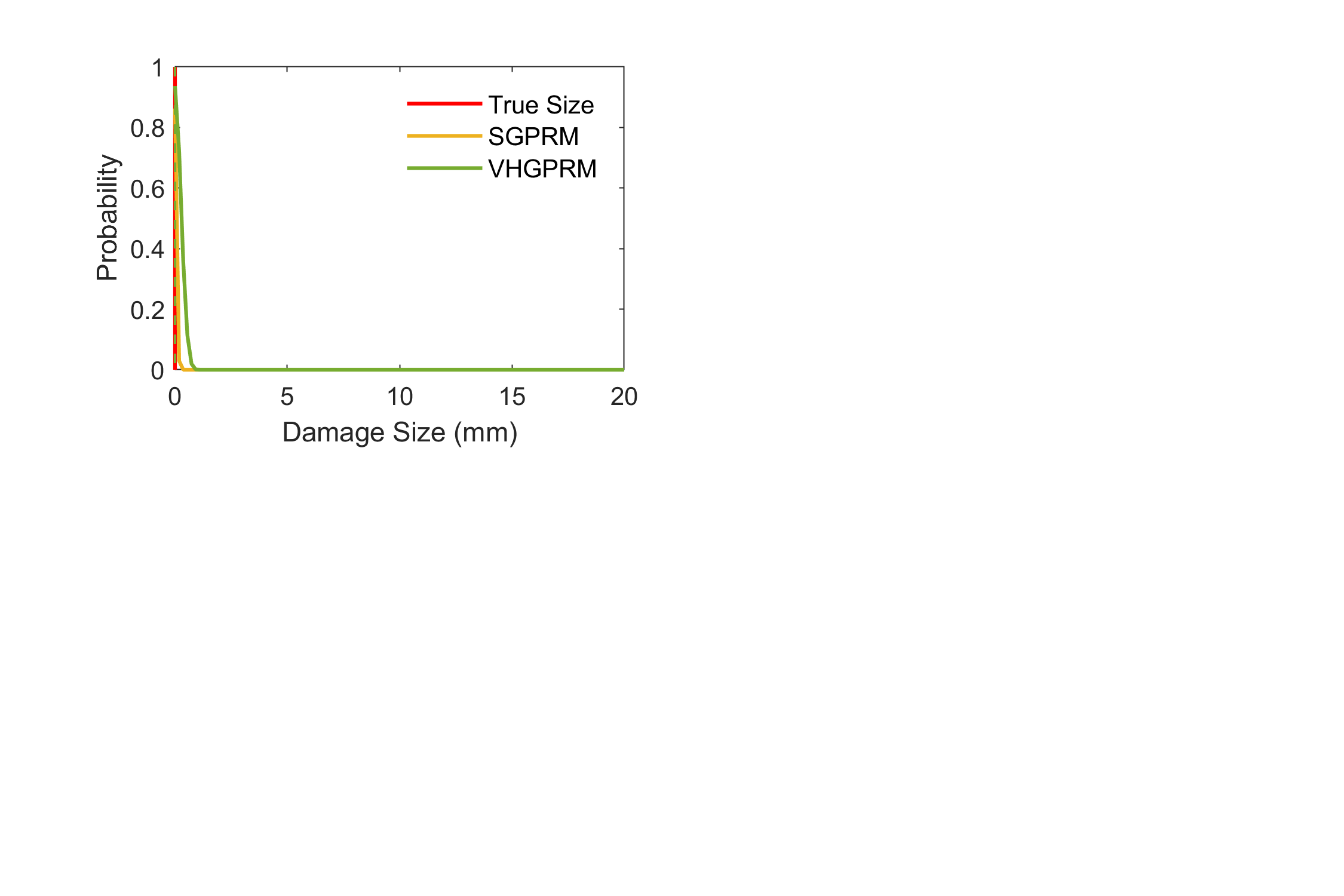}}
    \put(190,40){\includegraphics[trim = 20 0 20 15,clip,scale=0.8]{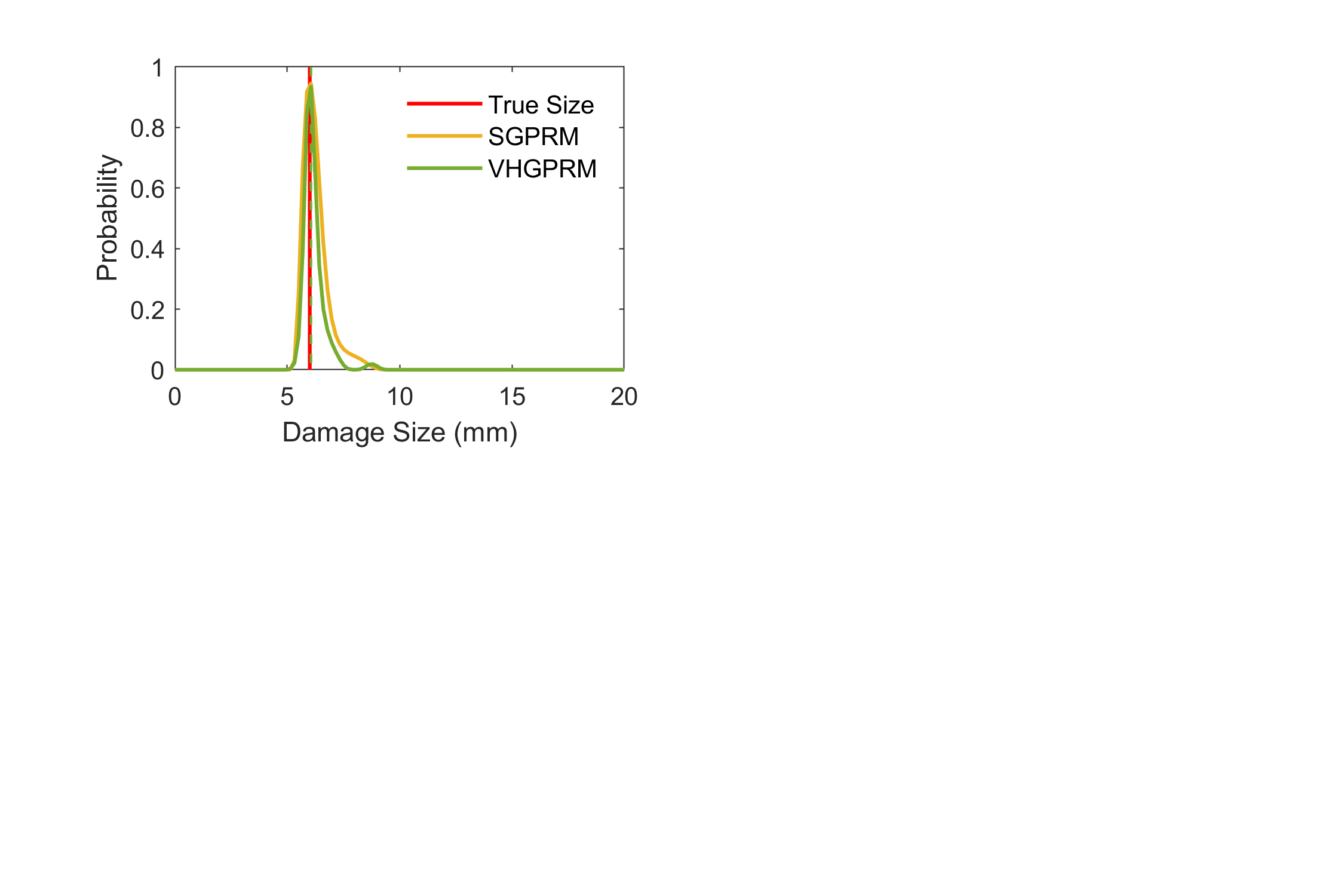}}
    \put(0,-95){\includegraphics[trim = 20 0 20 15,clip,scale=0.8]{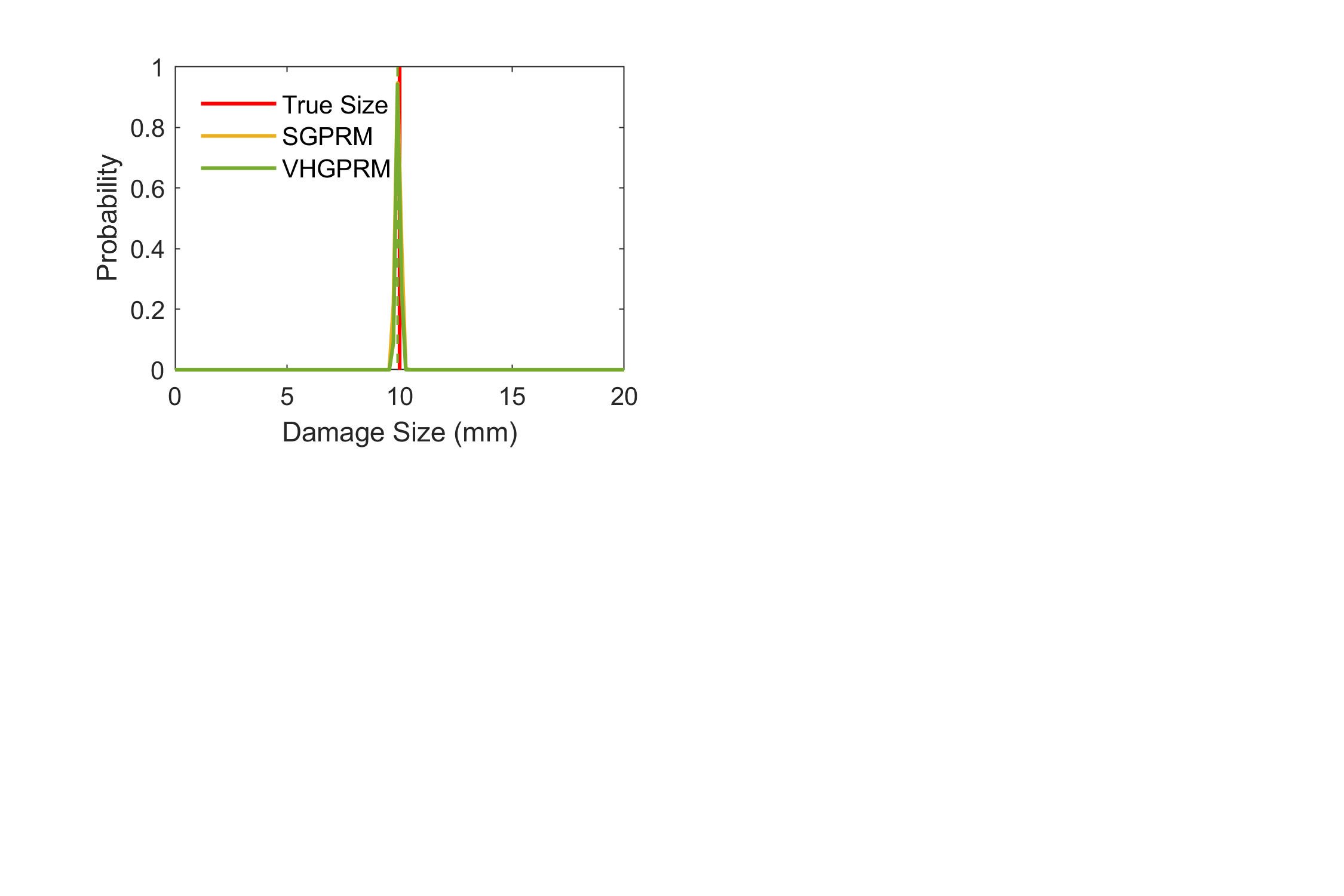}}
   \put(190,-95){\includegraphics[trim = 20 0 20 15,clip,scale=0.8]{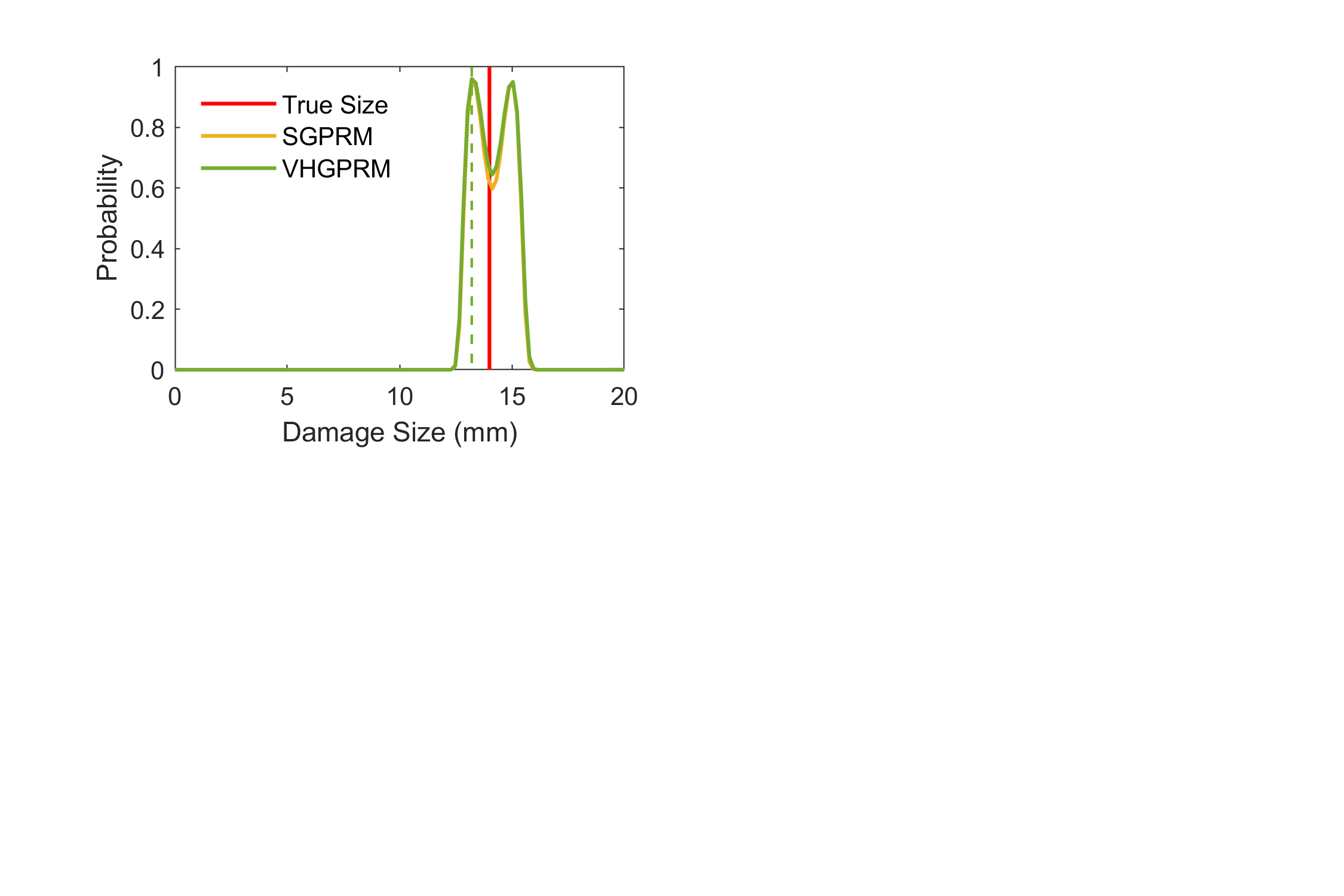}}
   \put(45,295){\color{black} \large {\fontfamily{phv}\selectfont \textbf{a}}}
    \put(235,295){\large {\fontfamily{phv}\selectfont \textbf{b}}}
   \put(45,160){\large {\fontfamily{phv}\selectfont \textbf{c}}} 
   \put(235,160){\large {\fontfamily{phv}\selectfont \textbf{d}}} 
    \end{picture} \vspace{-55pt}
    \caption{Notched Al coupon: damage size prediction results for path 6-1 based on the RMSD DI: (a) prediction probabilities for the healthy case; (b) prediction probabilities for a notch size of $6$ mm; (c) prediction probabilities for a notch size of $10$ mm; (d) prediction probabilities for a notch size of $14$ mm. Dashed vertical lines indicate the maximum probability corresponding to each model.} 
\label{fig:notch_prob_6-1_rmsd} \vspace{0pt}
\end{figure}

\begin{figure}[t!]
    \centering
    \begin{picture}(400,300)
    \put(0,40){\includegraphics[trim = 20 0 20 15
    ,clip,scale=0.8]{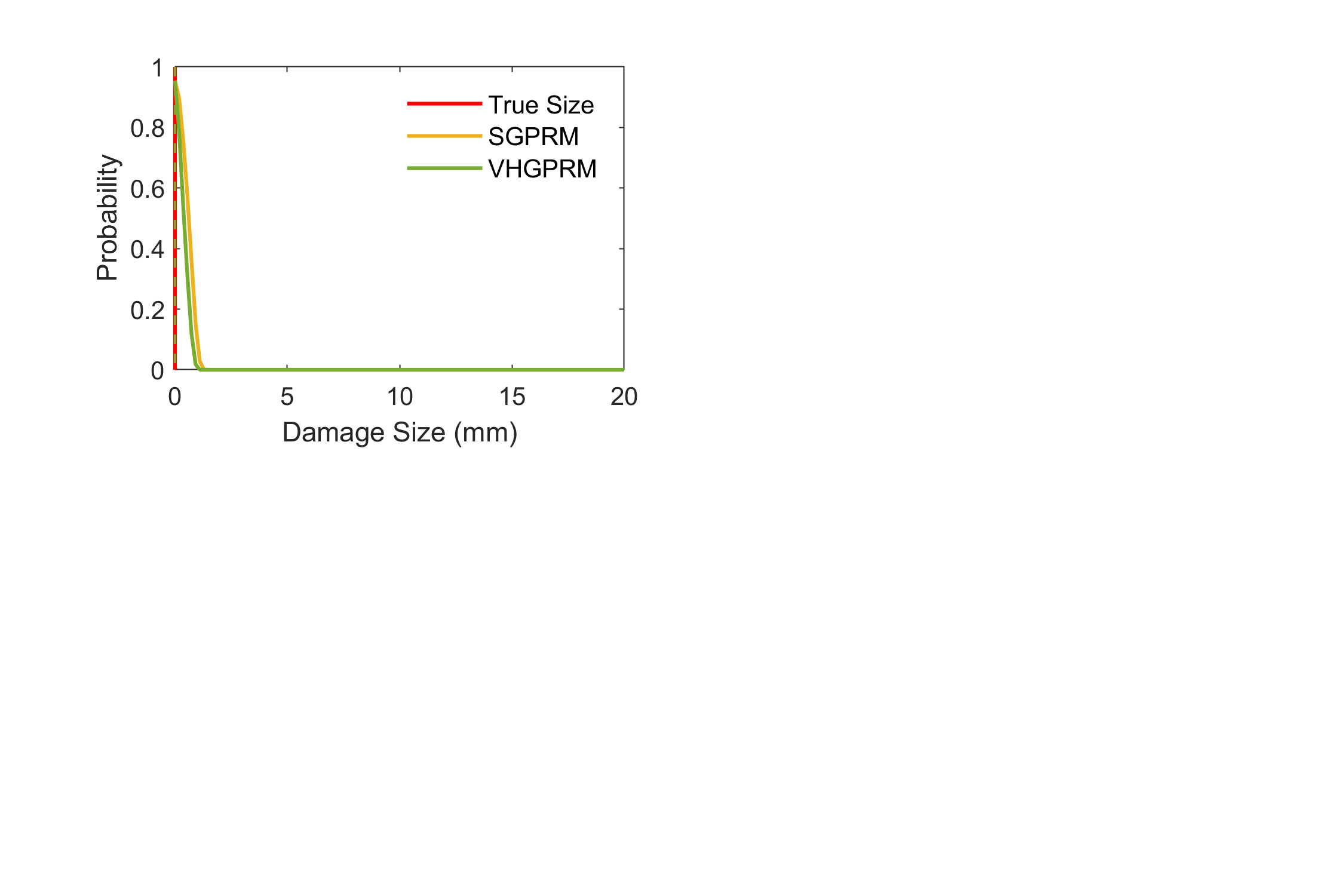}}
    \put(190,40){\includegraphics[trim = 20 0 20 15,clip,scale=0.8]{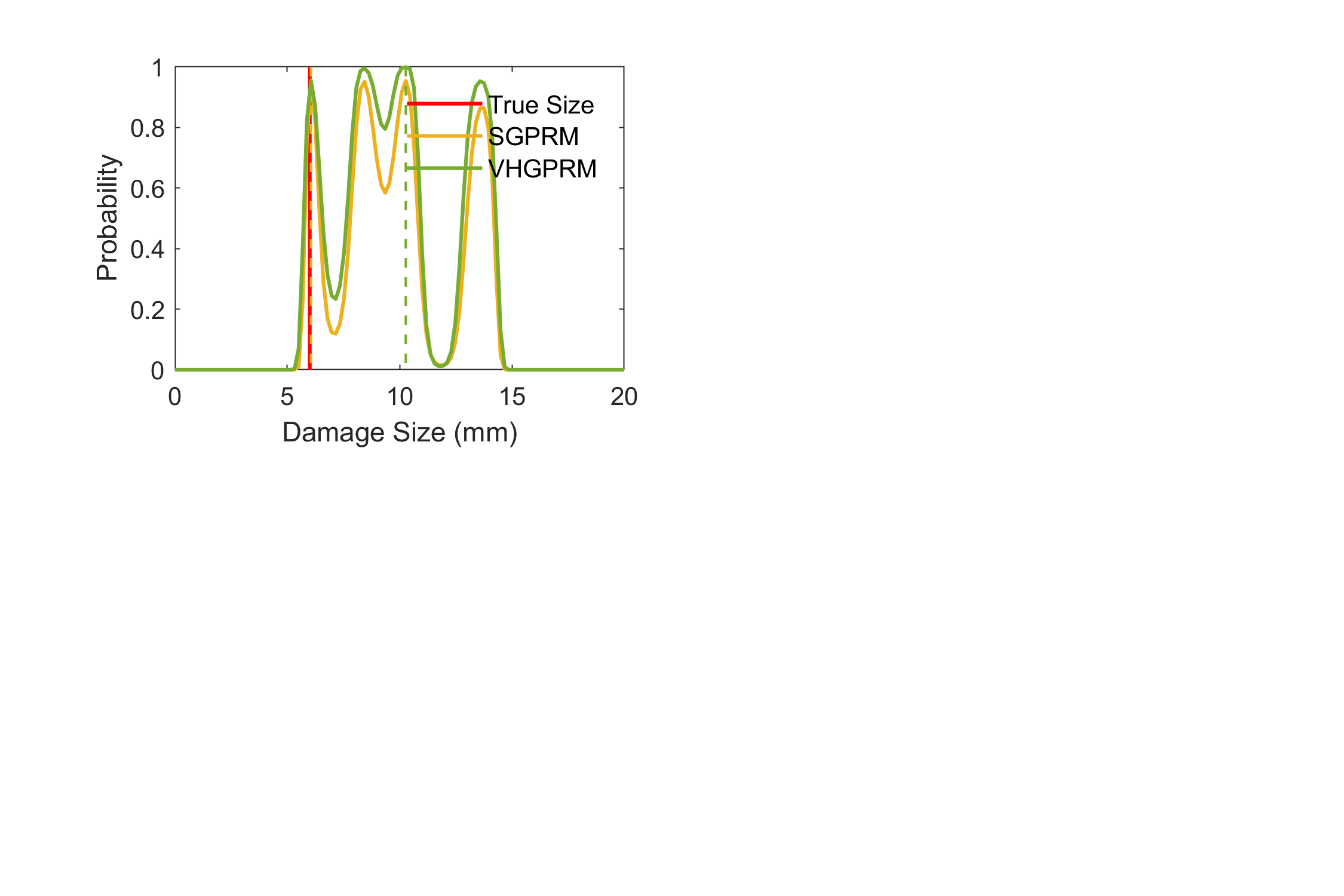}}
    \put(0,-95){\includegraphics[trim = 20 0 20 15,clip,scale=0.8]{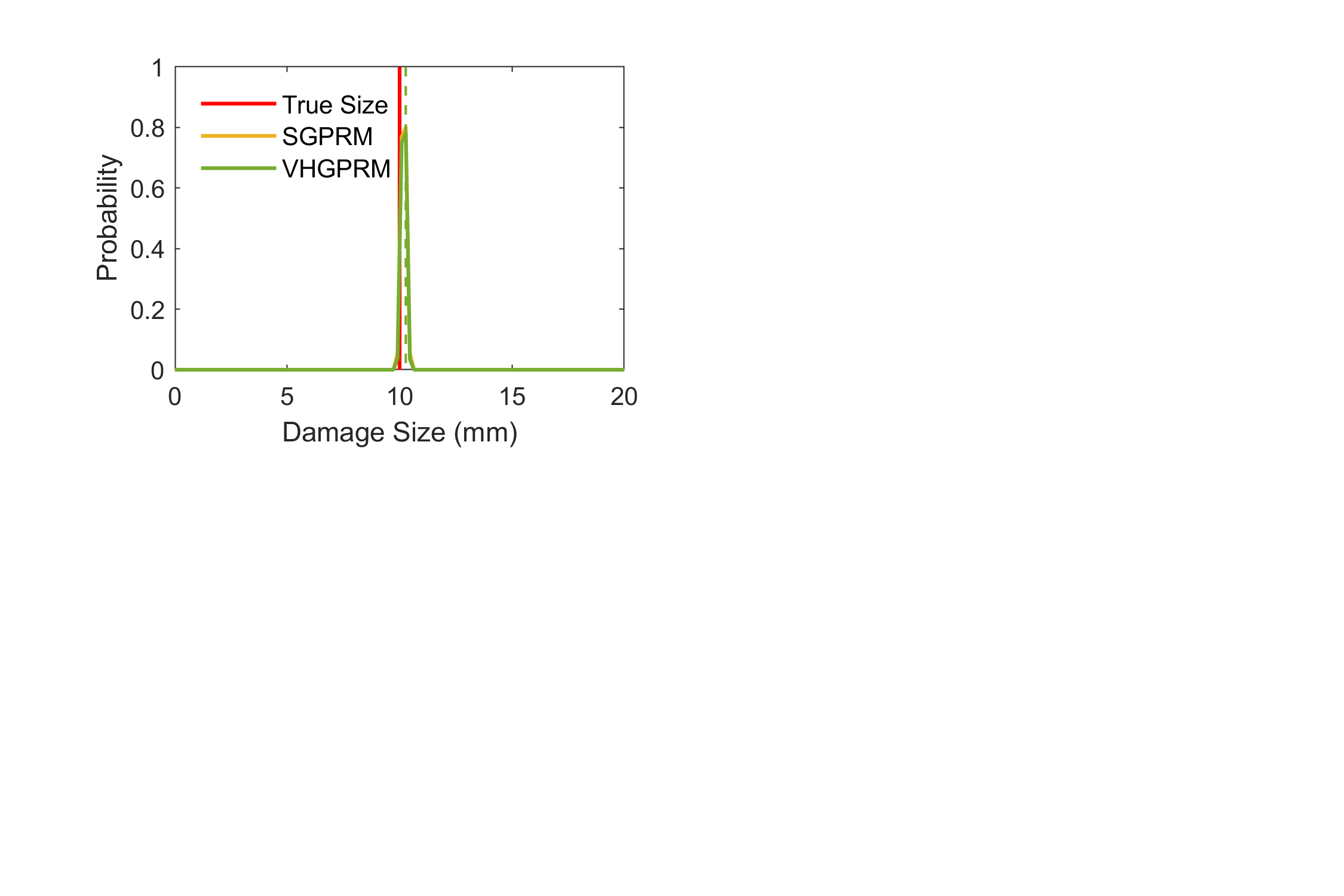}}
   \put(190,-95){\includegraphics[trim = 20 0 20 15,clip,scale=0.8]{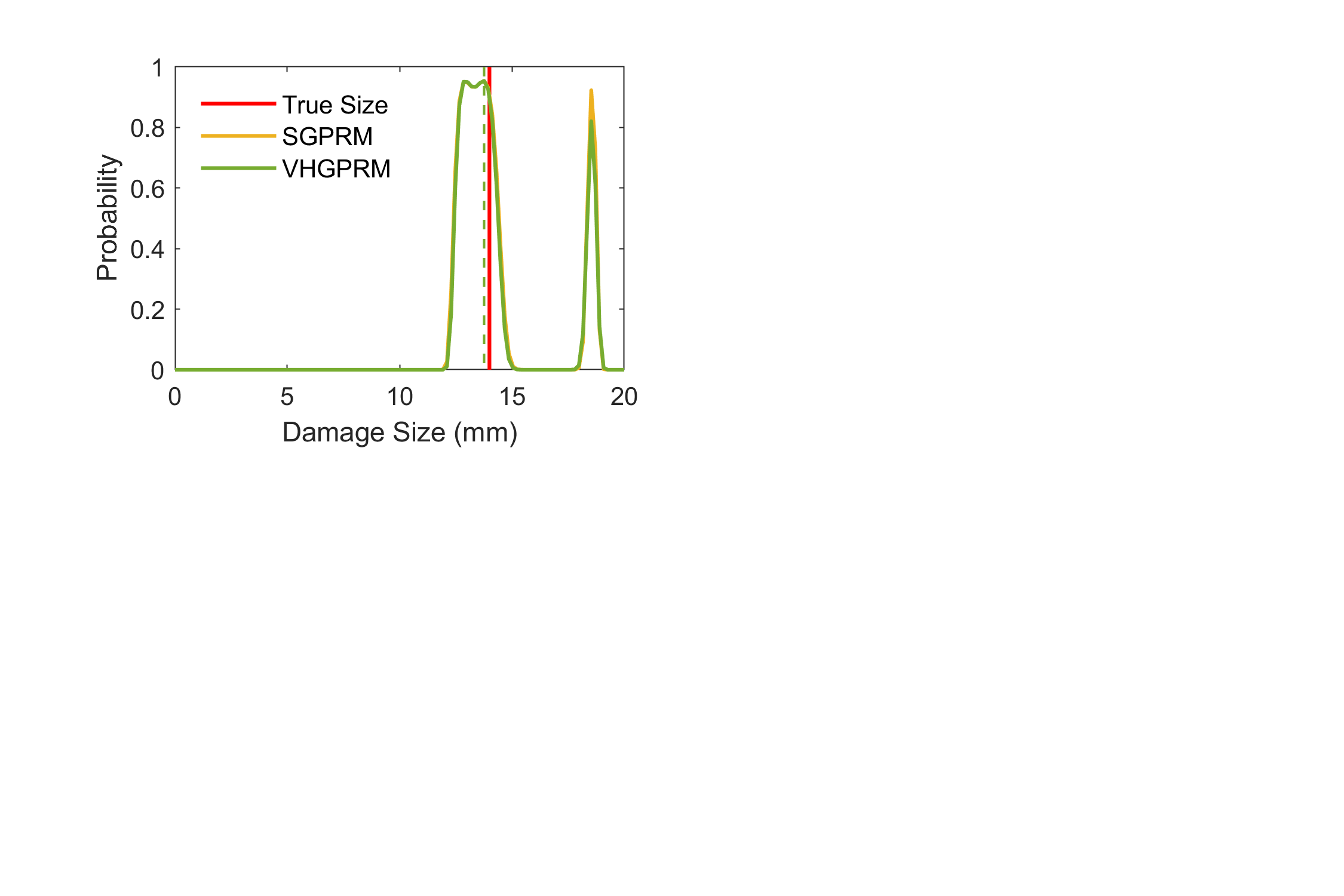}}
   \put(45,295){\color{black} \large {\fontfamily{phv}\selectfont \textbf{a}}}
    \put(235,295){\large {\fontfamily{phv}\selectfont \textbf{b}}}
   \put(45,160){\large {\fontfamily{phv}\selectfont \textbf{c}}} 
   \put(235,160){\large {\fontfamily{phv}\selectfont \textbf{d}}} 
    \end{picture} \vspace{-55pt}
    \caption{Notched Al coupon: damage size prediction results for path 6-1 based on the DI formulation in \cite{Janapati-etal16}: (a) prediction probabilities for the healthy case; (b) prediction probabilities for a notch size of $6$ mm; (c) prediction probabilities for a notch size of $10$ mm; (d) prediction probabilities for a notch size of $14$ mm. Dashed vertical lines indicate the maximum probability corresponding to each model.} 
\label{fig:notch_prob_6-1_janapati} \vspace{10pt}
\end{figure}

Figure \ref{fig:notch_prob_6-1_rmsd} shows 4 indicative predictions by the GPRMs trained using the RMSD DI data set from path 6-1. Next to the healthy case, three other points were selected based on their proximity in DI value again to test the model capability of differentiating between close test points coming from different states of the component. Coincidentally, because the development of the RMSD DI in the case of path 6-1 in this Al coupon is more or less monotonic, the three points were selected based on the prediction results of the GPRMs trained with the second DI formulation used in this study. As shown in all panels of Figure \ref{fig:notch_prob_6-1_rmsd}, both models accurately predict damage size, with a slight shift in the predictions of both models for the case of a $14$-mm notch. Examining the corresponding results from the GPRMs trained with the other DI formulation (Figure \ref{fig:notch_prob_6-1_janapati}), one can observe that both models again perform well except in the case of the selected DI test point at a notch size of $6$ mm, where the SGPRM outperforms the VHGPRM in prediction. All in all, as expected, the RMSD DI-trained GPRMs come on top compared to the GPRMs trained with the other DI formulation due to the nice evolution of the DI values with notch size. As with path 3-6, however, conclusions can only be withdrawn after examining the summary prediction results from each model.

\begin{figure}[t!]
    \centering
    \begin{picture}(400,300)
    \put(0,40){\includegraphics[trim = 20 0 20 15,clip,scale=0.8]{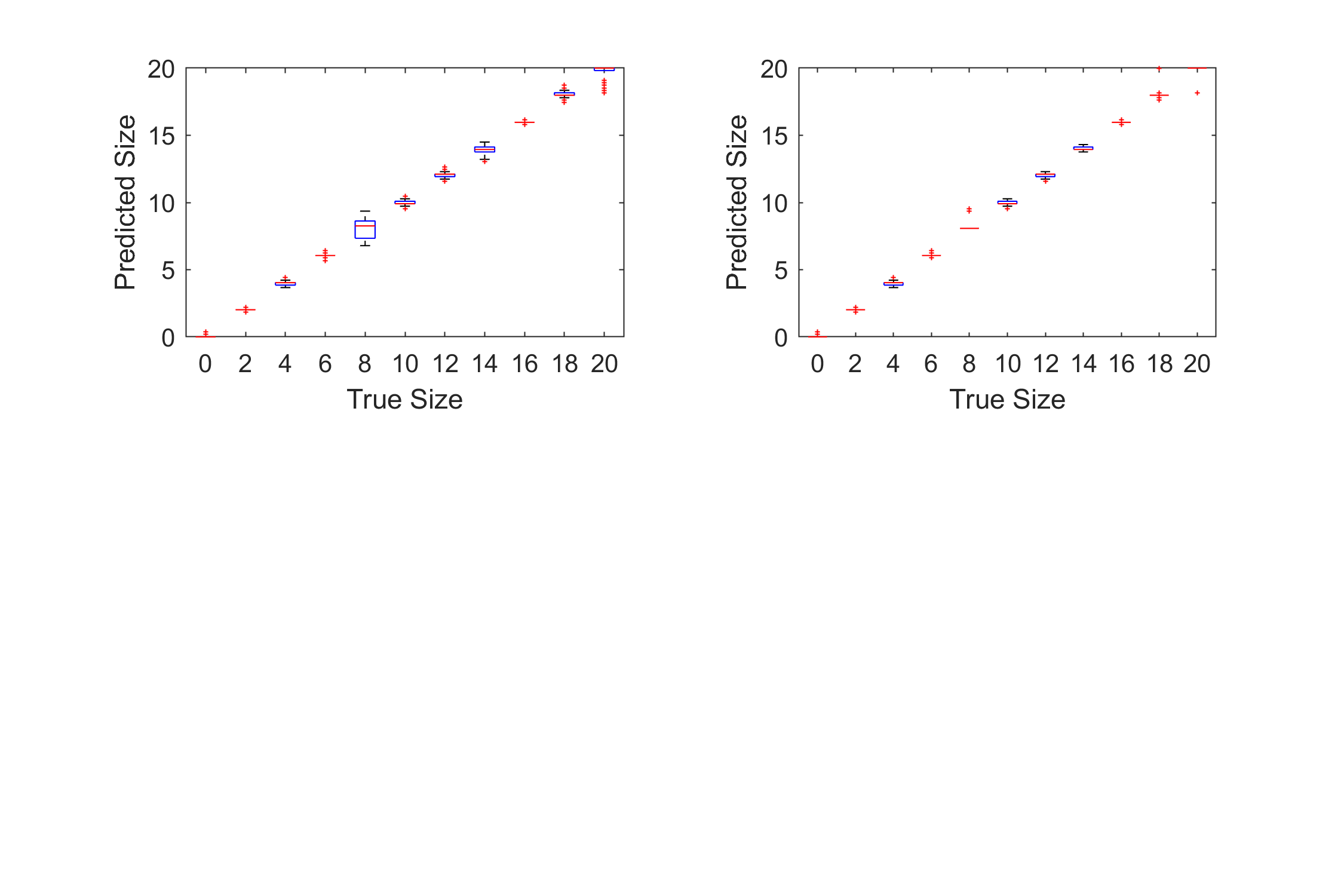}}
    \put(0,-85){\includegraphics[trim = 20 0 20 15,clip,scale=0.8]{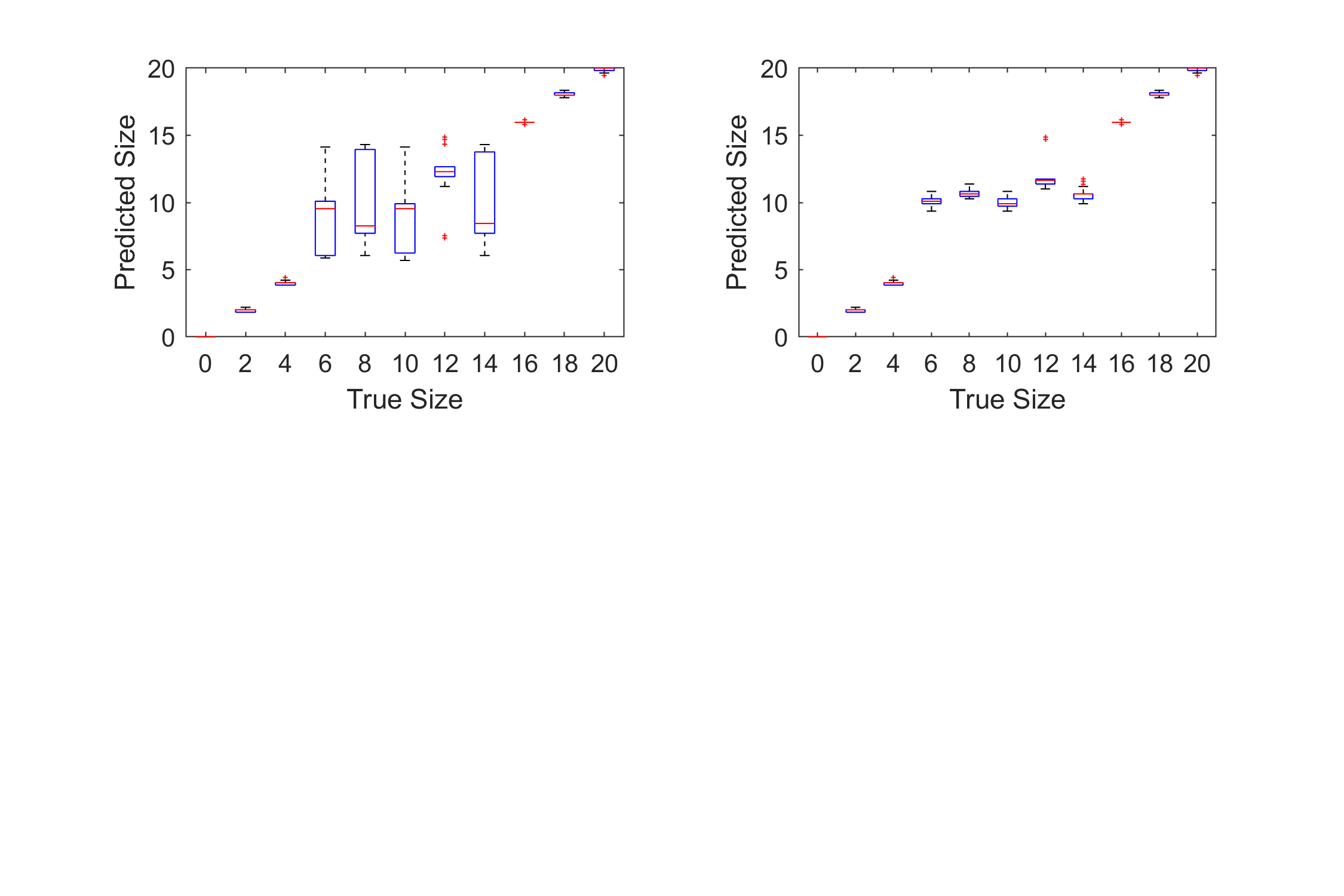}}
   \put(47,295){\color{black} \large {\fontfamily{phv}\selectfont \textbf{a}}}
    \put(245,295){\large {\fontfamily{phv}\selectfont \textbf{b}}}
   \put(47,170){\large {\fontfamily{phv}\selectfont \textbf{c}}} 
   \put(245,170){\large {\fontfamily{phv}\selectfont \textbf{d}}} 
    \end{picture} \vspace{-75pt}
    \caption{Notched Al coupon: true/predicted damage size box-plots for path 6-1: (a) SGPRM state prediction based on the RMSD DI; (b) VHGPRM state prediction based on the RMSD DI; (c) SGPRM state prediction based on the DI formulation from \cite{Janapati-etal16}; (d) VHGPRM state prediction based on the DI formulation from \cite{Janapati-etal16}.} 
\label{fig:notch_boxplot_6-1} \vspace{0pt}
\end{figure}

Figure \ref{fig:notch_boxplot_6-1} panels a and b present the summary prediction results for the rmsd DI-trained GPRMs for the standard and heteroscedastic models, respectively. As shown, overall, VHGPRMs show sharper predictions compared to SGPRMs. This phenomenon would generally result in sharper prediction probabilities, which has been observed for some test DI points (not shown here for brevity). Nonetheless, both models provide accurate predictions for each notch size. For the models trained using the second DI formulation (Figure \ref{fig:notch_boxplot_6-1} panels c and d), it can be observed that both model structures have trouble quantifying notch size between $6$ mm and $14$ mm, which can be expected given the saturation occurring in that range of notch sizes in the training DI values. This being said, a closer examination of the box-plots shows that the trained VHGPRM seems to give better damage size predictions at a notch size of $10$ and $12$ mm compared to the SGPRM, while the latter gives predictions that barely encompass the true notch sizes. In contrast, the broad box-plots shown in Figure \ref{fig:notch_boxplot_6-1}c at notch sizes of $6$, $8$, and $14$ mm do cover the true damage size, while the VHGPRM predictions completely miss the true state.

From the analysis of both paths presented herein, the following can concluded:
\begin{itemize}
\item GPRMs can accurately model both formulations of the DIs without the need for user experience. this can be seen in the relatively-small training and testing times required for GPRM training and validation processes.
\item Damage state prediction probabilities, extracted from model-based confidence bounds, give a richer representation of the state of the system, compared to the damage/no-damage paradigm of DIs.
\item Both variations of GPRMs used herein perform well in the damage quantification task in this simple Al coupon. This can be observed in the summary box-plots.
\item Neither the SGPRM nor the VHGPRM completely outperforms the other model in damage size quantification in the coupon presented herein.
\item The prediction performance of DI-trained GPRMs greatly depends on the evolution/trend of the DI values with damage size. This was evident in the case of path 6-1, where the RMSD DI-trained models performed almost perfectly in damage size quantification compared to the models trained with the other DI formulation, just because of the way the RMSD DI trend evolves with damage size.
\end{itemize}

For the sake of brevity, the analysis results for the remaining coupons presented in this study would be based on the DI formulation from \cite{Janapati-etal16}, and this formulation would be denoted hereon as the DI.


\section{Test Case II: CFRP Coupon with Simulated Damage} \label{Sec:cfrp}

\subsection{Test Setup}

\begin{figure}[t!]
\centering
\includegraphics[scale=0.35]{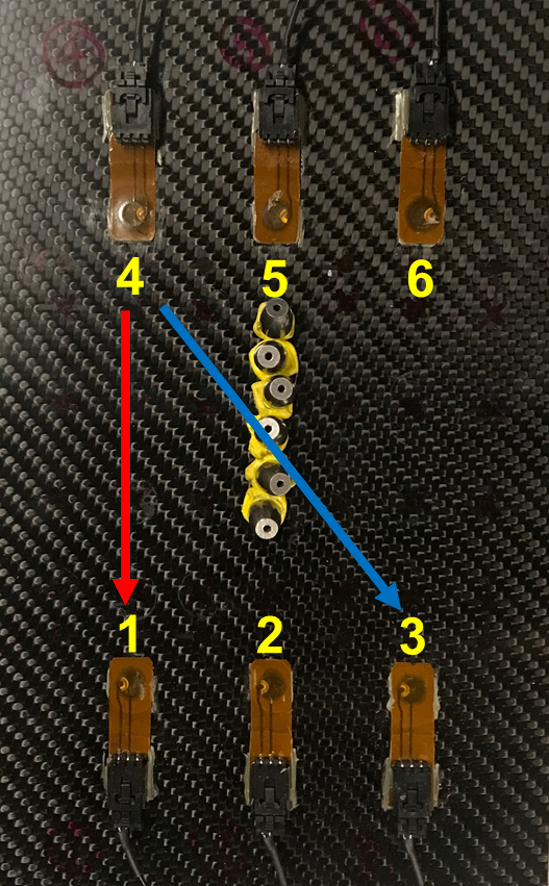}
\caption{The CFRP coupon used in this study shown here with 6 weights as simulated damage (largest damage size). The arrows show the signal paths presented herein.}
\label{fig:CFRP_plate} \vspace{10pt}
\end{figure}

The second coupon used in this study was a $152.4 \times  254$ mm ($6 \times 10$ in) CFRP plate ($2.36$ mm/$0.093$ in thickness; ACP Composites; $0/90^o$ unidirectional layup with Carbon fiber prepreg). As in the case of the Al coupon, the CFRP coupon was similarly fitted with six SMART Layers type PZT-5A using Hysol EA 9394 adhesive. In order to simulate damage, up to six three-gram weights were sequentially attached to the surface of the plate using tacky tape. Figure \ref{fig:CFRP_plate} shows the CFRP coupon with 6 weights attached.  Data acquisition and analysis was done in the same manner as in the Al coupon.

\subsection{Results \& Discussion}

\begin{figure}[t!]
    \centering
    \begin{picture}(400,300)
    \put(0,40){\includegraphics[trim = 20 0 20 12,clip,scale=0.8]{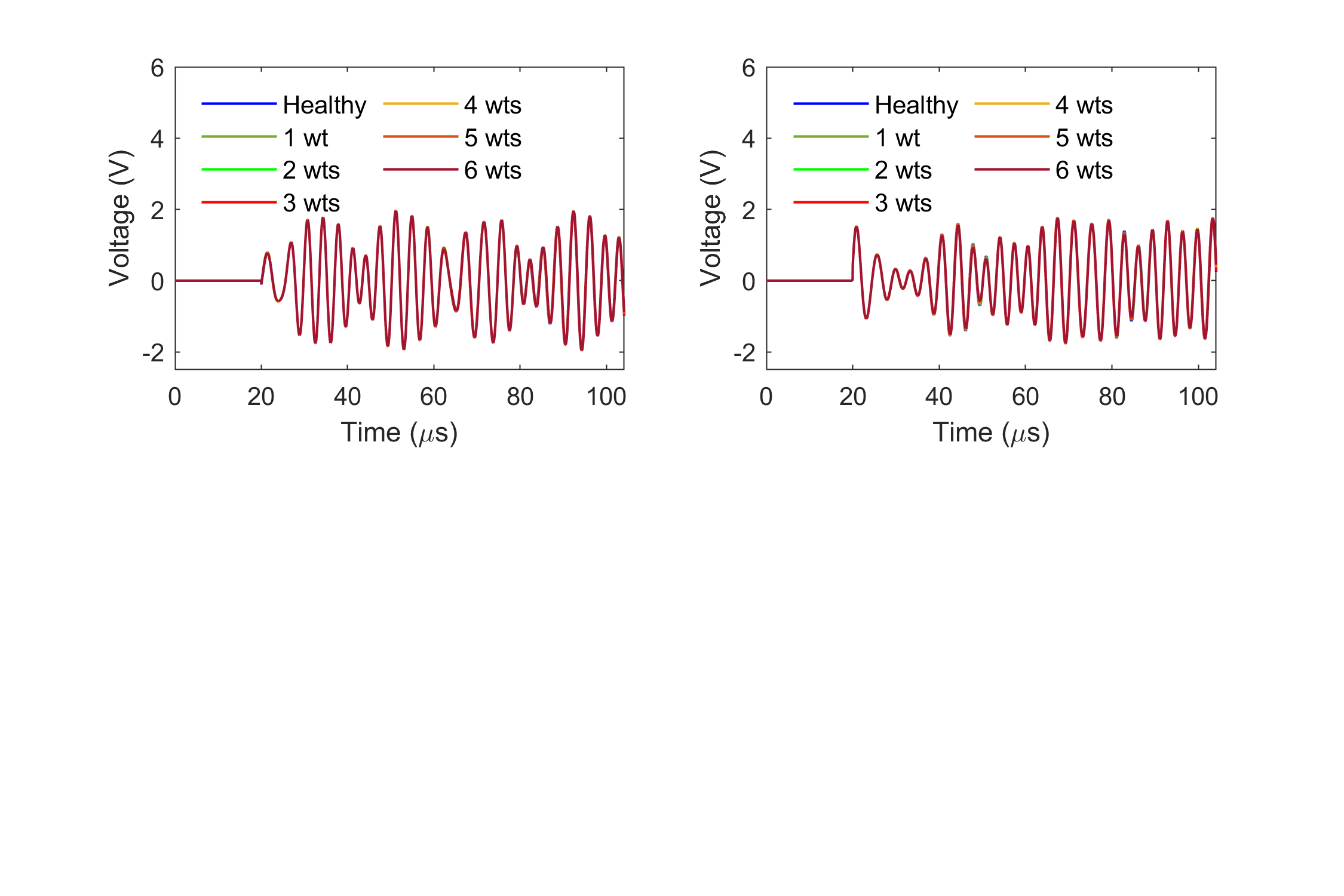}}
    \put(0,-95){\includegraphics[trim = 20 0 20 12,clip,scale=0.8]{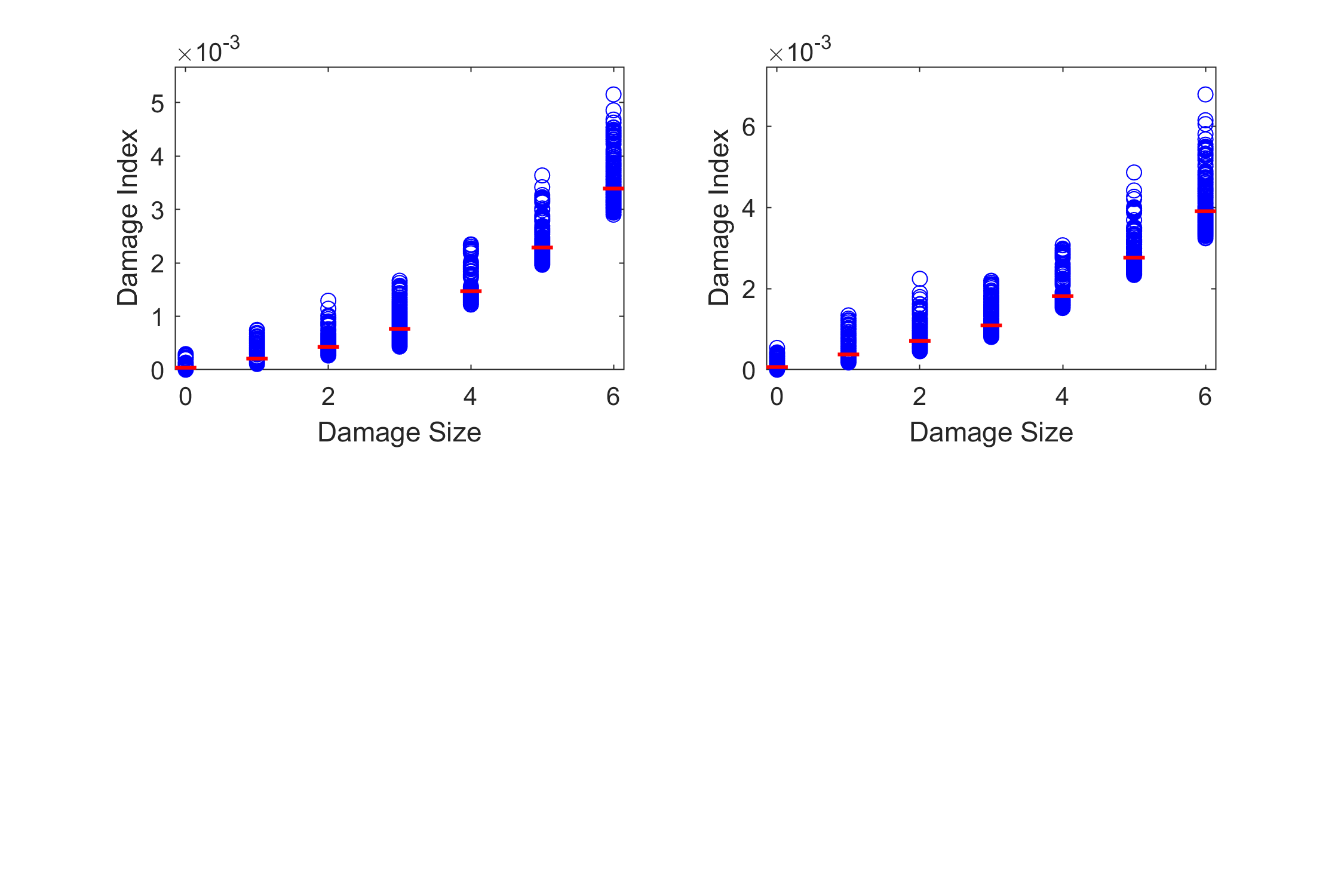}}
    \put(45,295){\color{black} \large {\fontfamily{phv}\selectfont \textbf{a}}}
    \put(235,295){\large {\fontfamily{phv}\selectfont \textbf{b}}}
   \put(45,160){\large {\fontfamily{phv}\selectfont \textbf{c}}} 
   \put(235,160){\large {\fontfamily{phv}\selectfont \textbf{d}}} 
    \end{picture} \vspace{-55pt}
    \caption{CFRP coupon: indicative signals and DI plots for the CFRP coupon: (a) signals from path 4-1; (b) signals from path 4-3; (c) DI values from path 4-1; (d) DI values from path 4-3. The red lines indicate the means of the DI values at every state.} 
\label{fig:cfrp_signals} \vspace{10pt}
\end{figure}

Figure \ref{fig:cfrp_signals} panels a and b show the signals received at sensors 1 and 3 in the CFRP coupon (see Figure \ref{fig:CFRP_plate} for sensor numbering), respectively, when sensor 4 was actuated under different levels of simulated damage. Note that the abrupt amplitude increase in panel b is due to the pre-processing of the signals where a set number of samples at the beginning of each signal was zeroed out to completely remove the effects of cross-talk. Figure \ref{fig:cfrp_signals} panels c and d show the corresponding DI evolution. The red dashed line indicates the $95\%$ confidence bound for the healthy case, as calculated from the standard deviation of the experimental baseline DI values. As shown, using this DI formulation, there seems to be substantial overlap between the DI values under multiple damage sizes, as well as a non-constant variance in the values with increasing damage size. These observations can be attributed to the nonlinearity exhibited by composites. Indeed, these paths in the CFRP coupon were specifically selected because of these two phenomena, which may prove interesting when quantifying damage size using GPRMs.

\begin{table}[b]
\centering
\caption{Summary of GPRM$^*$ information$^\dagger$ for the CFRP coupon based on the DI formulation from \cite{Janapati-etal16}}\label{tab:cfrp_janapati}
\renewcommand{\arraystretch}{1.2}
{\footnotesize
\begin{tabular}{|c|c|c|c|c|c|c|c|c|} 
\hline
Signal & \multicolumn{2}{c}{NMSE} & \multicolumn{2}{|c|}{RSS/SSS (\%)} & \multicolumn{2}{c}{Training Time (s)} & \multicolumn{2}{|c|}{Prediction Time (s)} \\ 
\cline{2-9}
 Path & SGPRM & VHGPRM & SGPRM & VHGPRM & SGPRM & VHGPRM & SGPRM & VHGPRM \\
\hline
4-1 & 0.0426 & 0.0425 & 2.036 & 2.035 & 4.31 & 9.82 & 0.0671 & 0.0996 \\
\hline
4-3 & 0.0611 & 0.0611 & 2.62 & 2.62 & 3.87 & 19.72 & 0.1006 & 0.0965 \\  
\hline
\multicolumn{7}{l}{$^*$22.5\% (990 points) of the data was used for training each model.} \\
\multicolumn{9}{l}{$^\dagger$Numbers approximated to the last quoted decimal place, and times estimated based on an Intel Core i3 laptop} \\
\multicolumn{8}{l}{  with 4 Gb of RAM.}
\end{tabular}} 
\end{table}

\begin{figure}[t!]
    \centering
    \begin{picture}(400,300)
    \put(0,40){\includegraphics[trim = 20 0 20 12,clip,scale=0.8]{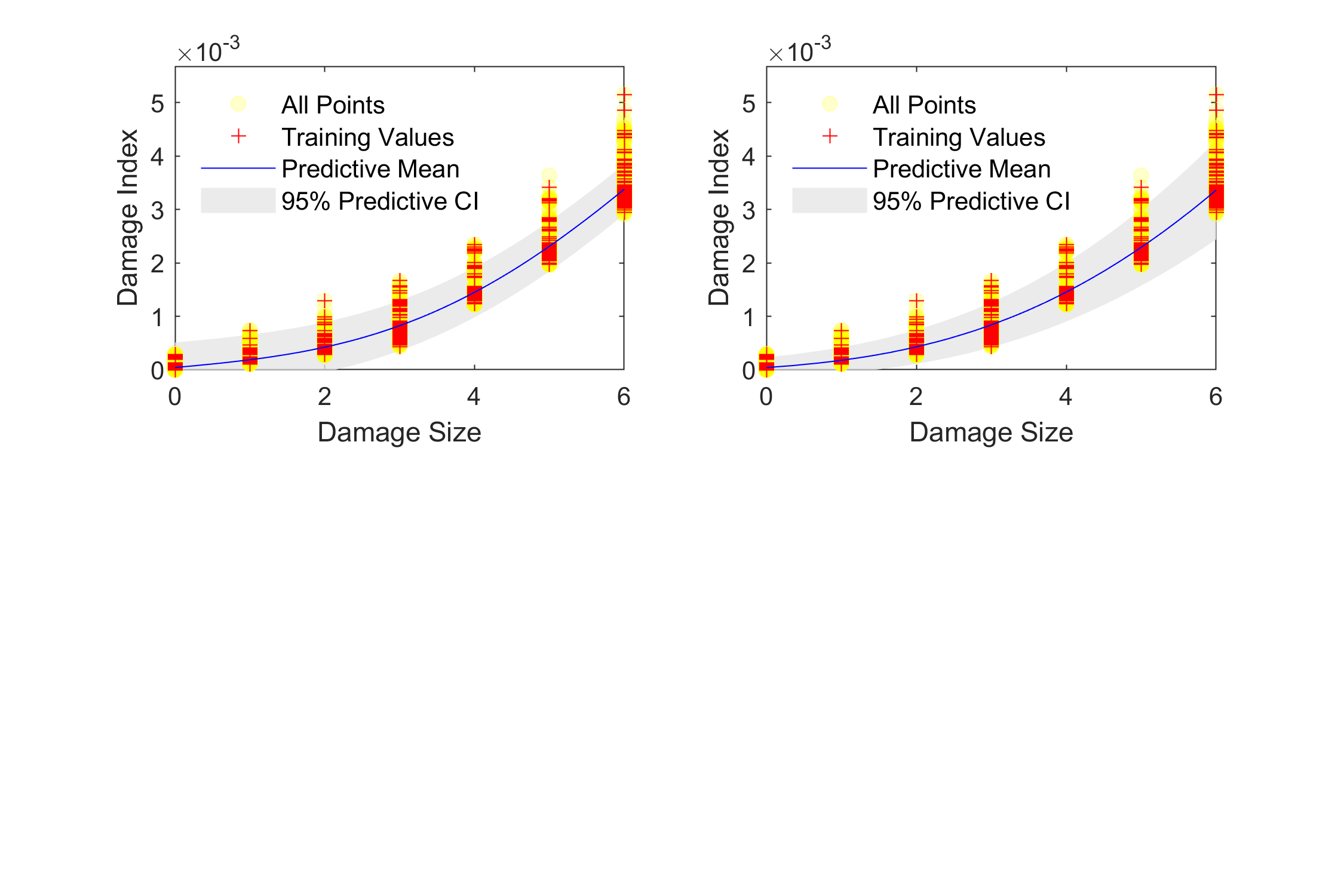}}
    \put(0,-95){\includegraphics[trim = 20 0 20 12,clip,scale=0.8]{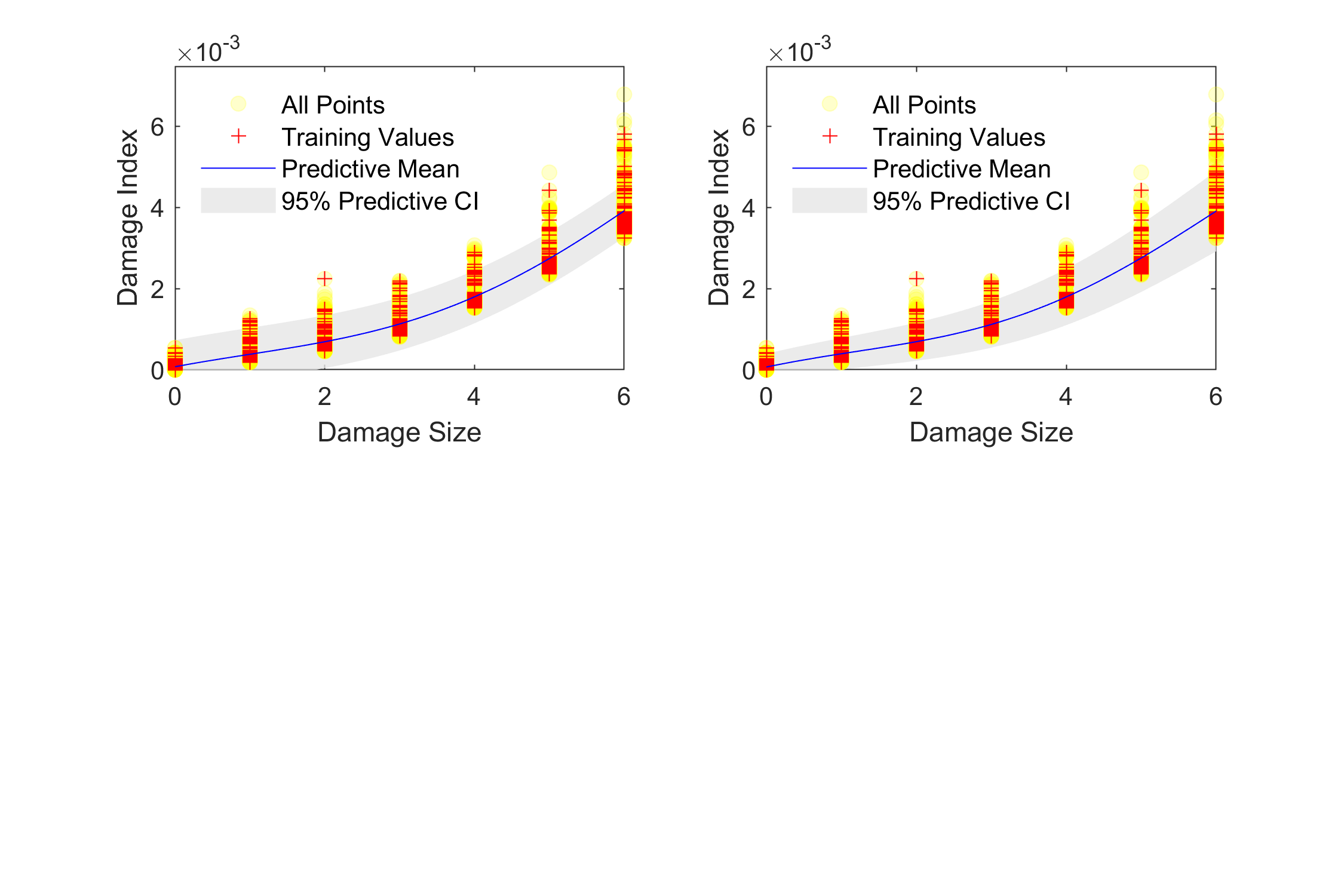}}
    \put(45,295){\color{black} \large {\fontfamily{phv}\selectfont \textbf{a}}}
    \put(235,295){\large {\fontfamily{phv}\selectfont \textbf{b}}}
   \put(45,160){\large {\fontfamily{phv}\selectfont \textbf{c}}} 
   \put(235,160){\large {\fontfamily{phv}\selectfont \textbf{d}}} 
    \end{picture} \vspace{-55pt}
    \caption{CFRP coupon: GPRM predictive mean and variance: (a) SGPRM results for path 4-1; (b) VHGPRM results for path 4-1; (c) SGPRM results for path 4-3; (d) VHGPRM results for path 4-3.} 
\label{fig:cfrp_gprm} \vspace{10pt}
\end{figure}

Similar to the case of the Al coupon, about $50\%$ of the available DI data points were used for training SGPRMs and VHGPRMs. Table \ref{tab:cfrp_janapati} presents summary information about the trained models. Figure \ref{fig:cfrp_gprm} panels a and b show the predictive means and confidence bounds of the trained SGPRM and VHGPRM, respectively, for path 4-1 under multiple damage cases. Figure \ref{fig:cfrp_gprm} panels c and d show the corresponding plots for path 4-3. As shown, both sets of models follow the evolution of the DI quite well, as can also be inferred from the low model criteria shown in Table \ref{tab:cfrp_janapati}. In addition, the VHGPRM in both paths exhibits a growing confidence interval, which follows the increasing variance in the DI values with damage size (number of attached weights). This widening in the confidence bounds should allow this model to estimate prediction probabilities more accurately compared to the trained SGPRM, especially since the predictive variance of the trained models is used in the prediction process, as described in Section \ref{sec:prediction}.

\begin{figure}[t!]
    \centering
    \begin{picture}(400,300)
    \put(0,40){\includegraphics[trim = 20 0 20 15
    ,clip,scale=0.8]{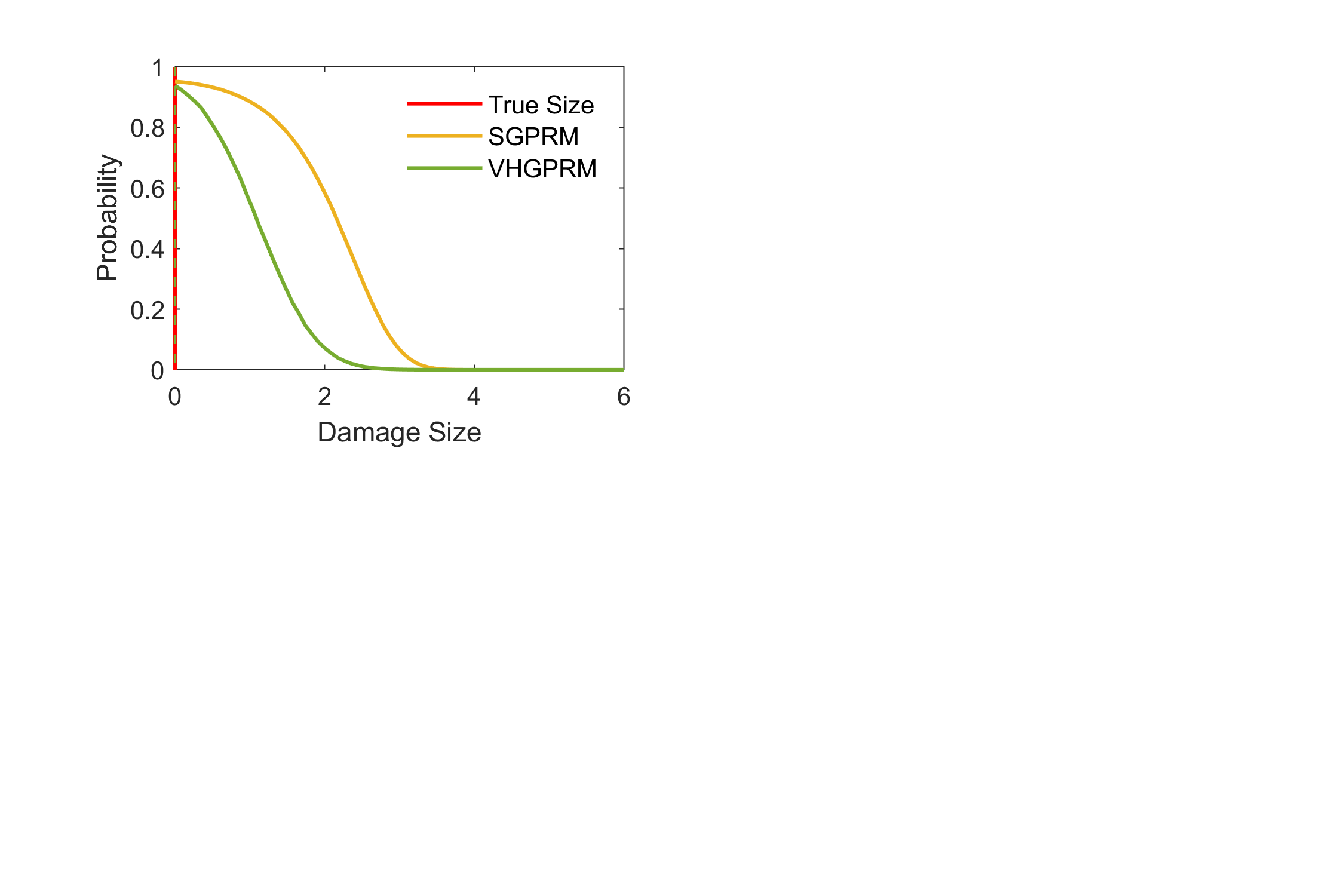}}
    \put(190,40){\includegraphics[trim = 20 0 20 15,clip,scale=0.8]{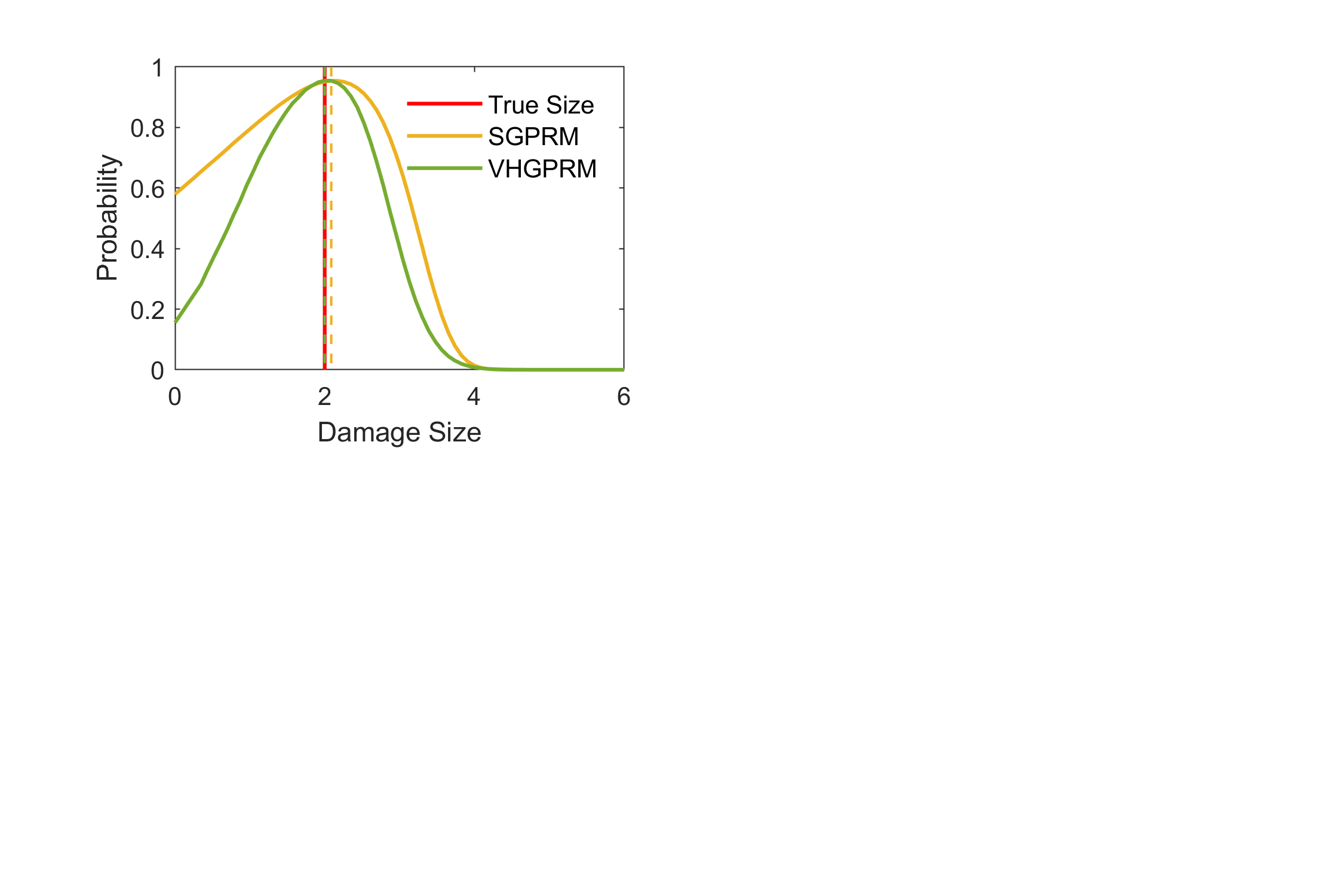}}
    \put(0,-95){\includegraphics[trim = 20 0 20 15,clip,scale=0.8]{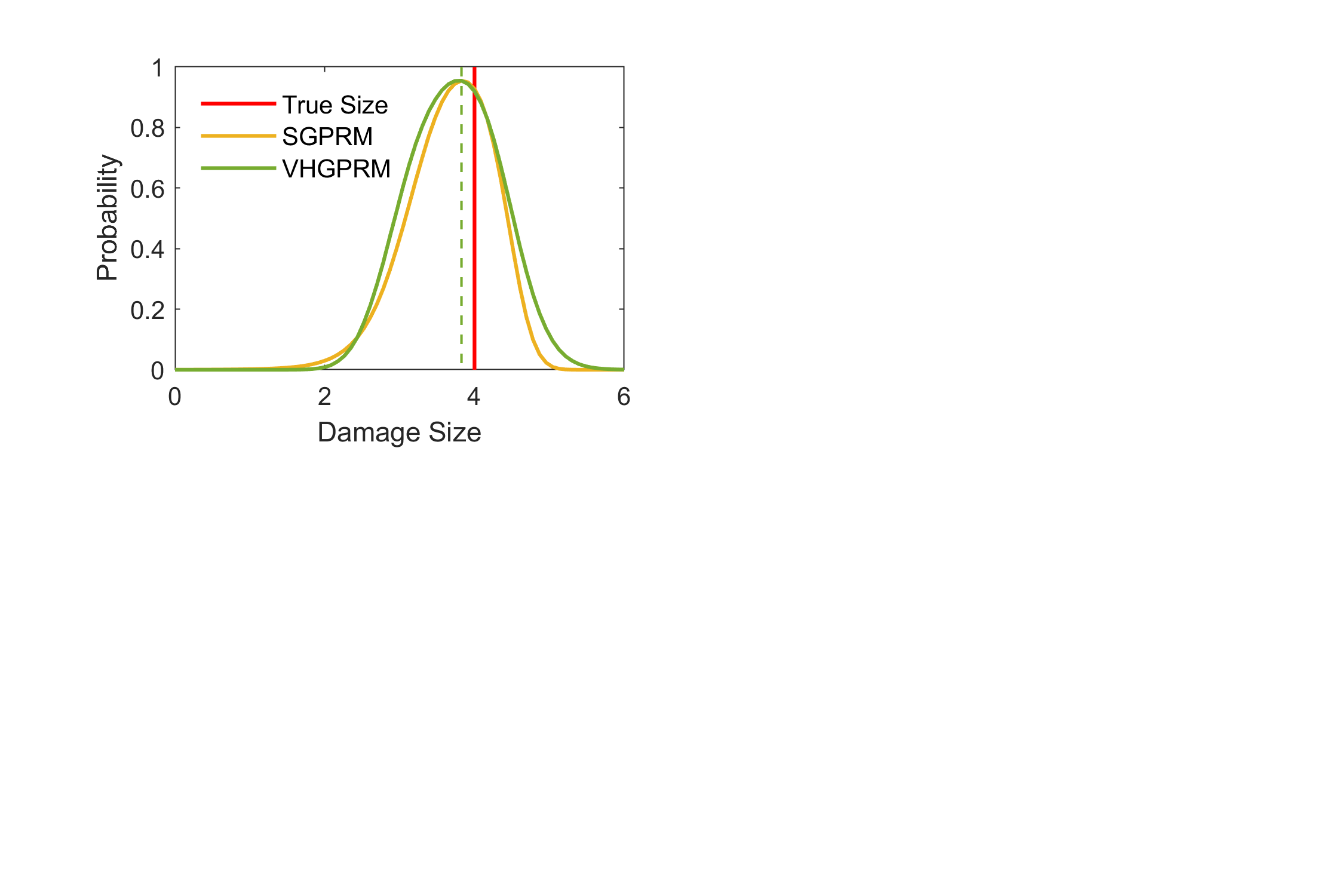}}
   \put(190,-95){\includegraphics[trim = 20 0 20 15,clip,scale=0.8]{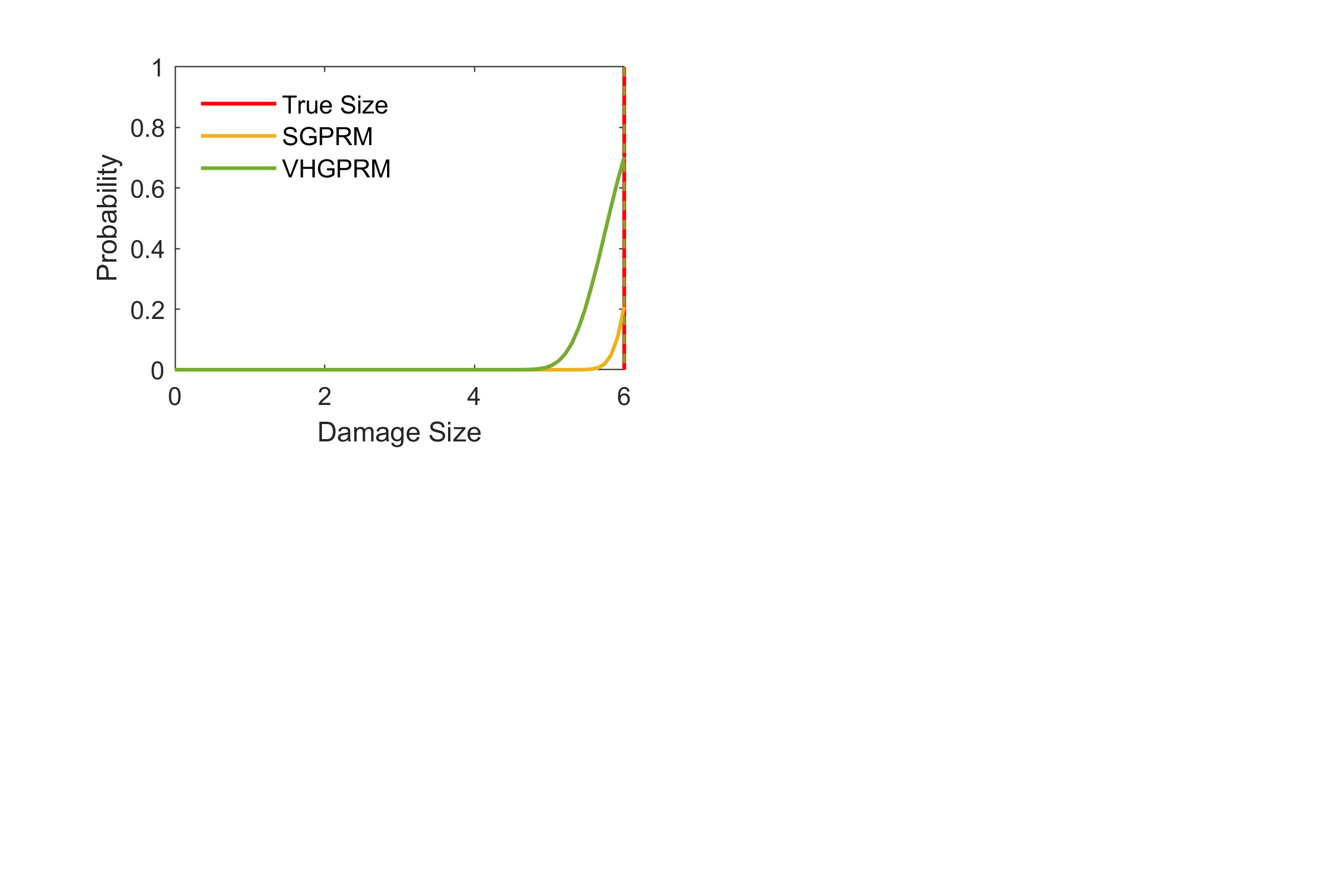}}
   \put(45,295){\color{black} \large {\fontfamily{phv}\selectfont \textbf{a}}}
    \put(235,295){\large {\fontfamily{phv}\selectfont \textbf{b}}}
   \put(45,160){\large {\fontfamily{phv}\selectfont \textbf{c}}} 
   \put(235,160){\large {\fontfamily{phv}\selectfont \textbf{d}}} 
    \end{picture} \vspace{-55pt}
    \caption{CFRP coupon: GPRM damage size prediction results for path 4-1: (a) prediction probabilities for the healthy case; (b) prediction probabilities for the case of 2 weights; (c) prediction probabilities for the case of 4 weights; (d) prediction probabilities for the case of 6 weights.} 
\label{fig:cfrp_prob_4-1_janapati} \vspace{10pt}
\end{figure}

In order to initially examine how the trained models perform in damage quantification, the prediction probabilities of a few indicative DI test points not used in the training process were plotted. Figure \ref{fig:cfrp_prob_4-1_janapati} shows such probabilities for the healthy (panel a), as well as 3 damaged cases. Although both model types accurately quantify damage size in all of the presented cases, it is worth noting that the VHGPRM prediction probability also closely follows the evolution of the variance; the probability is narrow towards the smaller damage sizes and broad towards the larger ones. Ont the other hand, the SGPRM prediction probabilities, owing to the SGPRM's constant noise term that is trained for the larger variance in the data, are broad towards the lower damage sizes, whereas they should be narrow because of the decreased variance in the DI values at these damage states. This shows the importance of utilizing heteroscedastic models, such as the VHGPRM, in order to properly estimate the prediction probability that accurately fits the variance in the data.

\begin{figure}[t!]
    \centering
    \begin{picture}(400,300)
    \put(0,40){\includegraphics[trim = 20 0 20 15
    ,clip,scale=0.8]{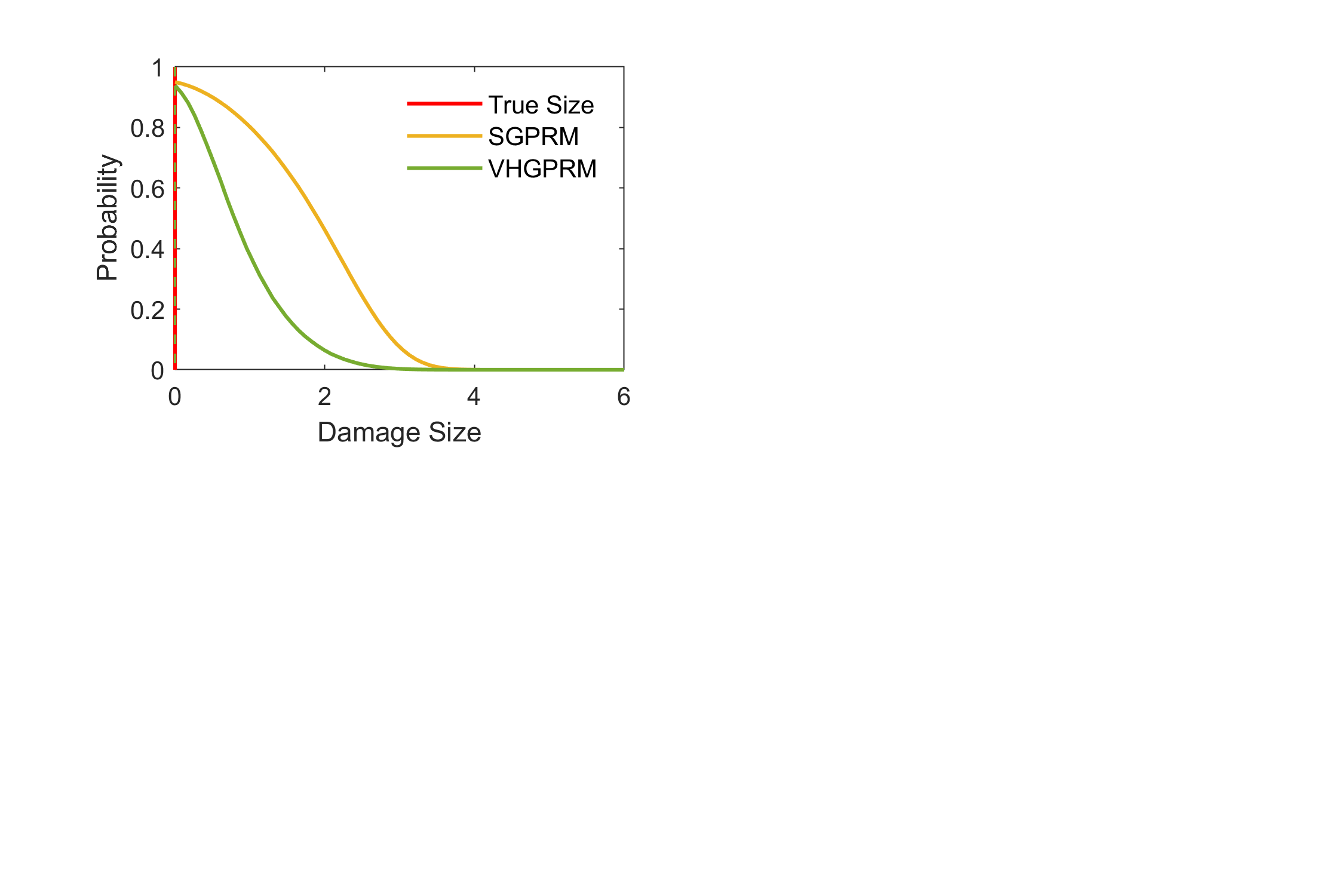}}
    \put(190,40){\includegraphics[trim = 20 0 20 15,clip,scale=0.8]{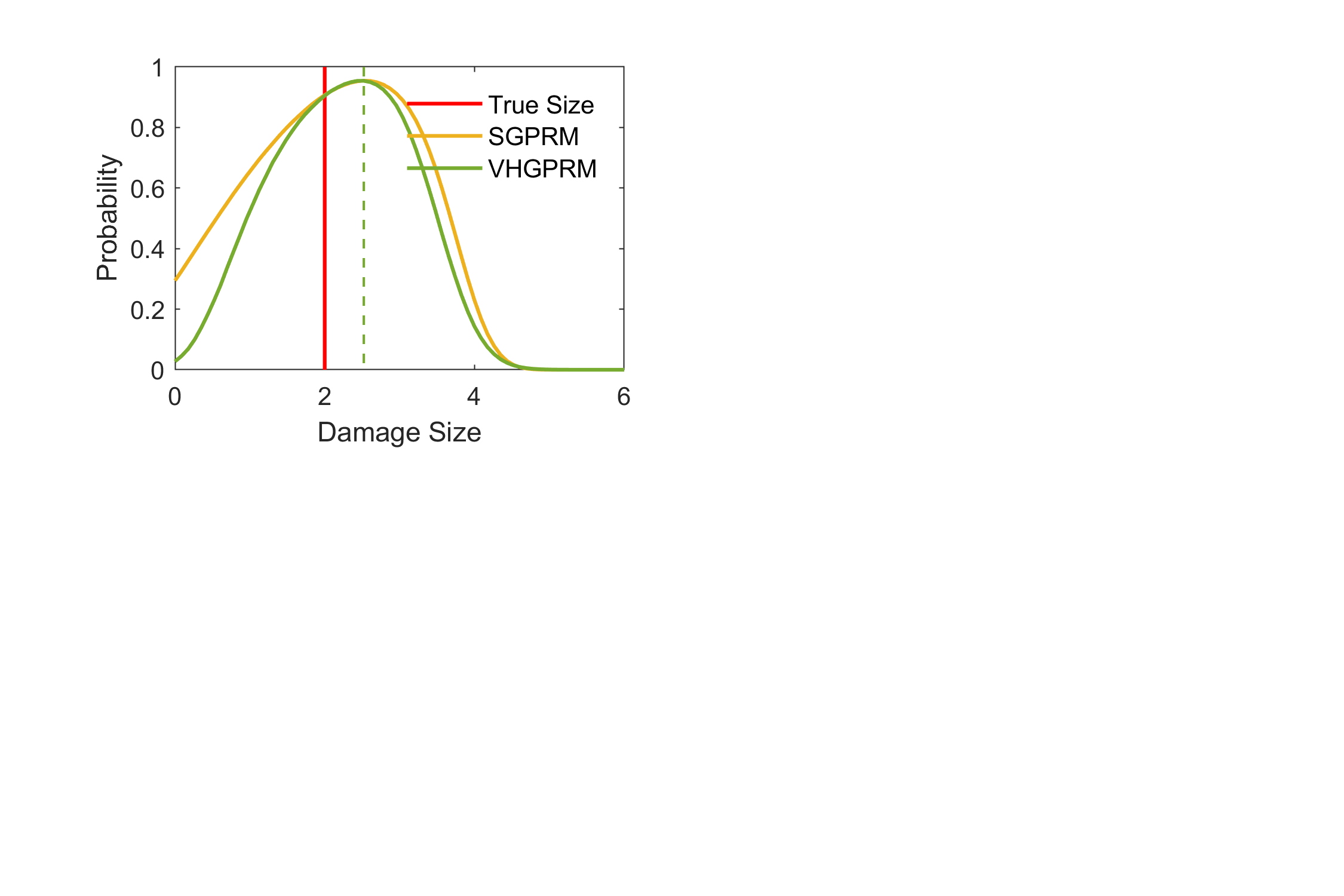}}
    \put(0,-95){\includegraphics[trim = 20 0 20 15,clip,scale=0.8]{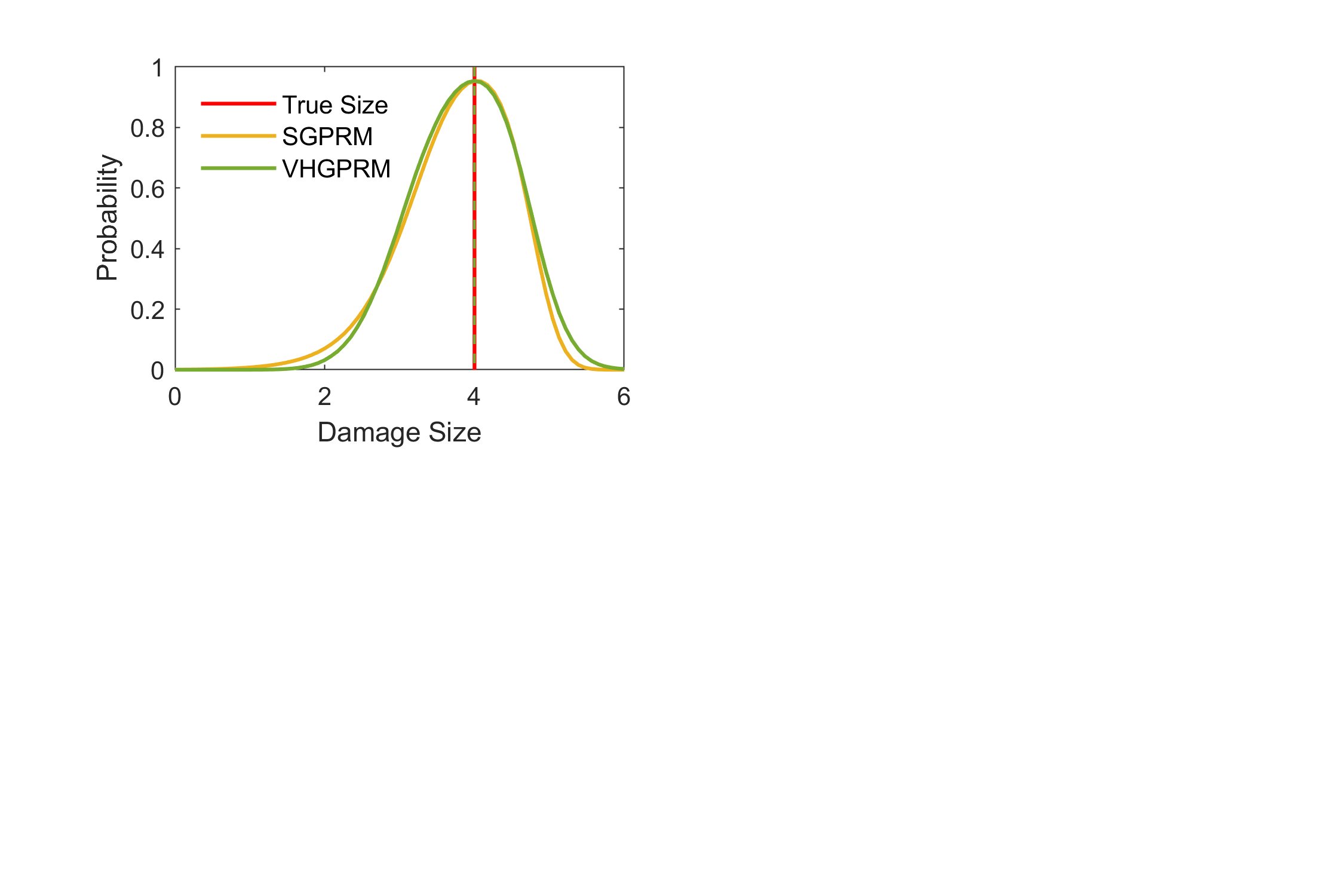}}
   \put(190,-95){\includegraphics[trim = 20 0 20 15,clip,scale=0.8]{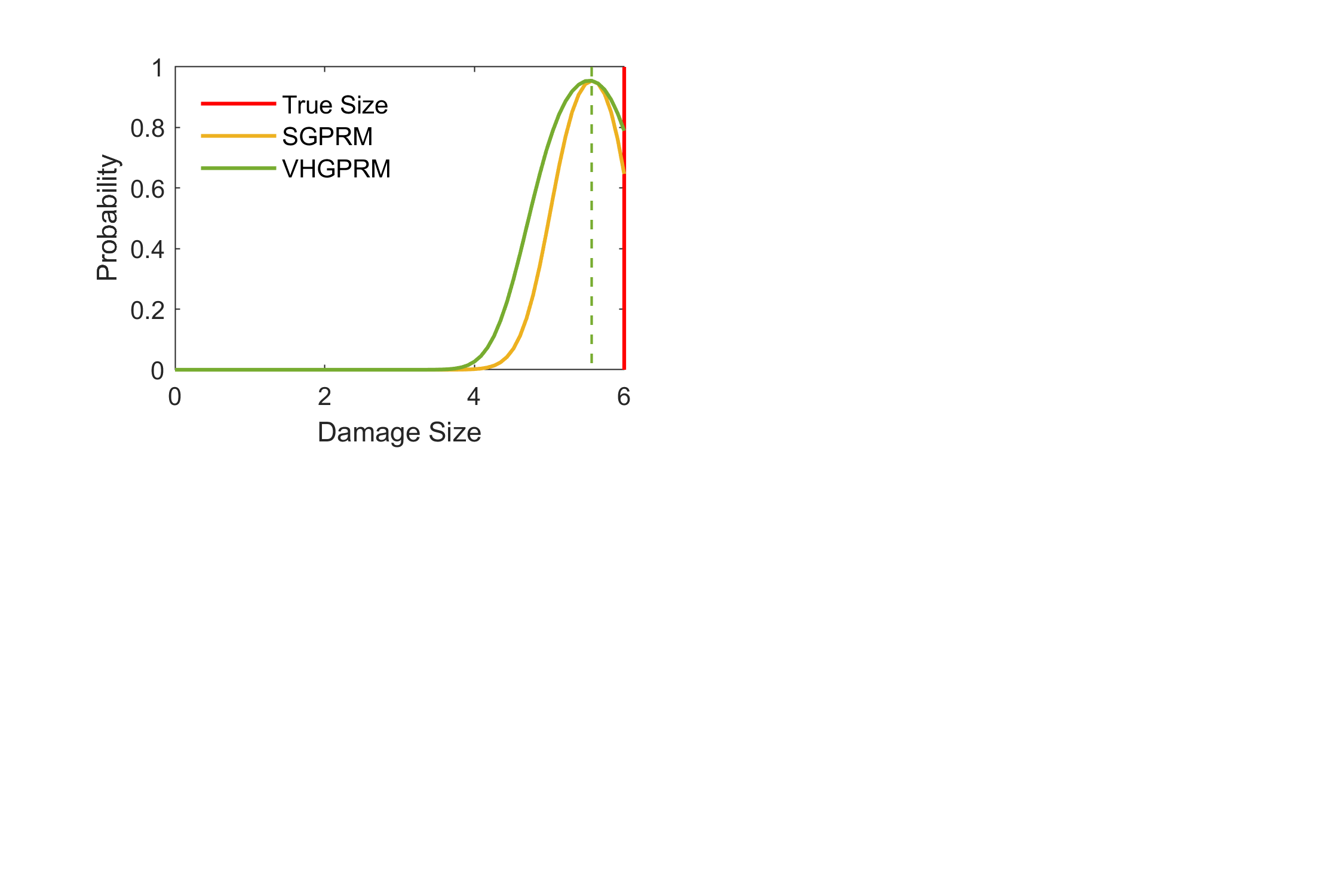}}
   \put(45,295){\color{black} \large {\fontfamily{phv}\selectfont \textbf{a}}}
    \put(235,295){\large {\fontfamily{phv}\selectfont \textbf{b}}}
   \put(45,160){\large {\fontfamily{phv}\selectfont \textbf{c}}} 
   \put(235,160){\large {\fontfamily{phv}\selectfont \textbf{d}}} 
    \end{picture} \vspace{-55pt}
    \caption{CFRP coupon: GPRM damage size prediction results for path 4-1: (a) prediction probabilities for the healthy case; (b) prediction probabilities for the case of 2 weights; (c) prediction probabilities for the case of 4 weights; (d) prediction probabilities for the case of 6 weights.} 
\label{fig:cfrp_prob_4-3_janapati} \vspace{5pt}
\end{figure}

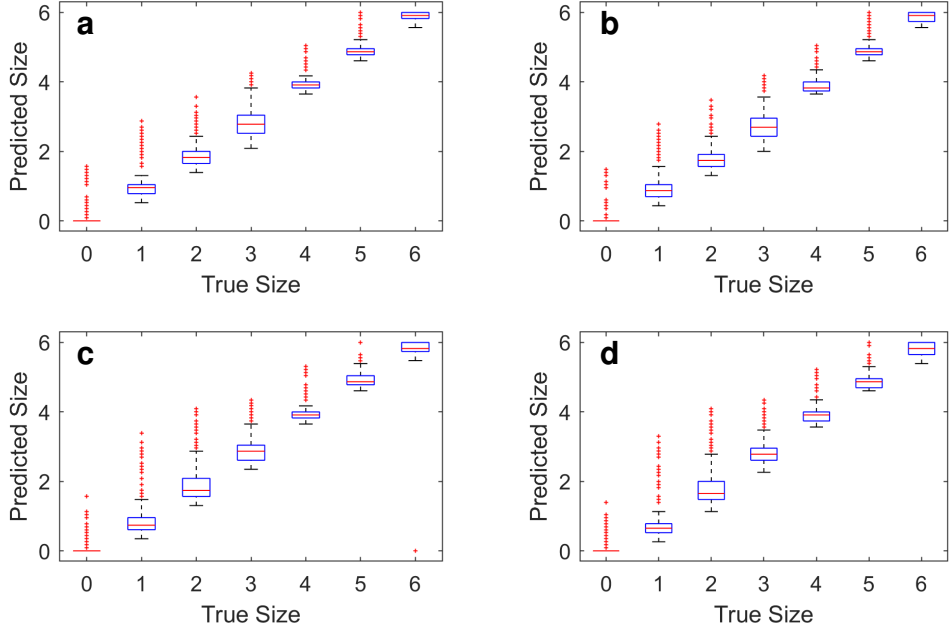
\begin{figure}[t!]
    \centering
    \begin{picture}(400,300)
    \put(0,40){\includegraphics[trim = 20 0 20 15,clip,scale=0.8]{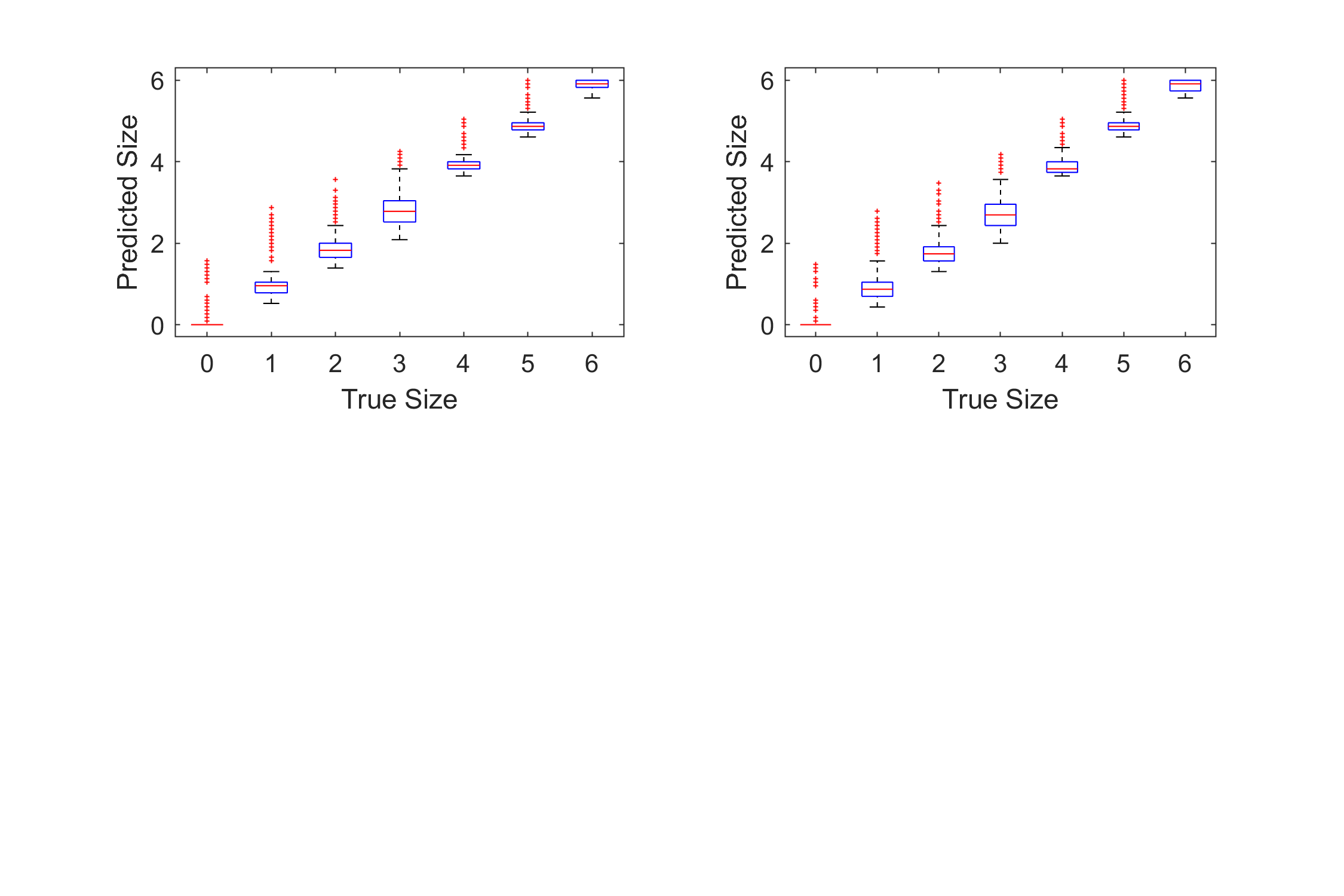}}
    \put(0,-85){\includegraphics[trim = 20 0 20 15,clip,scale=0.8]{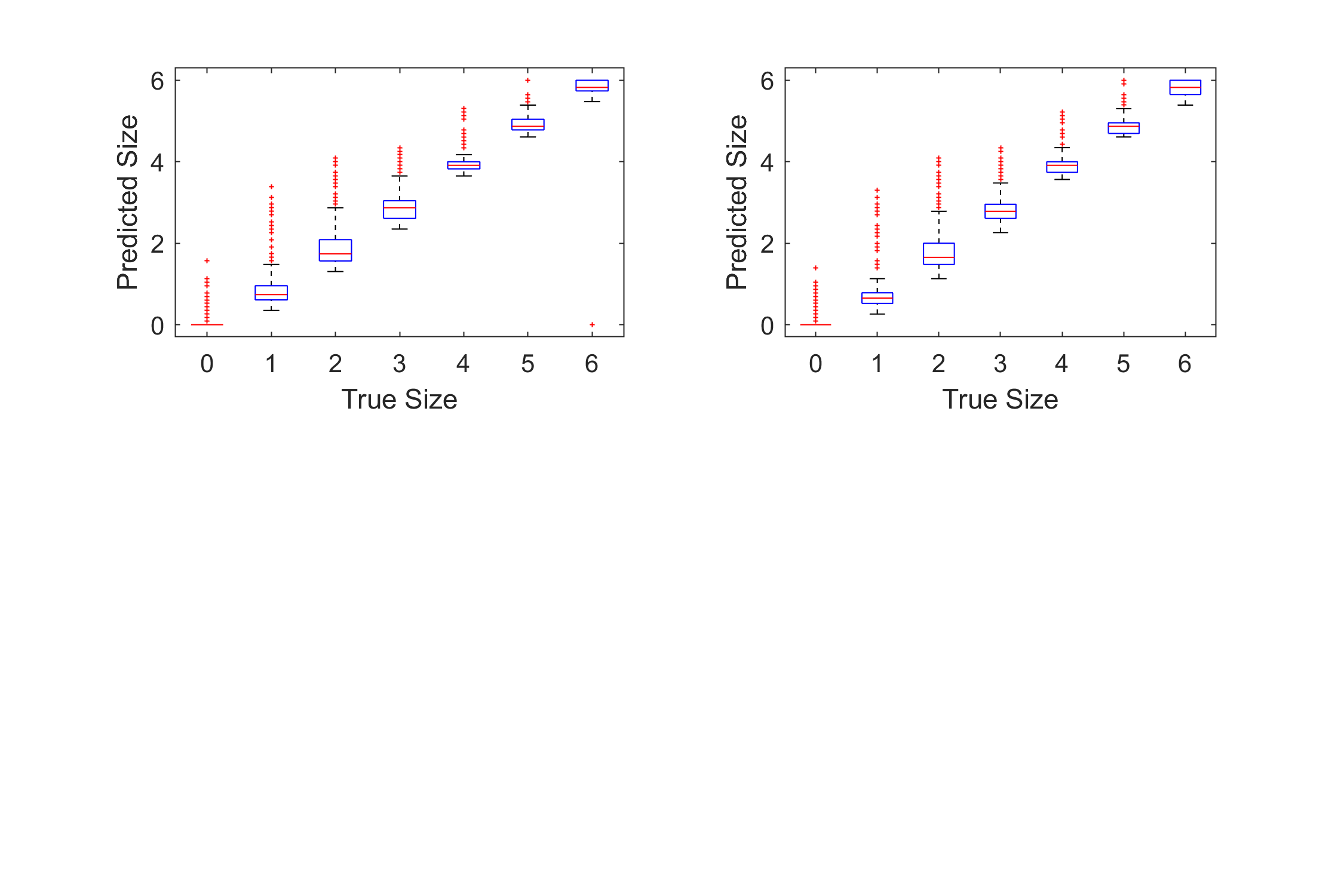}}
   \put(47,295){\color{black} \large {\fontfamily{phv}\selectfont \textbf{a}}}
    \put(245,295){\large {\fontfamily{phv}\selectfont \textbf{b}}}
   \put(47,170){\large {\fontfamily{phv}\selectfont \textbf{c}}} 
   \put(245,170){\large {\fontfamily{phv}\selectfont \textbf{d}}} 
    \end{picture} \vspace{-75pt}
    \caption{CFRP coupon: true/predicted damage size box-plots: (a) SGPRM state prediction for path 4-1; (b) VHGPRM state prediction for path 4-1; (c) SGPRM state prediction for path 4-3; (d) VHGPRM state prediction for path 4-3.} 
\label{fig:cfrp_boxplot_janapati} \vspace{0pt}
\end{figure}

Similarly, Figure \ref{fig:cfrp_prob_4-3_janapati} shows some indicative prediction probabilities from the trained models for path 4-3 in the CFRP coupon. Again, the adaptation of the VHGPRM-based prediction probability to the variance in the data can be clearly seen. In addition, both models show a slight deviation from the true damage size in the case of 2 and 6 weights. This deviation originates from the high overlap in the data sets between different damage sizes as can be clearly seen in Figure \ref{fig:cfrp_signals}d. Examining the summary results for this coupon (Figure \ref{fig:cfrp_boxplot_janapati}), one can observe that both models, overall, perform well in damage quantification, with the bulk of predictions per each damage size (the boxes) situated around the true damage size. The abundance of outliers in the box-plots originates from the large variation in DI values for each damage size. Again, these results indicate that GPRMs can accurately predict damage size in composites within an active-sensing, guided-wave SHM framework. In addition, the results presented in this section highlight the advantage of heteroscedastic GPRMs in accurately estimating the damage quantification probability, which should prove beneficial in real-life applications where the variation in the data can be large.

\section{Test Case III: Al Coupon Under Varying Simulated Damage \& Load States} \label{Sec:instron_Al}

\subsection{Test Setup}

\begin{figure}[t!]
\centering
\includegraphics[scale=0.6]{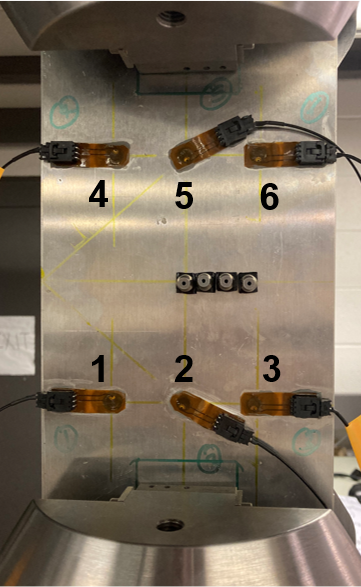} 
\caption{The third test case used in this study with four 3-gm weights simulating damage (largest damage size) shown here with the testing machine's grips.}  \label{fig:plate} \vspace{10pt}
\end{figure}

The third test case in this study was a $152.4\times304.8$ mm ($6\times12$ in) 6061 Aluminum coupon ($2.36$ mm/$0.093$ in thick). In a similar fashion to the two previous coupons, 6 PZT-5A sensors were attached to the plate. However, instead of leaving the adhesive to cure in ambient pressure for 7 days as with the two previous test cases, the adhesive was cured under vacuum for 24 hrs at room temperature. The plate was then installed onto a tensile testing machine (Instron, Inc). This allowed for the application of 4 static loading conditions consecutively: 0, 5, 10, and 15 kN. 1-4 three-gram weights were attached onto the surface of the plate during each loading state in the manner shown in Figure \ref{fig:plate}. Data acquisition and analysis was done in the same manner as in the other two coupons, with the difference that 20 signals/sensor/damage/loading condition (a total of 2400 signal data sets) were recorded in this test case.

\subsection{Results \& Discussion}

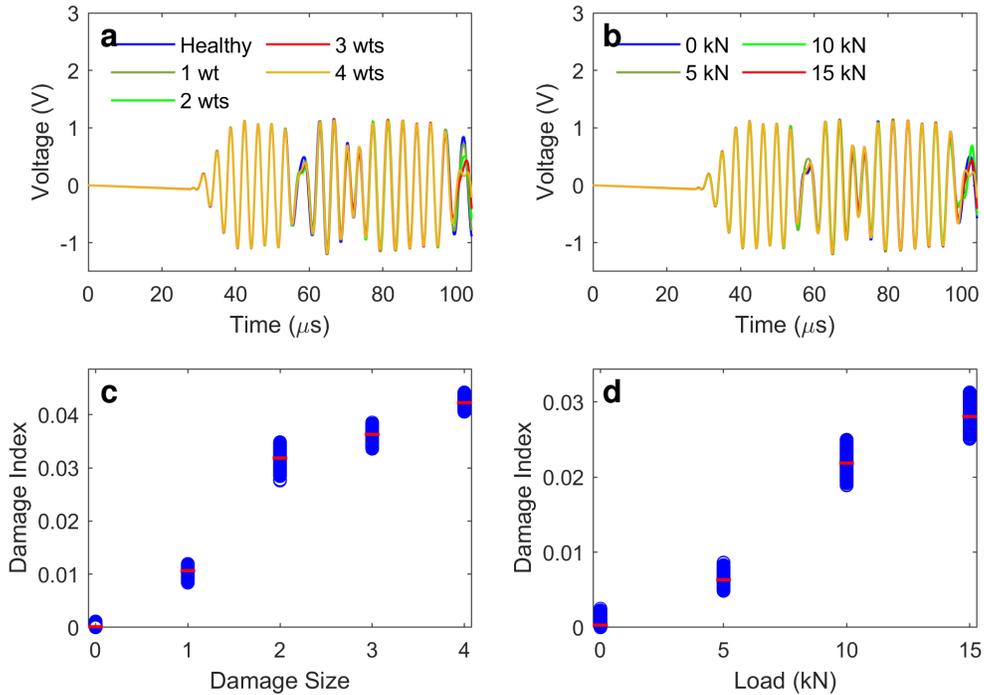
\begin{figure}[t!]
    \centering
    \begin{picture}(400,300)
    \put(0,40){\includegraphics[trim = 20 0 20 12,clip,scale=0.8]{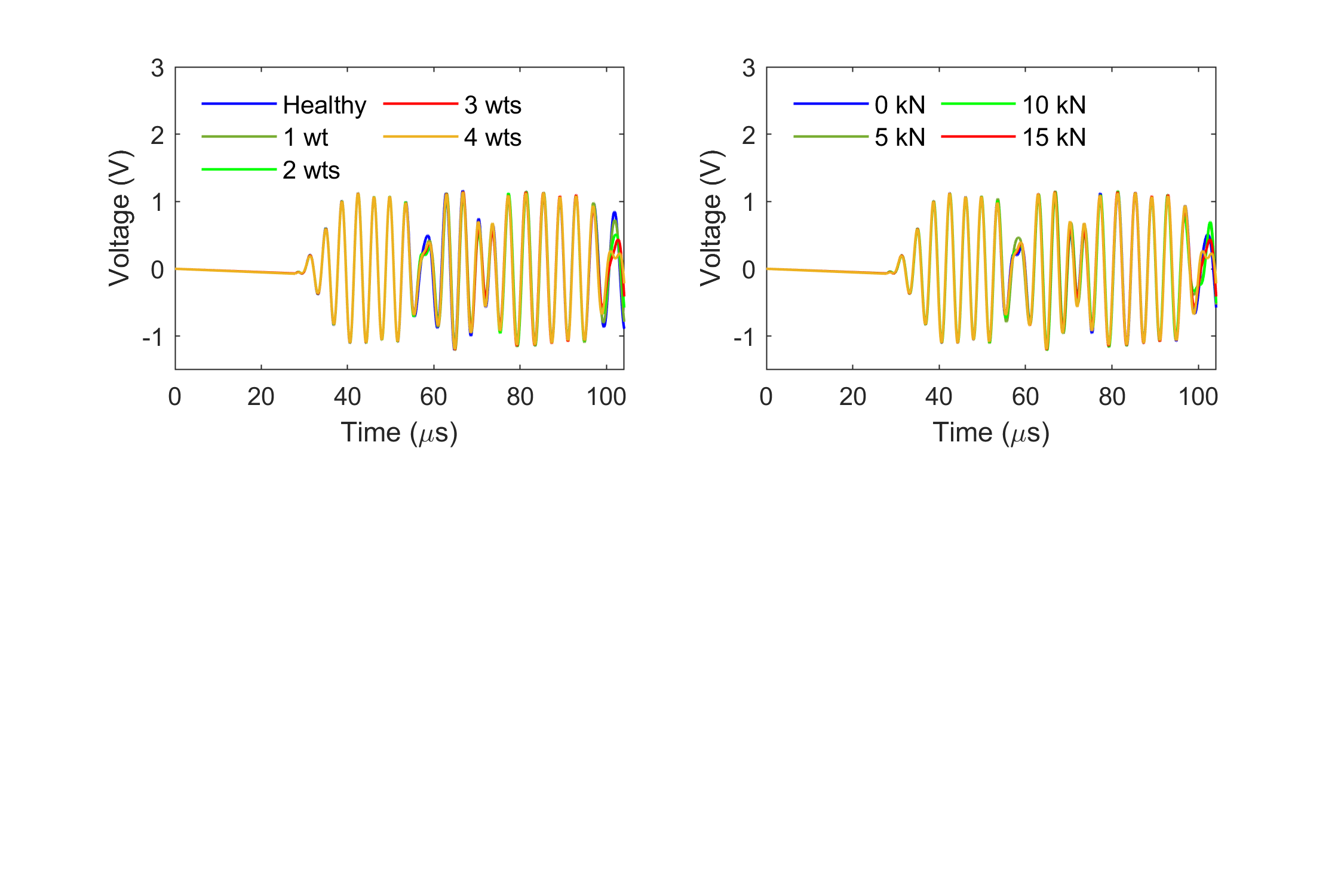}}
    \put(0,-95){\includegraphics[trim = 20 0 20 15,clip,scale=0.8]{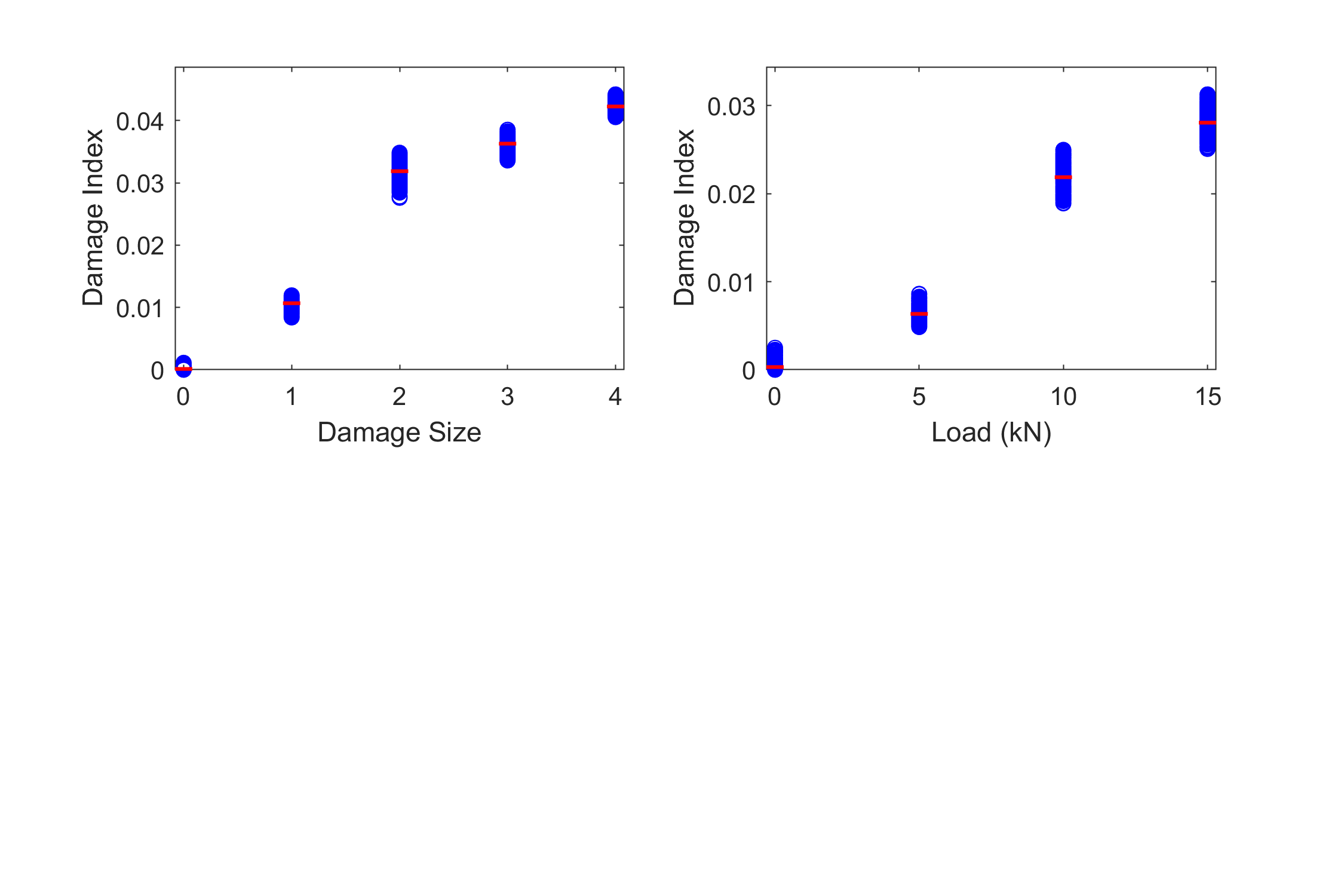}}
    \put(45,295){\color{black} \large {\fontfamily{phv}\selectfont \textbf{a}}}
    \put(235,295){\large {\fontfamily{phv}\selectfont \textbf{b}}}
   \put(45,160){\large {\fontfamily{phv}\selectfont \textbf{c}}} 
   \put(235,160){\large {\fontfamily{phv}\selectfont \textbf{d}}} 
    \end{picture} \vspace{-55pt}
    \caption{Al coupon with simulated damage: indicative signals and DI plots for path 1-6: (a) signals at unloaded condition; (b) signals with 2 weights attached; (c) DI values at unloaded condition; (d) DI values with 2 weights attached. The red lines indicate the means of the DI values at every state.} 
\label{fig:instron_1-6_signals} \vspace{0pt}
\end{figure}

\begin{figure}[t!]
    \centering
    \begin{picture}(400,300)
    \put(0,40){\includegraphics[trim = 20 0 20 12,clip,scale=0.8]{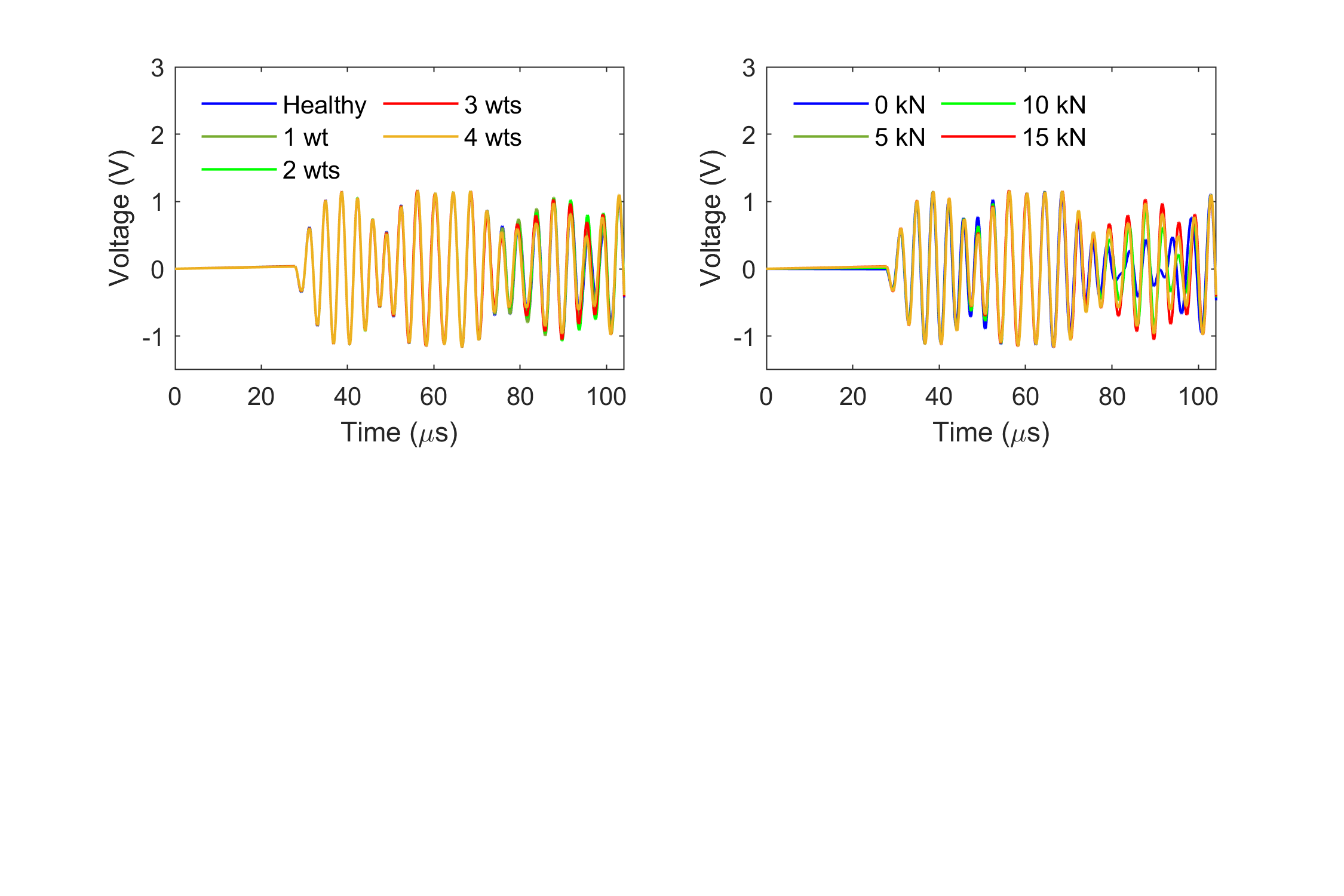}}
    \put(0,-95){\includegraphics[trim = 20 0 20 15,clip,scale=0.8]{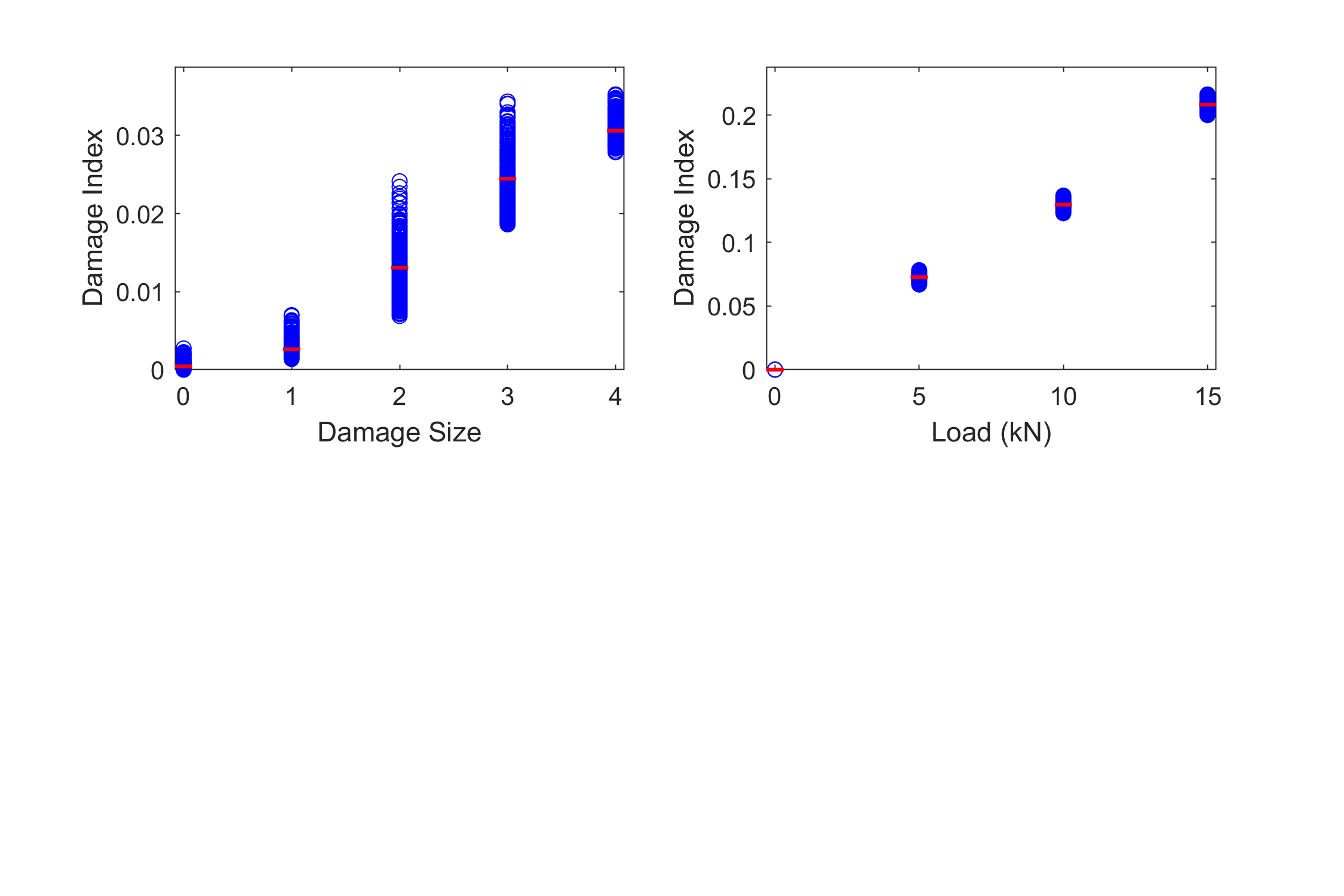}}
    \put(45,295){\color{black} \large {\fontfamily{phv}\selectfont \textbf{a}}}
    \put(235,295){\large {\fontfamily{phv}\selectfont \textbf{b}}}
   \put(45,160){\large {\fontfamily{phv}\selectfont \textbf{c}}} 
   \put(235,160){\large {\fontfamily{phv}\selectfont \textbf{d}}} 
    \end{picture} \vspace{-55pt}
    \caption{Al coupon with simulated damage: indicative signals and DI plots for path 3-6: (a) signals at 15 kN; (b) signals with 4 attached weights; (c) DI values at 15 kN; (d) DI values with 4 attached weights. The red lines indicate the means of the DI values at every state.} 
\label{fig:instron_3-6_signals} \vspace{0pt}
\end{figure}
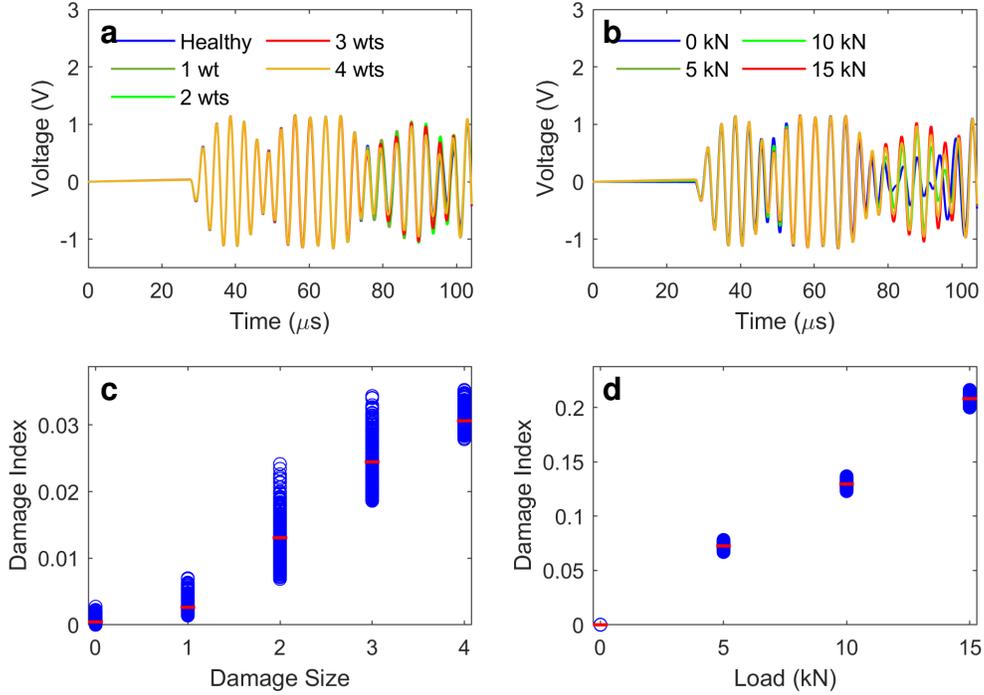

Figure \ref{fig:instron_1-6_signals}a shows the signal from path 1-6 in the third coupon in this study at the unloaded conditions under varying number of attached weights. Figure \ref{fig:instron_1-6_signals}b on the other hand shows the evolution of the signals with respect to load at a damage state of two attached weights. Figure \ref{fig:instron_1-6_signals} panels c and d show their corresponding DI plots, respectively. As shown, the DI formulation presented in \cite{Janapati-etal16} closely follows the evolution of damage size and load in this path. Also, a slight change in the variance of the DI values across multiple damage states or loads can be observed, with little-to-no overlap between te values in either DI plots. Examining a different path, Figure \ref{fig:instron_3-6_signals}a shows the signals from path 3-6 under varying damage states at a load of 15 kN, while Figure \ref{fig:instron_3-6_signals}b presents the signals from the same path at different loads under a damage state of 4 attached weights. Figure \ref{fig:instron_3-6_signals} panels c and d again show their respective DI plots. As shown in Figure \ref{fig:instron_3-6_signals}c, there exists a substantial overlap between the DI values, as well as a clearly-changing variance across the different damage states. On the other hand, Figure \ref{fig:instron_3-6_signals}d shows an almost linear evolution of the DI with loads, with no overlap at all, and an almost constant variance across the different loading states at a damage size of 4 attached weights. For a presentation of the full array of DI plots at all damage sizes and loading states, the readers are directed to Figures \ref{fig:instron_1-6_DI_dam}-\ref{fig:instron_3-6_DI_load} in the Appendix.

Because this coupon exhibits variation in both damage size and load, it is of interest to examine two cases in a typical state-quantification scenario. The first case is when the applied load on the component is known, and it is of interest to only quantify damage size. The second case, is when both the applied load and the damage state are unknown, and both need to be simultaneously quantified. The following sections present indicative results from both cases.

\subsubsection{The case of a known loading state: Single-input GPRMs}

For the first case, again $50\%$ of the available DI data points were used in training single-input SGPRMs and VHGPRMs under each loading state. Table \ref{tab:instron_1D} shows the model information for the trained models pertaining to the paths presented in Figures \ref{fig:instron_1-6_signals} and \ref{fig:instron_3-6_signals}. As shown in the table, again SGPRMs and VHGPRMs exhibit almost identical model criteria  (NMSE and RSS/SSS), with the substantial difference in training time. It is important to note here that, as presented in the case of the CFRP coupon, although the model criteria might be the same, the performance of both model types with respect to accurate estimation of the quantification probabilities can be different according to the nature of the training data.

\begin{table}[b!]
\centering
\caption{Summary of single-input GPRM$^*$ information$^\dagger$ for the Al coupon with simulated damage.}\label{tab:instron_1D}
\renewcommand{\arraystretch}{1.2}
{\footnotesize
\begin{tabular}{|c|c|c|c|c|c|c|c|c|c|} 
\hline
Signal & Load & \multicolumn{2}{c}{NMSE} & \multicolumn{2}{|c|}{RSS/SSS (\%)} & \multicolumn{2}{c}{Training Time (s)} & \multicolumn{2}{|c|}{Prediction Time (s)} \\
\cline{3-10}
Path & (kN) & SGPRM & VHGPRM & SGPRM & VHGPRM & SGPRM & VHGPRM & SGPRM & VHGPRM \\
\hline
\multirow{4}{*}{1-6} & 0 & 0.0034 & 0.0034 & 0.073 & 0.073 & 16.413 & 102.982 & 0.1913 & 0.3686 \\
& 5 & 0.0052 & 0.0052 & 0.1 & 0.1 & 17.294 & 86.772 & 0.1895 & 0.4157 \\
& 10 & 0.0012 & 0.0012 & 0.028 & 0.028 & 17.108 & 85.548 & 0.2558 & 0.3755 \\
& 15 & 0.0016 & 0.0016 & 0.036 & 0.036 & 16.561 & 108.205 & 0.2276 & 0.3589 \\
\hline
\multirow{4}{*}{3-6} & 0 & 0.0226 & 0.0226 & 0.97 & 0.97 & 10.348 & 23.044 & 0.1279 & 0.1126 \\
& 5 & 0.105 & 0.105 & 3.16 & 3.16 & 9.037 & 22.989 & 0.0957 & 0.112\\
& 10 & 0.0473 & 0.0473 & 1.77 & 1.77 & 9.25 & 25.79 & 0.1136 & 0.1082 \\
& 15 & 0.0379 & 0.0379 & 1.59 & 1.59 & 9.061 & 25.689 & 0.094 & 0.109 \\
\hline
\multicolumn{8}{l}{$^*$50\% (1000 points) of the data was used for training each model.} \\
\multicolumn{10}{l}{$^\dagger$Numbers approximated to the last quoted decimal place, and times estimated based on an Intel Core i3 laptop} \\
\multicolumn{8}{l}{  with 4 Gb of RAM.}
\end{tabular}} 
\end{table}

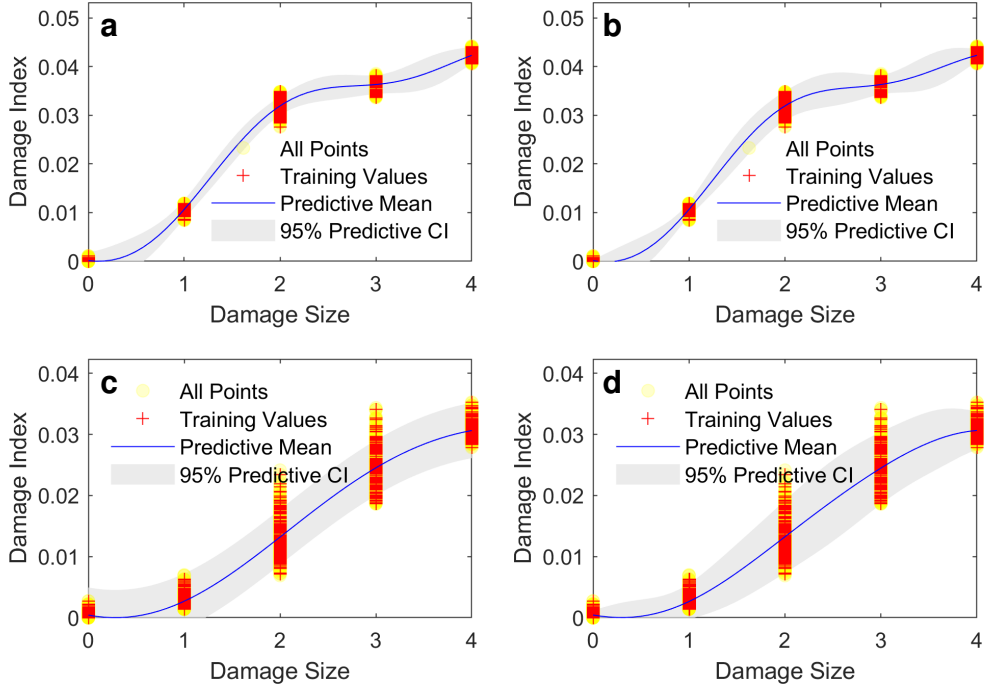
\begin{figure}[t!]
    \centering
    \begin{picture}(400,300)
    \put(0,40){\includegraphics[trim = 20 0 20 12,clip,scale=0.8]{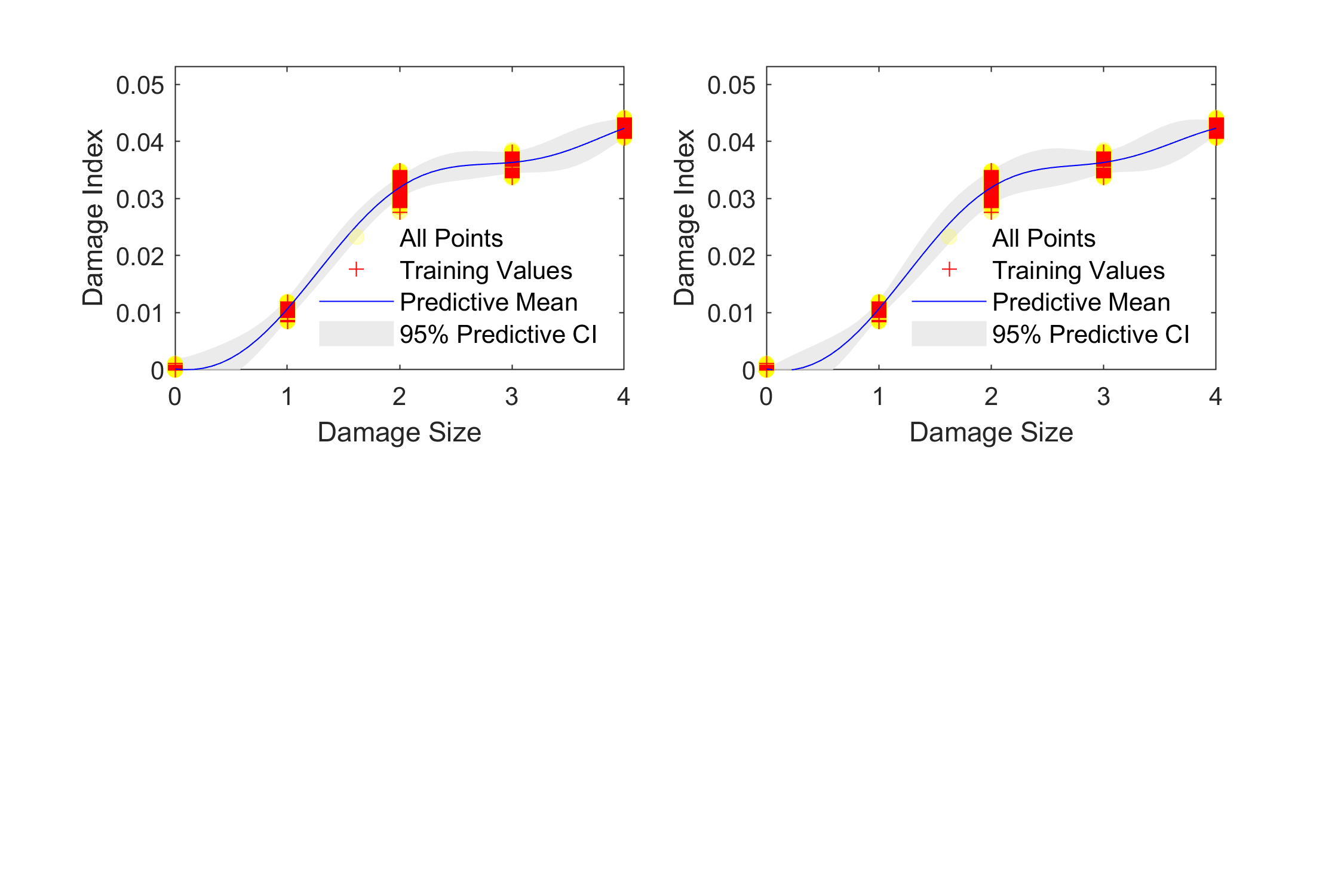}}
    \put(0,-95){\includegraphics[trim = 20 0 20 15,clip,scale=0.8]{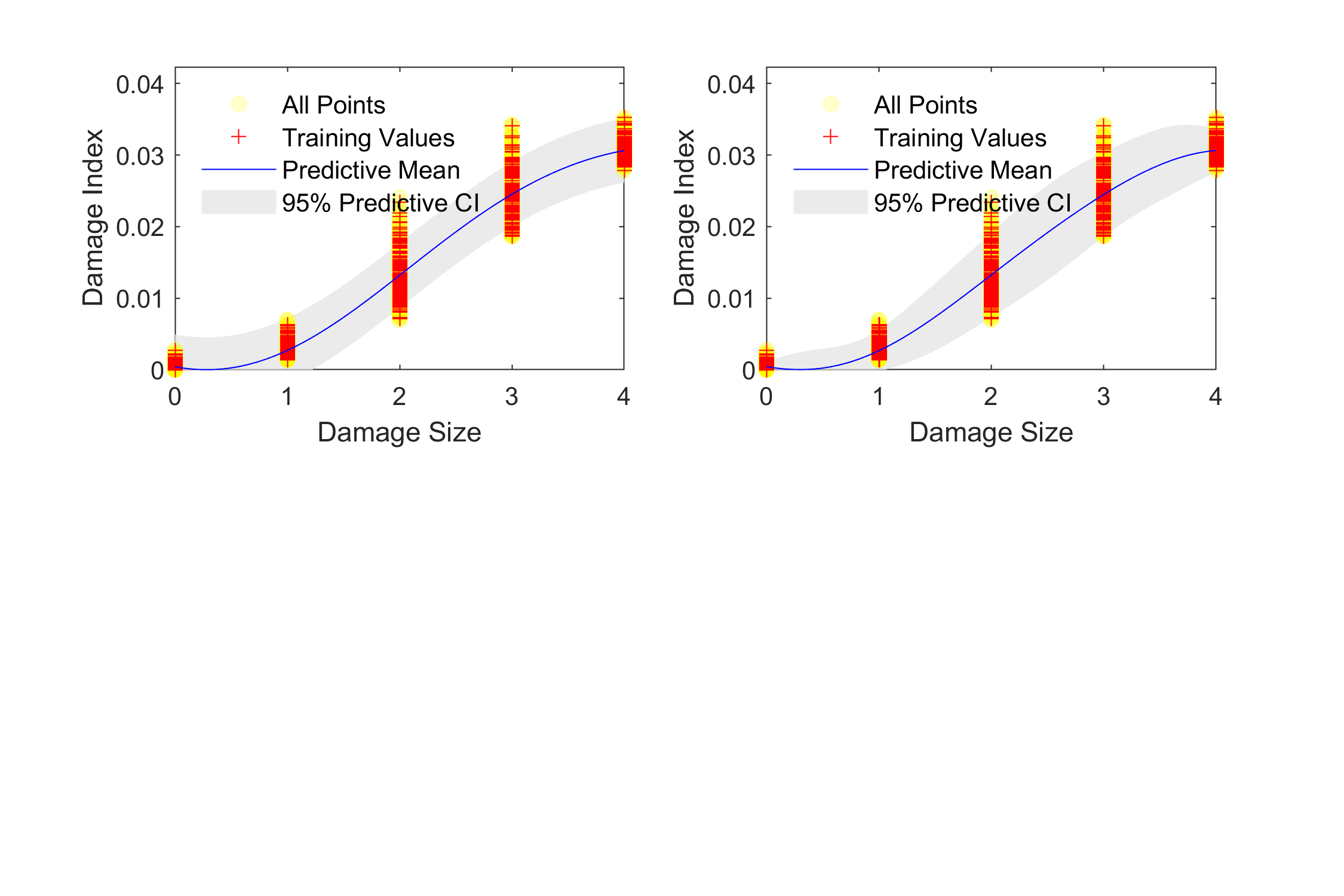}}
    \put(45,295){\color{black} \large {\fontfamily{phv}\selectfont \textbf{a}}}
    \put(235,295){\large {\fontfamily{phv}\selectfont \textbf{b}}}
    \put(45,160){\large {\fontfamily{phv}\selectfont \textbf{c}}} 
    \put(235,160){\large {\fontfamily{phv}\selectfont \textbf{d}}} 
    \end{picture} \vspace{-55pt}
    \caption{Al coupon with simulated damage: GPRM predictive mean and variance: (a) SGPRM for path 1-6 at unloaded conditions; (b) VHGPRM for path 1-6 at unloaded conditions; (c) SGPRM for path 3-6 at 15 kN; (d) VHGPRM for path 3-6 at 15 kN.} 
\label{fig:instron_gprm_janapati} \vspace{0pt}
\end{figure}

\begin{figure}[t!]
    \centering
    \begin{picture}(400,300)
    \put(0,40){\includegraphics[trim = 20 0 20 12,clip,scale=0.8]{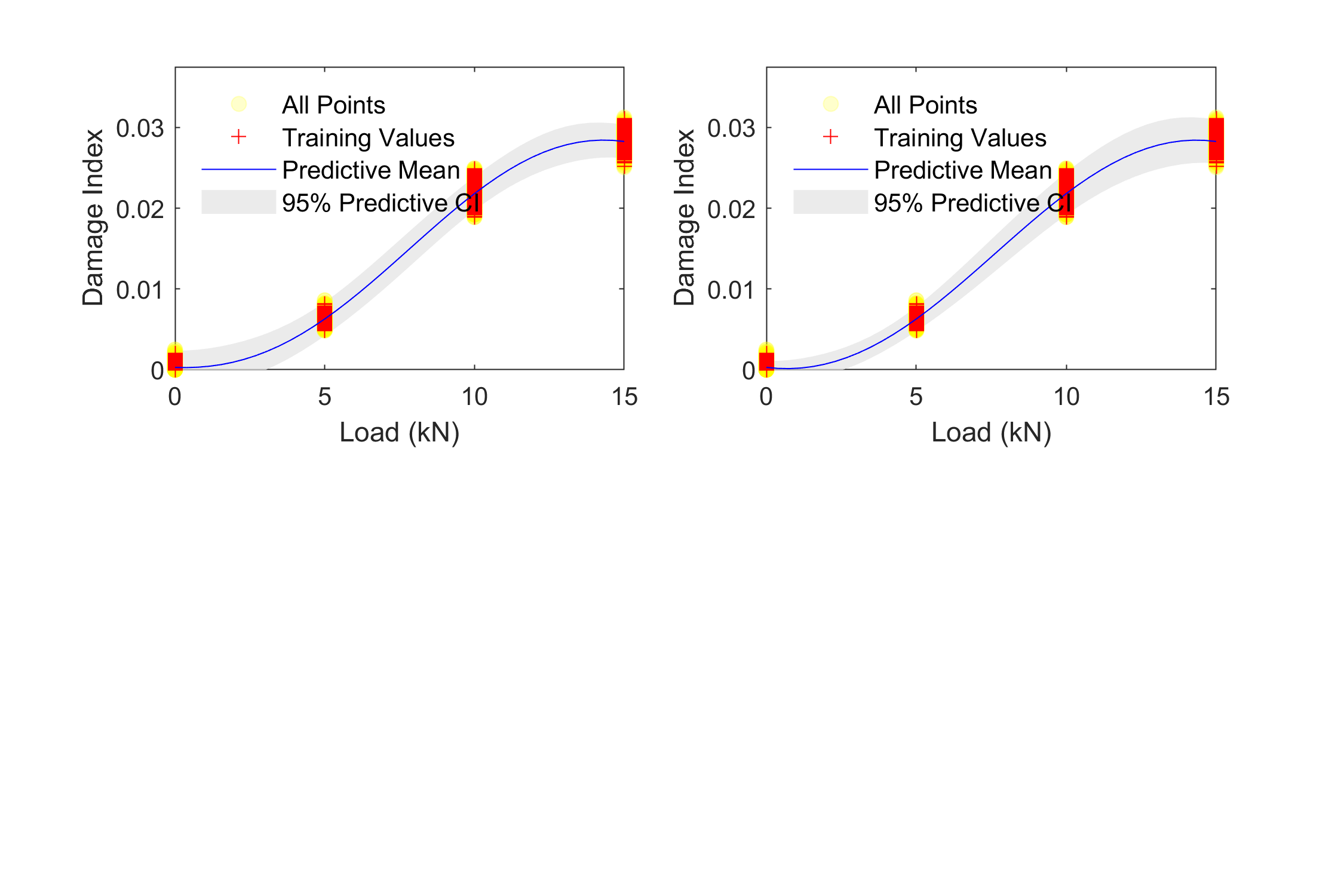}}
    \put(0,-95){\includegraphics[trim = 20 0 20 15,clip,scale=0.8]{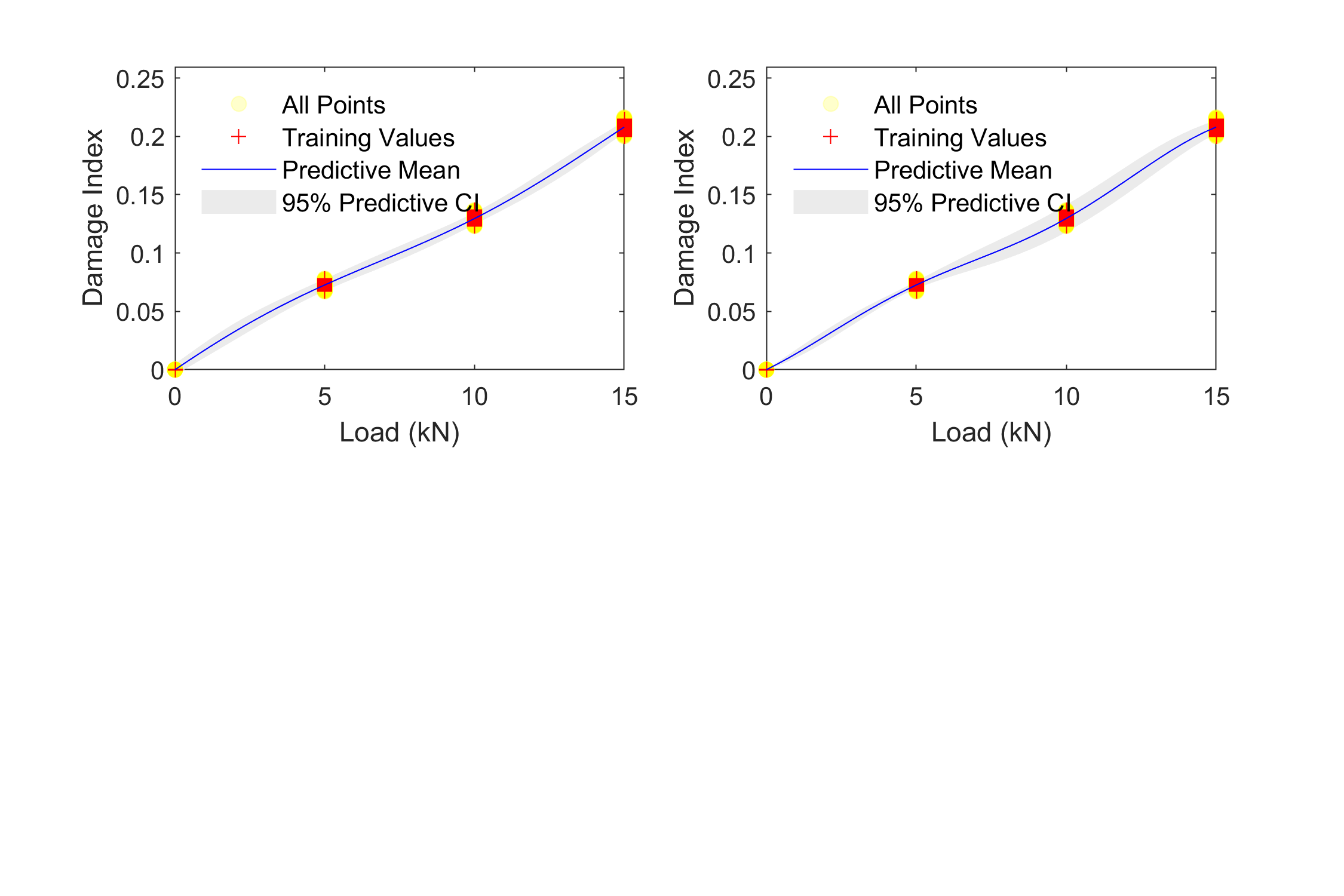}}
    \put(45,295){\color{black} \large {\fontfamily{phv}\selectfont \textbf{a}}}
    \put(235,295){\large {\fontfamily{phv}\selectfont \textbf{b}}}
    \put(45,160){\large {\fontfamily{phv}\selectfont \textbf{c}}} 
    \put(235,160){\large {\fontfamily{phv}\selectfont \textbf{d}}} 
    \end{picture} \vspace{-55pt}
    \caption{Al coupon with simulated damage: GPRM predictive mean and variance: (a) SGPRM for path 1-6 with 2 weights attached; (b) VHGPRM for path 1-6 with 2 weights attached; (c) SGPRM for path 3-6 with 2 weights attached; (d) VHGPRM for path 3-6 with 2 weights attached.} 
\label{fig:instron_gprm_janapati_load} \vspace{0pt}
\end{figure}
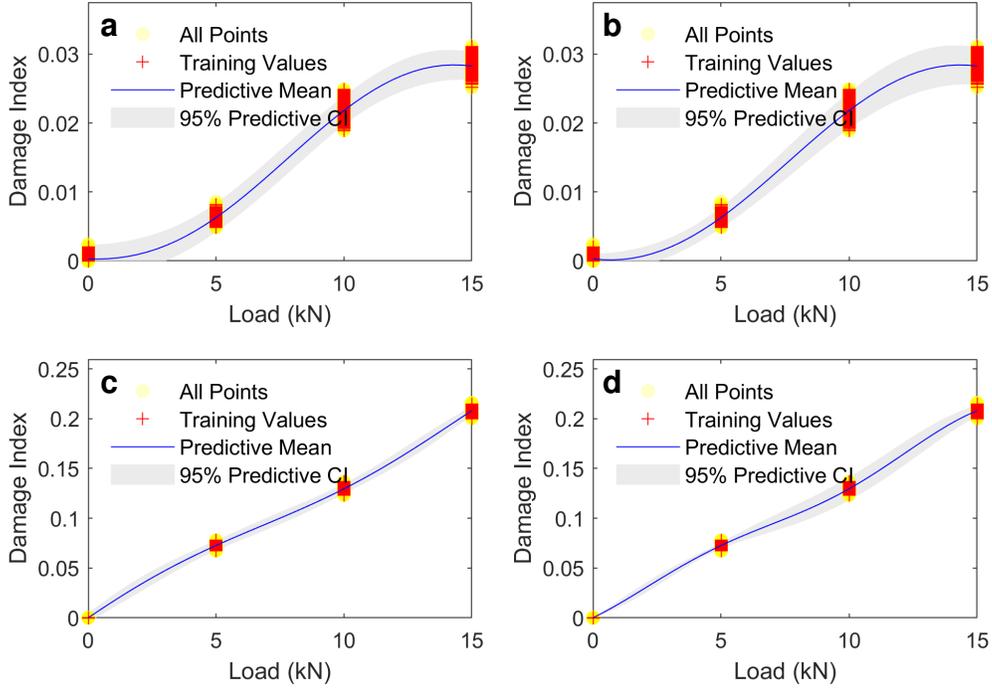

Figure \ref{fig:instron_gprm_janapati} panels a and b show the predictive means and confidence bounds from the GPRMs trained using the DI values presented in Figure \ref{fig:instron_1-6_signals}c for path 1-6. The difference between VHGPRMs and SGPRMs can be clearly seen at a damage size of 2 attached weights, where the VHGPRM can adapt its confidence bounds to fit the relatively-large variance in the training DI values at that specific damage size. The same observation can be persists in Figure \ref{fig:instron_gprm_janapati} panels a and b, which show the predictive means and confidence bounds from the GPRMs trained using the DI values presented in Figure \ref{fig:instron_3-6_signals}c for path 3-6 (a load of 15 kN). In both paths, though, each model type can clearly follow the evolution of the DI with damage size. Although this section is particular to the case where the load is known and the damage size is being quantified, for comparison, GPRMs trained at a specific damage size and unknown loading state are also presented herein. Figure \ref{fig:instron_gprm_janapati_load} shows the predictive means and confidence bounds from such GPRMs trained using DI values for paths 1-4 (panels a and b) and 3-6 (panels c and d) corresponding to the cases shown in Figure \ref{fig:instron_1-6_signals}d (two attached weights) and \ref{fig:instron_3-6_signals}d (four attached weights) respectively.

\begin{figure}[t!]
    \centering
    \begin{picture}(400,300)
    \put(0,40){\includegraphics[trim = 20 0 20 15,clip,scale=0.8]{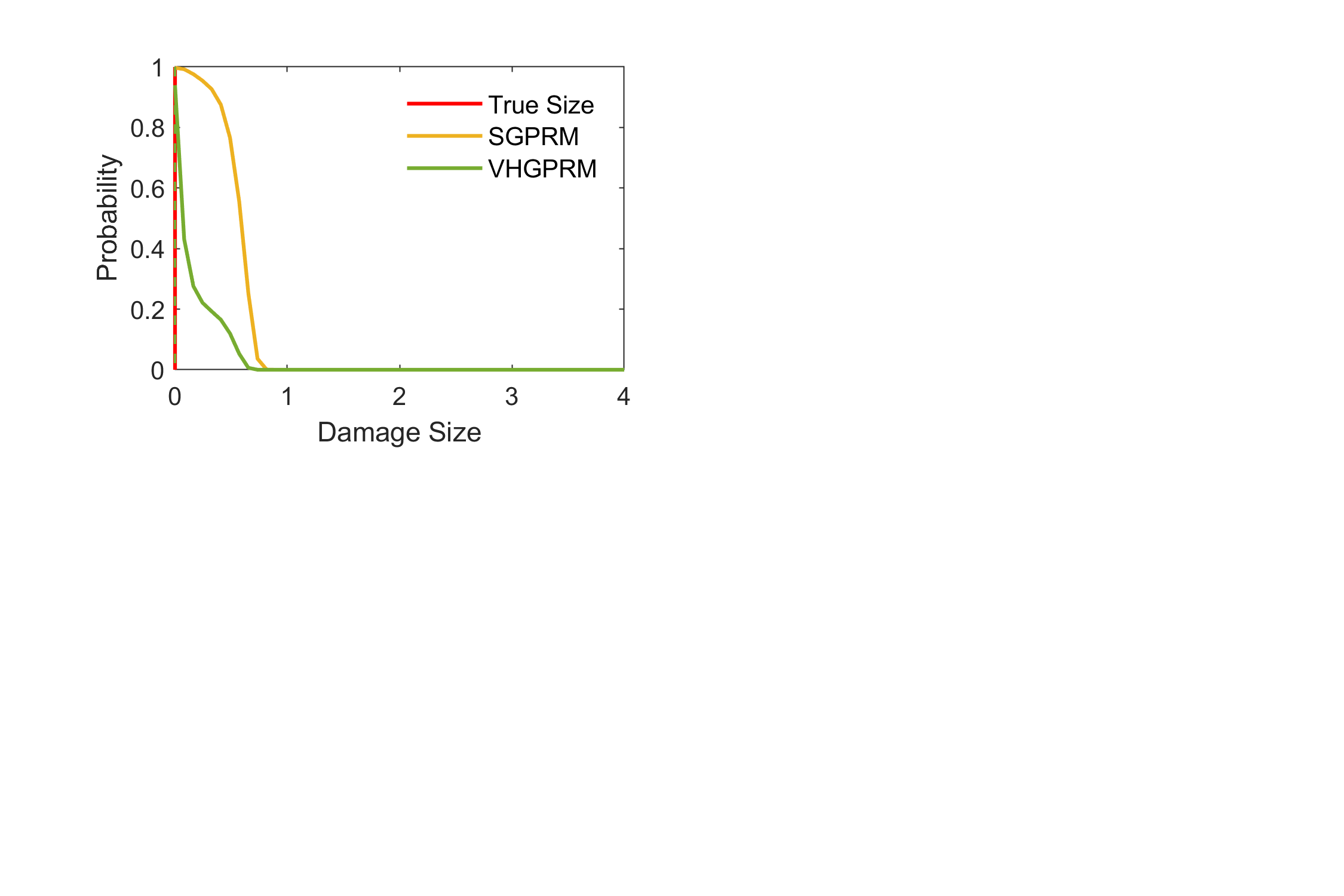}}
    \put(190,40){\includegraphics[trim = 20 0 20 15,clip,scale=0.8]{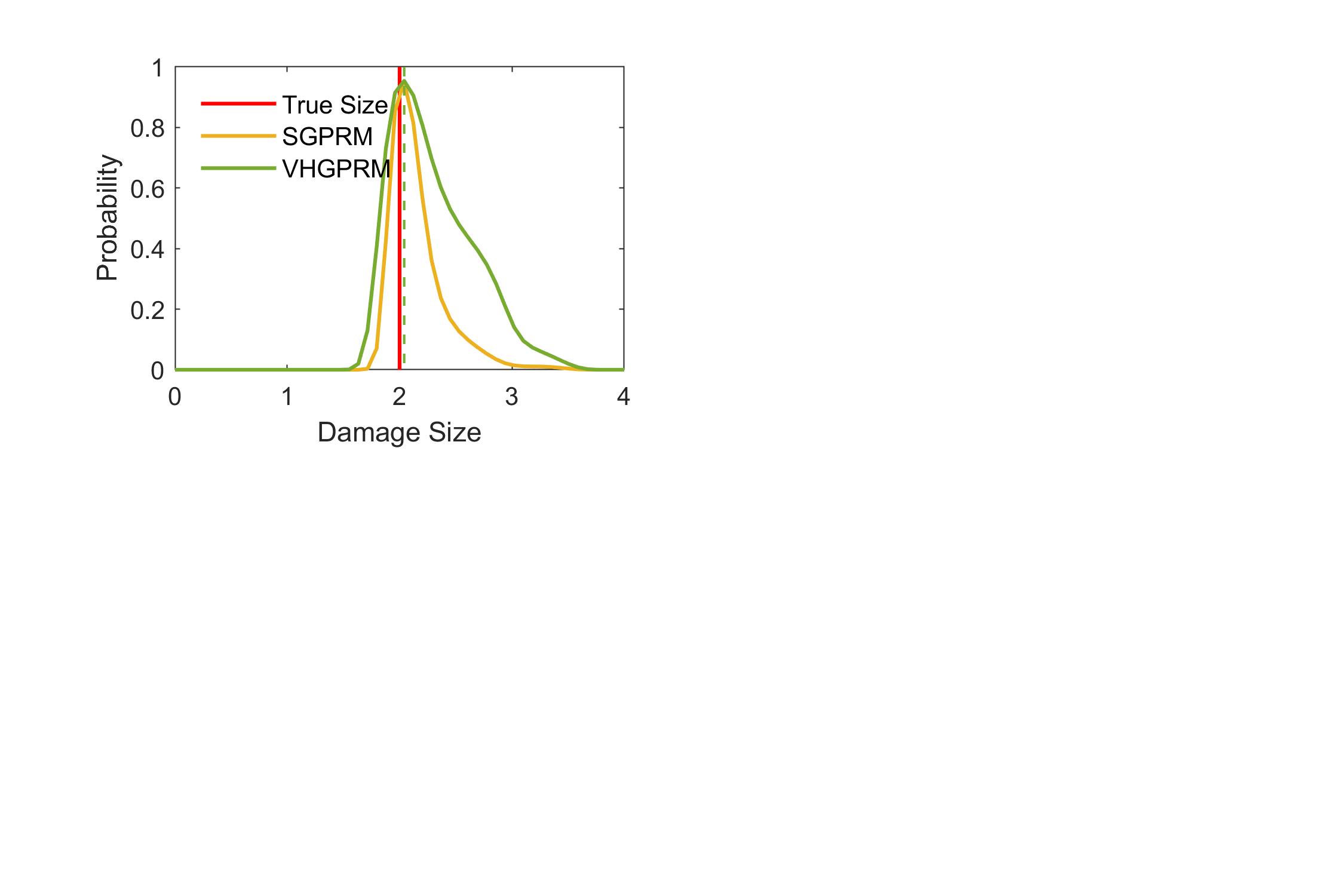}}
    \put(0,-95){\includegraphics[trim = 20 0 20 15,clip,scale=0.8]{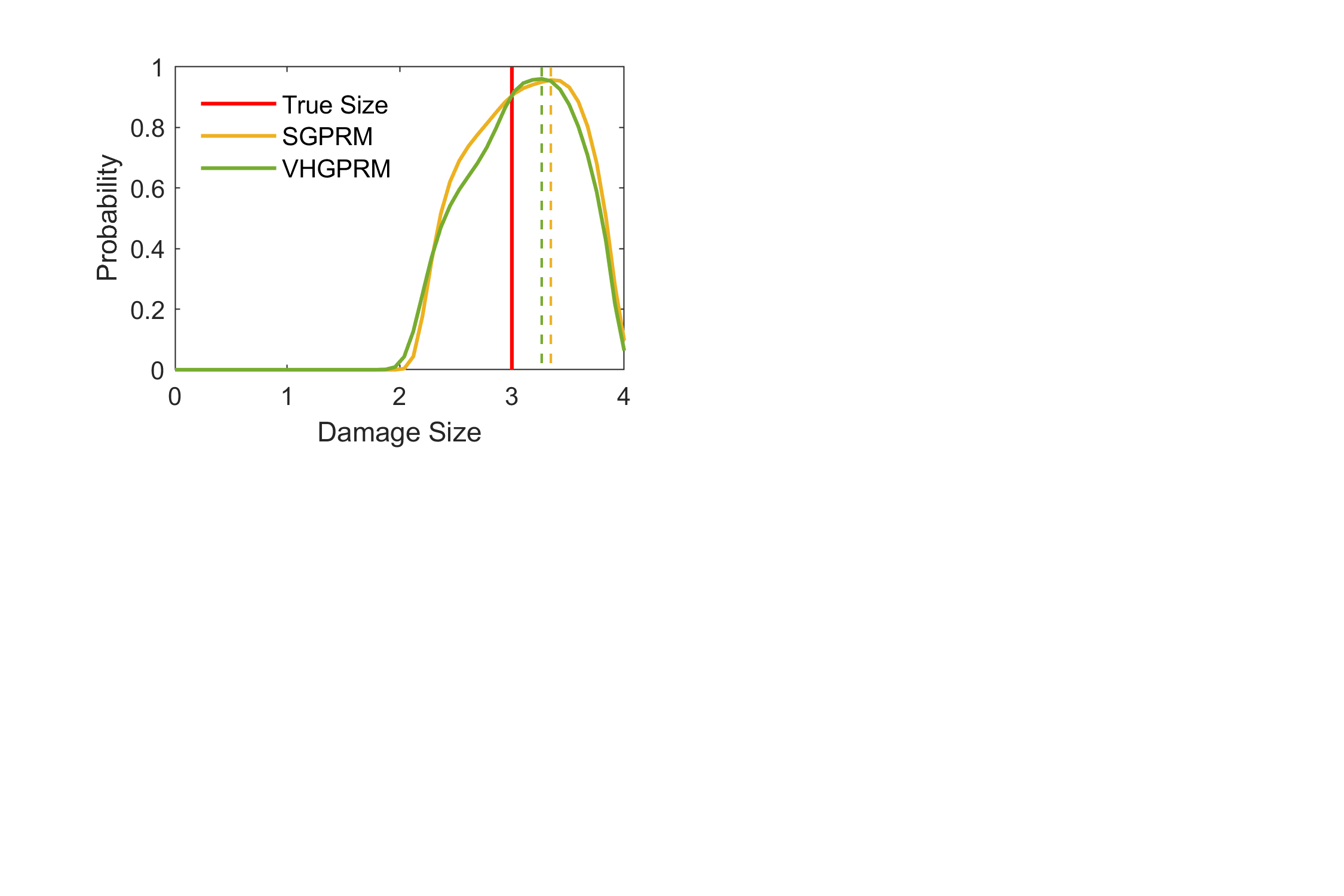}}
   \put(190,-95){\includegraphics[trim = 20 0 20 15,clip,scale=0.8]{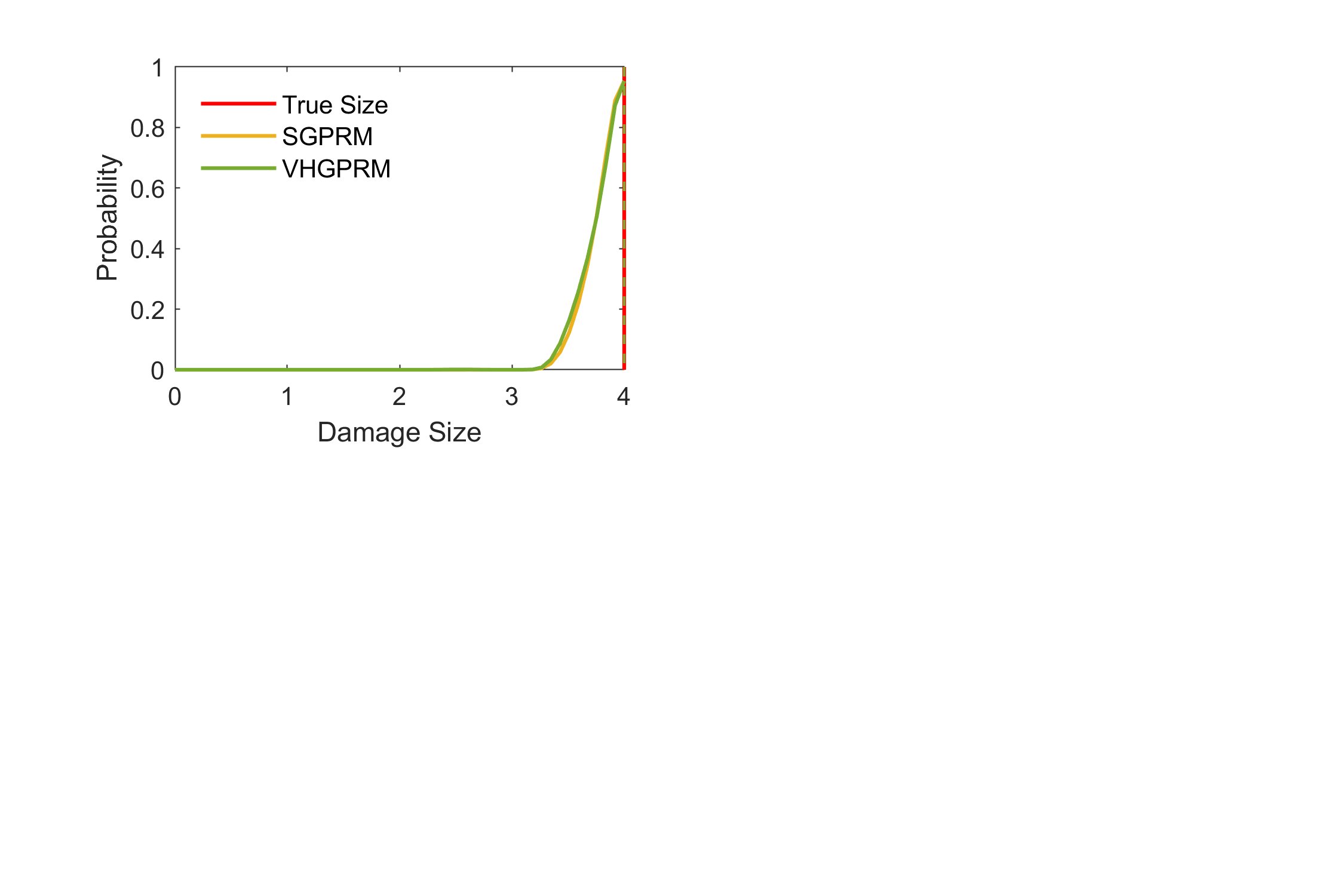}}
   \put(45,295){\color{black} \large {\fontfamily{phv}\selectfont \textbf{a}}}
    \put(235,295){\large {\fontfamily{phv}\selectfont \textbf{b}}}
   \put(45,160){\large {\fontfamily{phv}\selectfont \textbf{c}}} 
   \put(235,160){\large {\fontfamily{phv}\selectfont \textbf{d}}} 
    \end{picture} \vspace{-55pt}
    \caption{Al coupon with simulated damage: damage size prediction results for path 1-6 at the unloaded condition: (a) prediction probabilities for the healthy case; (b) prediction probabilities for the case of 1 attached weight; (c) prediction probabilities for the case of 3 attached weights; (d) prediction probabilities for the case of 4 attached weights.} 
\label{fig:instron_prob_1-6_janapati} \vspace{5pt}
\end{figure}
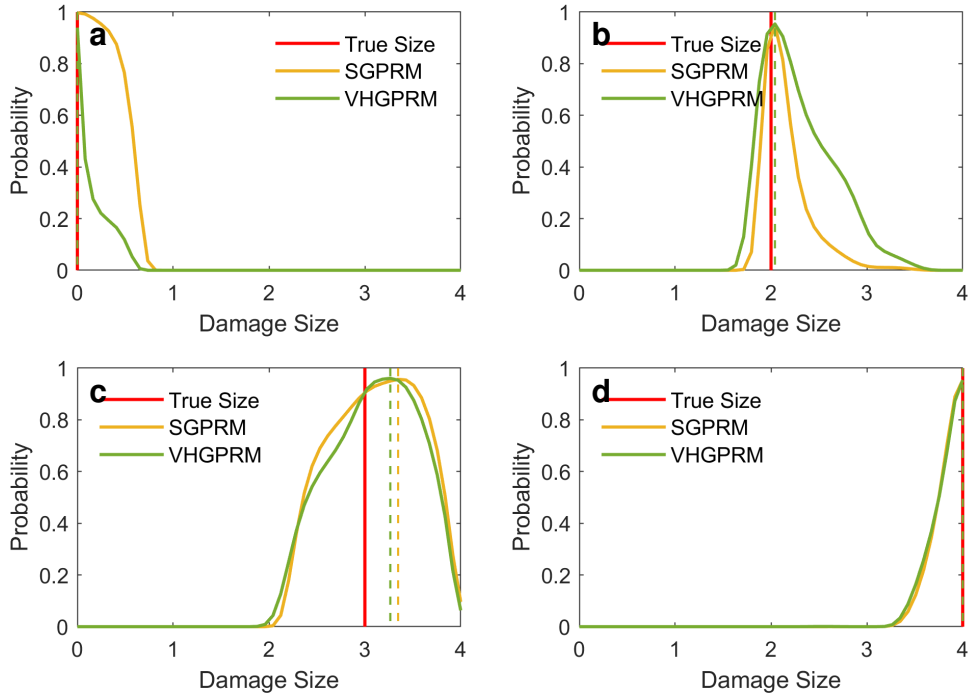

\begin{figure}[t!]
    \centering
    \begin{picture}(400,300)
    \put(0,40){\includegraphics[trim = 20 0 20 15,clip,scale=0.8]{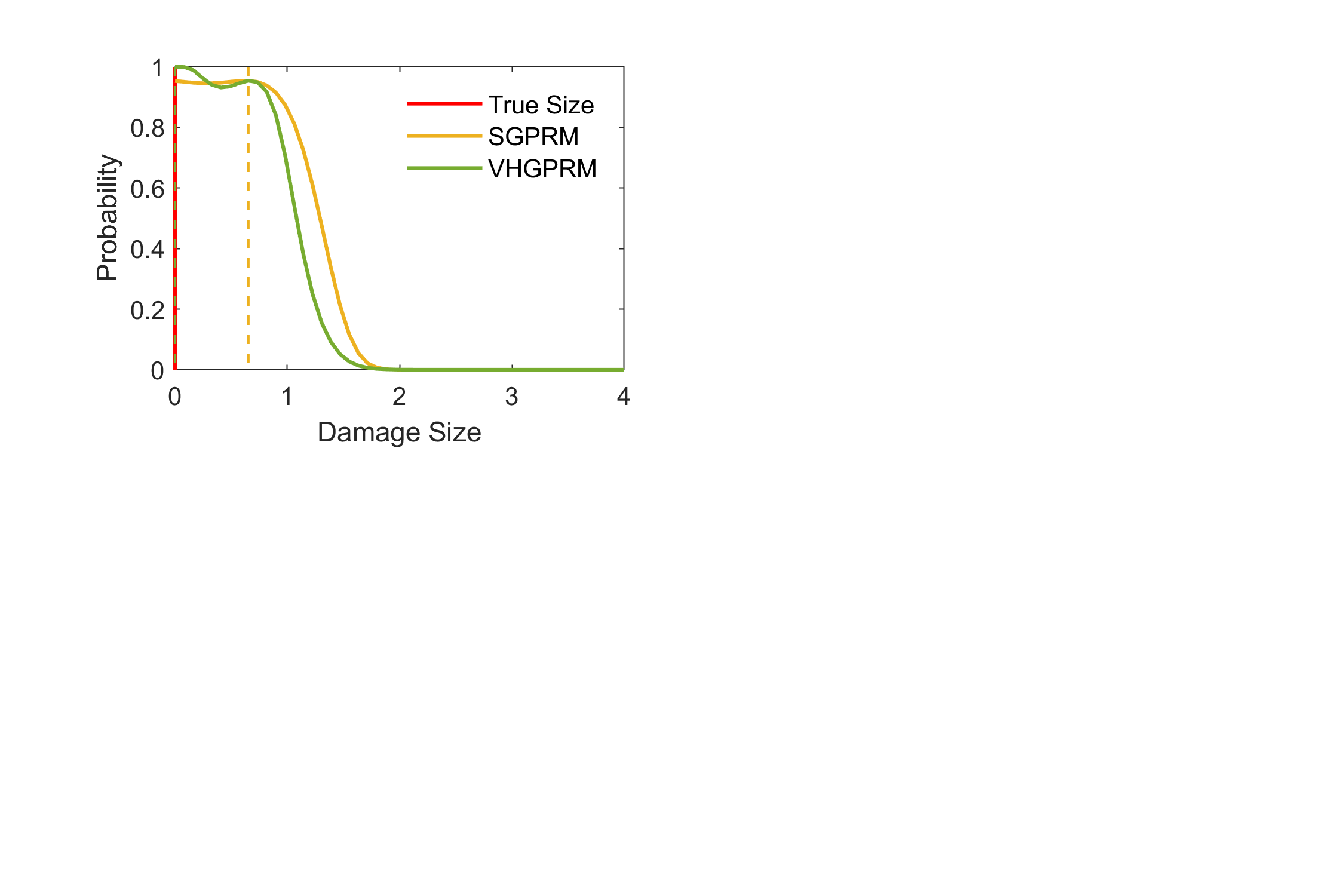}}
    \put(190,40){\includegraphics[trim = 20 0 20 15,clip,scale=0.8]{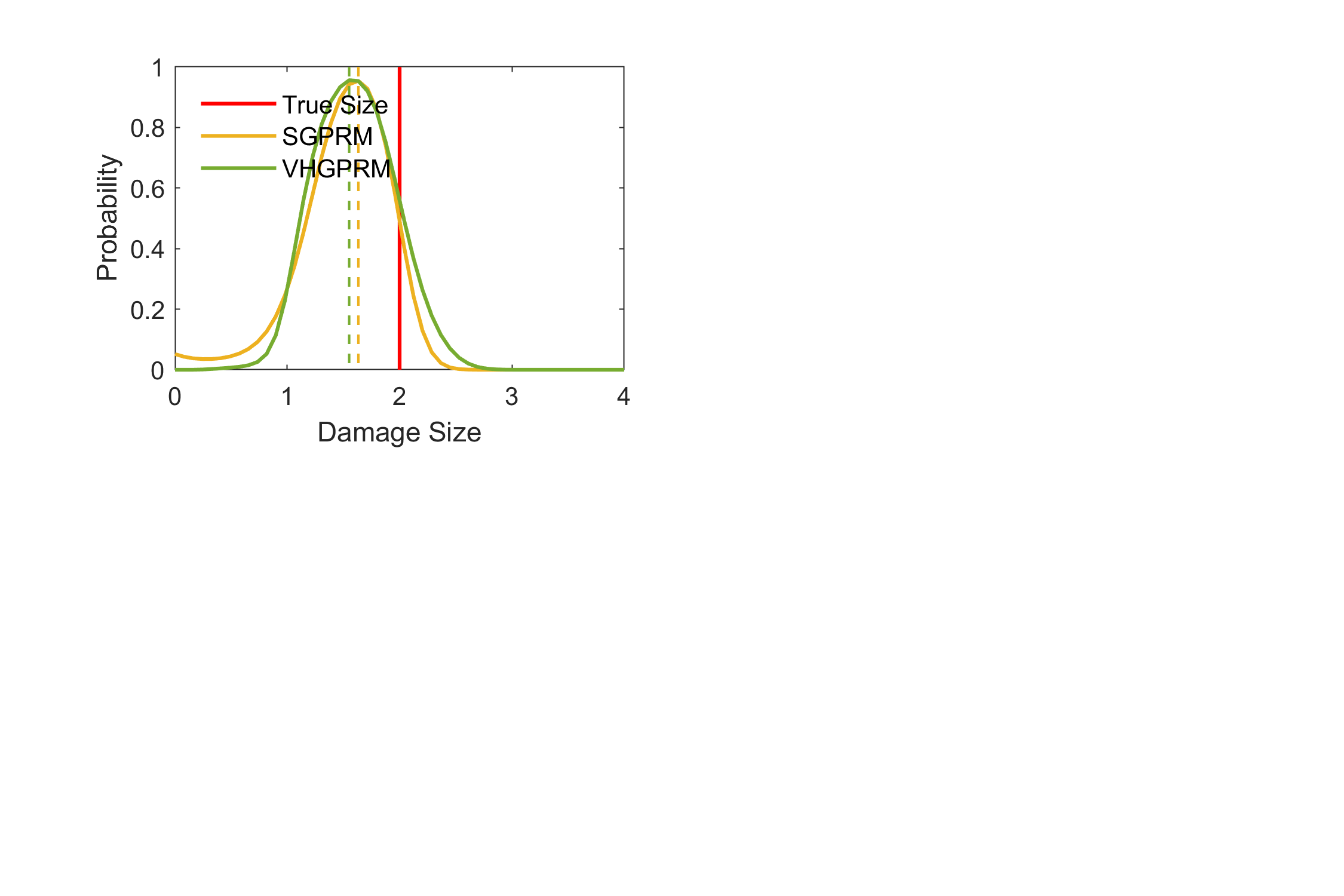}}
    \put(0,-95){\includegraphics[trim = 20 0 20 15,clip,scale=0.8]{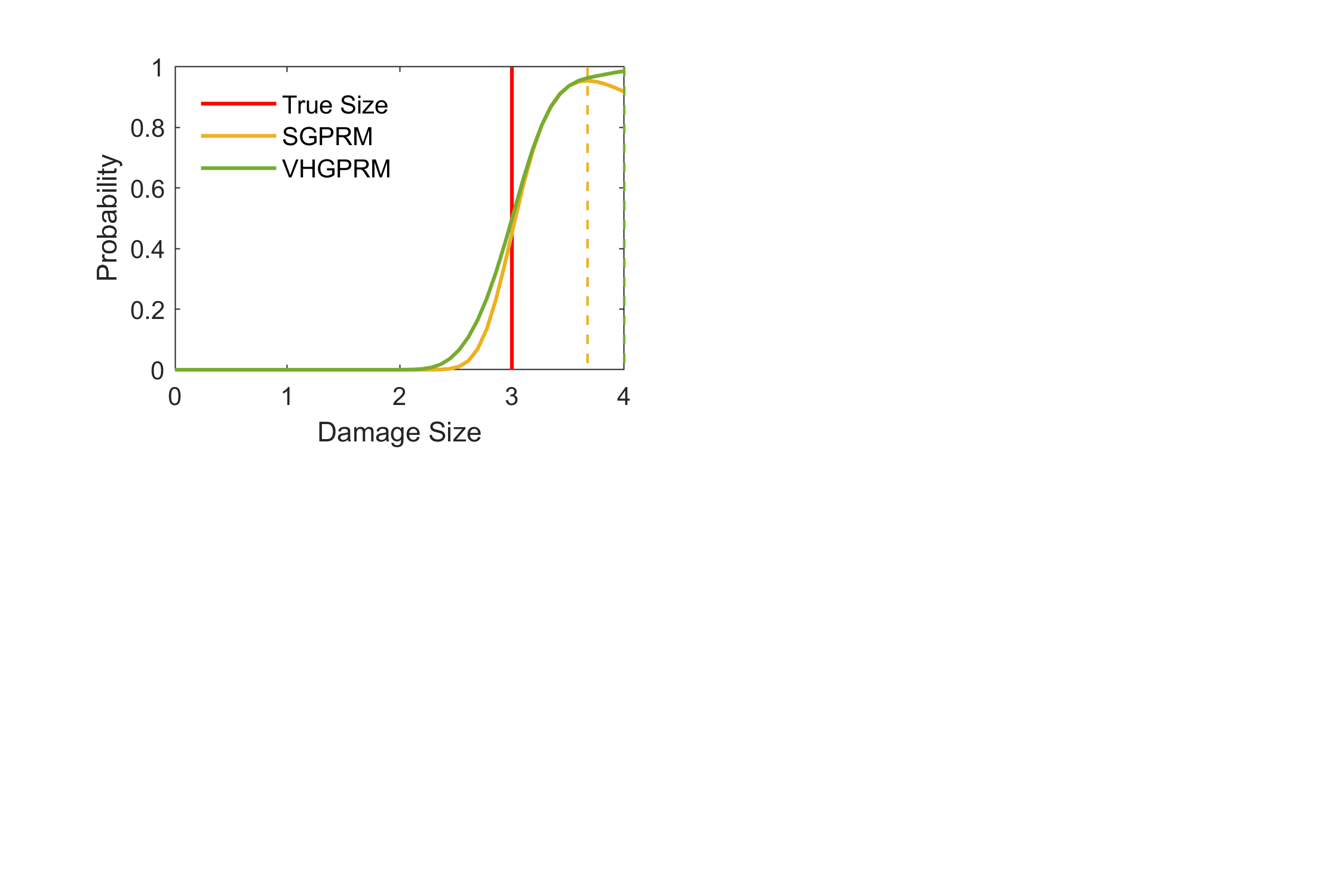}}
   \put(190,-95){\includegraphics[trim = 20 0 20 15,clip,scale=0.8]{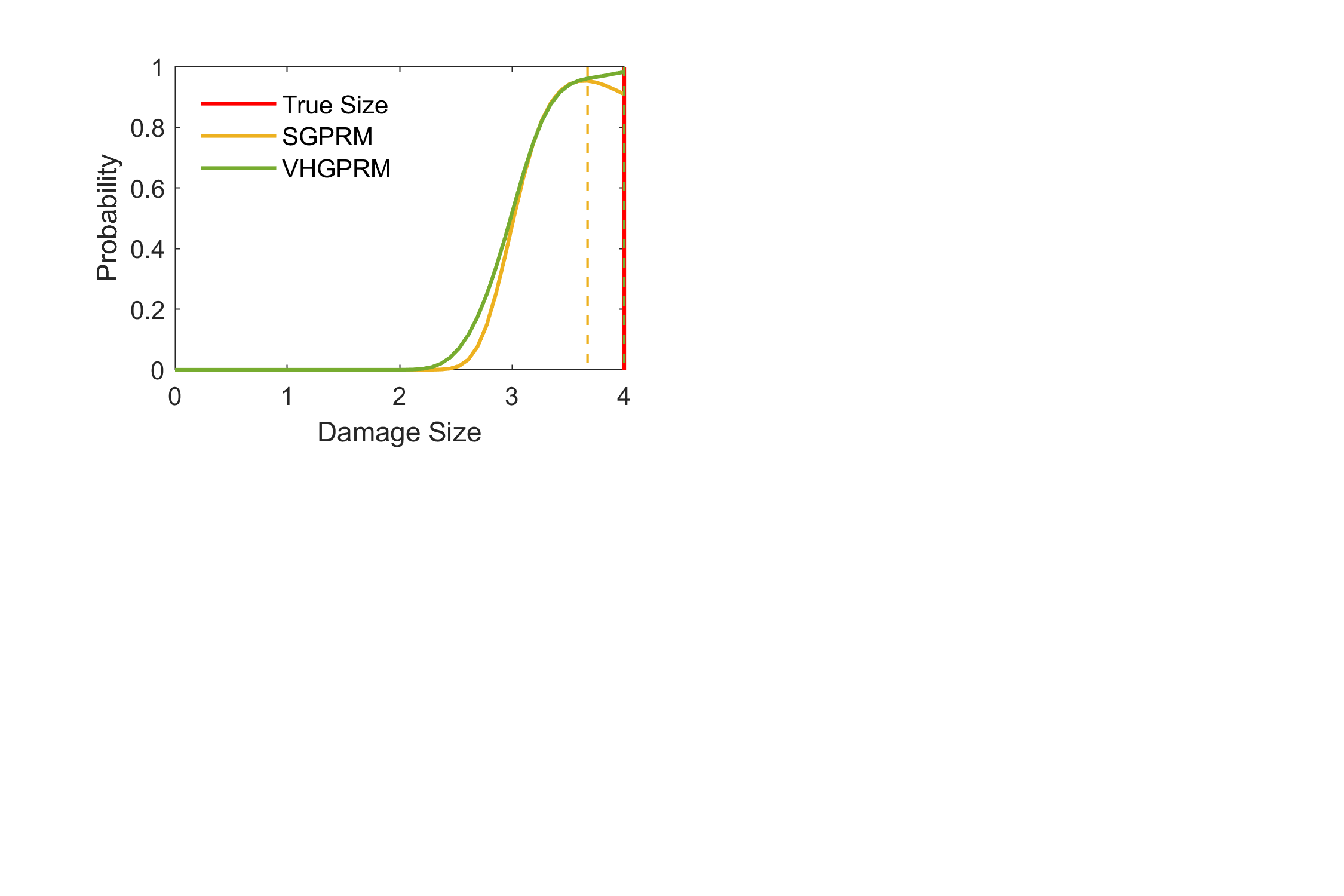}}
   \put(45,295){\color{black} \large {\fontfamily{phv}\selectfont \textbf{a}}}
    \put(235,295){\large {\fontfamily{phv}\selectfont \textbf{b}}}
   \put(45,160){\large {\fontfamily{phv}\selectfont \textbf{c}}} 
   \put(235,160){\large {\fontfamily{phv}\selectfont \textbf{d}}} 
    \end{picture} \vspace{-55pt}
    \caption{Al coupon with simulated damage: damage size prediction results for path 3-6 under 15 kN: (a) prediction probabilities for the healthy case; (b) prediction probabilities for the case of 1 attached weight; (c) prediction probabilities for the case of 3 attached weights; (d) prediction probabilities for the case of 4 attached weights.} 
\label{fig:instron_prob_3-6_janapati} \vspace{5pt}
\end{figure}

Figure \ref{fig:instron_prob_1-6_janapati} shows the prediction probabilities extracted from the GPRMs shown in Figure \ref{fig:instron_gprm_janapati} corresponding to 4 indicative test DI points not used in the training process from both SGPRMs and VHGPRMs for path 1-6. As shown, asside from a slight deviation in the prediction of both model types for the case of 3 attached weights, both models accurately predict damage size. In addition, as shown in Figure \ref{fig:instron_prob_1-6_janapati} panels a and b, the VHGPRM more accurately estimates the prediction probabilities as a bit narrow (panel a) or broad (panel b) given the respective change in variance in the training DI values at the healthy state, as well as the damage state of 2 attached weights. Examining the results for path 3-6 (Figure \ref{fig:instron_prob_3-6_janapati}, the superiority of VHGPRMs can be clearly shown here with more accurate predictions in the healthy case, as well as the case of 4 attached weights. Both models however fail to accurately quantify the case of three attached weights (Figure \ref{fig:instron_prob_3-6_janapati}c), while both of them fall just short of accurately predicting the case of 2 attached weights (Figure \ref{fig:instron_prob_3-6_janapati}b). This is attributed again to the immense overlap between the DI values at these two damage cases.

\begin{figure}[t!]
    \centering
    \begin{picture}(400,300)
    \put(0,70){\includegraphics[trim = 20 35 20 25,clip,scale=0.8]{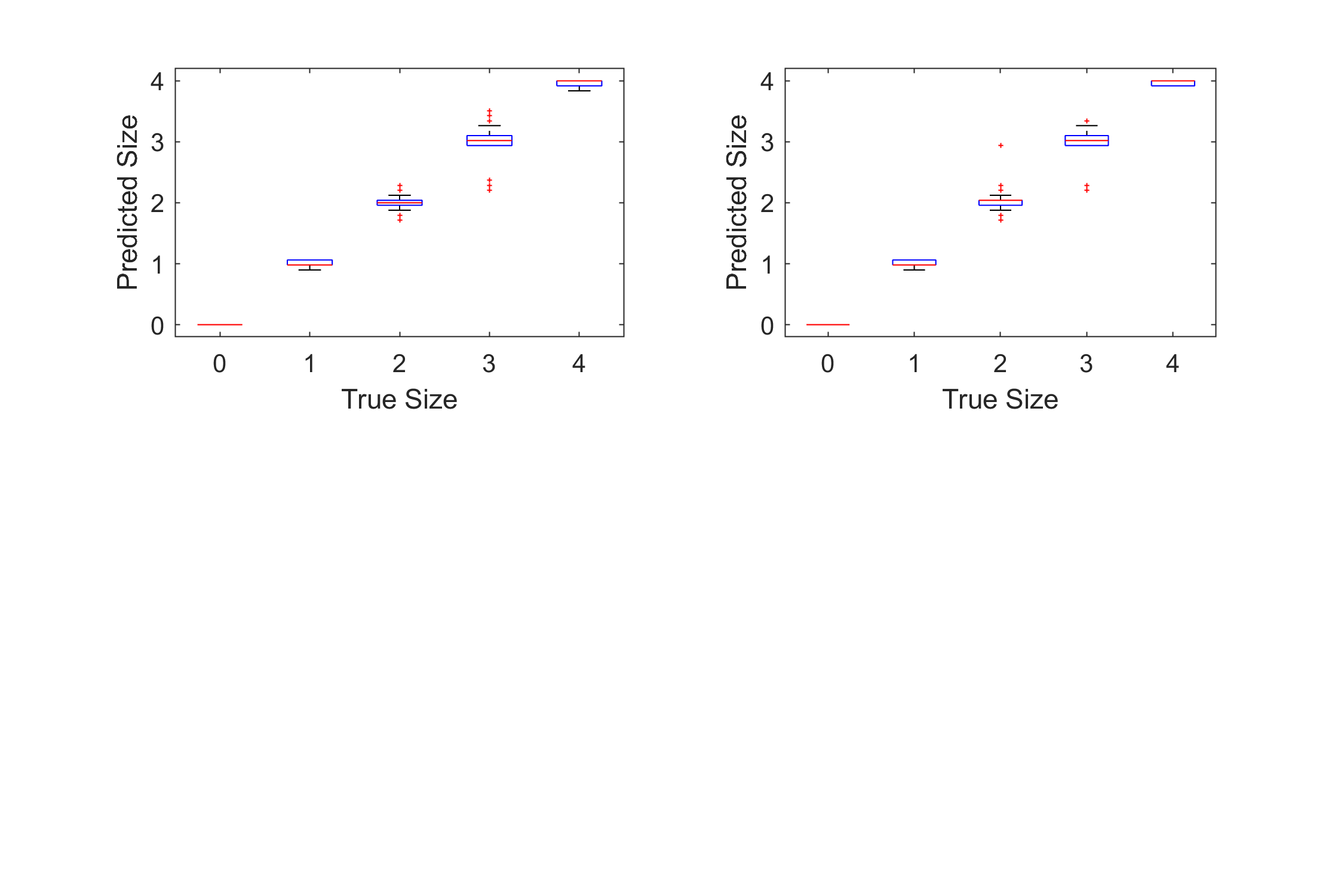}}
    \put(0,-55){\includegraphics[trim = 20 35 20 25,clip,scale=0.8]{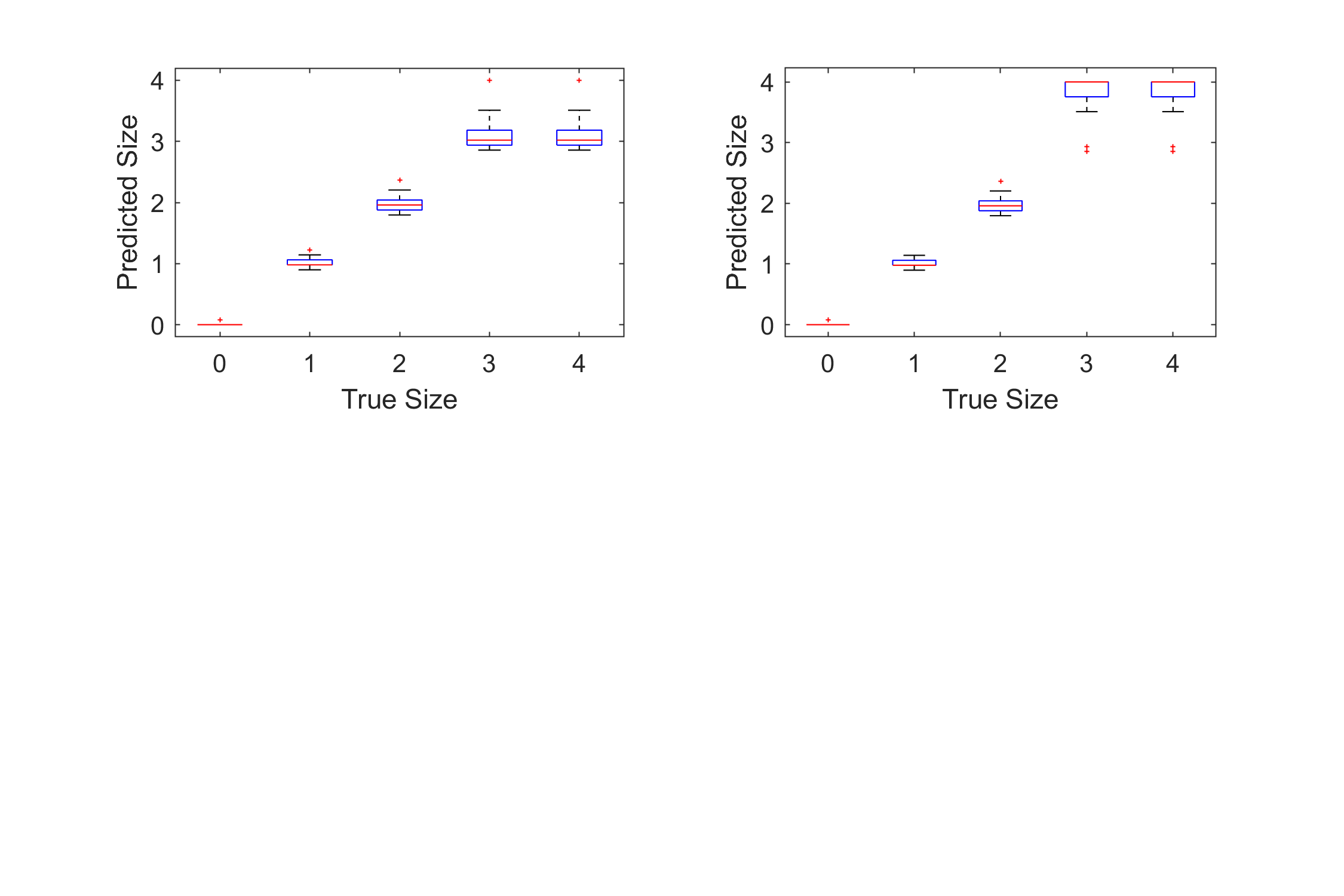}}
    \put(0,-180){\includegraphics[trim = 20 35 20 25,clip,scale=0.8]{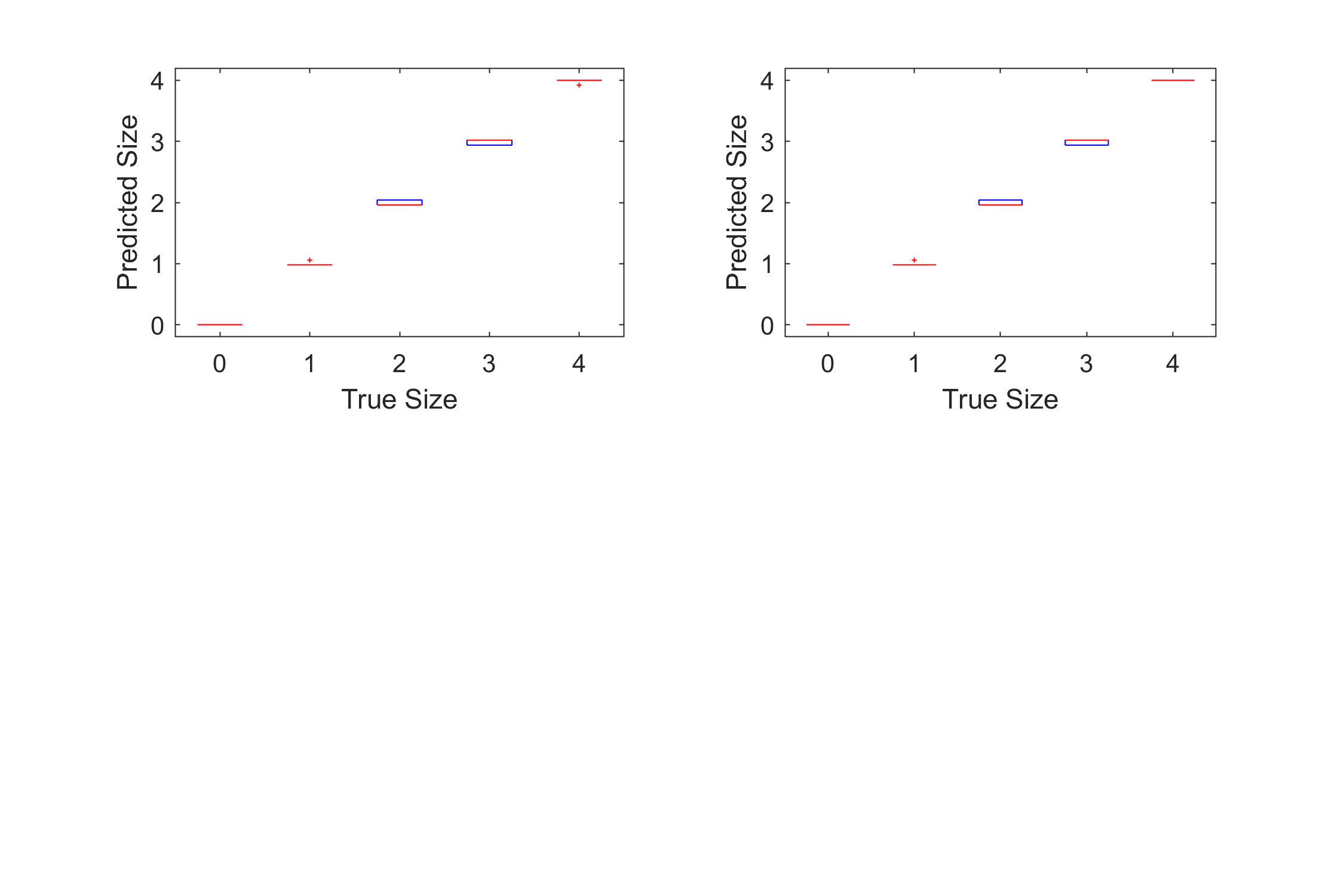}}
    \put(0,-305){\includegraphics[trim = 20 35 20 25,clip,scale=0.8]{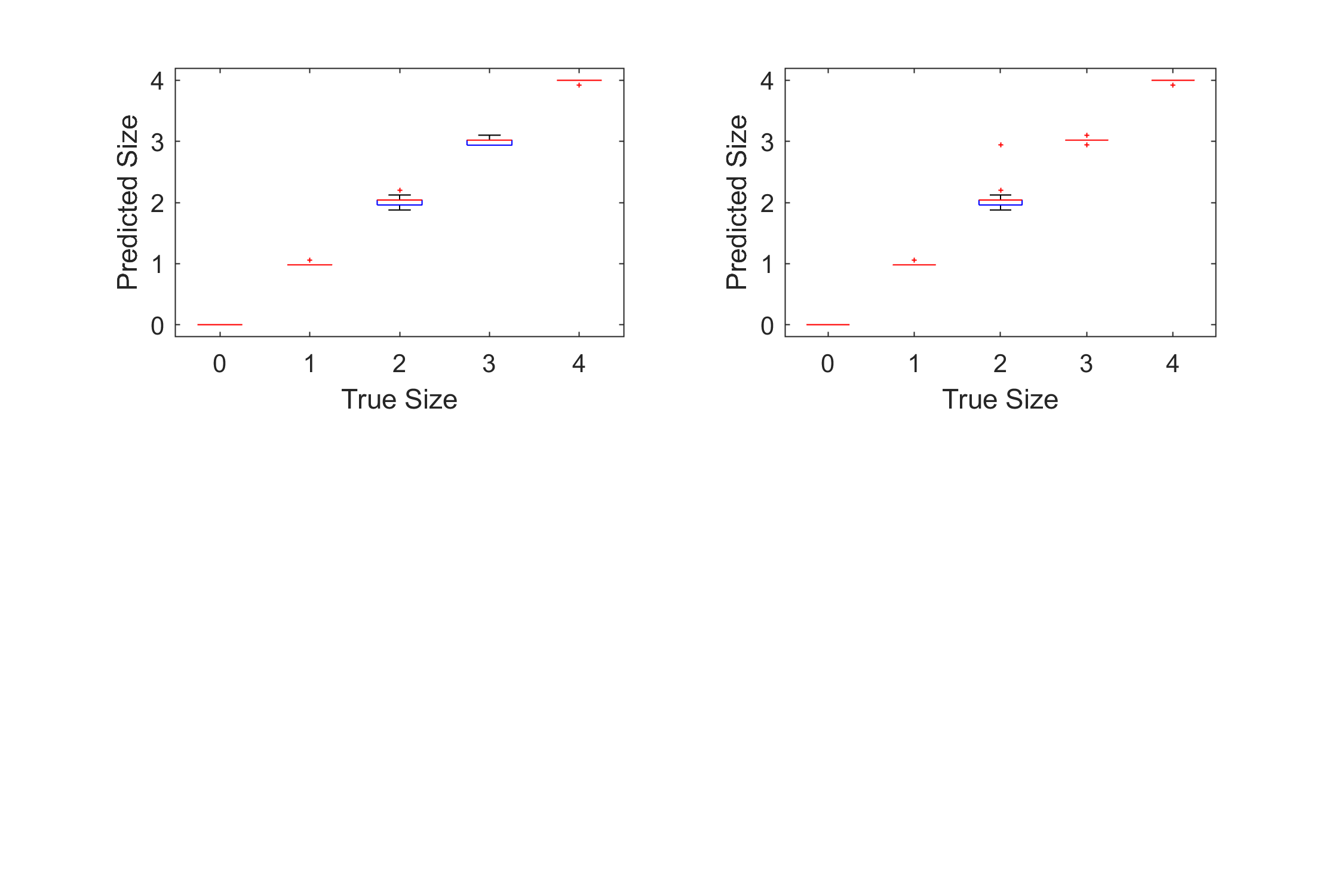}}
   \put(47,295){\color{black} \large {\fontfamily{phv}\selectfont \textbf{a}}}
    \put(245,295){\large {\fontfamily{phv}\selectfont \textbf{b}}}
   \put(47,170){\large {\fontfamily{phv}\selectfont \textbf{c}}} 
   \put(245,170){\large {\fontfamily{phv}\selectfont \textbf{d}}} 
   \put(47,45){\large {\fontfamily{phv}\selectfont \textbf{e}}} 
   \put(245,45){\large {\fontfamily{phv}\selectfont \textbf{f}}} 
   \put(47,-80){\large {\fontfamily{phv}\selectfont \textbf{g}}} 
   \put(245,-80){\large {\fontfamily{phv}\selectfont \textbf{h}}} 
    \end{picture} \vspace{170pt}
    \caption{Al coupon with simulated damage: true/predicted damage size boxplots for path 1-6: (a) SGPRM state predictions at 0 kN; (b) VHGPRM state predictions at 0 kN; (c) SGPRM state predictions at 5 kN; (d) VHGPRM state predictions at 5 kN; (e) SGPRM state predictions at 10 kN; (f) VHGPRM predictions at 10 kN; (g) SGPRM state predictions at 15 kN; (h) VHGPRM state predictions at 15 kN.} 
\label{fig:instron_1-6_boxplot_janapati} 
\end{figure}

In order to analyze the SGPRMs and VHGPRMs in damage quantification for the Al coupon with simulated damage, summary box-plots were plotted for both paths under the different loading states exhibited by the coupon in this study. Figure \ref{fig:instron_1-6_boxplot_janapati} presents the summary results for path 1-6. Examining the different panels across each loading state, it can be observed that both model types perform just as well in damage size quantification. In addition, all predictions for all test points under all loading states are accurate, except for the 5-kN case for the damage sizes of 3 and 4 attached weights, where it can be observed that the SGPRMs underestimate the latter damage size, while the VHGPRMs overestimate the former.

\begin{figure}[t!]
    \centering
    \begin{picture}(400,300)
    \put(0,70){\includegraphics[trim = 20 35 20 25,clip,scale=0.8]{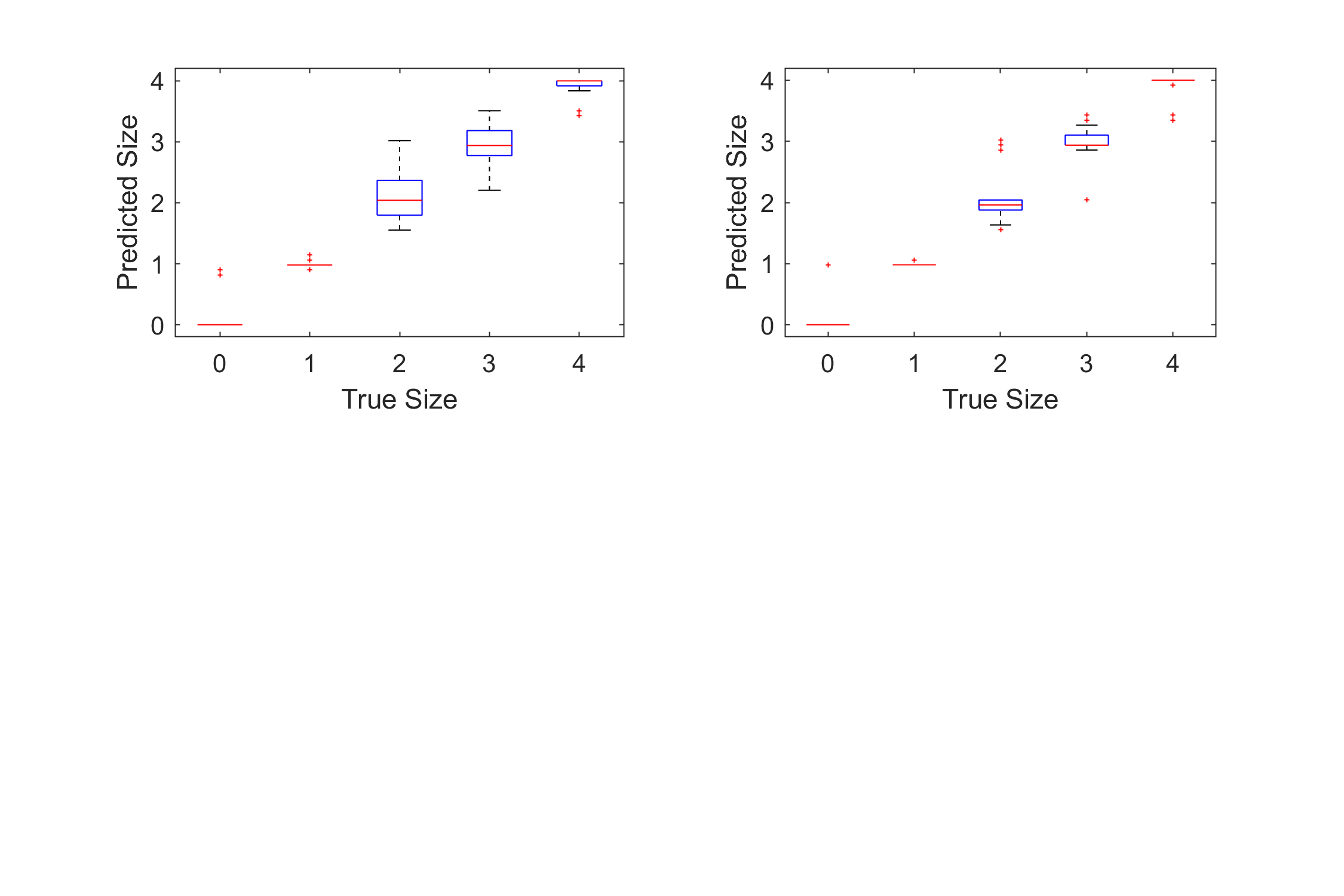}}
    \put(0,-55){\includegraphics[trim = 20 35 20 25,clip,scale=0.8]{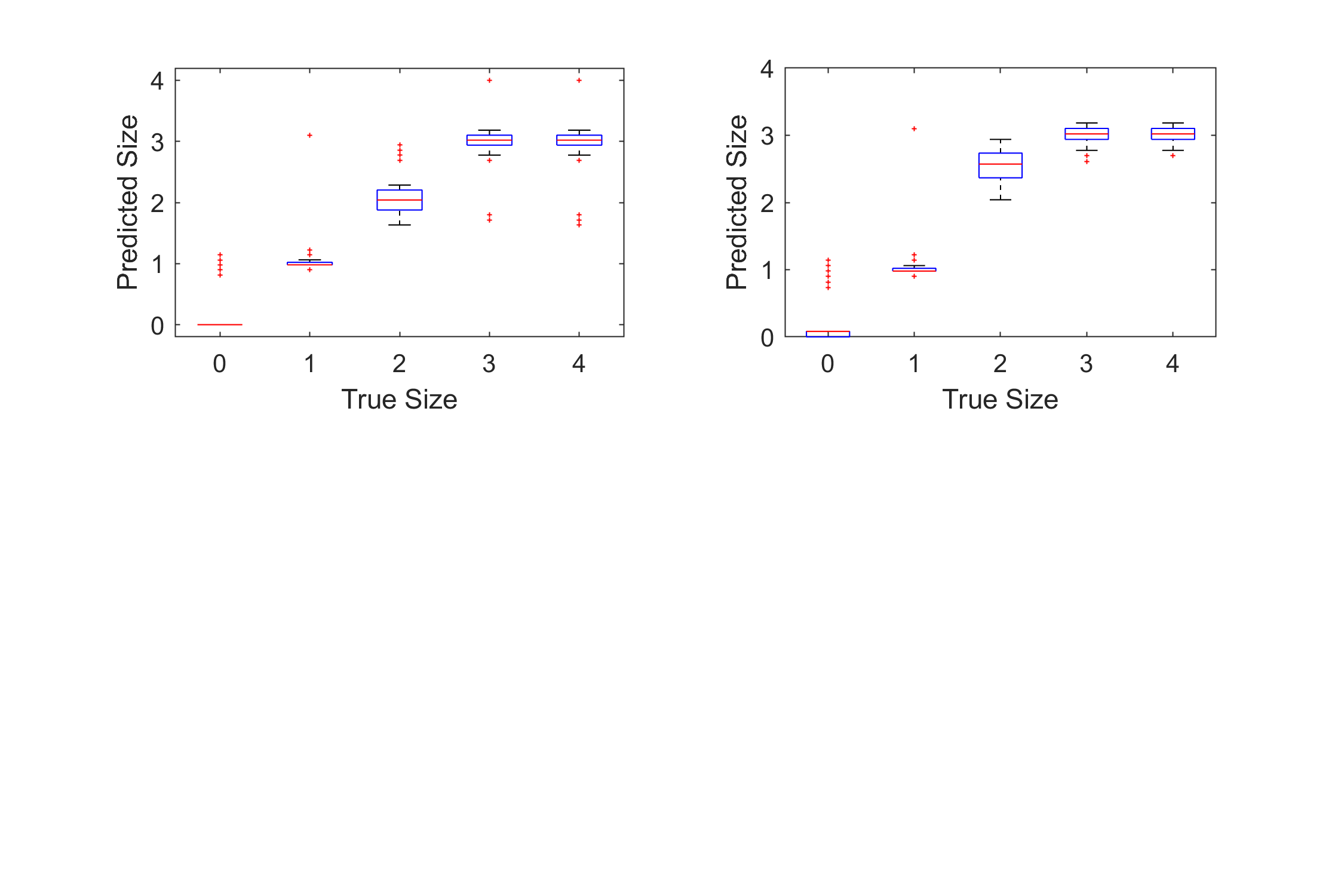}}
    \put(0,-180){\includegraphics[trim = 20 35 20 25,clip,scale=0.8]{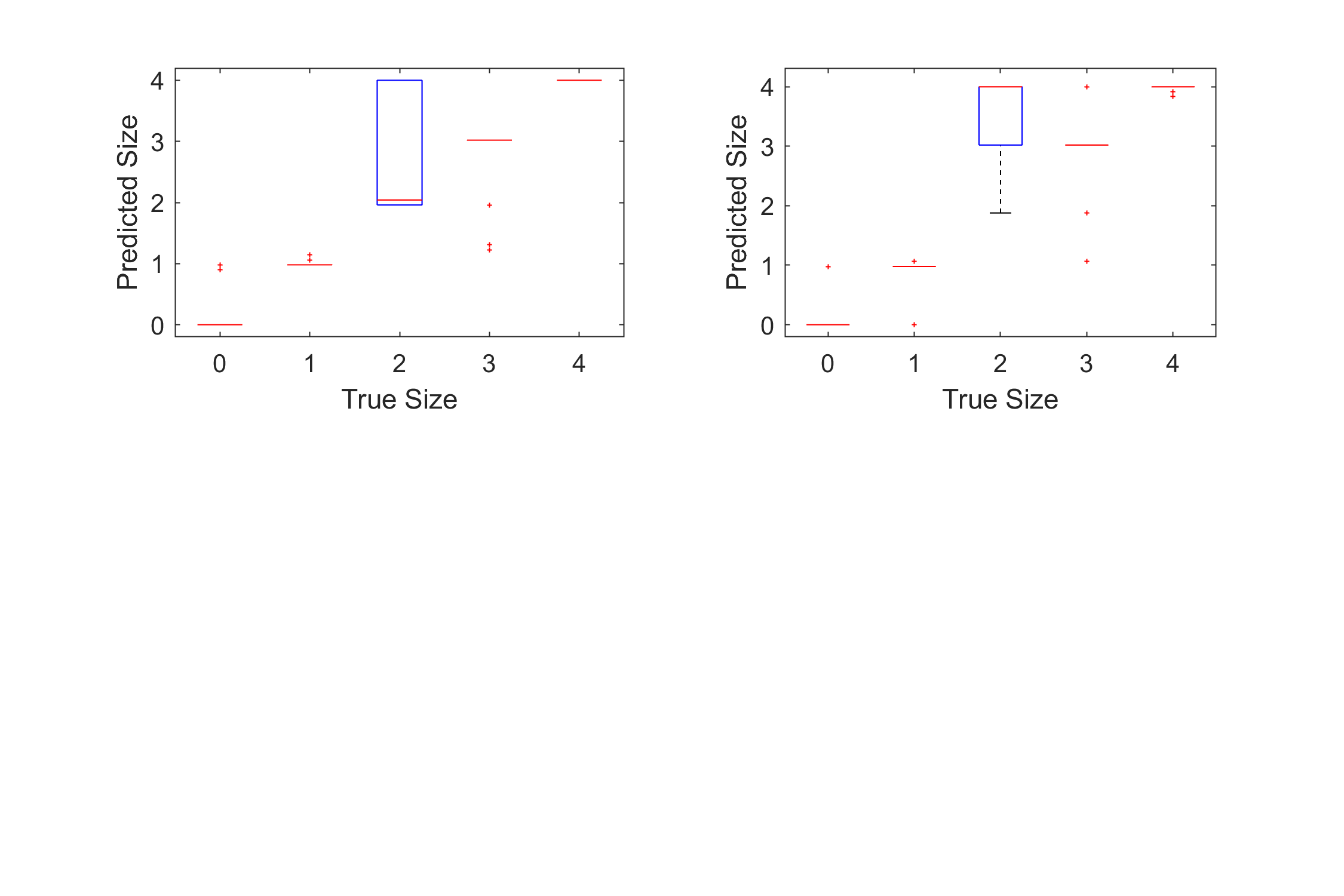}}
    \put(0,-305){\includegraphics[trim = 20 35 20 25,clip,scale=0.8]{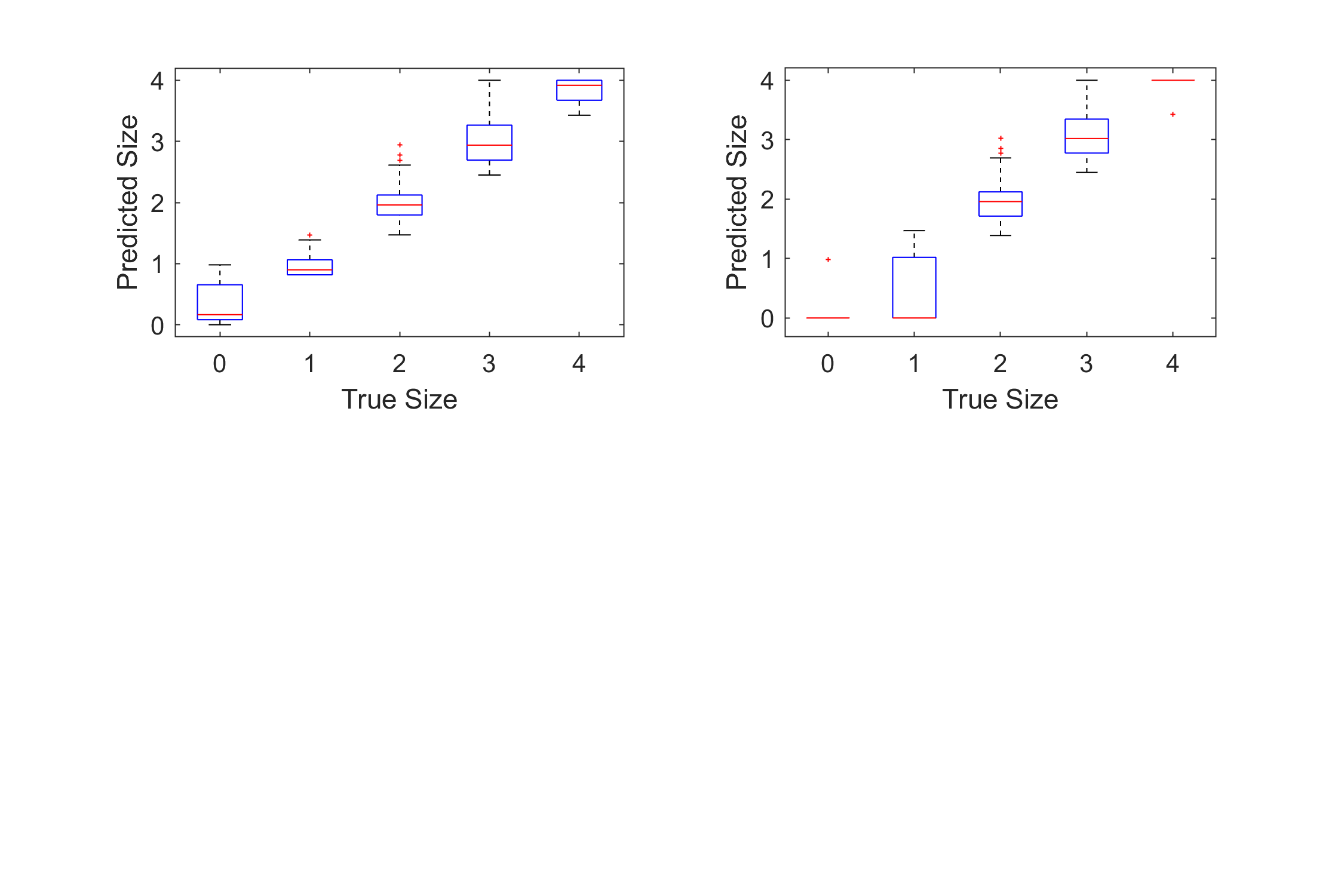}}
   \put(47,295){\color{black} \large {\fontfamily{phv}\selectfont \textbf{a}}}
    \put(245,295){\large {\fontfamily{phv}\selectfont \textbf{b}}}
   \put(47,170){\large {\fontfamily{phv}\selectfont \textbf{c}}} 
   \put(245,170){\large {\fontfamily{phv}\selectfont \textbf{d}}} 
   \put(47,45){\large {\fontfamily{phv}\selectfont \textbf{e}}} 
   \put(245,45){\large {\fontfamily{phv}\selectfont \textbf{f}}} 
   \put(47,-80){\large {\fontfamily{phv}\selectfont \textbf{g}}} 
   \put(245,-80){\large {\fontfamily{phv}\selectfont \textbf{h}}} 
    \end{picture} \vspace{170pt}
    \caption{Al coupon with simulated damage: true/predicted damage size boxplots for path 3-6: (a) SGPRM state predictions at 0 kN; (b) VHGPRM state predictions at 0 kN; (c) SGPRM state predictions at 5 kN; (d) VHGPRM state predictions at 5 kN; (e) SGPRM state predictions at 10 kN; (f) VHGPRM state predictions at 10 kN; (g) SGPRM state predictions at 15 kN; (h) VHGPRM state predictions at 15 kN.} 
\label{fig:instron_3-6_boxplot_janapati} 
\end{figure}

Looking at the summary results from path 3-6 (Figure \ref{fig:instron_3-6_boxplot_janapati}), panels a and b show the results for the unloadeding condition. It can be seen that the VHGPRM outperforms the SGPRM here with narrower prediction ranges for all damage sizes. Panels c and d show the prediction results for the 5-kN case. As shown, both models underestimate the case of 4 attached weights, with the SGPRM outperforming the VHGPRM in predicting the case of 2 attached weights. Examining the loading case of 10 kN, it can be observed that both models perform well in damage size quantification except when 2 weights are attached to the Al coupon, in which case the SGPRM seem to be a bit more accurate with the median prediction around the true damage size. Finally, when 15 kN are applied, although the VHGPRM outperforms the SGPRM in model accuracy in the healthy, as well as the case of 4 attached weights, the SGPRM outperforms the VHGPRM models when 1 weight is attached to the coupon. Again, in all cases, both models seem to perform well in the task of damage size quantification in this coupon.

\begin{table}[b] \vspace{-12pt}
\centering
\caption{Summary of multi-input GPRM$^*$ information$^\dagger$ for the Al coupon with simulated damage.}\label{tab:instron_2D}
\renewcommand{\arraystretch}{1.2}
{\footnotesize
\begin{tabular}{|c|c|c|c|c|c|c|c|c|} 
\hline
Signal & \multicolumn{2}{c}{NMSE} & \multicolumn{2}{|c|}{RSS/SSS (\%)} & \multicolumn{2}{c}{Training Time (s)} & \multicolumn{2}{|c|}{Prediction Time (s)} \\ 
\cline{2-9}
 Path & SGPRM & VHGPRM & SGPRM & VHGPRM & SGPRM & VHGPRM & SGPRM & VHGPRM \\
\hline
1-6 & 0.0026 & 0.0026 & 0.12 & 0.12 & 7.3113 & 36.5709 & 0.2571 & 0.8023 \\  
\hline
3-6 & 0.0023 & 0.0023 & 0.14 & 0.15 & 8.2146 & 47.5281 & 0.2943 & 0.5018 \\
\hline
\multicolumn{8}{l}{$^*$10\% (800 points) of the data was used for training each model.} \\
\multicolumn{9}{l}{$^\dagger$Numbers approximated to the last quoted decimal place, and times estimated based on an Intel Core i3 laptop} \\
\multicolumn{8}{l}{  with 4 Gb of RAM.}
\end{tabular}} 
\end{table}
\subsubsection{Multi-input GPRMs}
\begin{figure}[t!]
    \centering
    \begin{picture}(400,300)
    \put(0,40){\includegraphics[trim = 20 0 20 12,clip,scale=0.8]{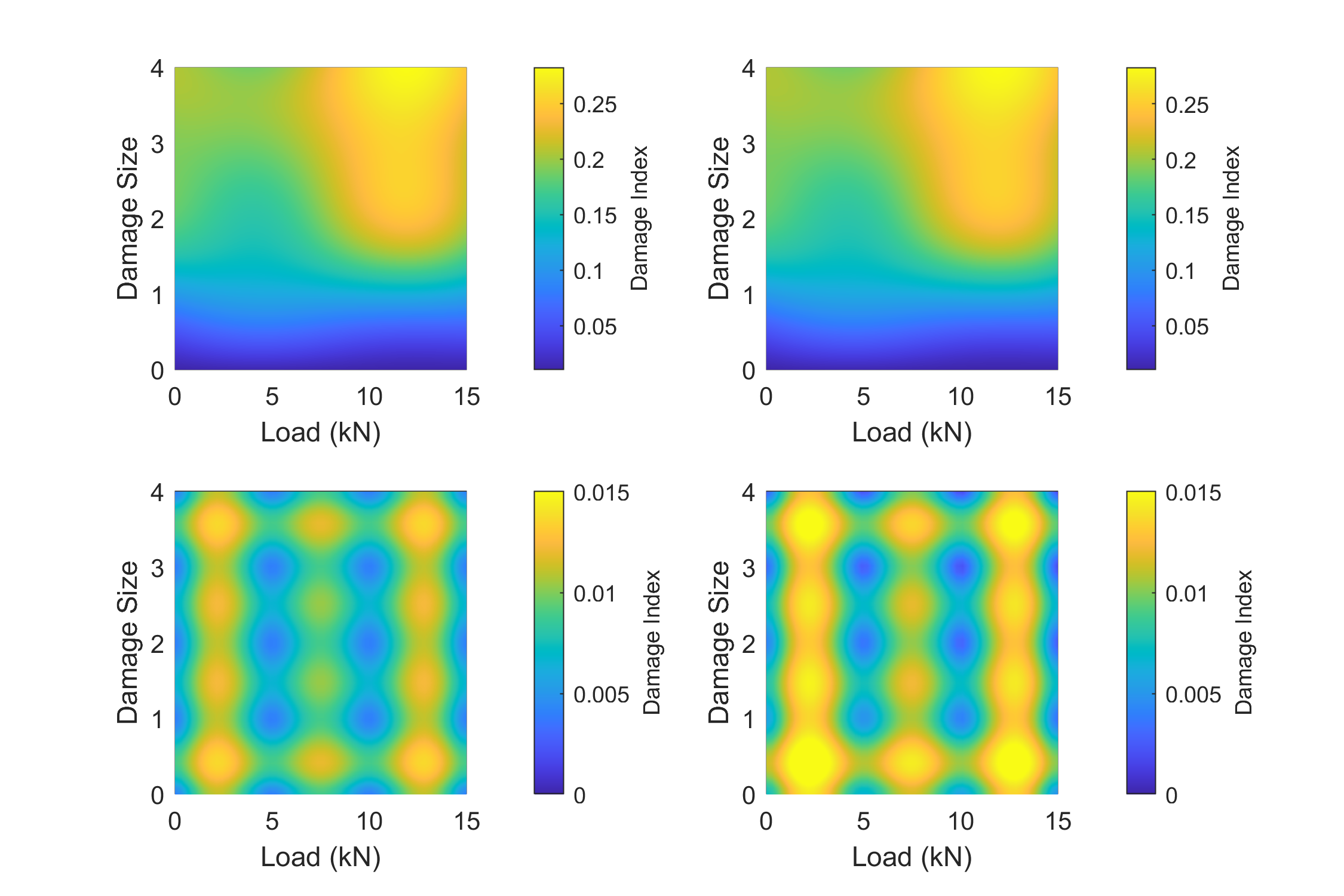}}
    \put(45,295){\color{black} \large {\fontfamily{phv}\selectfont \textbf{a}}}
    \put(235,295){\color{black} \large {\fontfamily{phv}\selectfont \textbf{b}}}
   \put(45,160){\color{black} \large {\fontfamily{phv}\selectfont \textbf{c}}} 
   \put(235,160){\color{black} \large {\fontfamily{phv}\selectfont \textbf{d}}} 
    \end{picture} \vspace{-55pt}
    \caption{Al coupon with simulated damage: Multi-input GPRM predictive mean and standard deviation for path 1-6: (a) SGPRM predictive mean; (b) VHGPRM predictive mean; (c) SGPRM predictive standard deviation; (d) VHGPRM predictive standard deviation.} 
\label{fig:instron_1-6_surf_janapati} \vspace{10pt}
\end{figure}

\begin{figure}[t!]
    \centering
    \begin{picture}(400,300)
    \put(0,40){\includegraphics[trim = 20 0 20 15,clip,scale=0.8]{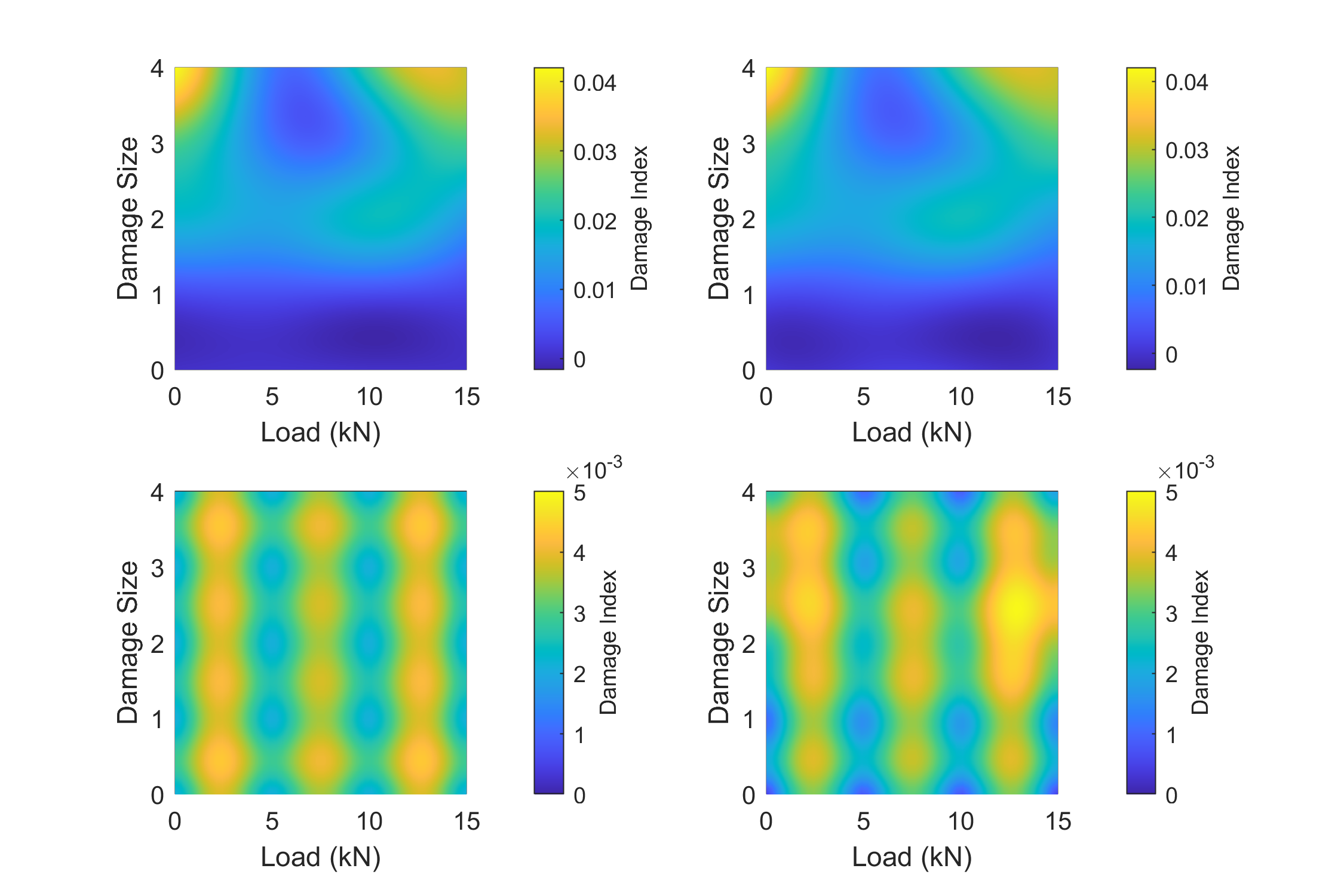}}
    \put(45,295){\color{black} \large {\fontfamily{phv}\selectfont \textbf{a}}}
    \put(235,295){\color{black} \large {\fontfamily{phv}\selectfont \textbf{b}}}
   \put(45,160){\color{black} \large {\fontfamily{phv}\selectfont \textbf{c}}} 
   \put(235,160){\color{black} \large {\fontfamily{phv}\selectfont \textbf{d}}} 
    \end{picture} \vspace{-55pt}
    \caption{Al coupon with simulated damage: Multi-input GPRM predictive mean and standard deviation for path 3-6: (a) SGPRM predictive mean; (b) VHGPRM predictive mean; (c) SGPRM predictive standard deviation; (d) VHGPRM predictive standard deviation..} 
\label{fig:instron_3-6_surf_janapati} \vspace{10pt}
\end{figure}

\begin{figure}[t!]
    \centering
    \begin{picture}(400,300)
    \put(0,40){\includegraphics[trim = 20 0 20 12,clip,scale=0.8]{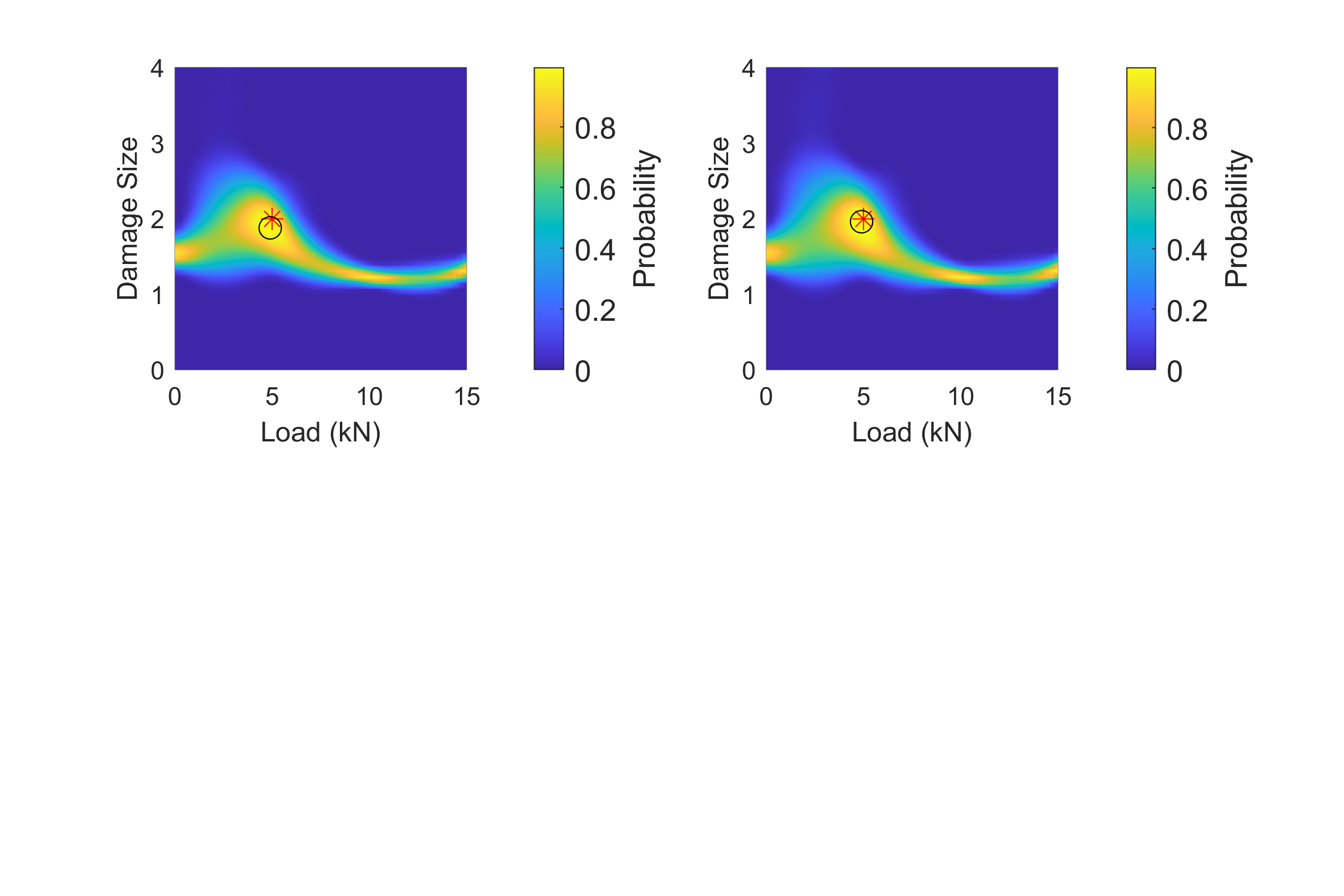}}
    \put(0,-95){\includegraphics[trim = 20 0 20 15,clip,scale=0.8]{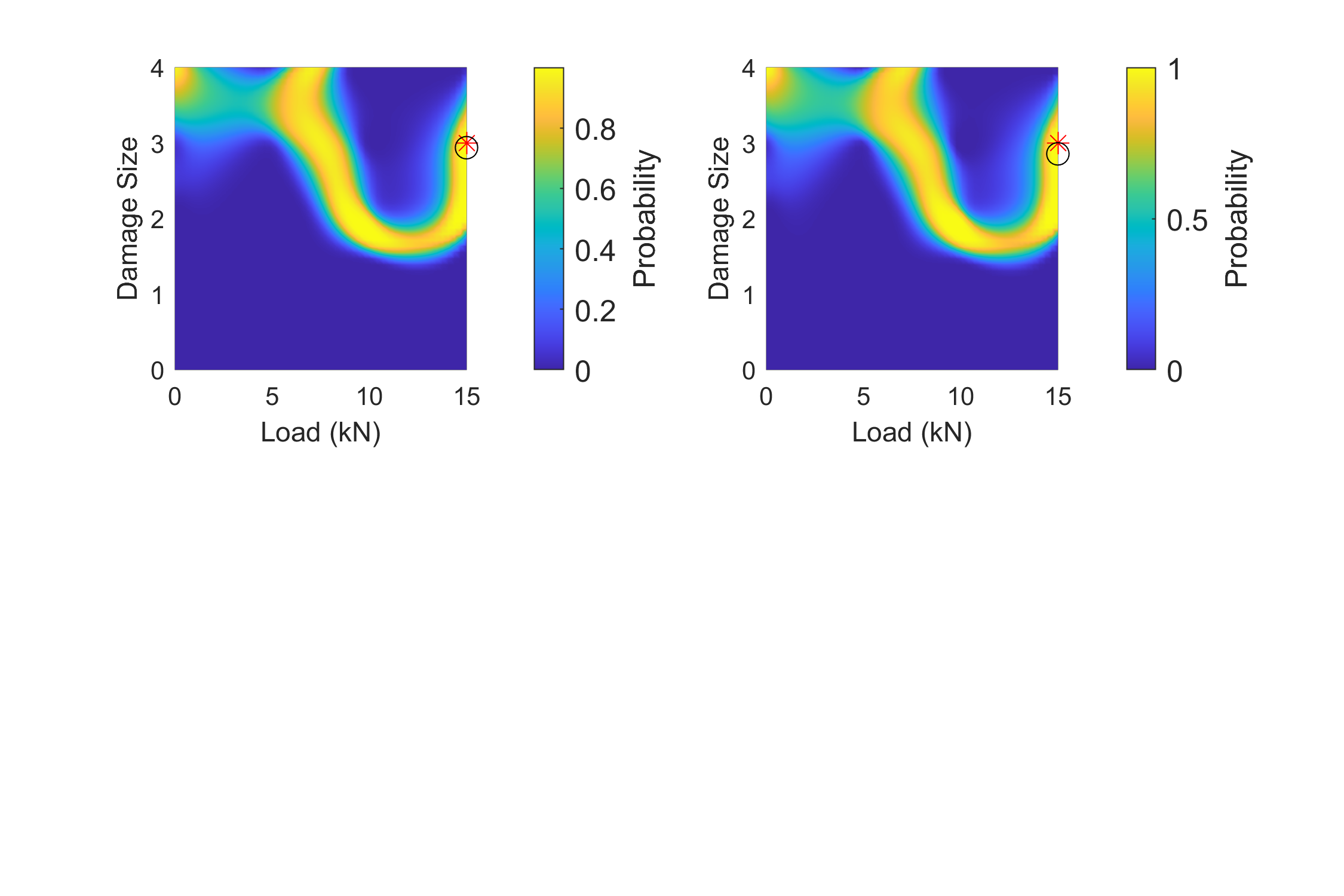}}
    \put(45,295){\color{white} \large {\fontfamily{phv}\selectfont \textbf{a}}}
    \put(235,295){\color{white} \large {\fontfamily{phv}\selectfont \textbf{b}}}
   \put(45,160){\color{white} \large {\fontfamily{phv}\selectfont \textbf{c}}} 
   \put(235,160){\color{white} \large {\fontfamily{phv}\selectfont \textbf{d}}} 
    \end{picture} \vspace{-55pt}
    \caption{Al coupon with simulated damage: Multi-input GPRM prediction results for path 1-6: (a) SGPRM prediction at 5 kN for 2 attached weights; (b) VHGPRM prediction at 5 kN for 2 attached weights; (c) SGPRM prediction at 15 kN for 3 attached weights; (d) VHGPRM prediction at 15 kN for 3 attached weights.} 
\label{fig:instron_multiprob_1-6_janapati} \vspace{10pt}
\end{figure}

\begin{figure}[t!]
    \centering
    \begin{picture}(400,300)
    \put(0,40){\includegraphics[trim = 20 0 20 12,clip,scale=0.8]{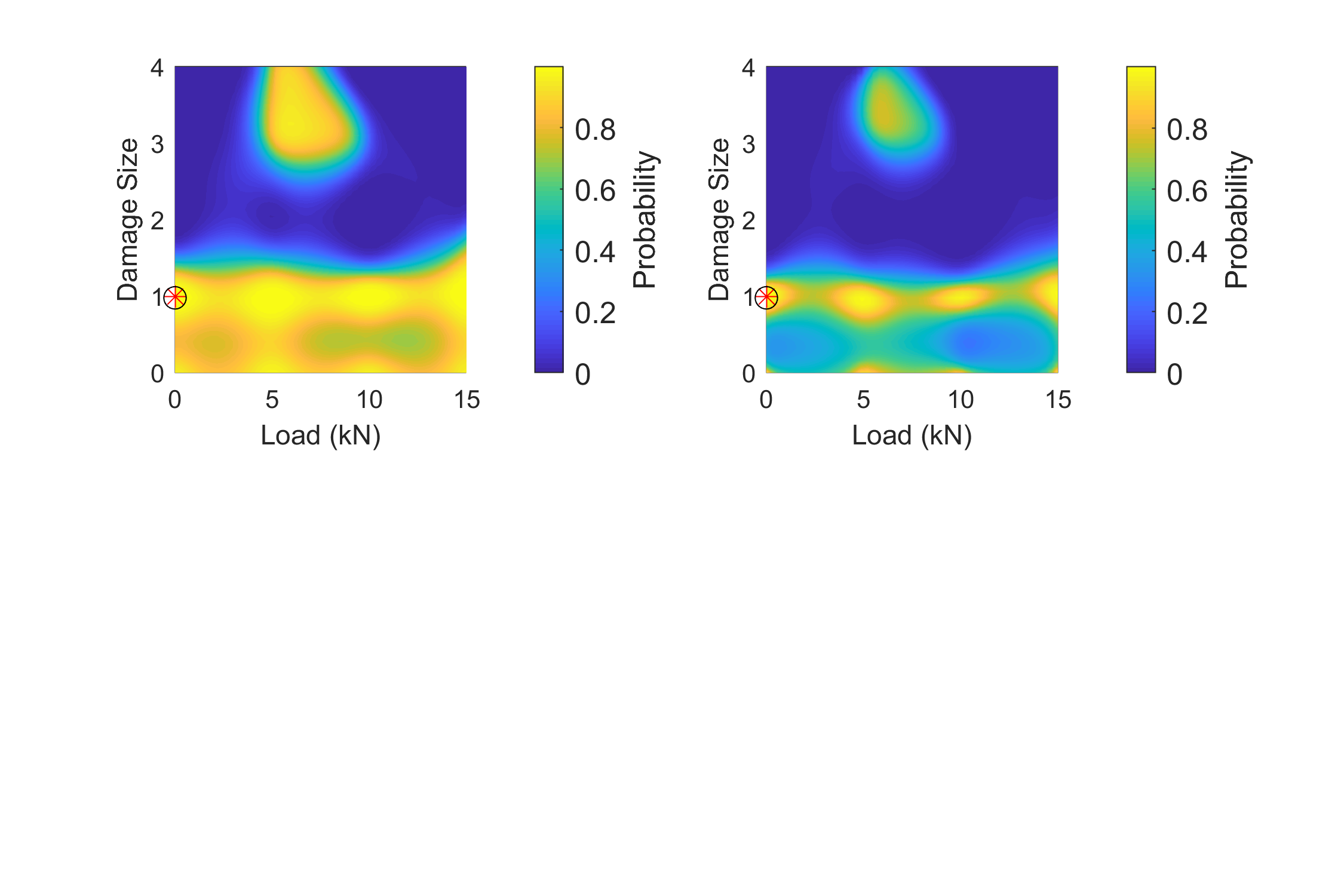}}
    \put(0,-95){\includegraphics[trim = 20 0 20 15,clip,scale=0.8]{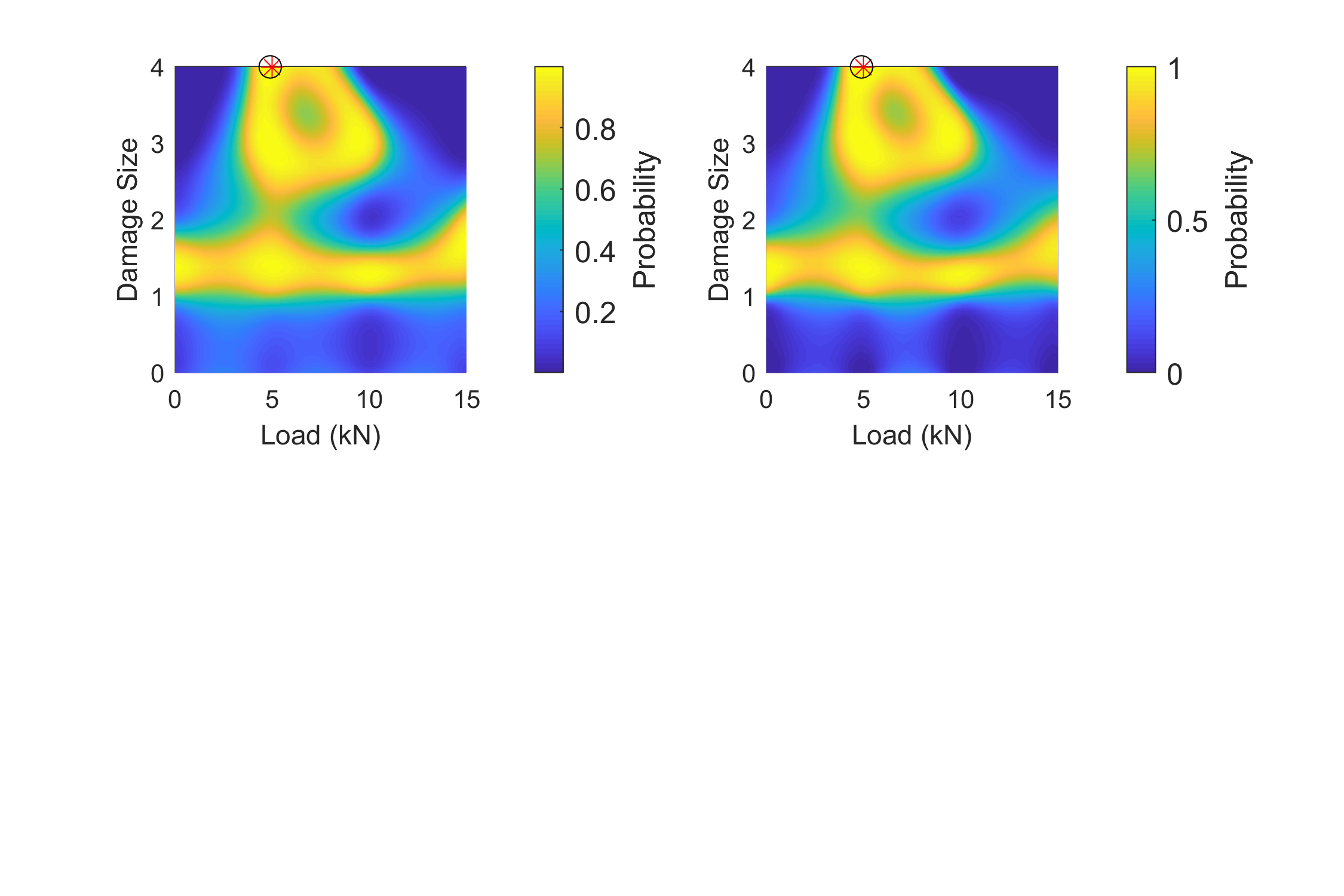}}
    \put(45,295){\color{white} \large {\fontfamily{phv}\selectfont \textbf{a}}}
    \put(235,295){\color{white} \large {\fontfamily{phv}\selectfont \textbf{b}}}
   \put(45,160){\color{white} \large {\fontfamily{phv}\selectfont \textbf{c}}} 
   \put(235,160){\color{white} \large {\fontfamily{phv}\selectfont \textbf{d}}} 
    \end{picture} \vspace{-55pt}
    \caption{Al coupon with simulated damage: Multi-input GPRM prediction results for path 3-6: (a) SGPRM prediction at 0 kN with 1 attached weight; (b) VHGPRM prediction at 0 kN with 1 attached weight; (c) SGPRM prediction at 5 kN for 4 attached weights; (d) VHGPRM prediction at 5 kN for 4 attached weights.} 
\label{fig:instron_multiprob_3-6_janapati} \vspace{-12pt}
\end{figure}

As aforementioned, a second scenario was looked at when studying this coupon, which is the case of simultaneously-unknown damage size and load. For that, multi-input GPRMs were trained using about as low as $0.05\%$ of all the available DI data sets under varying damage sizes and loads for the two paths presented herein. Table \ref{tab:instron_2D} presents some important aspects regarding the trained models. In order to visualize the trained GPRMs, the predictive means and standard deviations of the trained SGPRM and VGPRM were plotted for path 1-6 in Figure \ref{fig:instron_1-6_surf_janapati}. One interesting observation is that the VHGPRM predictive mean (panel a) looks almost identical to the SGPRM mean (panel b). Additionally, the predictive standard deviation of the VHGPRM is clearly more adaptive to the data, as can be seen from the high deviations in the regions where no training data was available (panel d - between weights and between loads), and as can also be seen with the slightly higher standard deviation at the regions where training data existed compared to the case of the SGPRM (panel c). This observation is clearer when looking at the same plots for path 3-6 (Figure \ref{fig:instron_3-6_surf_janapati} panels c and d), where the VHGPRM again shows more adaptability to the data, which is beneficial in capturing the true trends in the DI evolution with damage size and load. Figure \ref{fig:instron_multiprob_1-6_janapati} shows the prediction probabilities estimated by the trained multi-input models for path 1-6 at two indicative DI test points. As shown, both models perform excellently in simultaneously predicting both damage size and load state. These results show that multi-input GPRMs can accurately perform multi-state quantification for active-sensing, guided-wave SHM. Figure \ref{fig:instron_multiprob_3-6_janapati} also shows the prediction probabilities for 2 other indicative test DI points for path 3-6. Again, both models perform well in predicting both damage size and load. 

\begin{figure}[t!]
    \centering
    \begin{picture}(400,300)
    \put(0,40){\includegraphics[trim = 20 0 20 12,clip,scale=0.8]{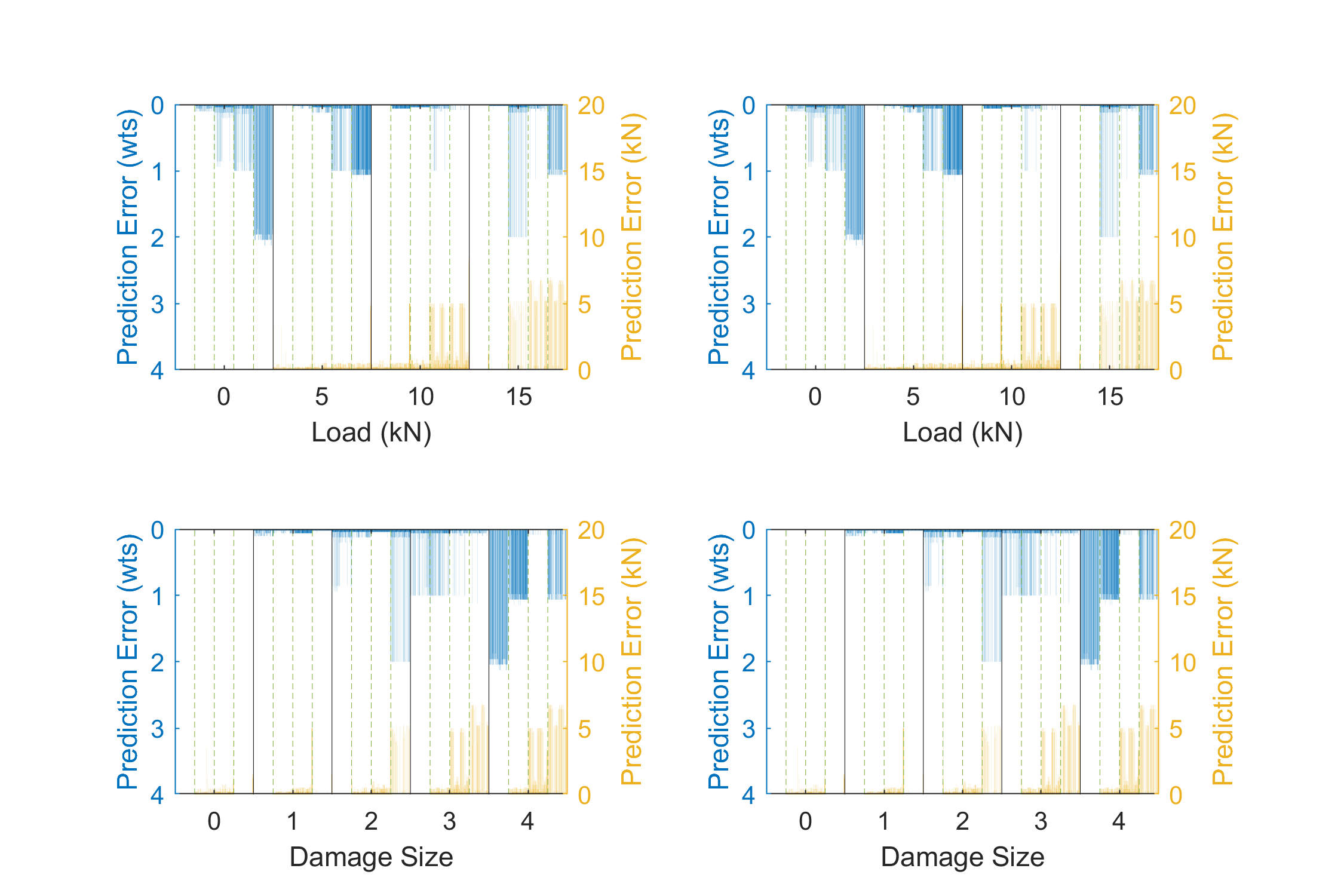}}
    \put(45,283){\color{black} \large {\fontfamily{phv}\selectfont \textbf{a}}}
    \put(235,283){\color{black} \large {\fontfamily{phv}\selectfont \textbf{b}}}
   \put(45,145){\color{black} \large {\fontfamily{phv}\selectfont \textbf{c}}} 
   \put(235,145){\color{black} \large {\fontfamily{phv}\selectfont \textbf{d}}} 
    \end{picture} \vspace{-55pt}
    \caption{Al coupon with simulated damage: Multi-input GPRM state prediction error results for 7500+ test DI values from path 1-6. Black horizontal lines divide regions of specific loads (panels a and b) or specific damage sizes (panels c and d): (a) SGPRM state prediction error with respect to load; (b) VHPRM state prediction error with respect to load; (c) SGPRM state prediction error with respect to damage size; (d) VHGPRM state prediction error with respect to damage size. In panels a and b (c and d), black lines separate bars of different loads (damage sizes) and green dashed lines separate bars of different damage sizes (loads).} 
\label{fig:instron_prederror_1-6_janapati} \vspace{10pt}
\end{figure}

In addition to the indicative results presented here, summary prediction error results were plotted In order to gain some insights into how prediction error evolves when using multi-input GPRMs. Figure \ref{fig:instron_prederror_1-6_janapati} panels a and b present the evolution of the load and damage size prediction errors plotted with respect to loading condition for SGPRM and VHGPRM of path 1-6, respectively, from all 7500+ test DI values not used in training. It is worth noting here that each pair of facing orange and blue bars corresponding to the load and damage size prediction errors, respectively, comes from the same test DI point i.e. there are just over 7500 bars with either color. Panels c and d show the like but plotted with respect to damage size. Doing an overall examination of where the models fall short in accurately predicting damage size and load (where the blue and yellow bars have high values) and comparing that with the DI plots in Figures \ref{fig:instron_1-6_DI_dam} and \ref{fig:instron_1-6_DI_load} in the Appendix, it can be observed that the GPRMs only fall short at areas where there is significant overlap between DI values across different states. Particularly, examining panels a and b of Figure \ref{fig:instron_prederror_1-6_janapati}, it can be seen that damage size miss-predictions come from the 0, 5, and 15 kN cases; examining panels c and d, it can be observed that most of these miss-predictions are related to the damage size of 4 weights, with some related to 3 and 2 weights also. This is expected since the 4-weight DI values in path 1-6 did exhibit significant overlap with smaller damage sizes (see Figure \ref{fig:instron_1-6_DI_dam}). The same can be said about the load miss-predictions, where most of them come form the 10 and 15-kN cases, as evident in panel a and b, and can be correlated to the damage sizes of 3 and 4 weights as shown in panels c and d. Again, examining the evolution of the DI values with load under multiple damage sizes as shown in figure \ref{fig:instron_1-6_DI_load}, the roots of these load miss-predictions can be well understood.

\begin{figure}[t!]
    \centering
    \begin{picture}(400,300)
    \put(0,40){\includegraphics[trim = 20 0 20 12,clip,scale=0.8]{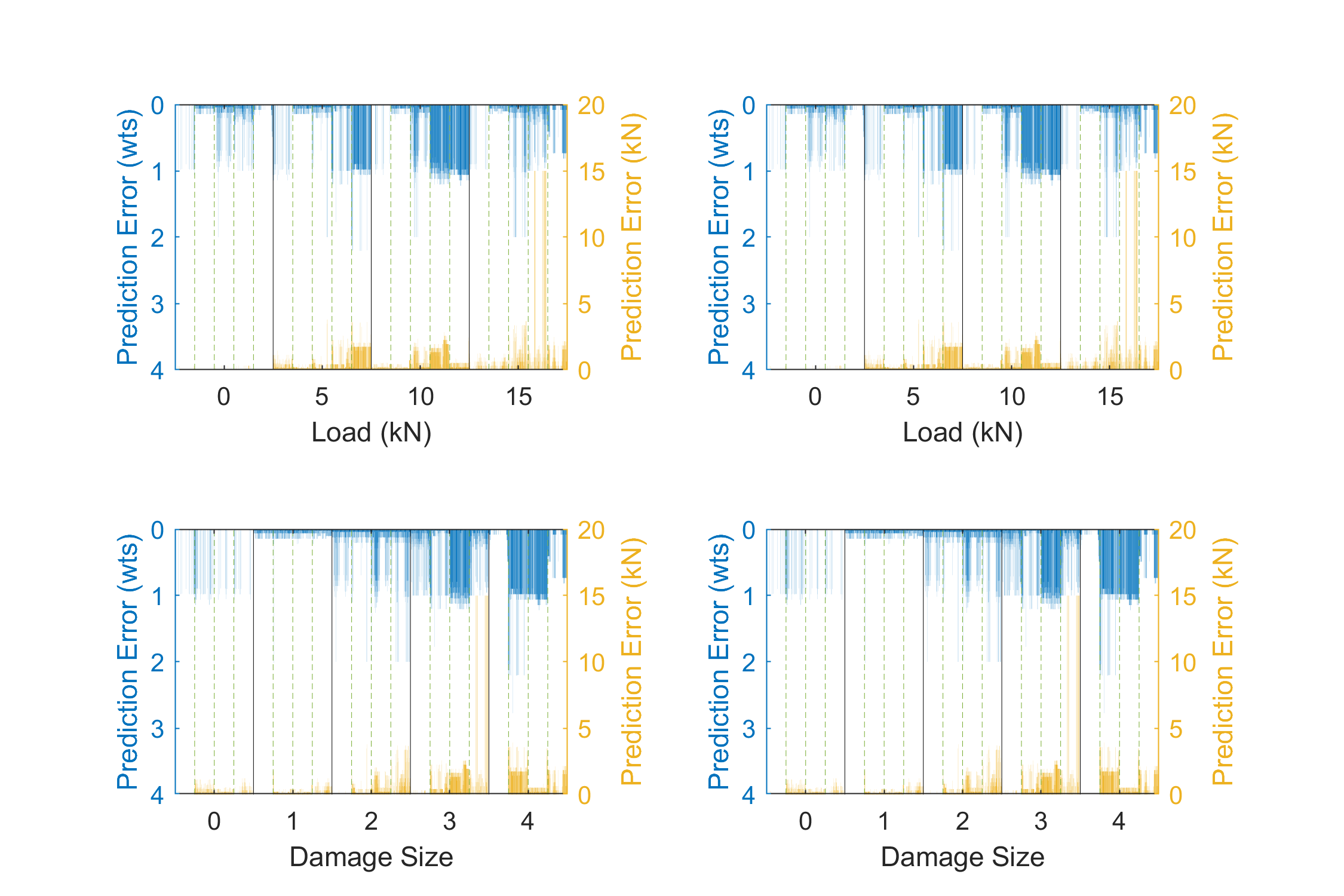}}
    \put(45,283){\color{black} \large {\fontfamily{phv}\selectfont \textbf{a}}}
    \put(235,283){\color{black} \large {\fontfamily{phv}\selectfont \textbf{b}}}
   \put(45,145){\color{black} \large {\fontfamily{phv}\selectfont \textbf{c}}} 
   \put(235,145){\color{black} \large {\fontfamily{phv}\selectfont \textbf{d}}} 
    \end{picture} \vspace{-55pt}
    \caption{Al coupon with simulated damage: Multi-input GPRM state prediction error results for 7500+ test DI values from path 3-6. The black horizontal lines divide regions of specific loads (panels a and b) or specific damage sizes (panels c and d): (a) SGPRM state prediction error with respect to load; (b) VHPRM state prediction error with respect to load; (c) SGPRM state prediction error with respect to damage size; (d) VHGPRM state prediction error with respect to damage size. In panels a and b (c and d), black lines separate bars of different loads (damage sizes) and green dashed lines separate bars of different damage sizes (loads).} 
\label{fig:instron_prederror_3-6_janapati} \vspace{10pt}
\end{figure}

Moving onto path 3-6, inspecting Figure \ref{fig:instron_prederror_3-6_janapati}, one can observe, again, a nice agreement between the miss-predictions in damage size and load and the overlap in DI values as evident in Figures \ref{fig:instron_3-6_DI_dam} and \ref{fig:instron_3-6_DI_load} in the Appendix. In particular, because this is a damage non-intersecting path, the evolution of the DI values with damage size across different loading states is not uniform, and thus there is significant overlap between DI values across the different damage sizes. This explains the higher damage size miss-classifications in this path compared to path 1-6. On the other hand, the overall miss-predictions in load are less than with path 1-6, which is again expected given the much more uniform evolution of the DI with load under different damage sizes for path 3-6 (see Figure \ref{fig:instron_3-6_DI_load}). Overall, both of the DI-trained SGPRM and VHGPRM perform well in simultaneously predicting damage size and load within the capabilities of the training DI data sets.

From all of the results presented in this test case, a few conclusions can be withdrawn as follows:

\begin{itemize}
    \item Single-input VHGPRMs outperform SGPRMs in accurately estimating prediction probabilities based on the CDF. This is due to the capability of the former models to adapt to the changing noise level in the data.
    \item DI-trained GPRMs can accurately simultaneously predict damage size and loading state in an active-sensing, guided-wave SHM framework.
    \item State miss-classification originates from the evolution of the DI with the states being quantified; if the evolution is uniform, state quantification is accurate. Otherwise, the trained models cannot provide an accurate quantification of the true state. Future work can target more robust SHM metrics that show a uniform evolution with respect to the different states for training GPRMs.
\end{itemize}

\section{Conclusion} \label{Sec:conc}

In this study, a novel method for damage/state quantification within the framework of active-sensing, guided-wave SHM was proposed and applied to three experimental test cases. The proposed method is based on damage index-trained Gaussian Process Regression Models (DI-trained GPRMs) in which the structural state is predicted based on the probability of an incoming test DI value from an unknown state of the structure originating from different structural states. The advantages of this framework lie in its simplicity, with only DI values needed for training a robust and accurate quantification model, as well as in the probability-based quantification process this method entertains. Three experimental test cases were presented in this study in which the proposed framework was applied for damage size quantification (Al coupon with notch and CFRP coupon with simulated damage), as well as for damage size and/or load state quantification (Al coupon with simulated damage under different loading conditions). In addition, two types of GPRMs were implemented within the proposed framework, namely: standard homoscedastic (SGPRMs) and variational heteroscedastic (VHGPRMs) models. Both types of GPRMs showed accurate damage size and/or load state quantification, with the VHGPRMs allowing for a better representation of the evolution of the DI with damage size and load, especially for signal paths that exhibited DIs with varying noise across different states. Also, the limit of accurate damage size and/or load state quantification seemed to be controlled by the training data; states at which training DI values overlapped were indistinguishable by the GPRMs when came to calculating the prediction probabilities. Finally, at least two interesting points remain open for exploration and are currently being explored by the authors. Firstly, although the proposed framework provides a simple route to accurate and robust state quantification in active-sensing, guided-wave SHM, it does involve multiple steps in some cases (2-state predictions for instance). Thus, other more straightforward, yet mathematically-involved, approaches, such as multi-output noisy-input GPRMs, might be a better alternative to the multi-step approach presented in this study. Secondly, because of the nature of how DI values can overlap across multiple states, using more robust and accurate SHM metrics, such as non-parametric and/or parametric time series representations, in training GPRMs might allow for limiting or totally avoiding state miss-classifications.

\section*{Acknowledgment}

This work is carried out at the Rensselaer Polytechnic Institute under the Vertical Lift Research Center of Excellence (VLRCOE) Program, grant number W911W61120012, with Dr. Mahendra Bhagwat and Dr. William Lewis as Technical Monitors. 


\vspace{-6pt}

\bibliographystyle{aiaa} 

\bibliography{References} 

\vspace{-6pt}

\clearpage
\appendix
\appendixpage
\addappheadtotoc

\setcounter{table}{0}
\renewcommand\thetable{\Alph{section}.\arabic{table}}

\setcounter{figure}{0}
\renewcommand\thefigure{\Alph{section}.\arabic{figure}}

\section{Additional results}

\begin{figure}[h!]
    \centering
    \begin{picture}(400,300)
        \put(0,40){\includegraphics[trim = 20 0 20 15,clip,scale=0.8]{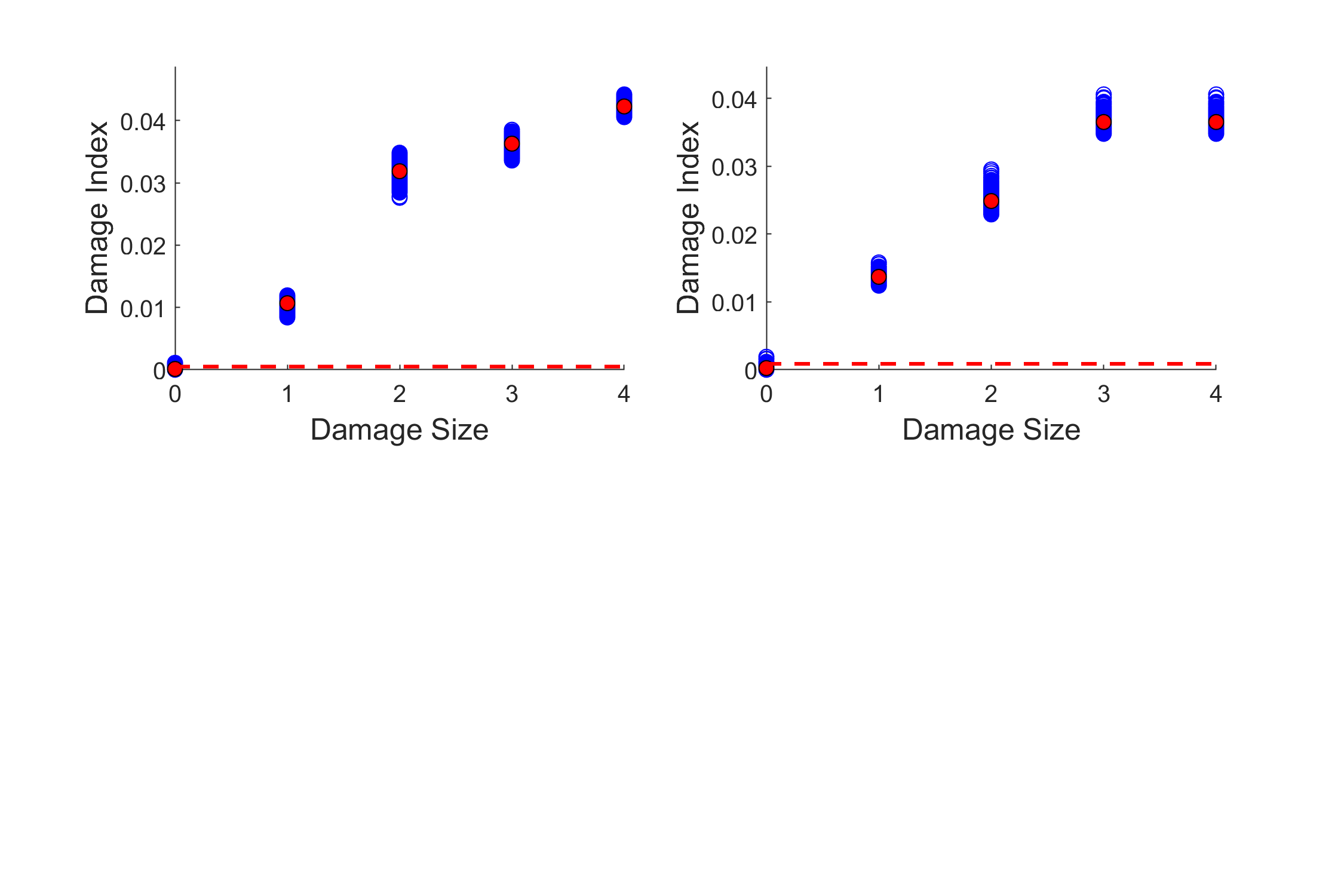}}
    \put(0,-95){\includegraphics[trim = 20 0 20 15,clip,scale=0.8]{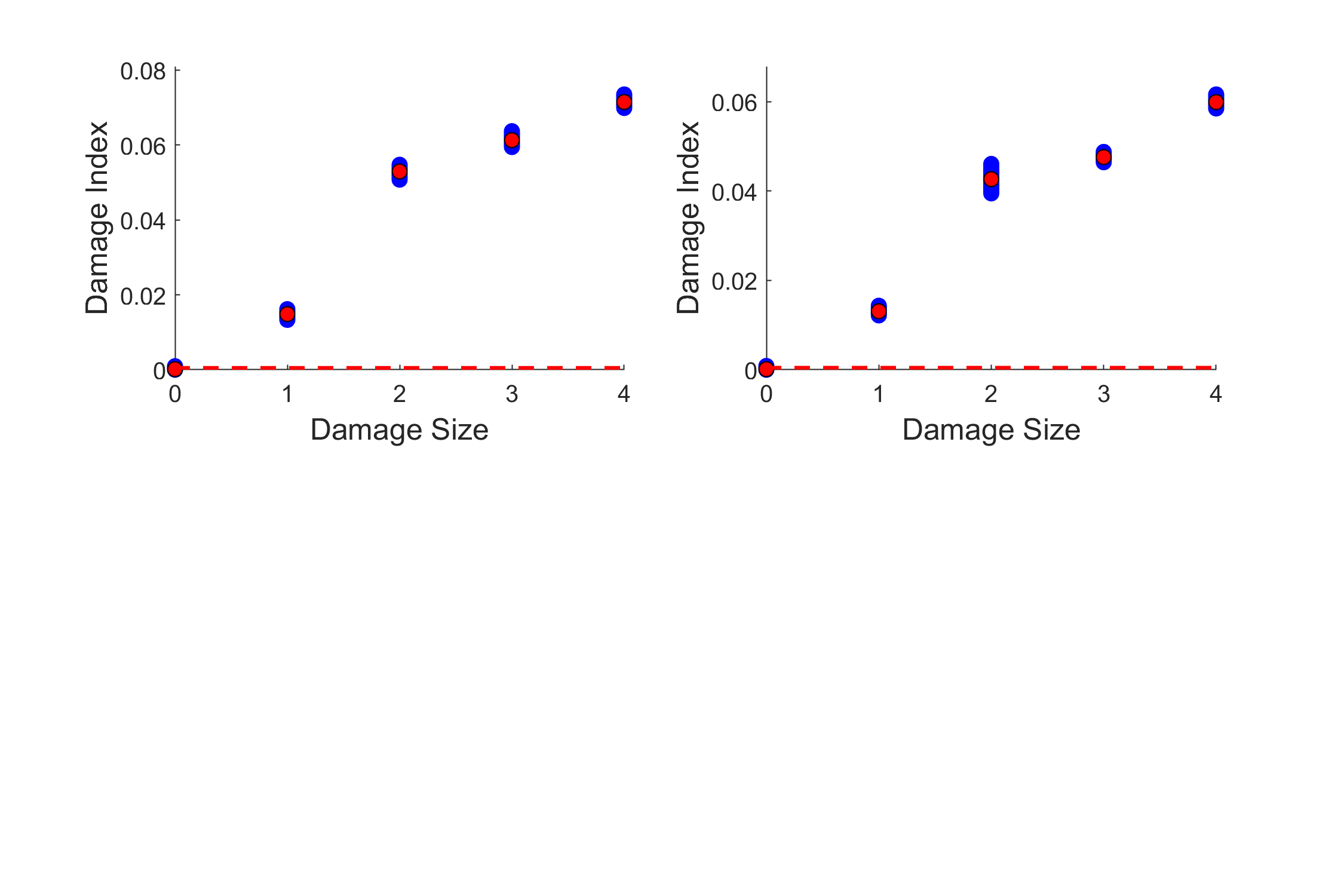}}
    \put(45,295){\color{black} \large {\fontfamily{phv}\selectfont \textbf{a}}}
    \put(235,295){\large {\fontfamily{phv}\selectfont \textbf{b}}}
    \put(45,160){\large {\fontfamily{phv}\selectfont \textbf{c}}} 
    \put(235,160){\large {\fontfamily{phv}\selectfont \textbf{d}}}
    \end{picture} \vspace{-55pt}
    \caption{Al coupon with simulated damage: DI plots from path 1-6 under multiple loading conditions: (a) 0 kN; (b) 5 kN; (c) 10 kN; (d) 15 kN. The red dashed lines indicate the healthy $95 \%$ confidence bounds, and the red circles indicate the mean DI values at each state.} 
\label{fig:instron_1-6_DI_dam} 
\end{figure}

\begin{figure}[t!]
    \centering
    \begin{picture}(400,300)
        \put(100,40){\includegraphics[trim = 20 0 20 15,clip,scale=0.8]{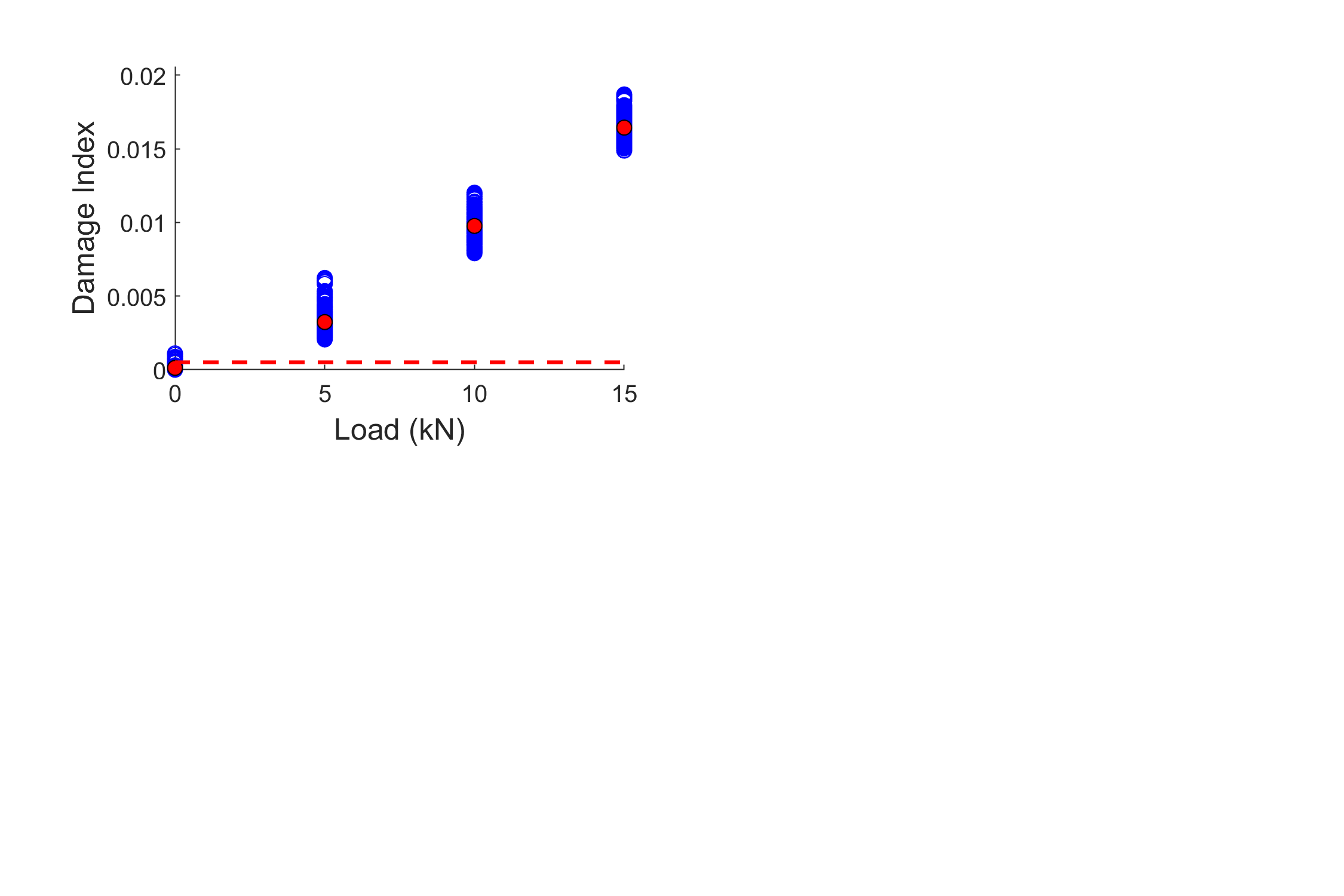}}
        \put(0,-95){\includegraphics[trim = 20 0 20 15,clip,scale=0.8]{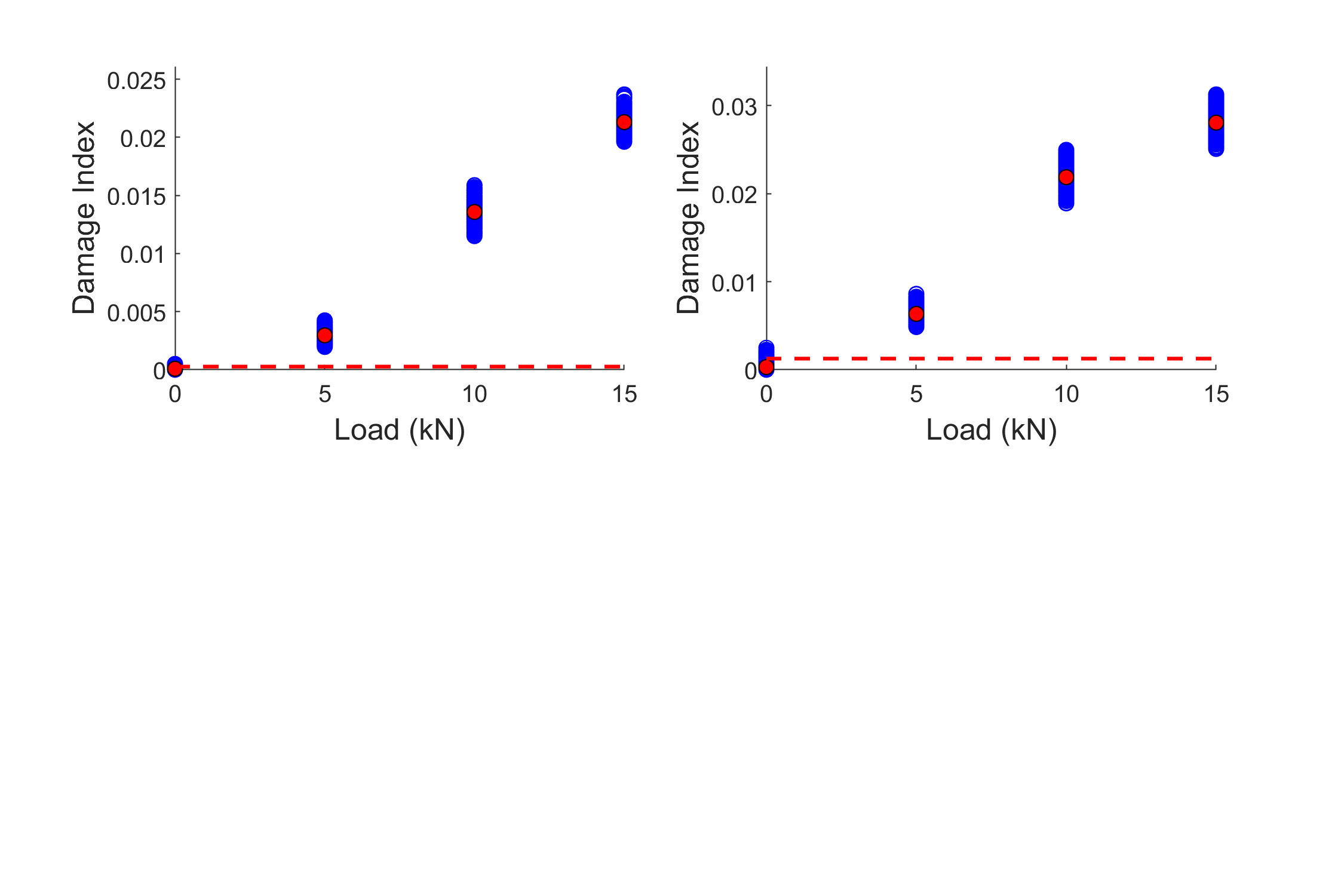}}
    \put(0,-240){\includegraphics[trim = 20 0 20 15,clip,scale=0.8]{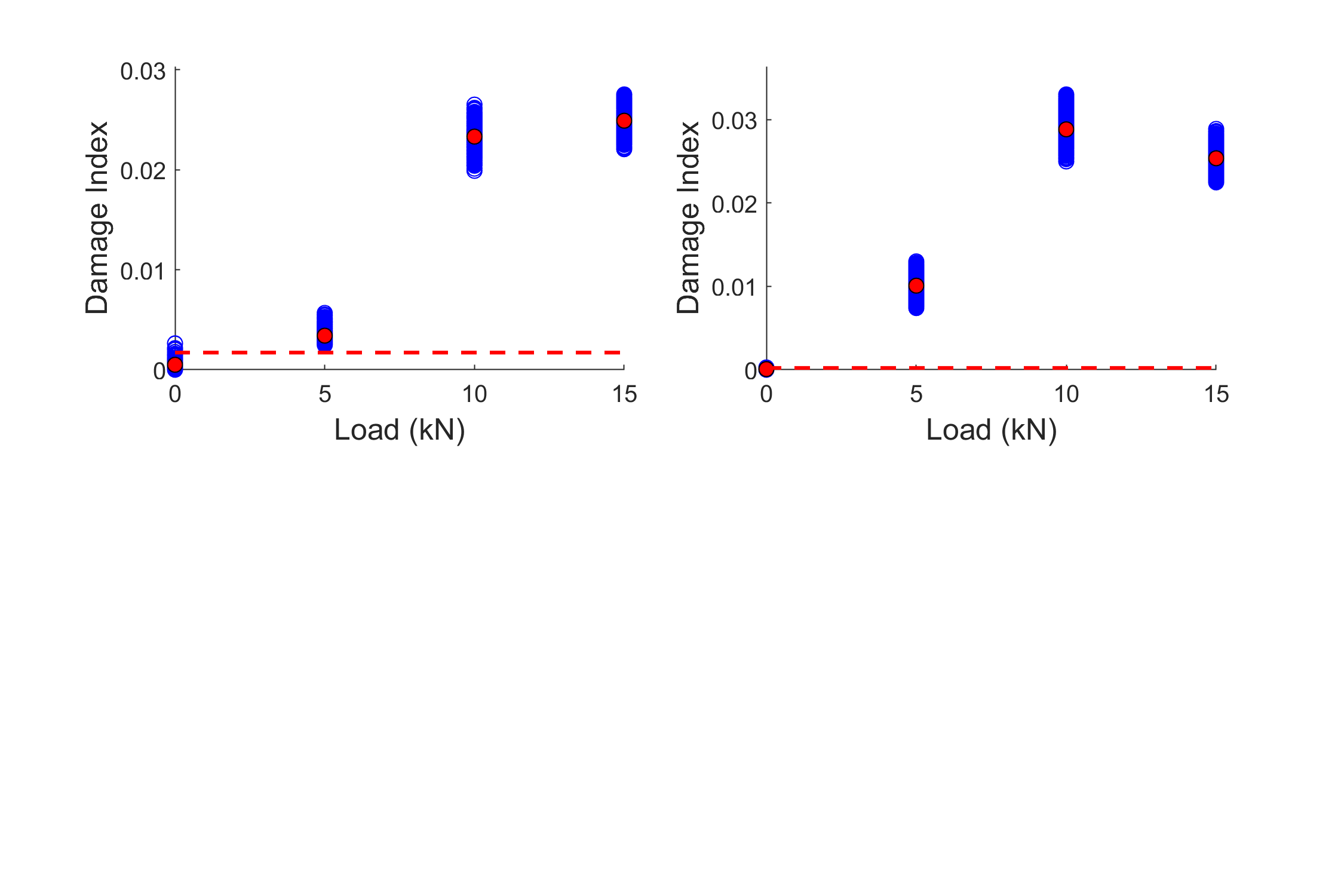}}
    \put(145,295){\color{black} \large {\fontfamily{phv}\selectfont \textbf{a}}}
    \put(45,160){\color{black} \large {\fontfamily{phv}\selectfont \textbf{b}}}
    \put(235,160){\large {\fontfamily{phv}\selectfont \textbf{c}}}
    \put(45,15){\large {\fontfamily{phv}\selectfont \textbf{d}}} 
    \put(235,15){\large {\fontfamily{phv}\selectfont \textbf{e}}}
    \end{picture} \vspace{90pt}
    \caption{Al coupon with simulated damage: DI plots from path 1-6 under multiple damage conditions: (a) Healthy; (b) One weight attached; (c) Two weights attached; (d) Three weights attached; (e) Four weights attached. The red dashed lines indicate the healthy $95 \%$ confidence bounds, and the red circles indicate the mean DI values at each state.} 
\label{fig:instron_1-6_DI_load} 
\end{figure}

\begin{figure}[t!]
    \centering
    \begin{picture}(400,300)
        \put(0,40){\includegraphics[trim = 20 0 20 15,clip,scale=0.8]{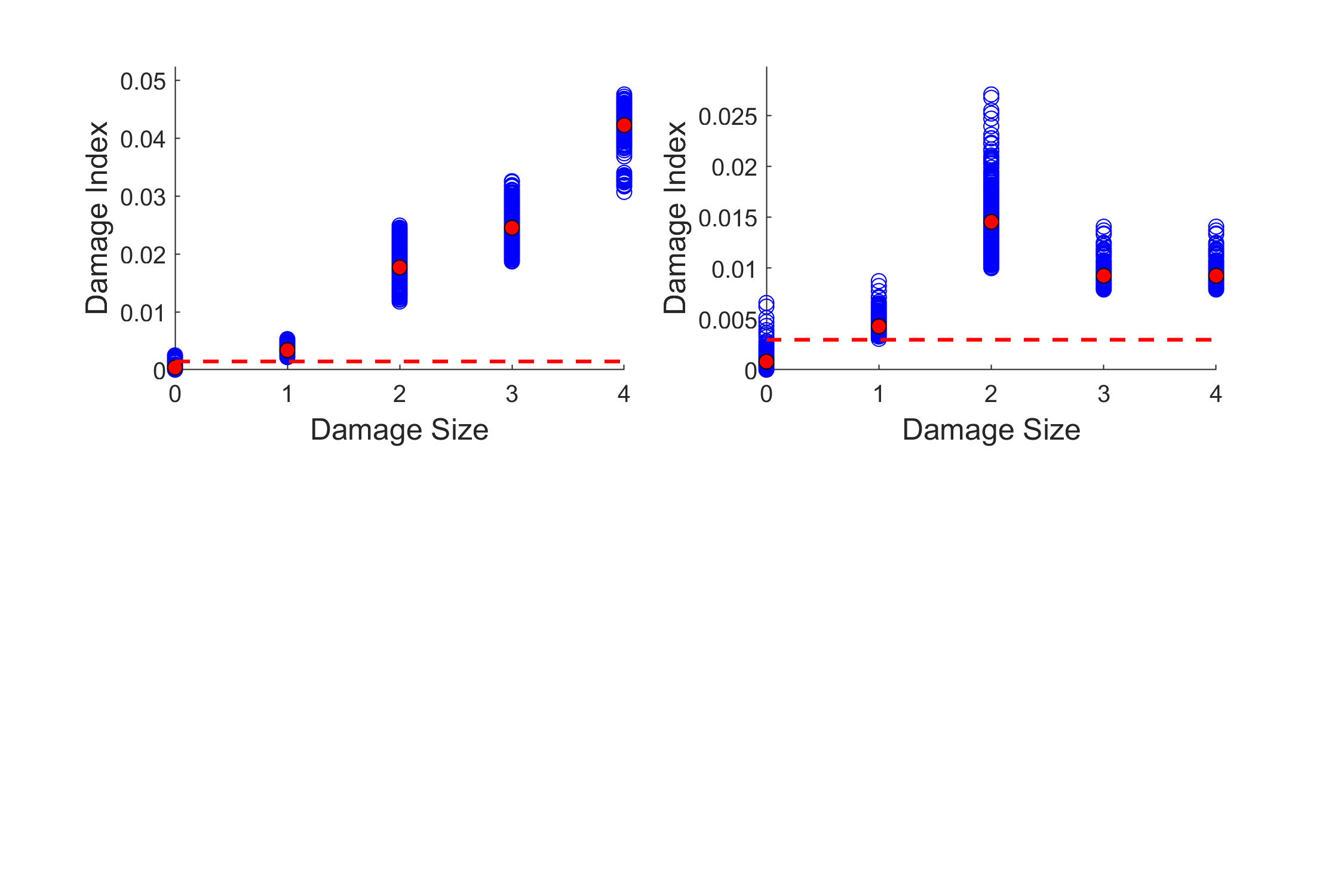}}
    \put(0,-95){\includegraphics[trim = 20 0 20 15,clip,scale=0.8]{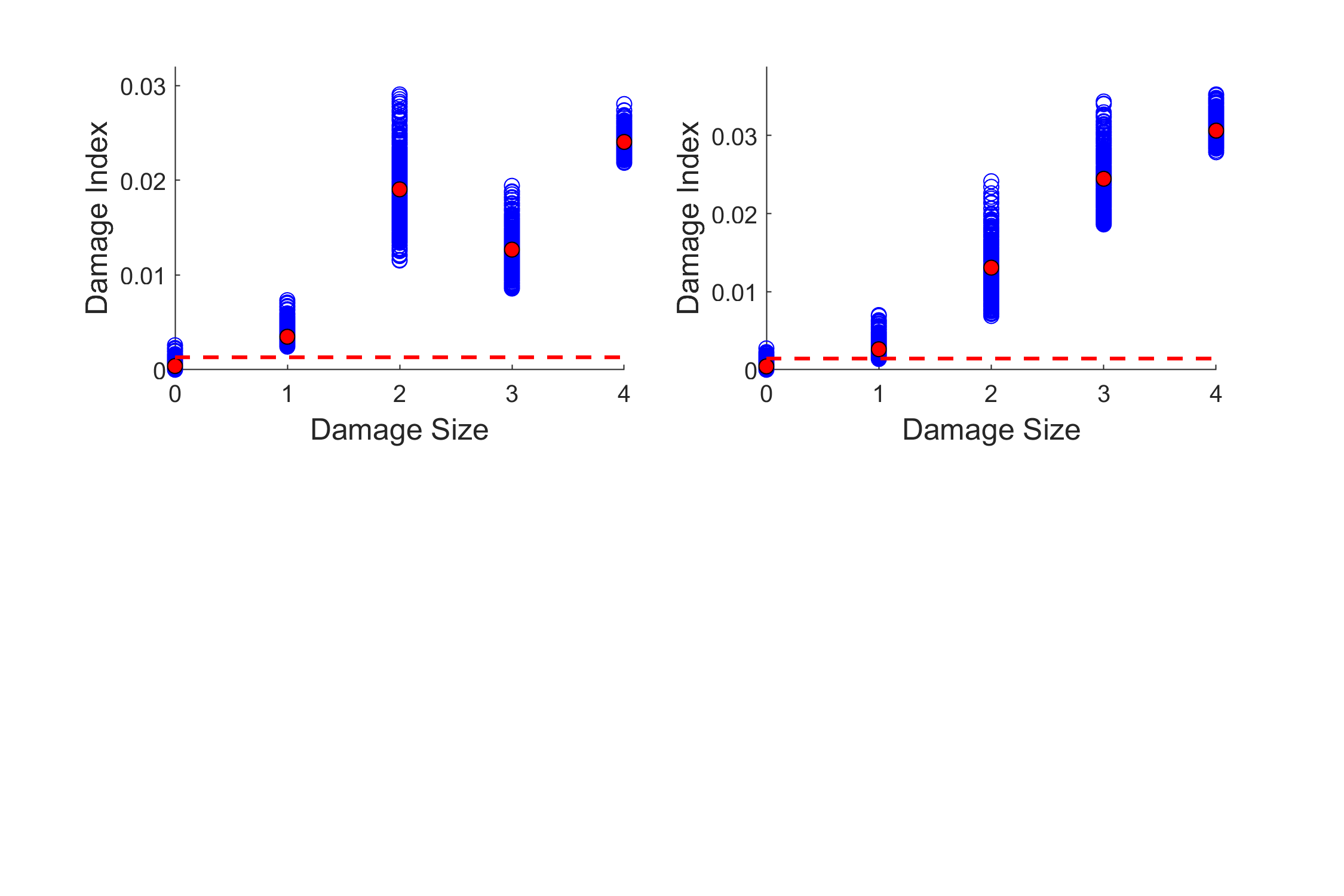}}
    \put(45,295){\color{black} \large {\fontfamily{phv}\selectfont \textbf{a}}}
    \put(235,295){\large {\fontfamily{phv}\selectfont \textbf{b}}}
    \put(45,160){\large {\fontfamily{phv}\selectfont \textbf{c}}} 
    \put(235,160){\large {\fontfamily{phv}\selectfont \textbf{d}}}
    \end{picture} \vspace{-55pt}
    \caption{Al coupon with simulated damage: DI plots from path 3-6 under multiple loading conditions: (a) 0 kN; (b) 5 kN; (c) 10 kN; (d) 15 kN. The red dashed lines indicate the healthy $95 \%$ confidence bounds, and the red circles indicate the mean DI values at each state.} 
\label{fig:instron_3-6_DI_dam} 
\end{figure}

\begin{figure}[t!]
    \centering
    \begin{picture}(400,300)
        \put(100,40){\includegraphics[trim = 20 0 20 15,clip,scale=0.8]{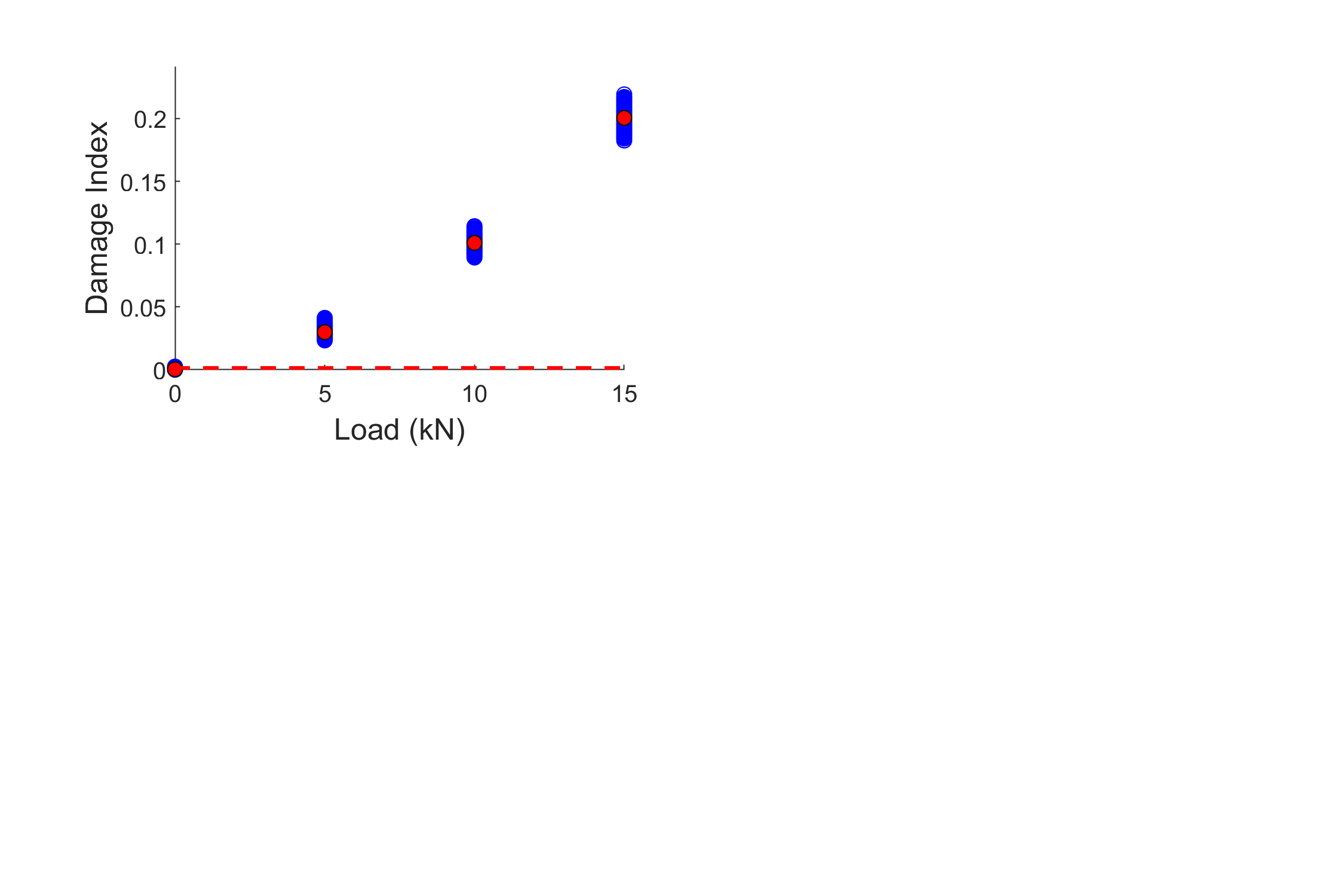}}
        \put(0,-95){\includegraphics[trim = 20 0 20 15,clip,scale=0.8]{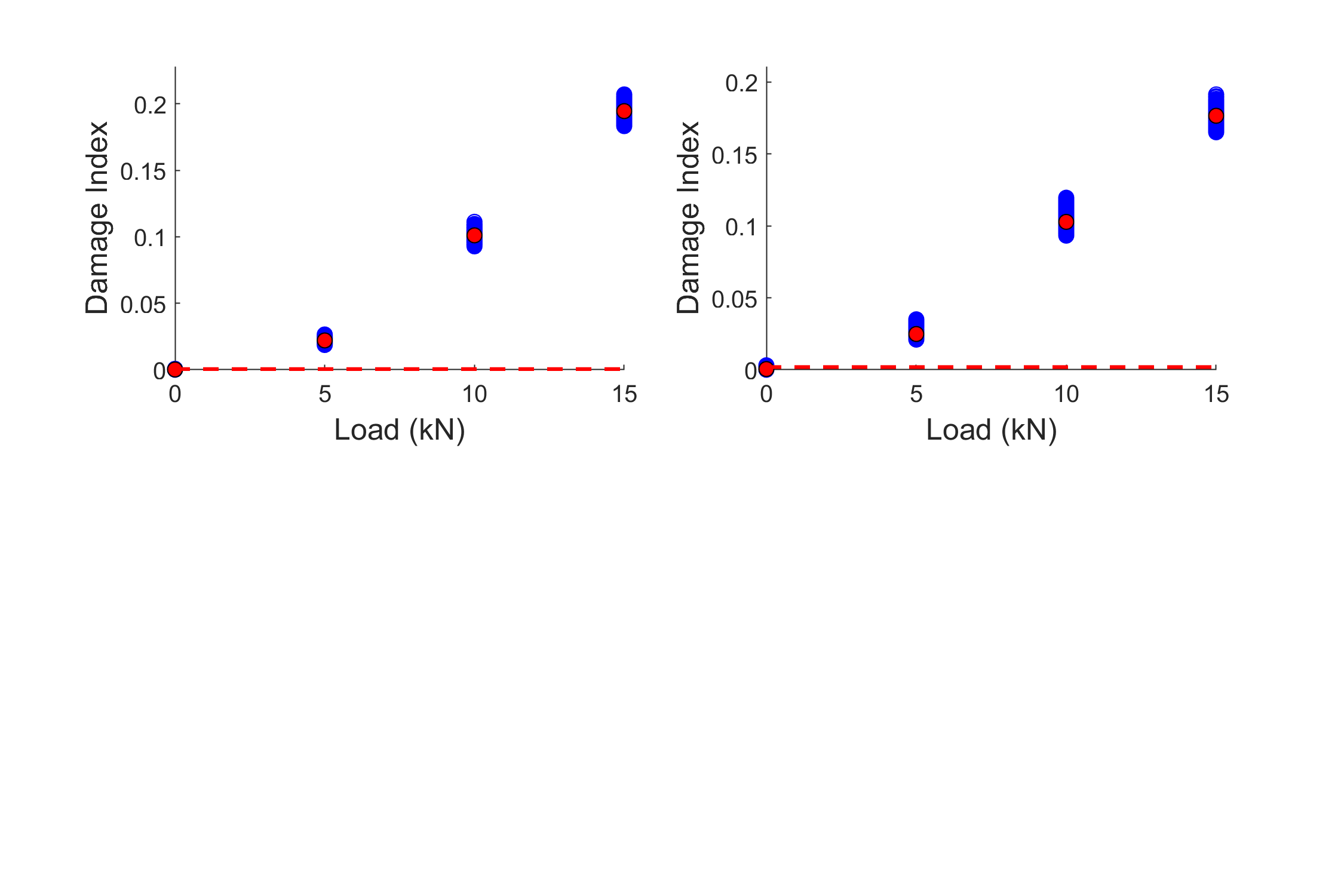}}
    \put(0,-240){\includegraphics[trim = 20 0 20 15,clip,scale=0.8]{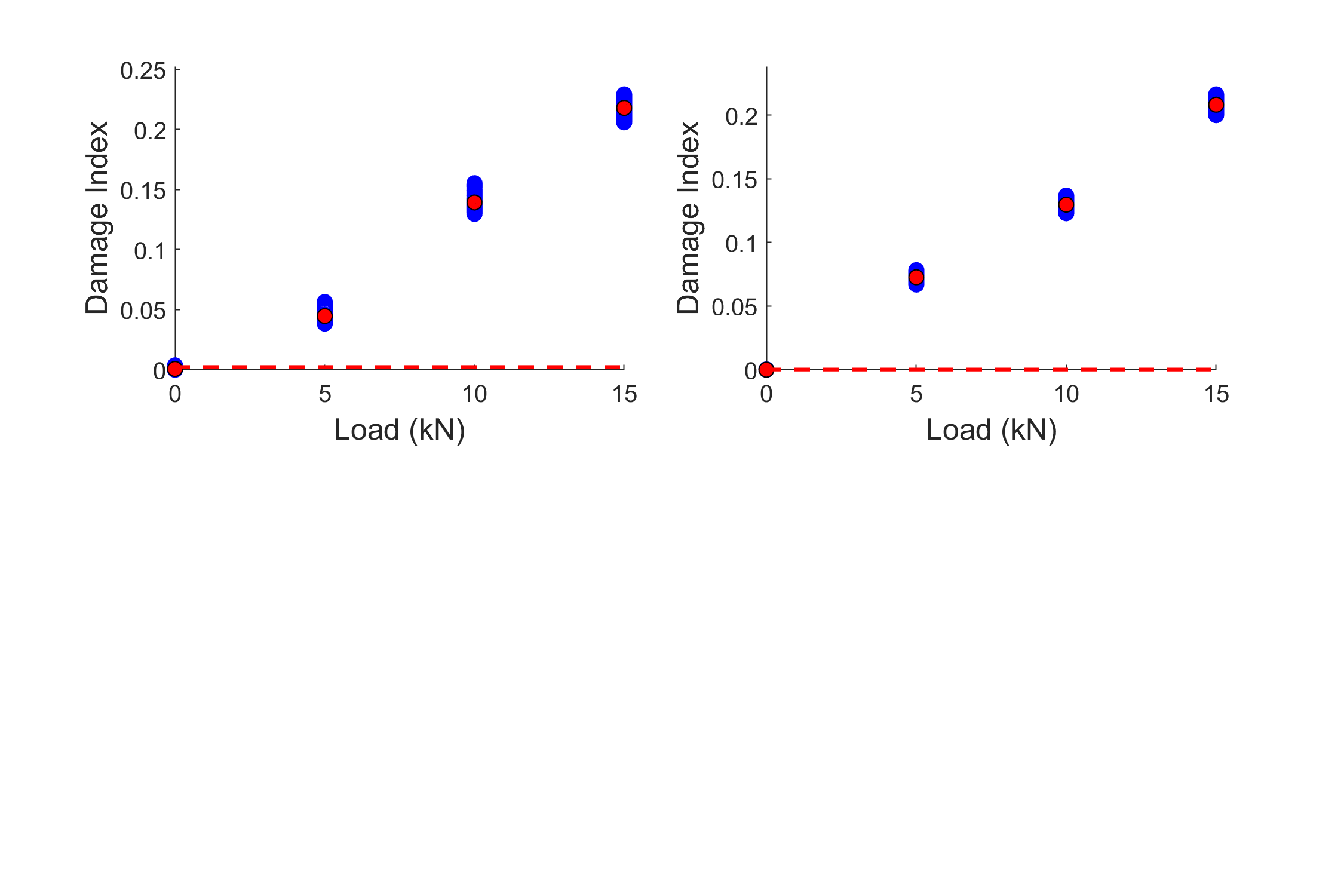}}
    \put(145,295){\color{black} \large {\fontfamily{phv}\selectfont \textbf{a}}}
    \put(45,160){\color{black} \large {\fontfamily{phv}\selectfont \textbf{b}}}
    \put(235,160){\large {\fontfamily{phv}\selectfont \textbf{c}}}
    \put(45,15){\large {\fontfamily{phv}\selectfont \textbf{d}}} 
    \put(235,15){\large {\fontfamily{phv}\selectfont \textbf{e}}}
    \end{picture} \vspace{90pt}
    \caption{Al coupon with simulated damage: DI plots from path 3-6 under multiple damage conditions: (a) Healthy; (b) One weight attached; (c) Two weights attached; (d) Three weights attached; (e) Four weights attached. The red dashed lines indicate the healthy $95 \%$ confidence bounds, and the red circles indicate the mean DI values at each state.} 
\label{fig:instron_3-6_DI_load} 
\end{figure}



\end{document}